\documentclass[11pt]{article}
\usepackage{a4}
\usepackage{amsmath}
\usepackage{amsfonts}
\usepackage{amssymb}
\usepackage[dvips]{graphicx}
\usepackage{enumitem}

\newlength{\PSB} 
\newcommand{\bm}[1]{\mbox{$\bf #1$}}

\newcommand{\argmax}{\operatornamewithlimits{argmax}} 
\newcommand{\argmin}{\operatornamewithlimits{argmin}} 
\newcommand{\moda}{\textrm{mod}}

\title{Report on Dirty Paper Coding Estimation with multidimensional lattices}
\author{Pedro Comesa\~na-Alfaro and Fernando P\'erez-Gonz\'alez}

\begin{document}

\pagenumbering{arabic}
\setcounter{page}{1}

\maketitle

\section{Notation}
Let us denote by $\bm x$ the host signal, $\bm y$ the watermarked
signal, $\bm d$ the dither vector, $\bm n$ the AWGN introduced
by the channel, $\bm z$ the received signal (all of them
taking values in $\mathbb{R}^n$) and $t_0$ the true real scaling
factor.

The watermark embedding can be written as
\begin{eqnarray}
  \bm y = (1-\alpha) \bm x + \alpha \left [
    Q_\Lambda\left (\bm x - \bm d \right ) + \bm d \right ] = 
  (1-\alpha) \bm x + \alpha \left [ \bm x - 
    \left (\bm x - \bm d \right )\moda \Lambda  \right ]  , \nonumber 
\end{eqnarray}
and the attack undergone by the watermarked signal as
\begin{eqnarray}
  \bm z = t_0 \bm y + \bm n.  \nonumber
\end{eqnarray}

The corresponding random vectors will be denoted by bold capital letters, with
$\bm X \sim \mathcal{N}(\bm 0, \sigma_X^2 \bm{I}_{n\times n})$, $\bm D \sim \mathcal{U}(\mathcal{V}(\Lambda))$, $\bm N \sim \mathcal{N}(\bm 0, \sigma_N^2 \bm{I}_{n\times n})$, $\sigma_N^2 > 0$, and $\bm{I}_{n\times n}$ the $n\times n$ identity matrix.

Our target will be to estimate $t_0$ from the received signal $\bm z$ and
the knowledge about the parameters of the system (i.e., $\alpha$, 
$\bm d$, $\Lambda$, $\sigma_X^2$, and $\sigma_N^2$) available at the decoder.

Throughout this report when we mention the {\it scalar quantizer} case
we will mean that the lattice $\Lambda = \Delta \mathbb{Z}^n$
is used. Similarly, when we say that a {\it low-dimensional lattice
quantizer} is used, we will mean that the used
lattice $\Lambda$ can be written as $\Lambda = \Lambda' \times \Lambda'
\cdots \times \Lambda'$, i.e., the considered lattice
is the Cartesian product of a low-dimensional lattice $\Lambda'$.
Finally, if we denote by $A$ the $n\times n$ generating matrix of $\Lambda$,
$A$ will verify that  for the lattice
it generates the ratio between covering radius and
packing radius is minimal among the lattices generated by the
matrix class
$\{\textrm{diag}(\bm v)\cdot A: \bm v \in \mathbb{R}^n\}$, where $\textrm{diag}(\bm v)$ denotes the diagonal matrix with elements those of $\bm v$. This last condition tries
to avoid cases as $\Lambda = \mathbb{Z}\times 10 \mathbb{Z}$, where
the large difference of the quantization region sizes in different directions
makes that the results developed in this report may not be applied.
%%%%%%%%%%%%%%%%%%%%%%%%%%%%%%%%%%%%%%%%%%%%%%%%%%%%%%%%%%%%%%%%%%%%%
%%%%%%%%%%%%%%%%%%%%%%%%%%%%%%%%%%%%%%%%%%%%%%%%%%%%%%%%%%%%%%%%%%%%%

\section{Hypotheses}\label{sec:hypotheses}

In this report we will consider the following hypotheses:
\begin{eqnarray}
  \textrm{Hypothesis 1:} & & \sigma_X^2>>\sigma_\Lambda^2 ,\nonumber \\
  \textrm{Hypothesis 2:} & & (1-\alpha)^2 t_0^2 \sigma_\Lambda^2   << \sigma_N^2,\nonumber \\
  \textrm{Hypothesis 3:} & & (1-\alpha)^2 t_0^2 \sigma_\Lambda^2 + \sigma_N^2 << t_0^2 \sigma_\Lambda^2, \nonumber
\end{eqnarray}
implying
\begin{eqnarray}
  \textrm{Hypothesis 1:} & & \textrm{HLR} \triangleq \alpha^2 \textrm{DWR} \rightarrow \infty,\nonumber \\
  \textrm{Hypothesis 2:} & & \textrm{SCR} \triangleq \frac{(1-\alpha)^2}{\alpha^2} \textrm{WNR} t_0^2 \rightarrow 0,\nonumber \\
  \textrm{Hypothesis 3:} & & \textrm{TNLR} \triangleq (1-\alpha)^2 + \frac{\alpha^2}{\textrm{WNR} t_0^2} \rightarrow 0, \nonumber
\end{eqnarray}
where HLR stands for {\it Host to Lattice Ratio}, SCR for {\it Self-noise to Channel-noise Ratio}, and TNLR for {\it Total-Noise to Lattice Ratio}.

Summarizing, the first hypothesis requires the variance of the host to be much larger than the quantization lattice second moment per dimension, the second one that the channel noise must be dominant over the self-noise (used in order to ensure that the
total noise is approximately Gaussian when low-dimensional lattices are used), and the third one establishes
that the variance of the total noise must be much smaller than
the quantization lattice second moment per dimension (defining in that sense a high-SNR
scenario).

All of these hypotheses, as well as $n\rightarrow \infty$, will be required
for the low-dimensional lattice case analysis.

Nevertheless, whenever the number of dimensions goes
to infinity and lattices with fundamental Voronoi region approaching a
hypersphere are considered (hereafter, {\it high-dimensional good lattices}):
\begin{itemize}
\item {\bf Hypothesis 2} can be dropped,
as in that case a random vector uniformly distributed over that fundamental Voronoi region will be asymptotically Gaussian \cite{Zamir96}.
\item  {\bf Hypothesis 3} can be relaxed, just requiring 
\begin{eqnarray}
(1-\alpha)^2 t_0^2 \sigma_\Lambda^2 + \sigma_N^2 < t_0^2 \sigma_\Lambda^2, \nonumber
\end{eqnarray}
or, equivalently,
\begin{eqnarray}
\textrm{TNLR} < 1. \nonumber
\end{eqnarray}
Be aware that the last condition can be verified for any $\frac{\sigma_W^2}
{\sigma_N^2}$, as one could play with $\alpha$;
specifically, TNLR can be checked to be smaller than $1$ whenever
\begin{eqnarray}
  \alpha < \frac{2\sigma_W^2 t_0^2}{\sigma_N^2 + \sigma_W^2 t_0^2}. \nonumber
\end{eqnarray}
Nevertheless, a very small value of $\alpha$ implies that
{\bf Hypothesis 1} is more difficult to be hold.
\end{itemize}
%%%%%%%%%%%%%%%%%%%%%%%%%%%%%%%%%%%%%%%%%%%%%%%%%%%%%%%%%%%%%%%%%%%%%
%%%%%%%%%%%%%%%%%%%%%%%%%%%%%%%%%%%%%%%%%%%%%%%%%%%%%%%%%%%%%%%%%%%%%

\section{Target function definition}\label{sec:target_function}

In order  to obtain the estimate $\hat{t}$ of the scaling factor, we use the Maximum
Likelihood (ML) criterion, which seeks the most likely value of $t$ given a
vector of observations $\bm z$ when there is not {\it a priori}  knowledge
about the distribution of the scaling factor, i.e.,
\begin{equation}
  \hat{t}(\bm z)  \triangleq  \argmax_t  \log \left (
  f_{\bm Z|T, \bm K}(\bm z|t, \bm d) \right ) = \argmin_t -2  \log \left (
  f_{\bm Z|T,\bm K}(\bm z|t, \bm d) \right ). \nonumber
\end{equation}

The hypotheses introduced above will be used to find
a mathematically tractable approximation  to $f_{\bm Z|T, \bm K}(\bm z|t, \bm d)$. The followed
approach will be based on the study of that probability density function (pdf) as
\begin{equation}
  f_{\bm Z|T, \bm K}(\bm z|t, \bm d) = \sum_{\lambda\in \Lambda}  p_{I|\bm K}(\lambda|\bm d) f_{\bm Z|T, \bm K, I}(\bm z|t,\bm d,\lambda),\label{eq:addGauss}
\end{equation}
where $p_{I|\bm K}(\lambda| \bm d)$
denotes the probability of the event that
a sample of the host signal $\bm X$ belongs to
the Voronoi region of lattice point $\lambda$ given that the dither takes value
$\bm d$; be aware that due to its definition, $p_{I|\bm K}(\lambda | \bm d)$ does not depend
on $t$. Indeed, it is straightforward to see that
\begin{equation}
  p_{I|\bm K}(\lambda |\bm d) = \int_{\lambda + \bm d + \mathcal{V}(\Lambda)}
  \frac{e^{-\frac{||\bm x||^2}{2 \sigma_X^2}}}{(2 \pi \sigma_X^2)^{n/2}} d \bm x.\nonumber
\end{equation}

Whenever the HLR goes to infinity (i.e., under {\bf Hypothesis 1}) one can approximate the previous probability mass function (pmf) by
\begin{equation}
  p_{I|\bm K}(\lambda |\bm d) \approx \frac{|\mathcal{V}(\Lambda)|  e^{-\frac{||\lambda + \bm d||^2}{2 \sigma_X^2 }}}{(2 \pi \sigma_X^2)^{n/2}}, \nonumber
\end{equation}
where $|\mathcal{V}(\Lambda)|$ denotes the volume of the
fundamental Voronoi region of $\Lambda$.

Also when the HLR goes to infinity we can approximate the self-noise
to be uniformly distributed within a scaled version of the fundamental Voronoi  region (and consequently independent of $\lambda$), yielding the following approximation
\begin{eqnarray}
  f_{\bm Z|T, \bm K}(\bm z|t,\bm d) & \approx & \sum_{\lambda \in \Lambda} \frac{p_{I|\bm K}(\lambda|\bm d)}{[ t (1 -
    \alpha)]^n|\mathcal{V}(\Lambda)|} \int_{(1-\alpha) \mathcal{V}(\Lambda)} \frac{e^{\frac{||\bm z/t - \lambda - \bm d + \tau||^2}{2 \sigma_N^2 / t^2}}}{(2\pi \sigma_N^2/t^2)^{n/2}}d \tau. \label{eq:FlatHost}
\end{eqnarray}

Since SCR $\rightarrow 0$ (i.e., {\bf Hypothesis 2}) in the scalar quantizer case
or low-dimensional lattice quantizer case, and based on the results in \cite{Zamir96}
for the high-dimensional good lattice quantizer case,
in (\ref{eq:FlatHost})
one can approximate the integral of the Gaussian in the scaled version
of the fundamental Voronoi region of the lattice
by a Gaussian distribution with variance the addition
of the variances of the self-noise and the channel noise. 
Using this approximation, the pdf of $\bm Z$ given $t$, that the centroid $\lambda$ was transmitted, and the dither $\bm d$, can be written like
\begin{equation}
  f_{\bm Z|T,\bm K,I}(\bm z|t,\bm d,\lambda) \approx \frac{ e^{- \frac{||\bm z - t \lambda  - t \bm d||^2}{2 \left ( \sigma_N^2 + (1-\alpha)^2 t^2 \sigma_\Lambda^2 \right ) } }}{\left (2 \pi \left ( \sigma_N^2 + (1-\alpha)^2 t^2 \sigma_\Lambda^2 \right )\right)^{n/2} }. \label{eq:iGauss}
\end{equation}
Taking into account the last expression, and TNLR $\rightarrow 0$ (i.e., {\bf Hypothesis 3}) for the low-dimensional lattice quantizer case,
or TNLR $< 1$ in the
high-dimensional good lattice quantizer case,
one can estimate, with a success probability
asymptotically (with the TNLR in the first case, and with the number of dimensions
in the second one)
close to $1$, the lattice point $\lambda$ that was used at the embedder as $\lambda = Q_\Lambda(\bm z/t - \bm d)$.
Therefore (\ref{eq:iGauss}) can be rewritten as
\begin{equation}
  f_{\bm Z|T,\bm K,I}(\bm z|t,\bm d,\lambda) \approx \frac{ e^{- \frac{||\bm z - t \lambda - t \bm d ||^2}{2 \left ( \sigma_N^2 + (1-\alpha)^2  t^2 \sigma_\Lambda^2 \right ) } }}{\left (2 \pi \left ( \sigma_N^2 + (1-\alpha)^2 t^2 \sigma_\Lambda^2 \right )  \right )^{n/2}} \delta \left ( \lambda - Q_\Lambda \left (\frac{\bm z}{t}-\bm d \right) \right ), \nonumber
\end{equation}
and
\begin{eqnarray}
  f_{\bm Z|T,  \bm K}(\bm z|t, \bm d) &\approx&  \sum_{\lambda \in \Lambda} \frac{|\mathcal{V}(\Lambda)| e^{-\frac{||\lambda + \bm d||^2}{2 \sigma_X^2 }}}{\left (2 \pi
\sigma_X^2\right)^{n/2}} \frac{ e^{- \frac{||\bm z - t\lambda - t \bm d||^2}{2 \left ( \sigma_N^2 + (1-\alpha)^2 t^2 \sigma_\Lambda^2 \right ) }}}{\left (2 \pi \left ( \sigma_N^2 + (1-\alpha)^2 t^2 \sigma_\Lambda^2 \right ) \right)^{n/2}  } \delta \left ( \lambda - Q_\Lambda \left (\frac{\bm z}{t}-\bm d \right )\right)  \nonumber \\
  &\approx& \frac{|\mathcal{V}(\Lambda)| e^{-\frac{||\bm z||^2}{2 \sigma_X^2 t^2}}}{(2 \pi
\sigma_X^2)^{n/2}} \frac{ e^{- \frac{||(\bm z - t \bm d )\moda (t \Lambda)||^2}{2 \left ( \sigma_N^2 + (1-\alpha)^2 t^2 \sigma_\Lambda^2 \right ) } }}{\left (2 \pi \left ( \sigma_N^2 + (1-\alpha)^2  t^2 \sigma_\Lambda^2\right )  \right)^{n/2}},  \label{eq:unidimen_natural}
\end{eqnarray}
where in the last approximation we have taken into account in the first Gaussian function that HLR $\rightarrow \infty$, this condition jointly with $n\rightarrow \infty$ in order to the terms in both exponentials to be independent, and TNLR $< 1$ jointly
with $n\rightarrow \infty$ in order to $\bm z - t\lambda - t \bm d$ to be in $\mathcal{V}(t\Lambda)$. Therefore,
\begin{eqnarray}
& -2  \log \left (
  f_{\bm Z|T, \bm K}(\bm z|t, \bm d) \right ) +K\approx \nonumber \\ &L(t, \bm z) \triangleq
\frac{||(\bm z - t\bm d )\moda (t \Lambda)||^2}{ \sigma_N^2 + (1-\alpha)^2 t^2 \sigma_\Lambda^2  } 
  + n \log \left ( 2 \pi \left ( \sigma_N^2 + (1-\alpha)^2 
  t^2 \sigma_\Lambda^2 \right ) \right )
  +  \frac{||\bm z||^2}{ \sigma_X^2 t^2} ,  \label{eq:SecEstimator} 
\end{eqnarray}
and
\begin{eqnarray}
  \hat{t}(\bm z) \approx \argmin_t L(t, \bm z). \nonumber
\end{eqnarray}
Be aware that for the sake of notational simplicity 
$K$ stands for $\frac{|\mathcal{V}(\Lambda)|}{(2 \pi
\sigma_X^2)^{n/2}}$; as it does not depend on $t$, the solution to the previous optimization problem will not be modified.

For the scalar quantizer case, i.e., $\Lambda = \Delta \mathbb{Z}^n$,
from (\ref{eq:unidimen_natural}) it is straightforward to write
\begin{eqnarray}
  f_{\bm Z|T, \bm K}(\bm z|t, \bm d) &\approx& 
  \frac{\Delta^n e^{-\frac{||\bm z||^2}{2 \sigma_X^2 t^2}}}{\left (2 \pi
\sigma_X^2 \right)^{n/2}} \frac{ e^{- \frac{||(\bm z - t\bm d )\moda (t \Delta)||^2}{2 \left ( \sigma_N^2 + (1-\alpha)^2 \Delta^2 t^2/12 \right ) } }}{\Big [ 2 \pi \left ( \sigma_N^2 + (1-\alpha)^2 \Delta^2 t^2/12 \right )  \Big]^{n/2}},  \label{eq:multidimen_natural}
\end{eqnarray}
so
\begin{eqnarray}
L(t, \bm z) =
\frac{||(\bm z - t\bm d )\moda (t \Delta)||^2}{ \sigma_N^2 + (1-\alpha)^2 \Delta^2 t^2/12  } 
  + n \log \left ( 2 \pi \left ( \sigma_N^2 + (1-\alpha)^2 \Delta^2
  t^2/12 \right ) \right )
  +  \frac{||\bm z||^2}{ \sigma_X^2 t^2} .  \label{eq:SecEstimator_escalar} 
\end{eqnarray}

In order to study (\ref{eq:SecEstimator}) we will first take into consideration
its first term. Applying basic properties of the modulo operation, one can write
\begin{eqnarray}
  \left ( \bm z - t \bm d \right ) \moda (t \Lambda) &=& \left [t_0 \left (  (1-\alpha) \bm x + \alpha \Big [ \bm x - 
    \left (\bm x - \bm d \right )\moda \Lambda  \Big ] \right) + \bm n - t \bm d \right] \moda (t \Lambda) \nonumber \\
&=& \left [t_0 \left (  \bm x - \alpha  
    \Big [\left (\bm x - \bm d \right ) \moda \Lambda \Big ]   \right) + \bm n - t \bm d \right] \moda (t \Lambda) \nonumber  \\
&=& \Bigg [(t_0-t) \bm x  - (t_0 - t) \alpha \Big [ \left ( \bm x - \bm d \right )
\moda \Lambda \Big] + \bm n \nonumber \\
&+& t \left (\bm x - \bm d - \alpha \Big [ \left ( \bm x - \bm d 
 \right ) \moda \Lambda \Big ] \right)  \Bigg]\moda(t\Lambda) \nonumber \\
&=& \Bigg [(t_0-t) \bm x  - (t_0 - t) \alpha \Big [ \left ( \bm x - \bm d \right )
\moda \Lambda \Big] + \bm n \nonumber \\
&+& t \left ((\bm x - \bm d)\moda\Lambda - \alpha \Big [ \left ( \bm x - \bm d 
 \right ) \moda \Lambda \Big ] \right)  \Bigg]\moda(t\Lambda) \nonumber \\
&=& \Bigg [(t_0-t) \bm x  - (t_0 - t) \alpha \Big [ \left ( \bm x - \bm d \right )
\moda \Lambda \Big] + \bm n \nonumber \\
&+& t (1-\alpha) \Big ( \left ( \bm x - \bm d 
 \right ) \moda \Lambda \Big )   \Bigg]\moda(t\Lambda) \nonumber \\
&=& \left [(t_0-t) \bm x  + (t - \alpha t_0)  \Big [ \left ( \bm x - \bm d \right )
\moda \Lambda \Big] + \bm n \right ]\moda(t\Lambda) \nonumber.
\end{eqnarray}

\subsection{Derivation of $f_{L(t, \bm z)|t = t_0}(x)$}\label{sec:target_function_t_0}

In the particular case where $t = t_0$, it is clear that
\begin{eqnarray}
  \left ( \bm z - t_0 \bm d \right ) \moda (t_0 \Lambda)
&=& \left [t_0 (1 - \alpha )  \Big [ \left ( \bm x - \bm d \right )
\moda \Lambda \Big] + \bm n \right ]\moda(t_0\Lambda) \nonumber. 
\end{eqnarray}
If the dimensionality of the problem $n$ is large enough one could use
the Central Limit Theorem (CLT), and approximate the pdf of $L(t, \bm Z)$
by a Gaussian with mean
\begin{eqnarray}
  \textrm{E}[L(t, \bm Z)|t = t_0] = n + n \log \left (2\pi\Big [\sigma_N^2
  +(1-\alpha)^2 t_0^2 \sigma_\Lambda^2 \Big ]\right ) + \frac{n\left[\left(\sigma_X^2+ \alpha^2 \sigma_\Lambda^2\right) t_0^2 + \sigma_N^2 \right ]}{\sigma_X^2 t_0^2}, \label{eq:mean_mom_t_0}
\end{eqnarray}
where we have used that both HLR $\rightarrow \infty$, and $n\rightarrow \infty$
in order to consider the Euclidean norm of the quantization error
in the first term
to be independent of the Euclidean norm of the received signal;
additionally, we have also consider TNLR $\rightarrow 0$ for the
low-dimensional lattice case, and TNLR $ < 1$ for the high-dimensional
good lattice case, in order to neglect the modulo reduction 
in the first term.

These assumptions are also taken into account in the derivation
of the variance, which 
in the scalar quantizer case can be written as
\begin{eqnarray}
\textrm{Var}[L(t, \bm Z)|t = t_0] &=& n \left [\frac{\frac{(1-\alpha)^4t_0^4 \Delta^4}{180} 
+2\sigma_N^4 + \frac{(1-\alpha)^2t_0^2\Delta^2\sigma_N^2}{3}}{\left (\sigma_N^2 + 
(1-\alpha)^2 \Delta^2 t^2/12 \right )^2} \right .\nonumber \\ &+&  \left . \frac{\frac{\alpha^4t_0^4 \Delta^4}{180} 
+2(t_0^2 \sigma_X^2 + \sigma_N^2)^2 + \frac{\alpha^2t_0^2\Delta^2(t_0^2 \sigma_X^2 + \sigma_N^2)}{3}}{\sigma_X^4 t_0^4} \right ], \label{eq:var_mom_t_0}
\end{eqnarray}
while in the high-dimensional good lattice case is
\begin{eqnarray}
\textrm{Var}[L(t, \bm Z)|t = t_0] &=& n \left [2+    \frac{2( \sigma_X^2 t_0^2 + \alpha^2 t_0^2 \sigma_\Lambda^2 + \sigma_N^2)^2 }{\sigma_X^4 t_0^4} \right ]. \label{eq:var_mom_t_0_mult}
\end{eqnarray}

Alternatively, in the following scenarios:
\begin{itemize}
\item low-dimensional lattice quantizer, 
HLR $\rightarrow \infty$, SCR $\rightarrow 0$, TNLR $\rightarrow 0$, and
$n\rightarrow \infty$, or
\item high-dimensional good lattice quantizer, HLR $\rightarrow \infty$ and
TNLR $<1$,
\end{itemize}
the first term
in (\ref{eq:SecEstimator}) will asymptotically follow a $\chi^2$ distribution
with $n$ degrees of freedom. Similarly, based on HLR $\rightarrow \infty$, and
TNLR $<1$
the third term in (\ref{eq:SecEstimator}) will also asymptotically follow a
$\chi^2$ distribution with $n$ degrees of freedom. Again,
based
on HLR $\rightarrow \infty$, and $n\rightarrow \infty$, we will assume that the Euclidean norm of the
quantization error considered
in the first term is independent of the Euclidean norm of the received signal.
Therefore, in the asymptotic
case described by the considered hypotheses, when $t = t_0$,
(\ref{eq:SecEstimator})
can be seen as the addition of a deterministic term which depends on $t$
and a $\chi^2$ random variable with $2n$ degrees of freedom.

Be aware that the Gaussian approximation derived from this $\chi^2$
distribution is asymptotically equivalent to that introduced at the
beginning of this section whenever the asymptotic framework
(i.e., HLR $\rightarrow \infty$, SCR $\rightarrow 0$, TNLR $\rightarrow 0$, and $n \rightarrow \infty$, for the low-dimensional lattice quantizer case, or
HLR $\rightarrow \infty$, and TNLR $<1$ for the high-dimensional good lattice quantizer case)
is applied to the mean and variance in (\ref{eq:mean_mom_t_0}) and (\ref{eq:var_mom_t_0}) or (\ref{eq:var_mom_t_0_mult}), respectively.

\subsection{Derivation of $f_{L(t, \bm z)|t \neq t_0}(x)$}\label{sec:pdf_no_eq}

On the other hand, whenever $t$ is not close to $t_0$ the vector in the
numerator of the first term in (\ref{eq:SecEstimator}) will be uniformly distributed
in $\mathcal{V}(t \Lambda)$, whereas the last term will follow a scaled $\chi^2$ distribution with $n$ degrees of freedom. In a quantitative way, one can
see that the quantization error considered in the first term of (\ref{eq:SecEstimator}) is asymptotically
uniform over the lattice fundamental Voronoi region whenever
\begin{eqnarray}
  \frac{(t_0-t)^2\sigma_X^2 + (t-\alpha t_0)^2 \sigma_\Lambda^2 + \sigma_N^2}{ t^2  \sigma_\Lambda^2} \rightarrow \infty. \nonumber
\end{eqnarray}
From HLR $\rightarrow \infty$, and TNLR $<1$  usually the dominant effect in the left-hand side
will be $(t_0-t)^2 \sigma_X^2$ and the uniform approximation can be taken
into account in practice for $(t_0-t)^2 \sigma_X^2 \geq K_{\textrm{unif}} t^2 \sigma_\Lambda^2$,
with $K_{\textrm{unif}}$ a positive real number.\footnote{In practice the uniform
approximation is pretty accurate for $K_{\textrm{unif}} \geq 5$ for the scalar quantizer case. In the high-dimensional good lattice quantizer case, it would be indeed enough that
$K_{\textrm{unif}} \geq 1$.}
Therefore, one will be able to use the uniform quantization error approximation
whenever $|t_0 - t| \geq \sqrt{\frac{K_{\textrm{unif}}  t^2 \sigma_\Lambda^2}{ \sigma_X^2}}$, or equivalently
\begin{eqnarray}
 t \leq \frac{t_0}{1 + \sqrt{\frac{K_{\textrm{unif}} \sigma_\Lambda^2}{ \sigma_X^2}}},
\textrm{   or  } t \geq \frac{t_0}{1 - \sqrt{\frac{K_{\textrm{unif}} \sigma_\Lambda^2}{ \sigma_X^2}}} ; \nonumber
\end{eqnarray}
consequently, taking into account that HLR $\rightarrow \infty$, the size of the interval around $t_0$
where the asymptotic uniform behavior can not be considered is rather
small. Concerning the last term of (\ref{eq:SecEstimator}), we will characterize it by
a $\chi^2$ distribution with $n$ degrees of freedom and scaled by 
\begin{eqnarray}
  \frac{(\sigma_X^2 + \alpha^2\sigma_\Lambda^2)t_0^2 + \sigma_N^2}{\sigma_X^2 t^2} . \nonumber
\end{eqnarray}
Taking into account the distribution of both the first and third terms and
using again the assumption of both terms being independent (based on HLR $\rightarrow \infty$, and $n\rightarrow \infty$), the distribution of
(\ref{eq:SecEstimator}) for values of $t$ which are not close to $t_0$ will
be approximately the convolution of the pdf of the scaled square
Euclidean norm of a random vector uniformly distributed on $\mathcal{V}(\Lambda)$, and a scaled $\chi^2$ with $n$ degrees of freedom. In order
to simplify the resulting pdf, we will assume that $n$ is large enough
to the CLT be applied, yielding that
whenever $t$ is not close to $t_0$ and a scalar quantizer is used,
(\ref{eq:SecEstimator}) will
approximately follow a
\begin{eqnarray}
\mathcal{N} &\Bigg (&\frac{n t^2 \sigma_\Lambda^2}{
\sigma_N^2 + (1-\alpha)^2t^2 \sigma_\Lambda^2} + n \log \left [ 2 \pi \left ( \sigma_N^2 + (1-\alpha)^2 t^2 \sigma_\Lambda^2
   \right ) \right ] + \frac{n [(\sigma_X^2 + \alpha^2\sigma_\Lambda^2) t_0^2 + \sigma_N^2 ]}{\sigma_X^2 t^2} ,  \nonumber \\
&&\frac{n 144 t^4 \sigma_\Lambda^4 /180 }{\left (\sigma_N^2 + (1-\alpha)^2t^2 \sigma_\Lambda^2\right )^2} + \frac{2n \left[(\sigma_X^2 + \alpha^2 \sigma_\Lambda^2) t_0^2 + \sigma_N^2\right]^2}{\sigma_X^4 t^4}\Bigg ). \label{eq:pdf_t_neq_t0}
\end{eqnarray}
On the other hand, if we are in the high-dimensional good lattice quantizer scenario, then
\begin{eqnarray}
\mathcal{N} &\Bigg (&\frac{n t^2 \sigma_\Lambda^2}{
\sigma_N^2 + (1-\alpha)^2t^2 \sigma_\Lambda^2} + n \log \left [ 2 \pi \left ( \sigma_N^2 + (1-\alpha)^2 t^2 \sigma_\Lambda^2
   \right ) \right ] + \frac{n [(\sigma_X^2 + \alpha^2\sigma_\Lambda^2) t_0^2 + \sigma_N^2 ]}{\sigma_X^2 t^2} ,  \nonumber \\
&&\frac{2 n t^4 \sigma_\Lambda^4 }{\left (\sigma_N^2 + (1-\alpha)^2t^22\sigma_\Lambda^2 \right )^2} + \frac{2n \left[(\sigma_X^2 + \alpha^2 \sigma_\Lambda^2) t_0^2 + \sigma_N^2\right]^2}{\sigma_X^4 t^4}\Bigg ). \label{eq:pdf_t_neq_t0_mult}
\end{eqnarray}

\subsection{Study of the behavior of $L(t, \bm z)$ under the
aymptotical work hypotheses}

As it was mentioned in Sect.~\ref{sec:hypotheses}, the analysis performed
in this report for the low-dimensional lattice case is based on HLR $\rightarrow \infty$, SCR $\rightarrow 0$, TNLR $\rightarrow 0$, and $n\rightarrow \infty$,
while the high-dimensional good lattice case relies on 
HLR $\rightarrow \infty$, TNLR $< 1$, and, of course, $n\rightarrow \infty$.
In this section we will analyze the considered target function $L(t, \bm z)$
in terms of its limits with respect to those parameters,
in order to check if the resulting
limit function is well defined, i.e., if it does not depend
on the order of the limits (as one would desire), or
if it is indeed ill-defined. In order to avoid the problems
due to $L(t,\bm z)$ diverging  when
$n\rightarrow \infty$ (as it is proportional to this parameter),
we will study $L(t, \bm z)/n$, without loss
of generality. Moreover, in order to avoid degenerated cases in our
 analysis, we will assume $\sigma_X^2 > 0$, $\sigma_\Lambda^2 > 0$, $\sigma_N^2 > 0$,
$t_0 \neq 0$, and $t \neq 0$. This will be proved to be
a sufficient condition for showing the result
we are interested in;  in any case,
practical cases are considered within this framework.

We will find useful to define $L(t, \bm z)$
in terms of the $4$ considered limits. Therefore, we will make
the following variable changes:
\begin{eqnarray}
  \sigma_X^2 &=& \textrm{HLR} \sigma_\Lambda^2, \nonumber \\
  \sigma_N^2& =& t_0^2 \sigma_\Lambda^2 \left [ \textrm{TNLR} - (1-\alpha)^2 \right ], \nonumber \\
  (1-\alpha)^2& =& \frac{\textrm{SCR} \cdot \textrm{TNLR}}{1 + \textrm{SCR}}, \nonumber
\end{eqnarray}
where due to its definition, TNLR $\geq (1-\alpha)^2$. Thus,
(\ref{eq:SecEstimator}) can be rewritten as
\begin{eqnarray}
\frac{L(t, \bm z)}{n}
&\triangleq& \varphi_n(\textrm{HLR}, \textrm{SCR}, \textrm{TNLR}) = 
\frac{||(\bm z - t\bm d )\moda (t \Lambda)||^2/n}{ \sigma_\Lambda^2 \textrm{TNLR}
\left [ \frac{t_0^2 + t^2 \textrm{SCR}}{\textrm{SCR}+1}
\right ]}  \nonumber
  \\&+&\log \left  [2 \pi \left ( \sigma_\Lambda^2 \textrm{TNLR}
\left [ \frac{t_0^2 + t^2 \textrm{SCR}}{\textrm{SCR}+1}
\right ] \right ) \right ]\nonumber \\
  &+ & \frac{||\bm z||^2/n}{ \textrm{HLR} \sigma_\Lambda^2 t^2} \nonumber.
\end{eqnarray}

Due to the nature of the considered functions, the rightmost term of the previous
equalities is obviously an analitic function with HLR $>0$ (i.e., $\sigma_X^2 > 0$), and it will
be also analitic with TNLR and SCR
as long as the denominator of the first
term were positive; a sufficient condition for this
to happen is that the involved variances ($\sigma_\Lambda^2$, and $\sigma_N^2$)
were positive,
and the scaling factor values ($t$, and $t_0$) were non-null.
Thus, under those conditions,
$\varphi(\textrm{HLR}, \textrm{SCR}, \textrm{TNLR}) \triangleq
 \lim_{n\rightarrow \infty} \varphi_n(\textrm{HLR}, \textrm{SCR}, \textrm{TNLR})$
will be also an analytic function.

Therefore, under the previous conditions $\frac{L(t, \bm z)}{n}$
will uniformly converge \cite{Courant}
with $n$ to an analytic function in  $(\textrm{HLR}, \textrm{SCR},
\textrm{TNLR}) \in
(0, \infty) \times (0, \infty) \times
((1-\alpha)^2, \infty)$, implying
that the order of the passages to the limit can be interchanged.
This result is sufficient for proving that the limit target
function is well-defined both for the high-dimensional good lattice case
and the low-dimensional lattice one.

\subsection{Derivation of the asymptotic value of $L(t, \bm z)$ when
the number of observations goes to infinity}\label{sec:L_assympt}

First, we will focus on the case where
$t\approx t_0$, quantified through the inequality $(t_0-t)^2\sigma_X^2 + 
(t-\alpha t_0)^2\alpha^2 \sigma_\Lambda^2 + \sigma_N^2 <<  t^2 \sigma_\Lambda^2$
(implying TNLR $\rightarrow 0$) for the low-dimensional lattice quantizer case,
and through its relaxed version $(t_0-t)^2\sigma_X^2 + 
(t-\alpha t_0)^2\alpha^2 \sigma_\Lambda^2 + \sigma_N^2 <  t^2 \sigma_\Lambda^2$ (implying TNLR $<1$) for the high-dimensional good lattice quantizer one. In that
scenario
\begin{eqnarray}
L(t, \bm z) &\approx &n \frac{(t_0-t)^2\sigma_X^2 + (t-\alpha t_0)^2\sigma_\Lambda^2 + 
\sigma_N^2}{\sigma_N^2 + (1-\alpha)^2t^2\sigma_\Lambda^2} + 
n \log\left(2\pi\left ( \sigma_N^2 + (1-\alpha)^2 t^2 \sigma_\Lambda^2\right) \right )
\nonumber \\ &+& n\frac{(\sigma_X^2 + \alpha^2 \sigma_\Lambda^2)t_0^2 + \sigma_N^2}{\sigma_X^2  t^2} \nonumber \\
& \approx & 2n + \frac{n (t_0-t)^2\sigma_X^2}{\sigma_N^2 + (1-\alpha)^2 t^2 \sigma_\Lambda^2}
+ n \log\left(2\pi\left ( \sigma_N^2 + (1-\alpha)^2 t^2 \sigma_\Lambda^2 \right) \right ), \label{eq:L_mean_close}
\end{eqnarray}
where we have used that HLR $\rightarrow \infty$, 
TNLR $\rightarrow 0$, and $n\rightarrow \infty$
for the low-dimensional lattice quantizer case, and that HLR $\rightarrow \infty$, and TNLR $<1$ for the high-dimensional good lattice quantizers.

On the other hand, if we study the case where $t$ is not close to $t_0$,
meaning $(t_0-t)^2\sigma_X^2 + 
(t-\alpha t_0)^2\alpha^2 \sigma_\Lambda^2 + \sigma_N^2 >>  t^2 \sigma_\Lambda^2$ for the low-dimensional lattice quantizer case and $(t_0-t)^2\sigma_X^2 + 
(t-\alpha t_0)^2\alpha^2 \sigma_\Lambda^2 + \sigma_N^2 >  t^2 \sigma_\Lambda^2$ for the high-dimensional good lattice quantizers,
$L(t, \bm z)$ can be approximated by 
\begin{eqnarray}
L(t, \bm z) &\approx &n \frac{ t^2 \sigma_\Lambda^2}{\sigma_N^2 + (1-\alpha)^2t^2 \sigma_\Lambda^2} + 
n \log\left(2\pi\left ( \sigma_N^2 + (1-\alpha)^2 t^2 \sigma_\Lambda^2\right) \right )
\nonumber \\ &+& n\frac{(\sigma_X^2 + \alpha^2 \sigma_\Lambda^2)t_0^2 + \sigma_N^2}{\sigma_X^2  t^2} \nonumber \\
&\approx &n \frac{ t^2 \sigma_\Lambda^2}{\sigma_N^2 + (1-\alpha)^2 t^2 \sigma_\Lambda^2} + 
n \log\left(2\pi\left ( \sigma_N^2 + (1-\alpha)^2 t^2 \sigma_\Lambda^2\right) \right )
+ n \frac{t_0^2}{t^2};\nonumber \\ \label{eq:L_mean_far}
\end{eqnarray}
again, we have used that HLR $\rightarrow \infty$,
TNLR $<1$, and $n \rightarrow \infty$.
%%%%%%%%%%%%%%%%%%%%%%%%%%%%%%%%%%%%%%%%%%%%%%%%%%%%%%%%%%%%%%%%%%%%%
%%%%%%%%%%%%%%%%%%%%%%%%%%%%%%%%%%%%%%%%%%%%%%%%%%%%%%%%%%%%%%%%%%%%%

\section{Optimization}

Due to the modulo operation in the first term of (\ref{eq:SecEstimator}) the
considered target function $L(t, \bm z)$ is not convex with $t$; indeed,
it will have
a large number of non-differentiable points, corresponding to those values 
of $t$ that move $\bm z/t$ to a different Voronoi region of the original
quantizer $Q_\Lambda$.

Taking into account these difficulties, we will use an initial estimation
of the scaling factor $t_0$, denoted by $t_1$, to
define a search interval where the solution to our problem 
can be guaranteed to lie on (if a deterministic approach
is followeed), or at least can be guaranteed to lie on
with a given probability (if a probabilistic
approach is followed); this search interval will be sampled finely enough
to be reasonably sure that the global minimum of the target function
(or a good approximation to it) will
be found 
by the optimization algorithms to be introduced.

%%%%%%%%%%%%%%%%%%%%%%%%%%%%%%%%%%%%%%%%%%%%%%%%%%%%%%%%%%%%%%%%%%%%%
%%%%%%%%%%%%%%%%%%%%%%%%%%%%%%%%%%%%%%%%%%%%%%%%%%%%%%%%%%%%%%%%%%%%%

\section{$t_1$ choices}\label{sec:t_1_computation}

\subsection{$L(t, \bm z)$ approximation}
As a first strategy for defining $t_1$,
we define an approximation to our target function which has
a single minimum in $t$; by doing so, one can ensure the convergence to the
minimum of that approximating function, which is hoped to be  close to
$t_0$.
Considering that for most of $t$ values (i.e., those that are not close to
$t_0$) the uniform approximation of the quantization error can be assumed,
our proposal to approximate the target function is
\begin{eqnarray}
L_1(t, \bm z) \triangleq \frac{n t^2 \sigma_\Lambda^2}{ \sigma_N^2 + (1-\alpha)^2 \sigma_\Lambda^2 t^2  } 
  + n \log \left ( 2 \pi \left ( \sigma_N^2 + (1-\alpha)^2 \sigma_\Lambda^2
  t^2 \right ) \right )
  +  \frac{||\bm z||^2}{ \sigma_X^2 t^2}. \nonumber
\end{eqnarray}
It is just mechanical to check that the first derivative of $L_1(t, \bm z)$ with
respect to $t^2$ is
\begin{eqnarray}
  \frac{\partial L_1(t, \bm z)}{\partial \left (t^2 \right )} &=& 
\frac{n \sigma_\Lambda^2 \left [ \left (1+\left (1-\alpha\right )^2 \right )\sigma_N^2 + \left (1-\alpha \right )^4 \sigma_\Lambda^2 t^2 \right ]}{ \left [\sigma_N^2 + \left(1-\alpha \right )^2 \sigma_\Lambda^2 t^2 \right ]^2} - \frac{||\bm z||^2}{\sigma_X^2 t^4}; \nonumber
\end{eqnarray}
it is straightforward to see that one can write
\begin{eqnarray}
  \left ( \frac{\partial L_1(t, \bm z)}{\partial \left (t^2 \right )} \right ) \left [\sigma_N^2 + \left ( 1-\alpha \right)^2 \sigma_\Lambda^2 t^2 \right ]^2
\sigma_X^2 t^4 = a_0 + a_1 t^2 + a_2 t^4 + a_3 t^6, \label{eq:der_mult_approx}
\end{eqnarray}
where,
\begin{eqnarray}
  a_0 &=& -||\bm z||^2 \sigma_N^4 \nonumber \\
  a_1 &=& -2(1-\alpha)^2\sigma_\Lambda^2 \sigma_N^2 ||\bm z||^2\nonumber \\
  a_2 &=& \sigma_\Lambda^2 \left [  n\left ( 1 + \left (1 - \alpha \right)^2  \right )
    \sigma_N^2 \sigma_X^2 - \left (1 - \alpha \right )^4 \sigma_\Lambda^2 ||\bm z||^2  \right ]\nonumber \\
  a_3 &=& n \left (1 - \alpha \right )^4 \sigma_\Lambda^4 \sigma_X^2. \nonumber 
\end{eqnarray}

Therefore,  considering that $L_1(t, \bm z) \rightarrow \infty$
both when $t\rightarrow 0$ and $t \rightarrow \infty$, and that according to
Descartes' sign rule the number of real roots of (\ref{eq:der_mult_approx}) is at most
one, it is evident that
$L_1(t, \bm z)$ will have a single local (and consequently global) minimum
for $t > 0$, proving that
$t_1 = \argmin_t L_1(\bm z, t)$
can be easily computed.

Interestingly, when $n\rightarrow \infty$, and for an arbitrary lattice,
the left term in  (\ref{eq:der_mult_approx}) goes to
\begin{eqnarray}
&-&n\left [ \left ( \sigma_X^2 + \alpha^2 \sigma_\Lambda^2 \right) t_0^2 + \sigma_N^2 \right ] \left [\sigma_N^2 + \left (1-\alpha \right)^2 \sigma_\Lambda^2 t^2 \right ]^2 \nonumber \\ &+& n \sigma_\Lambda^2 \sigma_X^2 t^4 \left [\left ( 1 + (1-\alpha)^2 \right) \sigma_N^2 + (1-\alpha)^4 \sigma_\Lambda^2 t^2 \right ], \nonumber
\end{eqnarray}
which, assuming that HLR $\rightarrow \infty$, SCR $\rightarrow 0$, TNLR $\rightarrow 0$, and that $t$ is not much larger than $t_0$,\footnote{This is the only place where SCR $\rightarrow 0$, and TNLR $\rightarrow 0$ are used
in order to prove a property of
the high-dimensional good lattice case. Additionally, assumptions
on the value of $t$ are made. Nevertheless, the objective
of considering these conditions is just to prove the
non-consistent nature of the resulting estimator, no affecting to the
calculation of $t_1$.} can be approximated by
\begin{eqnarray}
  n \sigma_N^2 \sigma_X^2 \left [ - \sigma_N^2 t_0^2 + \left (1 + \left (1-\alpha \right )^2 \right ) \sigma_\Lambda^2 t^4 \right]; \nonumber
\end{eqnarray}
consequently, under the work hypotheses and when the dimensionality of the problem
goes to infinity, the value of $t$ minimizing $L_1(t, \bm z)$ can be
approximated by
\begin{eqnarray}
  t_1 \triangleq  \left ( \frac{\sigma_N^2 t_0^2}{\left [1 + \left (1-\alpha\right )^2 \right] \sigma_\Lambda^2}\right )^{1/4}. \label{eq:approx_min_L1}
\end{eqnarray}
Obviously this value does not coincide with $t_0$, indicating that the
obtained estimator is not consistent.

\begin{figure}[t] 
  \begin{center}
    \includegraphics[width=0.75\linewidth]{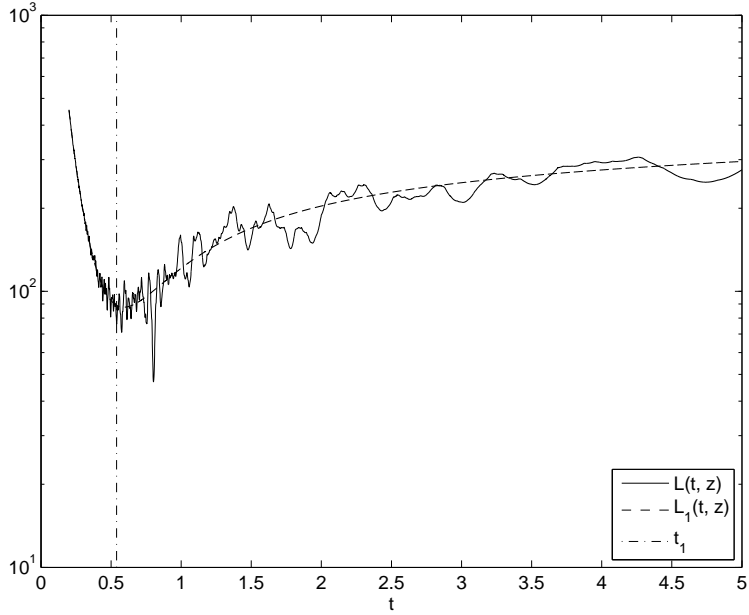}  
    \caption{Comparison between $L(t, \bm z)$ and $L_1(t, \bm z)$. The proposed
value of $t_1$ (dash-dot line) is not that exactly minimizing
$L_1(t, \bm z)$, but
that in (\ref{eq:approx_min_L1}). HLR $\approx 34.9765 $ dB, SCR $\approx -1.0618$ dB, TNLR $\approx -3.5736$ dB, $t_0 = 0.8$, 
$\alpha = \alpha_{\textrm{Costa}} \approx 0.5608$, $n = 40$, scalar
quantizer.}\label{fig:aprox_L1}
  \end{center}
\end{figure}

A graphical example of this approximation can be found in Fig.~\ref{fig:aprox_L1}.

\subsection{Variance-based estimators}\label{sec:variance_based_estimator}

The variance-based estimator for the i.i.d. Gaussian case
takes the form
\begin{eqnarray}
  \hat{t}_{\textrm{var}}(\bm z)= \sqrt{\frac{\frac{||\bm z||^2}{n} - \sigma_N^2}{\sigma_X^2 + 
      \alpha^2 \sigma_\Lambda^2}}. \label{eq:estimator_var_biased}
\end{eqnarray}
Furthermore, its Cramer Rao-Bound (CRB) is
\begin{eqnarray}
  \frac{\left [(\sigma_X^2 + \alpha^2 \sigma_\Lambda^2)t_0^2 + \sigma_N^2 \right]^2}{2 n \left (\sigma_X^2 + \alpha^2 \sigma_\Lambda^2 \right )^2 t_0^2}. \nonumber
\end{eqnarray}
Nevertheless, one must take into account that the previous estimator 
is biased (so the CRB is wrongly used); in order to prove it, we will consider
a simplified case, where $n=1$, $\sigma_N^2 = 0$, $\sigma_X^2 + \alpha^2 \sigma_\Lambda^2 =1$,
and $t_0 = 1$. In that scenario, it is obvious that
\begin{eqnarray}
  \textrm{E}\{\hat{T}_{\textrm{var}}(\bm Z)\} = \textrm{E}\{|Z|\} = 2\int_{0}^\infty \frac{\tau e^{\frac{-\tau^2}{2}}}{\sqrt{2\pi}} d\tau = \sqrt{\frac{2}{\pi}}.\nonumber
\end{eqnarray}

On the other hand, the following estimator could be used
\begin{eqnarray}
  \hat{t^2}_{\textrm{var}}(\bm z)= \frac{\frac{||\bm z||^2}{n} - \sigma_N^2}{\sigma_X^2 + 
      \alpha^2 \sigma_\Lambda^2}; \nonumber
\end{eqnarray}
the latter is indeed an  unbiased estimator of $t_0^2$, whose CRB
is easily proved to be
\begin{eqnarray}
  \frac{2 \left [ \left (\sigma_X^2 + \alpha^2 \sigma_\Lambda^2\right )t_0^2 + \sigma_N^2 \right]^2}{n\left (\sigma_X^2 + \alpha^2 \sigma_\Lambda^2 \right )^2}. \label{eq:crb_var_based_t2}
\end{eqnarray}

\subsubsection{Studying the links between  $\hat{T}_{\textrm{var}}(\bm Z)$ and 
$\hat{T^2}_{\textrm{var}}(\bm Z)$}\label{sec:var_bas_est_links}

In this section we will study the pdf of the square root
of $\hat{T^2}_{\textrm{var}}(\bm Z)$, derive its mean and variance, and analyze
in which asymptotic cases the statistics of $\hat{T}_{\textrm{var}}(\bm Z)$
obtained following this methodology converge to those previously derived
under the false premise that this estimator is unbiased.

Based on the assumption that the number of considered dimensions 
is very large, the pdf of $\hat{T^2}_{\textrm{var}}(\bm Z)$ given $t_0$
can be approximated
by $\mathcal{N}\left(t_0^2, \frac{2[(\sigma_X^2 + \alpha^2 \sigma_\Lambda^2)t_0^2 + \sigma_N^2]^2}{n (\sigma_X^2 + \alpha^2 \sigma_\Lambda^2)^2}\right )$.
%% , so
%% \begin{eqnarray}
%%   f_{\hat{T^2}_{\textrm{var}}|T_0^2}(t^2|t_0^2) \approx \frac{e^{\frac{- n (\sigma_X^2 + \alpha^2 \sigma_\Lambda^2)^2 (t^2 - t_0^2)^2}{4 [(\sigma_X^2 + \alpha^2 \sigma_\Lambda^2)t_0^2 + \sigma_N^2]^2}}}{\sqrt{\frac{4\pi [(\sigma_X^2 + \alpha^2 \sigma_\Lambda^2) t_0^2  + \sigma_N^2]^2}{n(\sigma_X^2 + \alpha^2 \sigma_\Lambda^2)^2}}}. \nonumber
%% \end{eqnarray}
In general, if we have a random variable $T^2$ with pdf $\mathcal{N}(\mu, \sigma^2)$, the mean of its square root $T$ is 
\begin{eqnarray}
  \textrm{E}\{T\} &=& \int_0^\infty \tau \left (\frac{2 \tau e^{\frac{-(\tau^2-\mu)^2}{2\sigma^2}}}{\sqrt{2\pi \sigma^2}} + j \frac{2 \tau e^{\frac{-(\tau^2+\mu)^2}{2\sigma^2}}}{\sqrt{2\pi \sigma^2}} \right ) d\tau  \nonumber \\
  &=&\frac{(1 + j)e^{\frac{-\mu^2}{2\sigma^2}}}{2^{3/4}\sqrt{\pi} \sigma^{1/2}}
  \left [ \sigma \Gamma \left (\frac{3}{4} \right ) 
    {}_1 F_1 \left (\frac{3}{4};\frac{1}{2}; \frac{\mu^2}{2\sigma^2} \right )
     - j \sqrt{2} \mu \Gamma \left (\frac{5}{4} \right ) 
    {}_1 F_1 \left (\frac{5}{4};\frac{3}{2}; \frac{\mu^2}{2\sigma^2} \right )
\right], \nonumber
\end{eqnarray}
where ${}_1 F_1 \left (\cdot;\cdot; \cdot \right )$ denotes the confluent
hypergeometric function of the first kind.
Similarly, the mean $|T|^2$ can be obtained as
\begin{eqnarray}
\textrm{E}\{|T|^2\} &=& \textrm{E}\{|T^2|\} = \int_0^\infty \tau^2 \left (\frac{2 \tau e^{\frac{-(\tau^2-\mu)^2}{2\sigma^2}}}{\sqrt{2\pi \sigma^2}} + \frac{2 \tau e^{\frac{-(\tau^2+\mu)^2}{2\sigma^2}}}{\sqrt{2\pi \sigma^2}} \right ) d\tau  \nonumber \\
&=& e^{\frac{-\mu^2}{2\sigma^2}} \sqrt{\frac{2\sigma^2}{\pi}} + \mu \textrm{Erf} 
\left (\frac{\mu}{\sqrt{2\sigma^2}} \right ). \nonumber
\end{eqnarray}

Replacing $\mu$ and $\sigma^2$ by  $t_0^2$ and (\ref{eq:crb_var_based_t2}),
respectively, it is straightforward to obtain the mean and variance
of $T$.

Nevertheless, it is also interesting to study the behavior of the
considered random variable when the mean of  $T^2$ is much larger than its standard deviation, i.e., $\frac{\mu^2}{\sigma^2} \rightarrow \infty$;
intuitively, one can see that in this case the effects of the tails
of the Gaussian taking negative values can be dismissed. Indeed, it
can be shown that
\begin{eqnarray}
  \lim_{\frac{\mu}{\sigma} \rightarrow \infty} \textrm{E}\{T\} &=& \sqrt{\mu}, \nonumber\\
  \lim_{\frac{\mu}{\sigma} \rightarrow \infty} \textrm{E}\{|T^2|\} \frac{\mu^2}{\sigma^2} &=& \frac{\mu}{4}. \nonumber
\end{eqnarray}
In this asymptotic case, if $\mu$ and $\sigma^2$ were replaced by
$t_0^2$ and (\ref{eq:crb_var_based_t2}), the mean and the
(wrongly calculated) CRB
for the estimator in (\ref{eq:estimator_var_biased}) are obtained.
This result seems to indicate that the estimator in
(\ref{eq:estimator_var_biased}) will be asymptotically unbiased, and
therefore its CRB correctly calculated, just
when the previously introduced condition on the ratio $\frac{\mu}{\sigma}$
is verified. For finite non-null values of $\sigma_X^2 + \alpha^2 \sigma_\Lambda^2$,
$\sigma_N^2$, and $t_0$ that condition holds, for example, when $n\rightarrow
\infty$.

Furthermore, when $n\rightarrow \infty$ both variance-based estimators
asymptotically
follow a Gaussian distribution.

\subsubsection{Additional comments on  $\hat{T^2}_{\textrm{var}}(\bm Z)$ and 
$\hat{T}_{\textrm{var}}(\bm Z)$}\label{sec:trunc_gauss_var_based_est}

At the sight of the definition and 
theoretical characterization of $\hat{T^2}_{\textrm{var}}(\bm Z)$
and $\hat{T}_{\textrm{var}}(\bm Z)$ provided in the previous sections,
it is obvious that the negative values of $\hat{T^2}_{\textrm{var}}(\bm Z)$ and
imaginary of $\hat{T}_{\textrm{var}}(\bm Z)$ have non-null probability.
 The causes of this problem
are basically two:
\begin{itemize}
  \item The approximate nature of the proposed pdf. Given that we are assuming
$||\bm Z||^2$ to follow a Gaussian distribution, the probability of that random
variable taking negative values will be non-null. This problem would be solved
if the real pdf (the convolution of two scaled $\chi^2$ distribution with
$n$ degrees of freedom) were considered. Nevertheless, the price to be paid
is a significant increase in the complexity of the resulting pdf, which
will make more hard to perform a theoretical analysis.
\item Even if one considers the exact pdf of $||\bm Z||^2$, the probability
of obtaining a negative value of $\hat{T^2}_{\textrm{var}}(\bm Z)$ or complex value
of $\hat{T}_{\textrm{var}}(\bm Z)$ will be still non-null. This is explained by the fact
that
{\it a priori} information is not used at the derivation of the proposed
estimators.
\end{itemize}

Therefore, if {\it a priori} information on the possible interval of
$t_0$ values were available, the ML optimization problem should be
redefined in order to consider those constraints.\footnote{Indeed, strictly speaking the resulting estimator is no longer ML, as {\it a priori} information is used; nevertheless, it is neither MAP, as the {\it a priori} knowledge
does not inform on the pdf of $T_0$, but just on its domain.}
 In our case, where the
{\it a priori} information constrains $t_0$ to be real, the ML
problem might be defined as
\begin{eqnarray}
  \hat{t^2}_{\textrm{var}}(\bm z) = \argmax_{t^2 \geq 0} f_{\bm Z|T^2}(\bm z| t^2). \nonumber
\end{eqnarray} 

In the case of i.i.d. Gaussian signals, as the derivative of the pdf with
respect to $t^2$ has a single local minimum, if that minimum were
located at a negative value of $t^2$ it is straightforward to see
that the value of $t^2$ maximizing $f_{\bm Z|T^2}(\bm z| t^2)$ will be $0$.
Therefore, the estimators obtained by following the above criterion are
\begin{eqnarray}
  \hat{t^2}_{\textrm{var}}(\bm z)= \left [\frac{\frac{||\bm z||^2}{n} - \sigma_N^2}{\sigma_X^2 + 
      \alpha^2 \sigma_\Lambda^2} \right ]^+, \label{eq:def_va_est2}
\end{eqnarray}
or 
\begin{eqnarray}
  \hat{t}_{\textrm{var}}(\bm z)= \sqrt{\left [\frac{\frac{||\bm z||^2}{n} - \sigma_N^2}{\sigma_X^2 + 
      \alpha^2 \sigma_\Lambda^2} \right ]^+}. \label{eq:def_va_est34}
\end{eqnarray}
Both of these estimators will be biased. Assuming
a large number of dimensions in order to: 1) the original non-truncated
estimators studied
at the beginning of this section be Gaussian, and 2)
for $\hat{T}_{\textrm{var}}(\bm Z)$ estimator, the
condition derived at the end of Sect.~\ref{sec:var_bas_est_links}
be hold, the pdfs of (\ref{eq:def_va_est2}) and (\ref{eq:def_va_est34}) are, respectively
\begin{eqnarray}
  f_{\hat{T^2}_{\textrm{var}}|T_0^2}(t^2|t_0^2) \approx \left \{
  \begin{array}{ll}
\frac{e^{\frac{- n (\sigma_X^2 + \alpha^2 \sigma_\Lambda^2)^2 (t^2 - t_0^2)^2}{4 [(\sigma_X^2 + \alpha^2 \sigma_\Lambda^2)t_0^2 + \sigma_N^2]^2}}}{\sqrt{\frac{4\pi [(\sigma_X^2 + \alpha^2 \sigma_\Lambda^2) t_0^2  + \sigma_N^2]^2}{n(\sigma_X^2 + \alpha^2 \sigma_\Lambda^2)^2}}} &
\textrm{   if   } t^2 > 0 \\
Q\left ( \frac{t_0^2}{\sqrt{\frac{2 [(\sigma_X^2 + \alpha^2 \sigma_\Lambda^2) t_0^2  + \sigma_N^2]^2}{n(\sigma_X^2 + \alpha^2 \sigma_\Lambda^2)^2}}} 
\right ) \delta(t^2) & \textrm {   otherwise} \end{array}
\right . , \label{eq:pdf_va_est2}
\end{eqnarray}
and
\begin{eqnarray}
  f_{\hat{T}_{\textrm{var}}|T_0}(t|t_0) \approx \left \{
  \begin{array}{ll}
\frac{e^{\frac{- n (\sigma_X^2 + \alpha^2 \sigma_\Lambda^2)^2 t_0^2 (t - t_0)^2}{ [(\sigma_X^2 + \alpha^2 \sigma_\Lambda^2)t_0^2 + \sigma_N^2]^2}}}{\sqrt{\frac{\pi [(\sigma_X^2 + \alpha^2 \sigma_\Lambda^2) t_0^2  + \sigma_N^2]^2}{n(\sigma_X^2 + \alpha^2 \sigma_\Lambda^2)^2 t_0^2}}} &
\textrm{   if   } t \in \mathbb{R}^+ \\
Q\left ( \frac{t_0}{\sqrt{\frac{ [(\sigma_X^2 + \alpha^2 \sigma_\Lambda^2) t_0^2  + \sigma_N^2]^2}{2n(\sigma_X^2 + \alpha^2 \sigma_\Lambda^2)^2 t_0^2}}} 
\right ) \delta(t) & \textrm {   otherwise} \end{array}
\right . , \nonumber
\end{eqnarray}
where $Q(x)$ is defined as
\begin{eqnarray}
  Q(x) \triangleq \int_{-\infty}^x \frac{e^{-x^2/2}}{\sqrt{2\pi}}. \nonumber
\end{eqnarray}

%%%%%%%%%%%%%%%%%%%%%%%%%%%%%%%%%%%%%%%%%%%%%%%%%%%%%%%%%%%%%%%%%%%%%
%%%%%%%%%%%%%%%%%%%%%%%%%%%%%%%%%%%%%%%%%%%%%%%%%%%%%%%%%%%%%%%%%%%%%
\section{Search interval lower-bound}\label{sec:lower_bound}
%%%%%%%%%%%%%%%%%%%%%%%%%%%%%%%%%%%%%%%%%%%%%%%%%%%%%%%%%%%%%%%%%%%%%
\subsection{General discussion on the methodologies proposed
for determining $t_{\textrm{lower}}$ and $t_{\textrm{upper}}$}

The methodologies proposed in Sect.~\ref{sec:lower_bound} and
Sect.~\ref{sec:upper_bound} for determining lower and upper-bounds
to the search interval, can be coarsely classified into three
categories:
\begin{itemize}
\item deterministic: Sects.~\ref{sec:lower_deter}, \ref{sec:upper_deter},
  and \ref{sec:deter_upper_2}. These approaches 
  ensure that
  $\hat{t}(\bm z)$ is contained in the derived search interval.
\item probabilistic: Sects.~\ref{sec:another_possible}, 
\ref{sec:upper_bound_prob}, and \ref{sec:another_possible2}.
These schemes are based on statistically
characterizing the target function, and determine an interval of
scaling factors such that if $t_0$ were out of
that interval, it would be very unlikely that the value of the target function 
at $t_0$ were smaller that $L(t_1, \bm z)$. Assuming the bias
of $\hat{t}(\bm z)$ to be small, these methods could be seen
as a relaxed version of the deterministic ones; as $t_0$ is just
ensured to be within the provided interval with a certain probability, the
width of that interval can be reduced. Links in this direction will
be established between the estimators proposed in 
Sects.~\ref{sec:lower_deter}, and \ref{sec:upper_deter}, and
those in Sects.~\ref{sec:another_possible}, and \ref{sec:another_possible2};
and between the estimators proposed in Sect.~\ref{sec:deter_upper_2}, and
Sect.~\ref{sec:upper_bound_prob}.
\item derived from the variance-based estimator: 
Sects.~\ref{eq:sec_lowerbound_var_based} and \ref{sec:upper_var_based}.
Although these approaches are also probabilistic, in this
case it is not the target function, but the
variance-based estimator which is statistically characterized in order
to define an interval of scaling factors such that if $t_0$
were out of that interval, it would be very unlikely
that the current variance-based estimator were obtained.
\end{itemize}

%%%%%%%%%%%%%%%%%%%%%%%%%%%%%%%%%%%%%%%%%%%%%%%%%%%%%%%%%%%%%%%%%%%%%
\subsection{Deterministic approach}\label{sec:lower_deter}

Defining $L_2(t, \bm z)$ as
\begin{eqnarray}
L_2(t, \bm z) \triangleq n \log \left ( 2 \pi \left ( \sigma_N^2 + (1-\alpha)^2 t^2 \sigma_\Lambda^2
   \right ) \right )
  +  \frac{||\bm z||^2}{ \sigma_X^2 t^2} , \label{eq:def_L2}
\end{eqnarray}
and
given that $L(\hat t(\bm z), \bm z) \leq L(t_1, \bm z)$ by definition,
and that the first term of $L(t, \bm z)$ is non-negative,
it is clear that $\hat t(\bm z)$ must verify
\begin{eqnarray}
L_2(\hat t(\bm z), \bm z) \leq L(\hat t(\bm z), \bm z) \leq L(t_1, \bm z). \nonumber
\end{eqnarray}
Furthermore, the first derivative of $L_2(t, \bm z)$ with respect to $t^2$ is 
\begin{eqnarray}
\frac{\partial L_2(t, \bm z)}{\partial (t^2)} =  \frac{n(1-\alpha)^2\sigma_\Lambda^2}{\sigma_N^2 + (1-\alpha)^2 t^2 \sigma_\Lambda^2} - \frac{||\bm z||^2}{\sigma_X^2 t^4},
\nonumber
\end{eqnarray}
which has $2$ roots in $t^2$, one of them negative; the only positive
one can be found at $t_2^2$
\begin{eqnarray}
  t_2^2 =  \frac{||\bm z||^2 + ||\bm z|| \sqrt{||\bm z||^2
 + \frac{4 n\sigma_N^2  \sigma_X^2}{(1-\alpha)^2 \sigma_\Lambda^2}}}{2 n\sigma_X^2}; \nonumber
\end{eqnarray}
additionally, since for $t \rightarrow 0$, $L_2(t, \bm z) \rightarrow \infty$,
we have that $L_2(t, \bm z)$ has a single minimum located at $t_2$.
Therefore, a lower-bound to the search interval can be calculated as
the largest $t\leq t_2$ such that $L_2(t, \bm z)$ is larger
than or equal to $L(t_1, \bm z)$, i.e.,
\begin{eqnarray}
  t_{\textrm{lower}} = \argmax_{t: t \leq t_2, L_2(t, \bm z) \geq L(t_1, \bm z) } t , \label{eq:compt_lower}
\end{eqnarray}
which due to the strictly monotonically decreasing nature of $L_2(t, \bm z)$
for $t < t_2$ and that for $t \rightarrow 0$, $L_2(t, \bm z) \rightarrow \infty$, will be well-defined. On the other hand $L_2(t_2, \bm z) \leq L_2(t_1, \bm z) \leq L(t_1, \bm z)$, holding both equalities only if $t_1 = t_2$ and if the
quantization error in the first term of $L(t_1, \bm z)$ is null.
Taking this into account, and
due to the continuity of $L_2(t, \bm z)$, is obvious that
$L_2(t_{\textrm{lower}}, \bm z) = L(t_1, \bm z)$, so we will focus on the study of
the solutions for $t \leq t_2$ to the equation
\begin{eqnarray}
&  n \log \left ( 2 \pi \left ( \sigma_N^2 + (1-\alpha)^2 
  t^2 \sigma_\Lambda^2 \right ) \right )
  +  \frac{||\bm z||^2}{ \sigma_X^2 t^2} = \nonumber \\ &\frac{||(\bm z - t_1\bm d )\moda (t_1 \Lambda)||^2}{ \sigma_N^2 + (1-\alpha)^2 t_1^2 \sigma_\Lambda^2  } 
  + n \log \left ( 2 \pi \left ( \sigma_N^2 + (1-\alpha)^2 
  t_1^2 \sigma_\Lambda^2 \right ) \right )
  +  \frac{||\bm z||^2}{ \sigma_X^2 t_1^2}. \label{eq:sol_t4}
\end{eqnarray}
Although to the best of authors' knowledge closed formulas for the solutions
of this equation are not available, due to the monotonically decreasing
nature of $L_2(t, \bm z)$ for $t \leq t_2$, and that $L_2(t_2, \bm z) \leq L(t_1, \bm z)$ the solution we are looking for is unique and can be found by
simple search methods.

The upper-bound counterpart of this method can be found in Sect.~\ref{sec:upper_deter}.
%%%%%%%%%%%%%%%%%%%%%%%%%%%%%%%%%%%%%%%%%%%%%%%%%%%%%%%%%%%%%%%%%%%%%
\subsubsection{Asymptotic value when the number of observations goes to infinity}\label{sec:lower_deter_asympt}

In this case
\begin{eqnarray}
  L_2(t, \bm z) \approx n \log \left ( 2 \pi \left ( \sigma_N^2 + (1-\alpha)^2 
  t^2 \sigma_\Lambda^2 \right ) \right )
  +  n\frac{(\sigma_X^2 + \alpha^2 \sigma_\Lambda^2)t_0^2  + \sigma_N^2}{ \sigma_X^2 t^2} , \label{eq:lower2_asym_1}
\end{eqnarray}
and
\begin{eqnarray}
  t_2 \approx  \sqrt{\frac{(\sigma_X^2 + \alpha^2 \sigma_\Lambda^2)t_0^2  + \sigma_N^2 + \sqrt{(\sigma_X^2 + \alpha^2 \sigma_\Lambda^2)t_0^2  + \sigma_N^2} \sqrt{(\sigma_X^2 + \alpha^2 \sigma_\Lambda^2)t_0^2  + \sigma_N^2
 + \frac{4 \sigma_N^2  \sigma_X^2}{(1-\alpha)^2 \sigma_\Lambda^2}}}{2 \sigma_X^2}}, \label{eq:lower2_asym_2}
\end{eqnarray}
which is obviously larger than or equal to $t_0$; whenever HLR $\rightarrow \infty$, and TNLR $<1$, the last expression can be approximated by
\begin{eqnarray}
  t_2 \approx  \sqrt{\frac{t_0^2 + t_0 \sqrt{t_0^2 + \frac{4\sigma_N^2}{(1-\alpha)^2\sigma_\Lambda^2}}}{2}}. \label{eq:lower2_asym_3}
\end{eqnarray}
Therefore, the remaining question to be answered is if $t_{\textrm{lower}}$ will
be smaller than or equal to $t_0$. In order to do so, we will first consider
the case $t_1 \approx t_0$, so, considering (\ref{eq:L_mean_close}) as
the second term in (\ref{eq:sol_t4}), one obtains
\begin{eqnarray}
&  n \log \left ( 2 \pi \left ( \sigma_N^2 + (1-\alpha)^2 
  t^2 \sigma_\Lambda^2 \right ) \right )
  + n \frac{ t_0^2}{t^2}= \nonumber \\ &2n + \frac{n (t_0-t_1)^2\sigma_X^2}{\sigma_N^2 + (1-\alpha)^2 t_1^2 \sigma_\Lambda^2}
+ n \log\left(2\pi\left ( \sigma_N^2 + (1-\alpha)^2 t_1^2 \sigma_\Lambda^2\right) \right ). \nonumber
\end{eqnarray}
If $t_0 \leq t_1$, then it is obvious that the first term evaluated
at $t = t_0$ is smaller than the
second one, implying that $t_{\textrm{lower}} < t_0$.
On the other hand, if $t_0 > t_1$, then,
given that both of them are smaller than or equal to $t_2$,
\begin{eqnarray}
  L_2(t_0, \bm z) < L_2(t_1, \bm z) \leq L(t_1, \bm z), \label{eq:cadena_mola}
\end{eqnarray}
proving that $t_{\textrm{lower}} < t_0$.

Finally, if $t_1$ is not close to $t_0$, (\ref{eq:L_mean_far}) will
replace to the second term in (\ref{eq:sol_t4}), yielding
\begin{eqnarray}
&  n \log \left ( 2 \pi \left ( \sigma_N^2 + (1-\alpha)^2
  t^2 \sigma_\Lambda^2 \right ) \right )
  +  n\frac{ t_0^2}{t^2} = \nonumber \\ &n \frac{ t_1^2 \sigma_\Lambda^2}{\sigma_N^2 + (1-\alpha)^2 t_1^2 \sigma_\Lambda^2} + 
n \log\left(2\pi\left ( \sigma_N^2 + (1-\alpha)^2 t_1^2 \sigma_\Lambda^2\right) \right )
+ n \frac{t_0^2}{t_1^2}. \label{eq:lotenemos}
\end{eqnarray}
If $t_0 \leq t_1$, then $\log \left ( 2 \pi \left ( \sigma_N^2 + (1-\alpha)^2 
  t_0^2 \sigma_\Lambda^2\right ) \right ) \leq \log \left ( 2 \pi \left ( \sigma_N^2 + (1-\alpha)^2 
  t_1^2\sigma_\Lambda^2 \right ) \right )$; additionally $\frac{ t^2\sigma_\Lambda^2}{\sigma_N^2 + (1-\alpha)^2 t^2 \sigma_\Lambda^2}$ is monotonically increasing with $t$,
implying that
\begin{eqnarray}
  \frac{ t_1^2 \sigma_\Lambda^2}{\sigma_N^2 + (1-\alpha)^2 t_1^2 \sigma_\Lambda^2} \geq 
  \frac{ t_0^2 \sigma_\Lambda^2}{\sigma_N^2 + (1-\alpha)^2 t_0^2 \sigma_\Lambda^2} > 1, \nonumber
\end{eqnarray}
due to TNLR $<1$. Therefore, at $t = t_0$ the first term in (\ref{eq:lotenemos}) will be smaller than the second one, proving that 
$t_{\textrm{lower}} < t_0$. For the case $t_0 > t_1$, (\ref{eq:cadena_mola})
can be applied.

%%%%%%%%%%%%%%%%%%%%%%%%%%%%%%%%%%%%%%%%%%%%%%%%%%%%%%%%%%%%%%%%%%%%%
%%%%%%%%%%%%%%%%%%%%%%%%%%%%%%%%%%%%%%%%%%%%%%%%%%%%%%%%%%%%%%%%%%%%%

\subsection{Partially probabilistic approach}\label{sec:another_possible}

A close look at (\ref{eq:SecEstimator}) allows us to see that the evaluation
of both its
second and third terms for different values of $t$ is computationally cheap.
Indeed, if we denote by $M_t$ the number of values of $t$ where
the target function has to be evaluated, the associated computational
cost will be $O(n + M_t)$.
Nevertheless, this is not the case for the first term, where the computational 
cost is linearly increased with the product of the dimensionality of the
problem and $M_t$, i.e., $O(n \cdot M_t)$.

Taking this into account, it could make sense to characterize the target
function by considering the computationally cheap second and third terms to be
deterministic (taking advantage of the knowledge of $||\bm z||^2$
available at the decoder), and the computationally expensive first term to be random, avoiding in this way the evaluation of the latter. Although the knowledge
of $\bm z$ at the decoder would allow to compute the exact value of the entire target
function, this methodology statistically characterizes the output of the target
function without explicitly computing its expensive first term, therefore
providing a valuable tool for analyzing the target function
output.

From a statistical point of view, one can justify
the approach presented in this section
by the asymptotic (when the number of
observations goes to infinity and HLR $\rightarrow \infty$)
independence between $||\bm Z||^2$,
and $||(\bm Z - t \bm d)\moda(t\Lambda)||^2$; the approximate independence
between these two terms has been already used in
Sect.~\ref{sec:target_function} for deriving $f_{\bm Z|T, \bm K}(\bm z|t, \bm d)$.

Formally, we will characterize $L(t, \bm Z)$ as
\begin{eqnarray}
  L(t, \bm Z ) = L_2(t, \bm z) + L_3(t, \bm Z), \nonumber
\end{eqnarray}
where
\begin{eqnarray}
  L_3(t, \bm Z) = \frac{||(\bm Z - t \bm d) \moda (t\Lambda)||^2}{\sigma_N^2 + (1-\alpha)^2t^2 \sigma_\Lambda^2}. \nonumber
\end{eqnarray}
Following this approach, it is straightforward to see that
whenever TNLR $<1$, the
pdf of $L_3(t, \bm Z)$ for $t= t_0$ would converge to $f_{\chi^2_{n}}(x)$ when
$n\rightarrow \infty$ and good lattices are used, and will approximately
follow
that distribution when the last conditions do not hold (i.e., the low-dimensional lattice case), but
SCR $\rightarrow 0$
is verified;
therefore,
\begin{eqnarray}
  \textrm{Pr} \left ( L(t, \bm Z) < L(t_1, \bm z) | t = t_0\right ) = 
\int_{-\infty}^{L(t_1, \bm z)} f_{\chi^2_{n}} \left (x - n \log \Big [ 2 \pi \left ( \sigma_N^2 + (1-\alpha)^2 
  t^2 \sigma_\Lambda^2 \right ) \Big ] - \frac{||\bm z||^2}{\sigma_X^2 t^2}\right ) dx, \label{eq:compt_prob2}
\end{eqnarray}
will not be monotonically decreasing with $t$, but will be null at
$t = 0$ and $t\rightarrow \infty$, and will have only one local maximum.
This behavior is inherited from 
$L_2(t, \bm z)$, defined at (\ref{eq:def_L2}) and contained in 
the argument of $f_{\chi^2_n}$ in (\ref{eq:compt_prob2}).
As it was proved in Sect.~\ref{sec:lower_deter},
the derivative of (\ref{eq:def_L2}) with
respect to $t$ will be null for just a positive value of $t$, denoted
by $t_2$.
From this result, we can see that no only
a lower-bound to the search interval can be found, calculated as
the largest $t\leq t_2$ such that the probability of
$L(t, \bm Z)$ being smaller than $L(t_1, \bm z)$ for $t=t_0$ is smaller
than or equal to $P_{e1}$, i.e.,
\begin{eqnarray}
  t_{\textrm{lower}} = \argmax_{t: t\leq t_2, \textrm{Pr} \left ( L(t, \bm Z) < L(t_1, \bm z) | t = t_0\right ) \leq P_{e1}} t , \nonumber
\end{eqnarray}
but also an upper-bound to the mentioned interval, defined as
\begin{eqnarray}
  t_{\textrm{upper}} = \argmin_{t: t\geq t_2, \textrm{Pr} \left ( L(t, \bm Z) < L(t_1, \bm z) | t = t_0\right ) \leq P_{e1}} t , \nonumber
\end{eqnarray}
i.e., the smallest $t\geq t_2$, such that the probability of
$L(t, \bm Z)$ being smaller than $L(t_1, \bm z)$ for $t=t_0$ is smaller
than or equal to $P_{e1}$.

As the analysis introduced in this section for the
lower-bound is strongly related to the corresponding one for the upper-bound,
we will jointly study both of them here, and Sect.~\ref{sec:another_possible2} will just
refer to the current section.

Whenever $L(t_1, \bm z) - L_2(t_2, \bm z) \leq F_{\chi^2_n}^{-1}(P_{e1})$,
both the upper and lower-bound derived above will coincide and will be equal
to $t_2$, i.e., $t_2$ is good enough for verifying the defined
constraint, so the subsequent optimization using the 
search interval is not longer required. Indeed, in that case other
points besides $t_2$ might be feasible solutions to both the
upper and lower-bounds definition problem; denoting those points by
$t_3$, they are defined by $L_2(t_3, \bm z) \geq L(t_1, \bm z) - F_{\chi^2_n}^{-1}(P_{e1})$.
In the extreme case where $P_{e1} \rightarrow 0$, the condition
for the upper and lower-bound coinciding can be rewritten
as 
\begin{eqnarray}
L_2(t_2, \bm z) \geq L(t_1, \bm z) ; \label{eq:const_55}
\end{eqnarray}
given that $L(t_1, \bm z) \geq L_2(t_1, \bm z)$, with equality only
if $||(\bm z - t\bm d )\moda (t \Lambda)||^2 =0$, and that $L_2(t_1, \bm z) 
\geq L_2(t_2, \bm z)$, with equality only if $t_1 = t_2$,
(\ref{eq:const_55}) will hold only when these two
conditions are simultaneously verified.

Otherwise, the upper and lower-bound  correspond to the two only solutions for $t>0$ to the equation
\begin{eqnarray}
 n \log \Big [ 2 \pi \left ( \sigma_N^2 + (1-\alpha)^2 
  t^2 \sigma_\Lambda^2 \right ) \Big ] + \frac{||\bm z||^2}{\sigma_X^2 t^2} +   F_{\chi^2_{n}}^{-1}(P_{e1}) = L(t_1, \bm z) . \label{eq:sec_acomer}
\end{eqnarray}
It is worth pointing out the similarities between the last formula
and (\ref{eq:sol_t4}); the only difference is the presence
of $F_{\chi^2_{n}}^{-1}(P_{e1})$ in (\ref{eq:sec_acomer}), reflecting
the relaxed nature  of the
lower-bound derived in this section (due to its probabilistic approach),
in comparison with the deterministic
strategy proposed in Sect.~\ref{sec:lower_deter}. Indeed, both formulas will
be exactly the same whenever the probabilistic nature is removed
from  (\ref{eq:sec_acomer}) by setting $P_{e1} = 0$.

To the best of the authors' knowledge closed formulas for the solutions to the previous equation
are not available, although due to the discussed properties of
$L_2(t, \bm z)$, the numerical search of those solutions can be performed
even using a dichotomy search algorithm.

Similarly to the analysis performed in Sect.~\ref{sec:target_function_t_0}, for 
values of $n$ large enough one could use the CLT approximation
for the pdf of $L_3(t, \bm Z)$, obtaining
a Gaussian distribution with mean 
\begin{eqnarray}
  \textrm{E}[L_3(t, \bm Z)|t = t_0] = n , \nonumber
\end{eqnarray}
and variance for the scalar quantizer case
\begin{eqnarray}
\textrm{Var}[L_3(t, \bm Z)|t = t_0] &=& n \left [\frac{\frac{144(1-\alpha)^4t^4 \sigma_\Lambda^4}{180} 
+2\sigma_N^4 + 4(1-\alpha)^2t^2 \sigma_\Lambda^2\sigma_N^2}{\left (\sigma_N^2 + 
(1-\alpha)^2  t^2 \sigma_\Lambda^2 \right )^2} \right ],
 \nonumber
\end{eqnarray}
while the variance for the high-dimensional good lattice quantizer case is
\begin{eqnarray}
\textrm{Var}[L_3(t, \bm Z)|t = t_0] &=& 2n; \nonumber
\end{eqnarray}
replacing these mean and variance by the asymptotic values achieved
by considering SCR $\rightarrow 0$ in the scalar quantizer case and taking the exact
values when the high-dimensional good lattice quantizer case is considered,
(\ref{eq:sec_acomer}) can be approximated by
\begin{eqnarray}
 n \log \Big [ 2 \pi \left ( \sigma_N^2 + (1-\alpha)^2 
  t^2 \sigma_\Lambda^2 \right ) \Big ] + \frac{||\bm z||^2}{\sigma_X^2 t^2} + n - \sqrt{2n}  Q^{-1}(P_{e1}) = L(t_1, \bm z). \nonumber
\end{eqnarray}

\begin{figure}[t] 
  \begin{center}
    \includegraphics[width=0.75\linewidth]{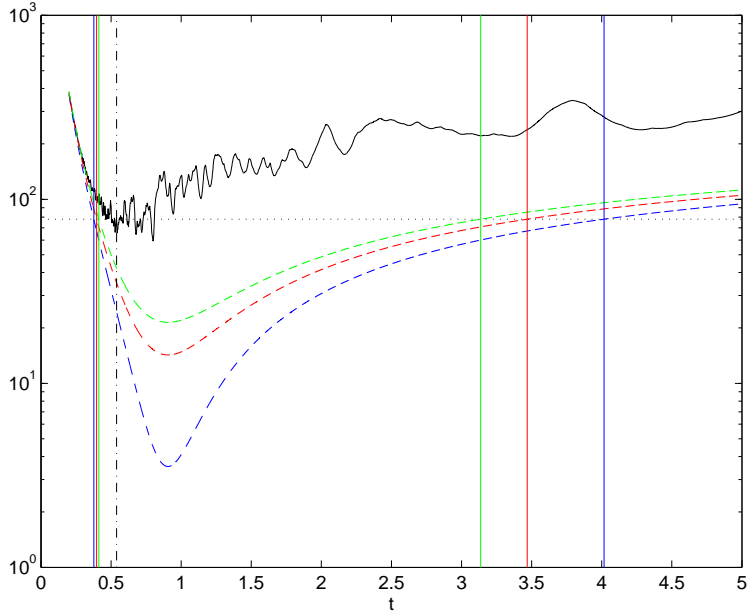}  
    \caption{Comparison between $L(t, \bm z)$ (solid black line),
      $L(t_1, \bm z)$ (dotted black line),
$L_2(t, \bm z)$ (dashed blue line), and the left term of (\ref{eq:sec_acomer})
for $P_{e1} = 10^{-6}$ (dashed red line), and $P_{e1} = 10^{-3}$ (dashed green line). The corresponding search interval bounds are plotted by using
vertical solid lines with the same color that the function used
for computing them. $t_1$ (vertical dash-dot black line) calculated by using (\ref{eq:approx_min_L1}).
HLR $\approx 34.9765 $ dB, SCR $\approx -1.0618$ dB, TNLR $\approx -3.5736$ dB, $t_0 = 0.8$, $\alpha = \alpha_{\textrm{Costa}} \approx 0.5608$, $n = 40$, scalar
quantizer.}\label{fig:bounds1}
  \end{center}
\end{figure}

In Fig.~\ref{fig:bounds1} we can see how the bounds derived by using
the approach proposed in this section are a relaxed version of those
obtained by using the deterministic method
proposed in Sects.~\ref{sec:lower_deter} and 
\ref{sec:upper_deter}. Specifically, the larger $P_{e1}$, the smaller
the obtained search interval.

\subsubsection{Asymptotic values when the number of observations goes to infinity}\label{sec:asym_large2}

Similarly to the previous section, here we will analyze the asymptotic
behavior of the search interval bounds when the number of considered
observations goes to infinity. In order to do that, we will consider
the asymptotic values of $L(t_1, \bm z)$ when $n\rightarrow \infty$,
both for $t_1 \approx t_0$, and
when $t_1$ is not  close to $t_0$ (both situations are quantitatively
described in Sect.~\ref{sec:L_assympt}), given by (\ref{eq:L_mean_close}) and 
(\ref{eq:L_mean_far}), respectively.

Note that in this case (\ref{eq:sec_acomer}) is still valid, but
one must consider that  $F_{\chi_n^2}^{-1}(P_{e1})$ will go $0$ if
$P_{e1} = 0$, and will go to $n$ if $P_{e1} > 0$ (assuming $P_{e1}$ to be
independent of $n$). 

If $P_{e1} = 0$,
we have exactly the deterministic version of the problem already studied in
Sect.~\ref{sec:lower_deter_asympt} (the upper-bound will
be studied in Sect.~\ref{sec:upper_deter_asympt}).

For $P_{e1} > 0$, we will separately study
the cases $t_1 \approx t_0$ and $t_1$ not close to $t_0$.

\begin{itemize}
 \item $t_1 \approx t_0$
\end{itemize}

Whenever $t_1 \approx t_0$ the upper and lower-bounds to the search
interval will be the solutions to 
\begin{eqnarray}
  \frac{(t_0 - t_1)^2\sigma_X^2}{\sigma_N^2 + (1-\alpha)^2 t_1^2 \sigma_\Lambda^2} + 1- \frac{t_0^2}{t^2} + \log \left (\frac{\sigma_N^2 + (1-\alpha)^2  t_1^2 \sigma_\Lambda^2}{\sigma_N^2 + (1-\alpha)^2  t^2 \sigma_\Lambda^2} \right ) = 0. \label{eq:equation_another}
\end{eqnarray}
Although a closed solution to the last equation does not exist,
one wonders
if $t_0$ will be always contained in the interval defined by the solutions
to that equation. In order to solve this question we will take
into account that the left side of (\ref{eq:equation_another}) goes to
$-\infty$ both when $t\rightarrow 0$ and $t\rightarrow \infty$; therefore,
a necessary and sufficient condition
for $t_0$ to be included in the resulting search interval is that the
result of evaluating 
(\ref{eq:equation_another}) at $t = t_0$, i.e.,
\begin{eqnarray}
  \frac{(t_0 - t_1)^2\sigma_X^2}{\sigma_N^2 + (1-\alpha)^2 t_1^2 \sigma_\Lambda^2}  + \log \left (\frac{\sigma_N^2 + (1-\alpha)^2  t_1^2 \sigma_\Lambda^2}{\sigma_N^2 + (1-\alpha)^2  t_0^2 \sigma_\Lambda^2} \right ) , \label{eq:asssdf}
\end{eqnarray}
should be non-negative for the considered $t_1$.
This is obviously the case for $t_1 \geq t_0$, confirming that in this scenario
$t_0$ is contained in the search interval; in particular, for $t_1 = t_0$
(\ref{eq:asssdf}) is null.

In order to study the case $t_1 < t_0$ we will take
into account that the derivative of (\ref{eq:asssdf}) 
with respect to $t_1$ is given by
\begin{eqnarray}
  \frac{2\left [-\sigma_N^2 \sigma_X^2 t_0 + \sigma_N^2\left ( \left (1-\alpha\right )^2 \sigma_\Lambda^2 + \sigma_X^2 \right)t_1 + (1-\alpha)^2\sigma_\Lambda^2 t_1
    \left (\left (1-\alpha\right)^2\sigma_\Lambda^2 t_1^2 + \sigma_X^2 t_0 (t_1 - t_0 ) \right) \right ]}{\left [ \sigma_N^2 + \left (1 - \alpha \right )^2 t_1^2\sigma_\Lambda^2  \right ]^2}. \nonumber
\end{eqnarray}
The numerator of the last formula can be written like $a_0 + a_1 t_1 + a_2 t_1^2 + a_3 t_1^3$, where
\begin{eqnarray}
  a_0 &=& -2\sigma_N^2 \sigma_X^2 t_0, \nonumber \\
  a_1 &=& 2 \left [\sigma_N^2 \sigma_X^2 + (1-\alpha)^2\sigma_\Lambda^2 \left (\sigma_N^2 - \sigma_X^2 t_0^2 \right ) \right ], \nonumber \\
  a_2 &=& 2(1-\alpha)^2 t_0 \sigma_\Lambda^2 \sigma_X^2, \nonumber \\
  a_3 &=& 2 (1-\alpha)^4 \sigma_\Lambda^4, \nonumber
\end{eqnarray}
proving, using Descartes' sign rule, that the derivative of (\ref{eq:asssdf})
will be null for at most a positive $t_1$. Additionally if that
derivative is evaluated at $t_1 = t_0$, the result is
\begin{eqnarray}
  \frac{2(1-\alpha)^2 \sigma_\Lambda^2 t_0}{\sigma_N^2 + (1-\alpha)^2 t_0^2
    \sigma_\Lambda^2},\nonumber
\end{eqnarray}
proving that there exists an interval, open in its upper-bound $t_0$, such that (\ref{eq:asssdf}) is negative for the values
of $t_1$ contained in it. Consequently, the question to be answered
is how large that interval is.

If one evaluates
(\ref{eq:asssdf}) at 
\begin{eqnarray}
  t_1 = t_0 - \sqrt{\frac{\sigma_N^2 + (1-\alpha)^2 t_0^2 \sigma_\Lambda^2}{\sigma_X^2}}, \label{eq:t1_asymp}
\end{eqnarray}
the result
can be written like $\log(x_1) + \frac{1}{x_1}$, where
\begin{eqnarray}
x_1 = \frac{\sigma_N^2 + (1-\alpha)^2\sigma_\Lambda^2 \left (t_0 - \sqrt{\frac{\sigma_N^2 + (1-\alpha)^2t_0^2\sigma_\Lambda^2 }{\sigma_X^2}} \right )^2}{\sigma_N^2 + (1-\alpha)^2t_0^2\sigma_\Lambda^2}. \nonumber
\end{eqnarray}
Thus, given that $x_1 \geq 0$, the result in App.~\ref{sec:app3} proves
that (\ref{eq:asssdf}) evaluated at the right hand of (\ref{eq:t1_asymp})
is non-negative. Furthermore, whenever HLR $\rightarrow \infty$,
the right term in (\ref{eq:t1_asymp}) asymptotically goes to $t_0$, proving that
whenever $\textrm{HLR} \rightarrow \infty$, the interval
where (\ref{eq:asssdf}) takes negative values is asymptotically
small.

Therefore, whenever $\textrm{HLR} \rightarrow \infty$, $t_0$ will be contained
in or will be asymptotically close to the search interval, irrespectively
of $t_1$.

Additionally, one can conclude that
for $t_1 = t_0$, (\ref{eq:equation_another}) will have a single solution
corresponding to $t=t_0$.

\begin{figure}[t] 
  \begin{center}
    \includegraphics[width=0.75\linewidth]{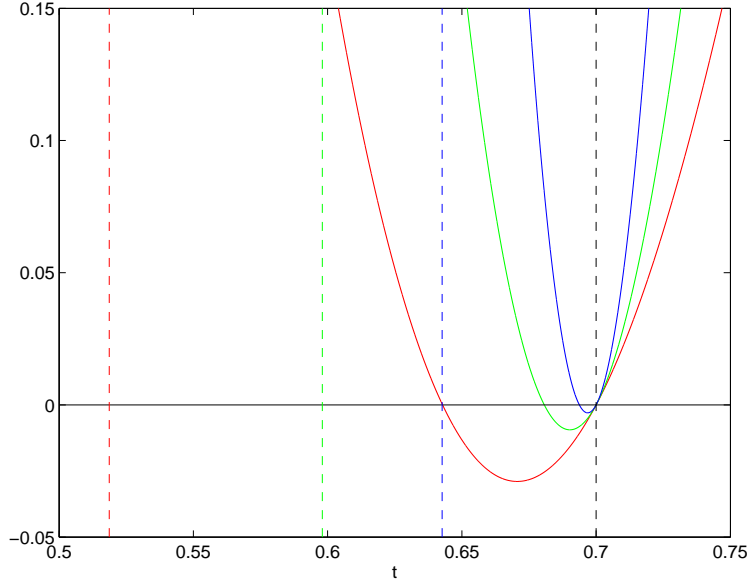}  
    \caption{Comparison  of (\ref{eq:asssdf}) for different values of 
      HLR (red $10$ dB, green $15$ dB, blue $20$ dB). SCR $\approx 3.0980$ dB, TNLR $\approx -1.7319$ dB, $t_0 = 0.7$, 
      $\alpha = \alpha_{\textrm{Costa}} \approx 0.3289$. The vertical colored dashed
    lines stand for the values of $t_1$ derived by using
    (\ref{eq:t1_asymp}), illustrating that they converge to $t_0$ (vertical black
dashed line) as HLR $\rightarrow \infty$.}\label{fig:aprox_semiprob_close}
  \end{center}
\end{figure}

Fig.~\ref{fig:aprox_semiprob_close} illustrates the conclusions derived
for the case $t_1\approx t_0$.

\begin{itemize}
\item $t_1$ not close to $t_0$
\end{itemize}

On the other hand, for the case where $t_1$ is not close to $t_0$, one will look
for two solutions to
\begin{eqnarray}
  \frac{t_0^2}{t_1^2} - \frac{t_0^2}{t^2} -1 + \frac{ t_1^2 \sigma_\Lambda^2}{\sigma_N^2 + (1-\alpha)^2 t_1^2 \sigma_\Lambda^2} + \log \left (\frac{\sigma_N^2 + (1-\alpha)^2  t_1^2 \sigma_\Lambda^2}{\sigma_N^2 + (1-\alpha)^2  t^2 \sigma_\Lambda^2} \right ) = 0. \label{eq:eq_another2}
\end{eqnarray}
Similarly to the case $t_1 \approx t_0$, the left-hand side of the previous
equation will go to $-\infty$ as $t \rightarrow 0$ and $t\rightarrow \infty$.
Therefore, for $t_0$ to be contained in the search interval,
(\ref{eq:eq_another2}) evaluated at $t = t_0$, i.e.,
\begin{eqnarray}
  f(t_1, \sigma_\Lambda^2) \triangleq \frac{t_0^2}{t_1^2}  -2 + \frac{ t_1^2 \sigma_\Lambda^2}{\sigma_N^2 + (1-\alpha)^2 t_1^2 \sigma_\Lambda^2} + \log \left (\frac{\sigma_N^2 + (1-\alpha)^2 t_1^2 \sigma_\Lambda^2}{\sigma_N^2 + (1-\alpha)^2  t_0^2 \sigma_\Lambda^2} \right )\label{eq:eq_another3}
\end{eqnarray}
should be non-negative.

In order to prove that this is the case, we will find useful to define
\begin{eqnarray}
    \sigma_{\Lambda_0}^2 \triangleq \frac{\sigma_N^2}{t_0^2 \left [1 - (1-\alpha)^2 \right]}, \nonumber
\end{eqnarray}
where
\begin{eqnarray}
  t_0^2 \sigma_{\Lambda_0}^2 = \sigma_N^2 + (1-\alpha)^2 t_0^2 \sigma_{\Lambda_0}^2, \nonumber
\end{eqnarray}
i.e., the threshold of the {\bf relaxed version of Hypothesis 3} (TNLR $=1$)
is achieved.

The proof is splitted in the next steps: 
\begin{enumerate}
\item Prove that
\begin{eqnarray}
  \frac{\partial f(t_1, \sigma_{\Lambda_0}^2)}{\partial t_1} \left \{
  \begin{array}{ll}
    = 0, & \textrm{  if  } t_1 = t_0\\
    \neq 0,& \textrm{  otherwise} 
  \end{array} \right . ,\nonumber
\end{eqnarray}
and
\begin{eqnarray}
  f(t_0, \sigma_{\Lambda_0}^2) &=& 0, \nonumber \\
  f(t_1, \sigma_{\Lambda_0}^2) &>& 0, t_1 \neq t_0. \nonumber 
\end{eqnarray}
\end{enumerate}

{\bf Proof}

\begin{eqnarray}
  f(t_1, \sigma_{\Lambda_0}^2) = \frac{t_0^2}{t_1^2}  -2 + \frac{ t_1^2 }{t_1^2 + \alpha(\alpha-2)(t_1^2 - t_0^2)} + \log \left (\alpha(2-\alpha) + \frac{(1-\alpha)^2 t_1^2}{t_0^2} \right ),
\label{eq:core_asymp}
\end{eqnarray}
which is obviously null at $t_1 = t_0$. Additionally, the derivative
of the last formula with respect to $t_1$ is
\begin{eqnarray}
  \frac{ -2  (\alpha-2)^2\alpha^2 t_0^6 - 4(2-\alpha)(1-\alpha)^2\alpha t_0^4
t_1^2 - 2 \left [1-2 (2-\alpha )\alpha \left (2 - (2-\alpha)\alpha \right ) \right ]t_0^2t_1^4   + 2(1-\alpha)^4 t_1^6 }{t_1^3 \left (t_1^2 - (2-\alpha)\alpha \left (t_1^2 - t_0^2 \right ) \right )^2}, \label{eq:num_der_aa}
\end{eqnarray} 
whose numerator can be written like $a_0 + a_1 t_1^2 + a_2 t_1^4 + a_3 t_1^6$, where
\begin{eqnarray}
  a_0 &= & -2  (\alpha-2)^2\alpha^2 t_0^6, \nonumber \\
  a_1 &=& - 4(2-\alpha)(1-\alpha)^2\alpha t_0^4, \nonumber  \\
  a_2 &=& - 2 \left [1-2 (2-\alpha )\alpha \left (2 - (2-\alpha)\alpha \right ) \right ]t_0^2, \textrm {  and  }\nonumber \\
  a_3 &=& 2(1-\alpha)^4. \nonumber
\end{eqnarray}
Using Descartes' sign rule, it is evident that (\ref{eq:num_der_aa})
will be null just for at most a positive $t_1$, which can be shown to be $t_0$
just by replacing $t_1$ by $t_0$ in (\ref{eq:num_der_aa}). Given
that (\ref{eq:core_asymp}) goes to infinity when $t_1 \rightarrow 0$ and 
$t_1 \rightarrow \infty$,
Step $1$ is proved.

\begin{figure}[t] 
  \begin{center}
    \includegraphics[width=0.75\linewidth]{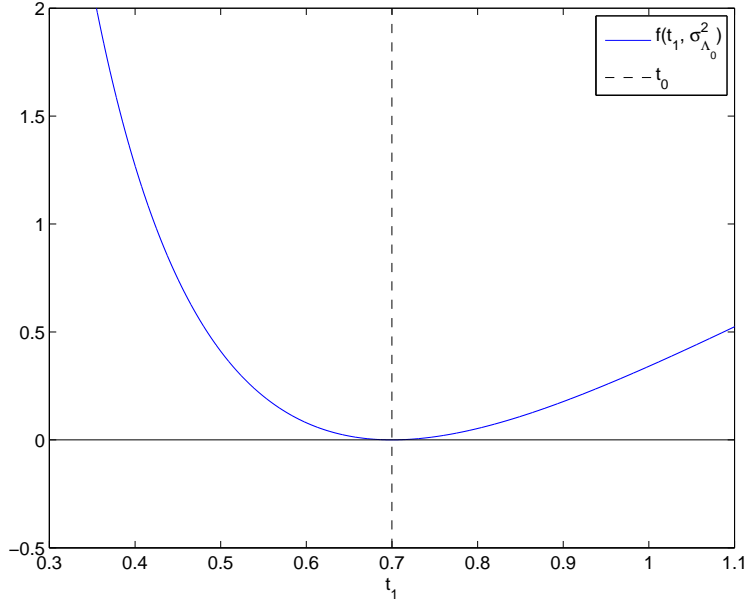}  
    \caption{$f(t_1, \sigma_{\Lambda_0}^2)$ as a function of $t_1$. $\sigma_N^2 = 1$, $\alpha = 0.5$, $t_0 = 0.7$.}\label{fig:demo_tocha_step1}
  \end{center}
\end{figure}

Fig.~\ref{fig:demo_tocha_step1} illustrates Step 1.

\begin{enumerate}[start=2]
\item Let us define
\begin{eqnarray}
  \xi_1 \triangleq \sqrt{\frac{(1-\alpha)^2t_0^2}{1 + (1-\alpha)^2}}. \nonumber
\end{eqnarray}
For each $t_1 \leq \xi_1$, there exists a single $\sigma_{t_1}^2 \geq 0$ such that
\begin{eqnarray}
  \left . \frac{\partial f(t_1, \sigma_\Lambda^2)}{\partial \sigma_\Lambda^2}\right |_{(t_1, \sigma_{t_1}^2)} = 0. \nonumber
\end{eqnarray}
Additionally, in that case
\begin{eqnarray}
  \left . \frac{\partial^2 f(t_1, \sigma_\Lambda^2)}{(\partial \sigma_\Lambda^2)^2}\right |_{(t_1, \sigma_{t_1}^2)} > 0. \nonumber
\end{eqnarray}

If $t_1 > \xi_1$, then
\begin{eqnarray}
  \left . \frac{\partial f(t_1, \sigma_\Lambda^2)}{\partial \sigma_\Lambda^2}\right |_{(t_1, \sigma^2)} > 0, \textrm{  for any  } \sigma^2 \geq 0. \nonumber
\end{eqnarray}
\end{enumerate}

{\bf Proof}
\begin{eqnarray}
  \frac{\partial f(t_1, \sigma_\Lambda^2)}{\partial \sigma_\Lambda^2} &= &
  \sigma_N^2 \left [-(1-\alpha)^2\sigma_N^2t_0^2 + \Big (1 + (1-\alpha)^2
\Big )\sigma_N^2 t_1^2 + (1-\alpha)^2\sigma_\Lambda^2 t_1^2 \Big ( \left(2-\alpha\right) \alpha \left (t_0^2 - t_1^2\right) + t_1^2 \Big ) \right ]
\nonumber \\
&&\left [ \Big (\sigma_N^2 + (1-\alpha)^2t_0^2 \sigma_\Lambda^2 \Big )
\Big (\sigma_N^2 + (1-\alpha)^2t_1^2 \sigma_\Lambda^2 \Big )^2 \right ]^{-1}.
\nonumber
\end{eqnarray}
It is easy to see that the last formula nullifies just when
\begin{eqnarray}
  \sigma_\Lambda^2 = \sigma_{t_1}^2 \triangleq \frac{\sigma_N^2 \left [(1-\alpha)^2t_0^2  - \left(1 + (1-\alpha)^2 \right )t_1^2 \right ]}{(1-\alpha)^2 t_1^2 \left ( (2-\alpha)\alpha(t_0^2 - t_1^2) + t_1^2
    \right )}, \nonumber
\end{eqnarray}
which is non-negative (therefore providing a valid variance value)
if and only if $t_1 \leq \xi_1$ (indeed, $\sigma_{\xi_1}^2 = 0$); in any case, 
\begin{eqnarray}
  \left . \frac{\partial^2 f(t_1, \sigma_\Lambda^2)}{\left ( \partial \sigma_\Lambda^2 \right )^2}\right |_{(t_1, \sigma_{t_1}^2)} =  \frac{t_1^4 \left (2\alpha(t_0^2- t_1^2) + t_1^2
     + \alpha^2(t_1^2 - t_0^2) \right )^4}{\sigma_N^4 (t_0^2 - t_1^2)^4}, \nonumber
\end{eqnarray}
proving that for each $t_1 \leq \xi_1$, the curve $f(t_1, \sigma_\Lambda^2)$
has a minimum at $\sigma_\Lambda^2 = \sigma_{t_1}^2 \geq 0$, and that
for $t_1 > \xi_1$, 
$\left . \frac{\partial f(t_1, \sigma_\Lambda^2)}{\partial \sigma_\Lambda^2}\right |_{ (t_1, \sigma^2)} > 0$,   for any  $\sigma^2 \geq 0$.

\begin{figure}[t] 
  \begin{center}
    \includegraphics[width=0.75\linewidth]{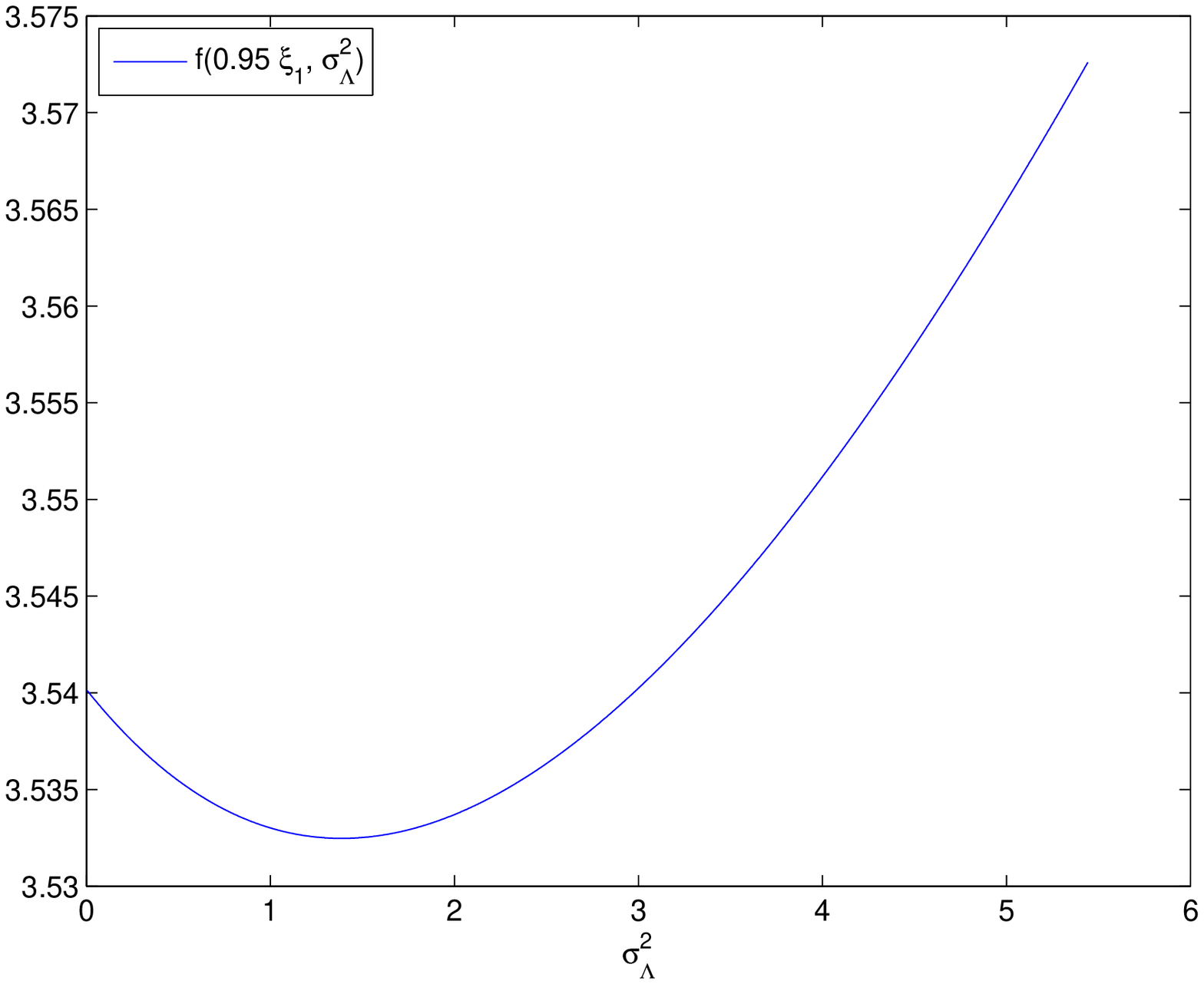}  
    \caption{$f(t_1, \sigma_{\Lambda}^2)$ as a function of $\sigma_{\Lambda}^2$. $\sigma_N^2 = 1$, $\alpha = 0.5$, $t_0 = 0.7$, $t_1 = 0.95\xi_1$.}\label{fig:demo_tocha_step21}
  \end{center}
\end{figure}

\begin{figure}[t] 
  \begin{center}
    \includegraphics[width=0.75\linewidth]{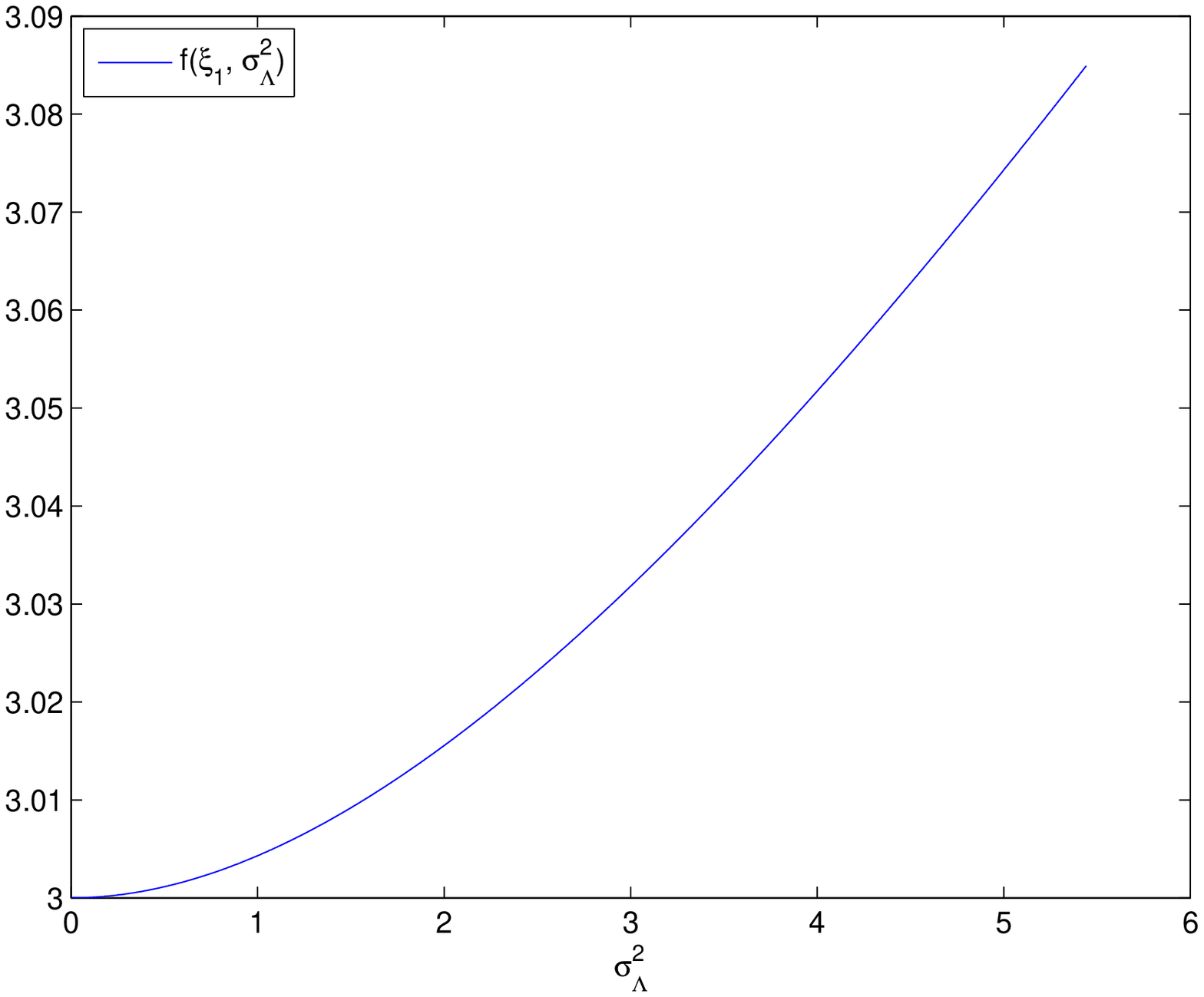}  
    \caption{$f(t_1, \sigma_{\Lambda}^2)$ as a function of $\sigma_{\Lambda}^2$. $\sigma_N^2 = 1$, $\alpha = 0.5$, $t_0 = 0.7$, $t_1 = \xi_1$.}\label{fig:demo_tocha_step22}
  \end{center}
\end{figure}

\begin{figure}[t] 
  \begin{center}
    \includegraphics[width=0.75\linewidth]{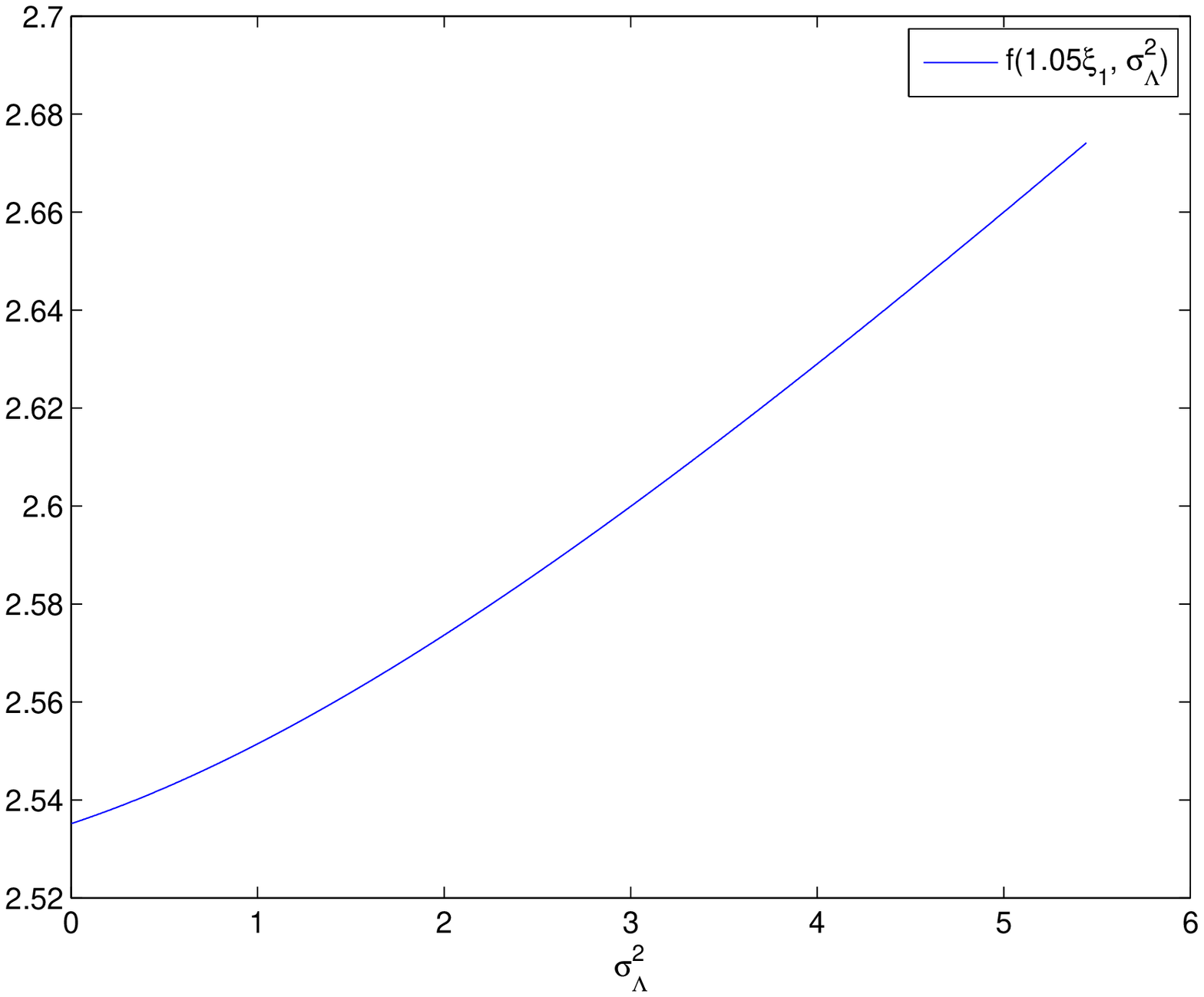}  
    \caption{$f(t_1, \sigma_{\Lambda}^2)$ as a function of $\sigma_{\Lambda}^2$. $\sigma_N^2 = 1$, $\alpha = 0.5$, $t_0 = 0.7$, $t_1 = 1.05\xi_1$.}\label{fig:demo_tocha_step23}
  \end{center}
\end{figure}

Figs.~\ref{fig:demo_tocha_step21}, \ref{fig:demo_tocha_step22},
and \ref{fig:demo_tocha_step23} illustrate Step 2 for the cases
$t_1 < \xi_1$, $t_1 = \xi_1$ and $t_1 > \xi_1$, respectively.

\begin{enumerate}[start=3]
\item Prove that 
\begin{eqnarray}
 \frac{\partial \left (\sigma_{t_1}^2 \right )}{\partial t_1} \leq 0, \textrm{    } \forall t_1 \in [0, t_0]; \nonumber
\end{eqnarray}
additionally $\sigma_{t_1}^2|_{t_1 = 0} = \infty$, and
$\sigma_{t_1}^2|_{t_1 = \xi_1} = 0$.
\end{enumerate}

{\bf Proof}
\begin{eqnarray}
 \frac{\partial \left (\sigma_{t_1}^2 \right )}{\partial \left (t_1^2 \right )}
 = \frac{\sigma_N^2\left [\left (-2 + \alpha \right )\alpha\left ( t_0^2 - t_1^2 \right ) - 2 t_1^2 \right ] \left (t_0^2 - t_1^2 \right)}{t_1^4 \left  (
   (2-\alpha)\alpha (t_0^2 - t_1^2) + t_1^2 \right)^2}, \label{eq:der_sigmat1}
\end{eqnarray}
has $4$ roots on $t_1$, located at $\pm t_0$ and
\begin{eqnarray}
  \pm \sqrt{\frac{-(2-\alpha)\alpha t_0^2}{1 + (1-\alpha)^2}}. \nonumber
\end{eqnarray}
Given that the last two are imaginary, the only real non-negative root
is that located at $t_0$. Therefore, (\ref{eq:der_sigmat1}) will
have the same sign for any $t_1 \in [0, t_0]$. Given that
\begin{eqnarray}
\lim_{t_1 \rightarrow 0}   \frac{\partial \left (\sigma_{t_1}^2 \right )}{\partial \left (t_1^2 \right )} = -\infty, \nonumber
\end{eqnarray}
it is straightforward that
\begin{eqnarray}
  \frac{\partial \left (\sigma_{t_1}^2 \right )}{\partial \left (t_1^2 \right )} \leq 0, \; \; \forall t_1 \in [0, t_0], \nonumber
\end{eqnarray}
so
\begin{eqnarray}
  \frac{\partial \left (\sigma_{t_1}^2 \right )}{\partial t_1} \leq 0, \; \; \forall t_1 \in [0, t_0].\nonumber
\end{eqnarray}

\begin{enumerate}[start=4]
\item Due to Step 3, it is possible to uniquely define $\xi_2 \triangleq \arg_{t_1: \sigma_{t_1}^2 = 
\sigma_{\Lambda_0}^2}$. 
\item Prove that $\xi_2 \leq \xi_1 \leq t_0$.
\end{enumerate}

{\bf Proof}

At the sight of the definition of $\xi_1$ it is obvious that  $\xi_1 \leq t_0$.
On the other hand, based on the monotonically decreasing nature of 
$\sigma_{t_1}^2$ with $t_1$ proved at Step 3, and the fact that
$\sigma_{\xi_1}^2 = 0$ and $\sigma_{\xi_2}^2 = \sigma_{\Lambda_0}^2$, it
is straightforward that $\xi_2 \leq \xi_1$.

\begin{figure}[t] 
  \begin{center}
    \includegraphics[width=0.75\linewidth]{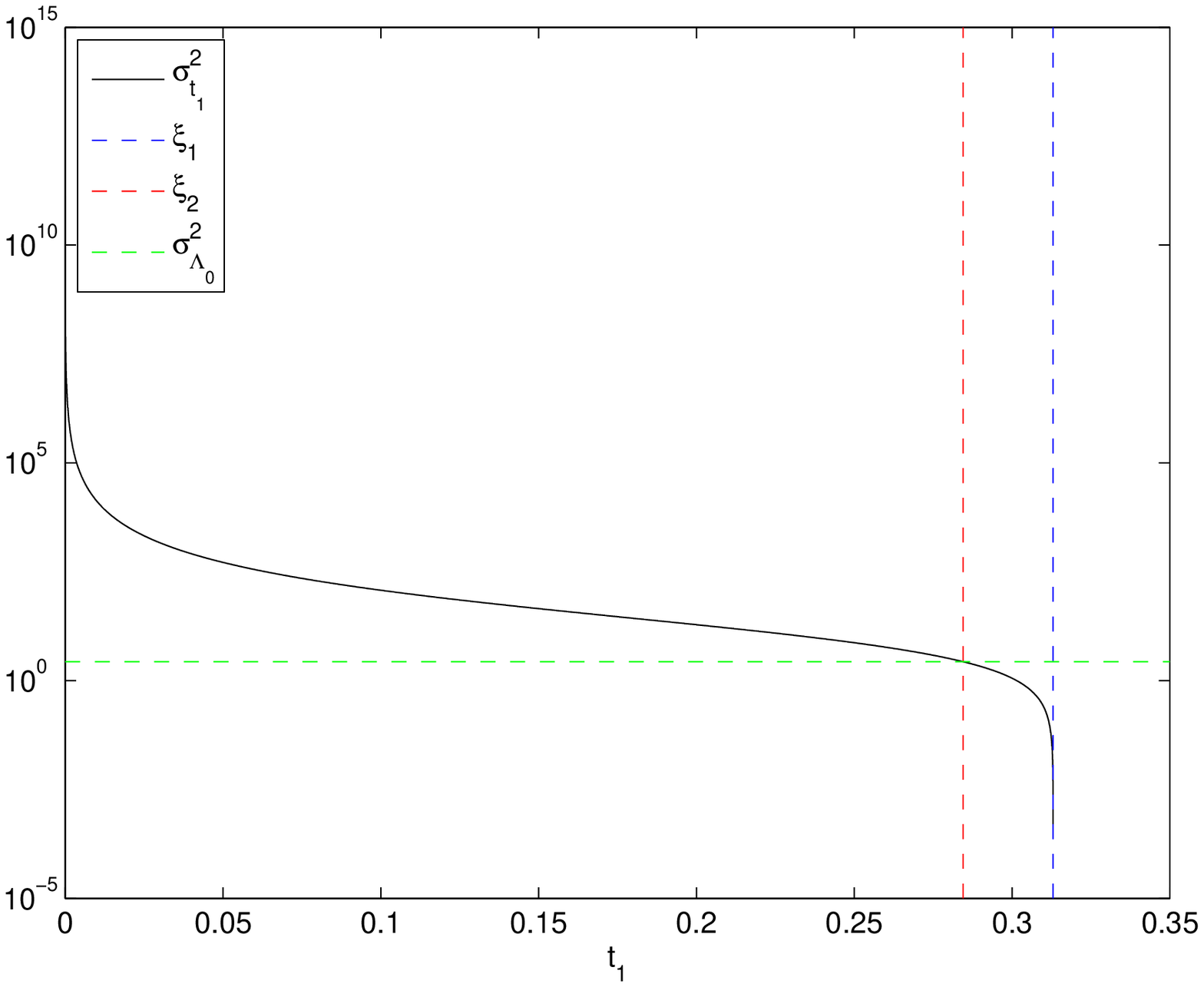}  
    \caption{$\sigma_{t_1}^2$ as a function of $t_1$. $\sigma_N^2 = 1$, $\alpha = 0.5$, $t_0 = 0.7$. Furthermore, one can check that $\sigma_{t_1}^2|_{t_1=0} = \infty$, $\sigma_{t_1}^2|_{t_1=\xi_1} = 0$, and $\sigma_{t_1}^2|_{t_1 = \xi_2} = \sigma_{\Lambda_0}^2$.}\label{fig:steps345}
  \end{center}
\end{figure}

Steps 3, 4, and 5 are illustrated in Fig.~\ref{fig:steps345}.

\begin{enumerate}[start=6]
\item Prove that  if $t_1 \geq \xi_2$, then
\begin{eqnarray}
  \left . \frac{\partial f(t_1, \sigma_\Lambda^2)}{\partial \sigma_\Lambda^2}\right |_{(t_1, \sigma^2)} \geq 0, \textrm{  for any  } \sigma^2 \geq \sigma_{\Lambda_0}^2. \nonumber
\end{eqnarray}
\end{enumerate}

{\bf Proof}

From Step 2, it is clear that the former statement is true for $t_1 > \xi_1$.
From Steps 2, 3 and 4, for $\xi_2 \leq t_1 \leq \xi_1$, $\sigma_{t_1}^2 \in
[0, \sigma_{\Lambda_0}^2]$; also from Step 2
\begin{eqnarray}
  \left . \frac{\partial^2 f(t_1, \sigma_\Lambda^2)}{(\partial \sigma_\Lambda^2)^2} \right |_{(t_1, \sigma_{t_1}^2)} > 0, \nonumber
\end{eqnarray}
which jointly with the fact that for a particular $t_1$ the derivative
nullifies just at $\sigma_{t_1}^2$ implies that for any
$t_1 \in [ \xi_2, \xi_1]$, then
\begin{eqnarray}
  \left . \frac{\partial f(t_1, \sigma_\Lambda^2)}{\partial \sigma_\Lambda^2}\right |_{(t_1, \sigma^2)} \geq 0, \textrm{  for any  } \sigma^2 \geq \sigma_{\Lambda_0}^2. \nonumber
\end{eqnarray}

\begin{enumerate}[start=7]
\item Prove that
\begin{eqnarray}
  f(t_1, \sigma_\Lambda^2) \geq 0, \; \; \; \forall t_1 \geq \xi_2, \forall \sigma_\Lambda^2 \geq \sigma_{\Lambda_0}^2. \nonumber
\end{eqnarray}
\end{enumerate}

{\bf Proof}

This result comes from the combination of Step 1, as $f(t_1, \sigma_{\Lambda_0}^2) \geq 0$, and Step 6, as for all $t_1 \geq \xi_2$
\begin{eqnarray}
\left .  \frac{\partial f(t_1, \sigma_\Lambda^2)}{\partial \sigma_\Lambda^2} \right |_{(t_1, \sigma^2)} \geq 0, \;\;\; \forall \sigma^2 \geq \sigma_{\Lambda_0}^2. \nonumber
\end{eqnarray}

\begin{enumerate}[start=8]
\item Prove that
\begin{eqnarray}
  f(t_1, \sigma_\Lambda^2) \geq 0, \; \; \; \forall t_1 \in [0, \xi_2], \forall \sigma_\Lambda^2 \geq 0. \nonumber
\end{eqnarray}
\end{enumerate}

{\bf Proof}

For all $t_1 \leq \xi_1$ we have that $\sigma_{t_1}^2 \geq 0$, i.e., it is
well defined as a variance. Therefore, in that scenario we can define
\begin{eqnarray}
  g(t_1) \triangleq f(t_1 ,\sigma_{t_1}^2), \nonumber
\end{eqnarray}
verifying that
\begin{eqnarray}
  g(t_1) = \min_{\sigma^2 \geq 0} f(t_1, \sigma^2). \nonumber
\end{eqnarray}
On the other hand, the derivative of $g(t_1)$ with respect to $t_1^2$ is
\begin{eqnarray}
  \frac{\partial g(t_1)}{\partial (t_1^2)} = \frac{-t_0^2}{(1-\alpha)^2(t_0^2 - t_1^2)^2} + \frac{1}{t_0^2 - t_1^2 } - \frac{t_0^2}{t_1^4} + \frac{1}{t_1^2}, \nonumber
\end{eqnarray}
which is null just at
\begin{eqnarray}
t_1 =   \pm \sqrt{\frac{-(1-\alpha) \left ( -3(1-\alpha) \pm \sqrt{(\alpha - 3)(1+\alpha)} \right) t_0^2}{6 -4(2-\alpha)\alpha}}. \nonumber
\end{eqnarray}
Given that the $4$ roots are complex, the derivative will take the same
sign in all $t_1 \geq 0$. In particular,
\begin{eqnarray}
\lim_{t_1 \rightarrow 0}  \frac{\partial g(t_1)}{\partial (t_1^2)} = -\infty, \nonumber
\end{eqnarray}
proving that $g(t_1)$ is monotonically decreasing with $t_1$ in $[0, \xi_1]$.
Finally, from the definition of $g(\cdot)$, the definition of $\xi_2$, and
Step 7, one obtains that
\begin{eqnarray}
  g(\xi_2) = f(\xi_2, \sigma_{\xi_2}^2) =
  f(\xi_2, \sigma_{\Lambda_0}^2) \geq 0. \nonumber
\end{eqnarray}

\begin{figure}[t] 
  \begin{center}
    \includegraphics[width=0.75\linewidth]{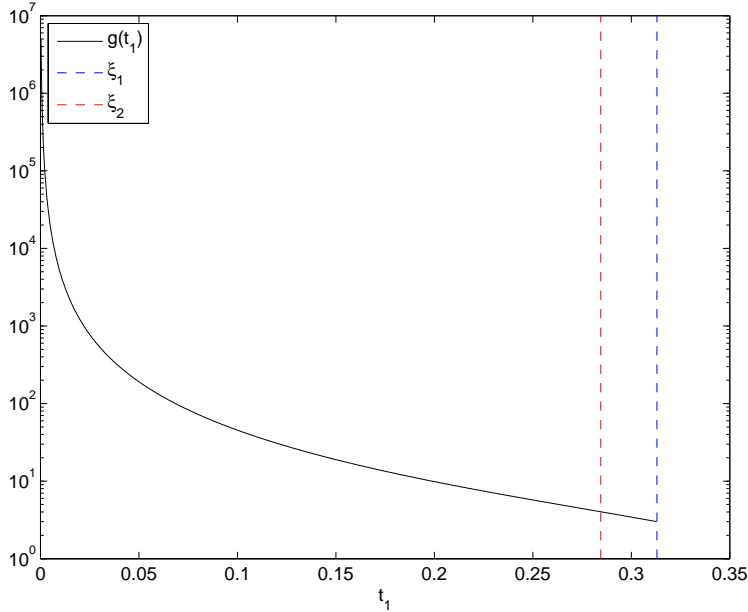}  
    \caption{$g(t_1)$. $\sigma_N^2 = 1$, $\alpha = 0.5$, $t_0 = 0.7$.}\label{fig:step8}
  \end{center}
\end{figure}

Step 8 is illustrated in Fig.~\ref{fig:step8}.

%%%%%%%%%%%%%%%%%%%%%%%%%%%%%%%%%%%%%%%%%%%%%%%%%%%%%%%%%%%%%%%%%%%%%
\subsection{Search interval lower-bound calculated
from the variance-based estimator}\label{eq:sec_lowerbound_var_based}

The previous strategies for defining the search interval lower-bound
are based on characteristics of the target function
$L(t, \bm z)$, so they are valid for any initial estimator $t_1$.
Nevertheless, a different approach for
defining the search interval could be based on the statistical characterization
of the variance-based
 estimator when this were used. In this section we will exploit some of the results derived in
Sect.~\ref{sec:variance_based_estimator} for obtaining a lower-bound
of the search interval. Given that throughout this report
we are assuming the scaling factor to be real, we will use the
estimators and pdfs derived in Sect.~\ref{sec:trunc_gauss_var_based_est};
additionally, as the probability of the upper tail of a Gaussian with mean and
variance those of $\hat{T}_{\textrm{var}}(\bm Z)$ is not monotonically increasing
with $t_0$,
$\hat{T^2}_{\textrm{var}}(\bm Z)$, proposed
in (\ref{eq:def_va_est2}) and whose  pdf was studied in
(\ref{eq:pdf_va_est2}),  will be used.

Taking these considerations into account, one choice for defining
$t_{\textrm{lower}}$ could be
\begin{eqnarray}
  t_{\textrm{lower}} &=& \sqrt{\argmax_{t^2: t^2\geq 0, \textrm{Pr} \left ( \hat{T^2}_{\textrm{var}} =t_1^2 | t_0^2 \leq t^2  \right ) \leq P_{e1}} t^2} , \nonumber
\end{eqnarray}
i.e., the largest $t$ such that the probability of obtaining the
considered variance-based estimator when the real scaling
factor is smaller than or equal to $t$, is smaller than or equal
to $P_{e1}$.
Nevertheless, given that the pdf we hace access to is $f_{\hat{T^2}_{\textrm{var}}|T_0^2}$, no having knowledge about $f_{\hat{T^2}_{\textrm{var}}}$ or $f_{T_0}$, the probability
in the constraint of the last formula can not be evaluated and a different
strategy must be followed. In that sense, we propose to use
\begin{eqnarray}
  t_{\textrm{lower}} &=& \sqrt{\argmax_{t^2: t^2\geq 0, \forall (t')^2 \leq t^2, \textrm{Pr} \left ( \hat{T^2}_{\textrm{var}} \geq t_1^2 | t_0^2 = (t')^2   \right ) \leq P_{e1}} t^2} , \label{eq:lower_var_est} 
\end{eqnarray}
which is nothing but the largest $t$ such that for every
$(t')^2 \leq t^2$, if 
the square real scaling factor is $(t')^2$, then
the probability of obtaining a variance-based estimator
larger than or equal to $t_1$,  is smaller than
or equal to $P_{e1}$.

It is important to note that
whenever $P_{e1} < 0.5$,  $t_{\textrm{lower}} < t_1$, so
one might just consider the corresponding interval in the optimizations
described in (\ref{eq:lower_var_est}).

Based on (\ref{eq:pdf_va_est2}), the probability in (\ref{eq:lower_var_est})
can be calculated as
\begin{eqnarray}
  \textrm{Pr} \left (\hat{T^2}_{\textrm{var}} \geq t_1^2|t_0^2 \right ) \approx \int_{t_1^2}^\infty
\frac{e^{\frac{- n (\sigma_X^2 + \alpha^2 \sigma_\Lambda^2)^2 (\tau - t_0^2)^2}{4 [(\sigma_X^2 + \alpha^2 \sigma_\Lambda^2)t_0^2 + \sigma_N^2]^2}}}{\sqrt{\frac{4\pi [(\sigma_X^2 + \alpha^2 \sigma_\Lambda^2) t_0^2  + \sigma_N^2]^2}{n(\sigma_X^2 + \alpha^2 \sigma_\Lambda^2)^2}}} d\tau
=
Q\left ( \frac{t_1^2 - t_0^2}{\sqrt{\frac{2 [(\sigma_X^2 + \alpha^2 \sigma_\Lambda^2) t_0^2  + \sigma_N^2]^2}{n(\sigma_X^2 + \alpha^2 \sigma_\Lambda^2)^2}}} 
\right ).
\label{eq:chorizo_Pe}
\end{eqnarray}
Given that $Q(x)$ is monotonically decreasing with $x$, we will prove that
its argument in (\ref{eq:chorizo_Pe}) is also monotonically decreasing with $t_0^2$ 
in order to show that (\ref{eq:lower_var_est})
has, at most, a single solution. Indeed,
\begin{eqnarray}
  \frac{\partial}{\partial t_0^2}\left (
  \frac{t_1^2 - t_0^2}{\sqrt{\frac{2 [(\sigma_X^2 + \alpha^2 \sigma_\Lambda^2) t_0^2  + \sigma_N^2]^2}{n(\sigma_X^2 + \alpha^2 \sigma_\Lambda^2)^2}}}  \right ) = 
  \frac{ - \sqrt{n}\left ((\sigma_X^2 + \alpha^2 \sigma_\Lambda^2) t_1^2  + \sigma_N^2 \right)\sigma_X^2}{\sqrt{2} \left ((\sigma_X^2 + \alpha^2 \sigma_\Lambda^2) t_0^2  + \sigma_N^2 \right )^2}.\nonumber
\end{eqnarray}
Therefore, based on $\textrm{Pr} \left ( \hat{T^2}_{\textrm{var}} \geq t_1^2 | t_0^2   \right )$ being monotonically increasing with $t_0^2$, (\ref{eq:lower_var_est})
is equivalent to 
\begin{eqnarray}
  t_{\textrm{lower}} &=& \argmax_{t:  \textrm{Pr} \left ( \hat{T^2}_{\textrm{var}} \geq t_1^2 | t_0^2=t^2   \right ) \leq P_{e1}} t.\nonumber 
\end{eqnarray}

Note that the minimum value of $P_{e1}$ one would expect to obtain is 
\begin{eqnarray}
  Q\left ( \sqrt{\frac{n}{2}} \frac{t_1^2 (\sigma_X^2 + \alpha^2 \sigma_\Lambda^2)}{\sigma_N^2}
\right ), \label{eq:chorizo_Pe2}
\end{eqnarray}
as $t_0^2 \geq 0$. A smaller value can not be obtained by the followed
approach due to the same reasons exposed at
Sect.~\ref{sec:trunc_gauss_var_based_est}. Nevertheless, in order
to achieve a trade-off between  simplicity and accuracy we will pursue
the current approximation. In any case, it is worth pointing out that
whenever the number of dimensions is very large (as it was assumed at the
derivation of the used approximation), HLR $\rightarrow \infty$, and
TNLR $<1$, the problem of (\ref{eq:chorizo_Pe})
being lower-bounded by (\ref{eq:chorizo_Pe2}) can be dismissed,
as the argument of (\ref{eq:chorizo_Pe}) is proportional to $\sqrt{n}$ and
$(\sigma_X^2 + \alpha^2 \sigma_\Lambda^2)/\sigma_N^2$, so
(\ref{eq:chorizo_Pe2}) goes to $0$.

Denoting by $\xi = Q^{-1}(P_{e1})$, and assuming that $P_{e1}$ is larger than or equal to (\ref{eq:chorizo_Pe2}), the mentioned unique solution  is
\begin{eqnarray}
  t_{\textrm{lower}}^2 = \left [\frac{2 \xi^2\sigma_N^2 + n (\sigma_X^2 + \alpha^2 \sigma_\Lambda^2) t_1^2 - \sqrt{2  n} \xi 
  [(\sigma_X^2 + \alpha^2 \sigma_\Lambda^2) t_1^2  + \sigma_N^2]}{(n-2\xi^2)
    (\sigma_X^2+ \alpha^2 \sigma_\Lambda^2)} \right ]^+. \nonumber
\end{eqnarray}

\begin{figure}[t] 
  \begin{center}
    \includegraphics[width=0.75\linewidth]{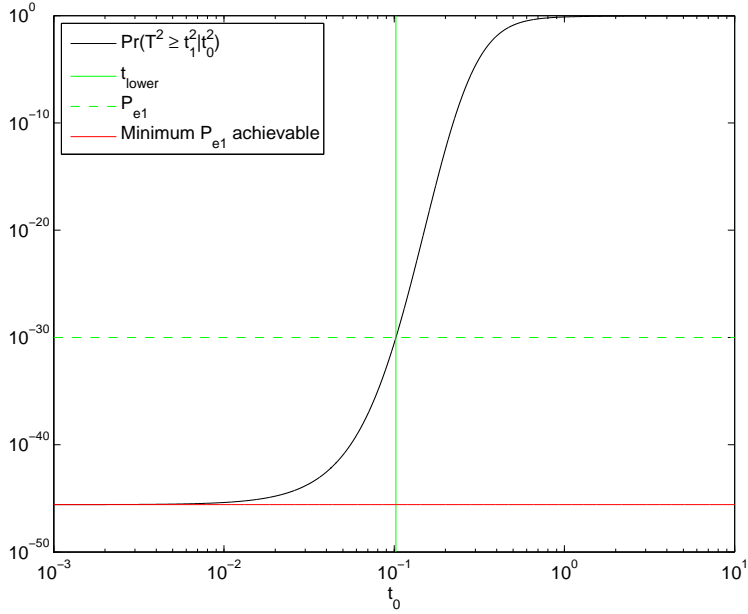}  
    \caption{$\textrm{Pr} \left (\hat{T^2}_{\textrm{var}} \geq t_1^2|t_0^2 \right )$, (\ref{eq:chorizo_Pe2}), and $t_{\textrm{lower}}$. HLR $= 10$ dB, WNR $=-3$ dB, $t_1 = 0.7$, $\alpha = 0.5$, $n = 4$, $P_{e1} = 10^{-30}$. In this case we use the WNR instead 
of the SCR and TNLR as the latter quantities depend
on $t_0$, while the WNR does not depend on it.}\label{fig:var_based_lower}
  \end{center}
\end{figure}

The probability studied at (\ref{eq:chorizo_Pe}), the lower-bound on it
derived in (\ref{eq:chorizo_Pe2}), and the value of $t_{\textrm{lower}}$
obtained in the last formula are illustrated in Fig.~\ref{fig:var_based_lower}.

The upper-bound counterpart of this method can be found in Sect.~\ref{sec:upper_var_based}.

%%%%%%%%%%%%%%%%%%%%%%%%%%%%%%%%%%%%%%%%%%%%%%%%%%%%%%%%%%%%%%%%%%%%%
%%%%%%%%%%%%%%%%%%%%%%%%%%%%%%%%%%%%%%%%%%%%%%%%%%%%%%%%%%%%%%%%%%%%%
\section{Search interval upper-bound}\label{sec:upper_bound}
%%%%%%%%%%%%%%%%%%%%%%%%%%%%%%%%%%%%%%%%%%%%%%%%%%%%%%%%%%%%%%%%%%%%%
\subsection{Deterministic approach I}\label{sec:upper_deter}
As it was proved in Sect.~\ref{sec:lower_deter}, $L_2(t, \bm z)$ first derivative
with respect to $t^2$ has a single positive root at $t_2$, corresponding
to a global minimum. Consequently, an upper-bound to the search interval can be
calculated similarly to (\ref{eq:compt_lower}), specifically
as the smallest $t\geq t_2$ such that 
the value of $L_2(t, \bm z)$ is larger
than or equal to $L(t_1, \bm z)$, i.e.,
\begin{eqnarray}
  t_{\textrm{upper}} = \argmin_{t: t \geq t_2,  L_2(t, \bm z) \geq L(t_1, \bm z) } t , \label{eq:compt_upper2}
\end{eqnarray}
which due to the strictly monotonically increasing nature of $L_2(t, \bm z)$
for $t > t_2$, and that for $t \rightarrow \infty$, $L_2(t, \bm z) \rightarrow \infty$, will be well-defined. Similarly to Sect.~\ref{sec:lower_deter}, $t_{\textrm{upper}}$
will verify $L_2(t_{\textrm{upper}}, \bm z) = L(t_1, \bm z)$, although in this
case we will focus on the study of the solutions for $t \geq t_2$ to 
(\ref{eq:sol_t4}). Again, despite we think that closed formulas are not available, due to the monotonically increasing nature of $L_2(t,\bm z)$ in the
range of interest, and that $L_2(t_2, \bm z) \leq L(t_1, \bm z)$ the solution
can be found by simple search methods.

The lower-bound counterpart of this method can be found in Sect.~\ref{sec:lower_deter}.
%%%%%%%%%%%%%%%%%%%%%%%%%%%%%%%%%%%%%%%%%%%%%%%%%%%%%%%%%%%%%%%%%%%%%
\subsubsection{Asymptotic value when the number of observations goes to infinity}\label{sec:upper_deter_asympt}

Taking into account that (\ref{eq:lower2_asym_1}), (\ref{eq:lower2_asym_2}), and
(\ref{eq:lower2_asym_3}) are still verified, is trivial
to note that $t_{\textrm{upper}}$ will be larger than or equal to $t_0$,
as this is smaller than or equal to $t_2$.

%%%%%%%%%%%%%%%%%%%%%%%%%%%%%%%%%%%%%%%%%%%%%%%%%%%%%%%%%%%%%%%%%%%%%
\subsection{Deterministic approach II}\label{sec:deter_upper_2}

For the upper-bound, simpler bounding functions can be used, allowing the closed calculation of $t_{\textrm{upper}}$; this
is the case, for example of
\begin{eqnarray}
L_4(t, \bm z) \triangleq n \log \left ( 2 \pi \left ( \sigma_N^2 + (1-\alpha)^2 
  t^2 \sigma_\Lambda^2 \right ) \right ), \nonumber
\end{eqnarray}
where we have removed the last term in $L_2(t, \bm z)$, and
which allows to introduce the following upper-bound to the search
interval
\begin{eqnarray}
  t_{\textrm{upper}} = \sqrt{\frac{1}{(1-\alpha)^2 \sigma_\Lambda^2} \left [\frac{e^{\frac{L(t_1, \bm z)}{n}}}{2\pi} - \sigma_N^2 \right]}. \label{eq:upper_3approx}
\end{eqnarray}

Furthermore, when the number of observations goes to infinity and $t_1 \approx t_0$, one can use the approximation derived in Sect.~\ref{sec:L_assympt}; therefore, replacing (\ref{eq:L_mean_close}) in (\ref{eq:upper_3approx}),
one obtains
\begin{eqnarray}
  t_{\textrm{upper}} &\approx & \sqrt{\frac{1}{(1-\alpha)^2 \sigma_\Lambda^2} \left [ \Big (
\sigma_N^2 + (1-\alpha)^2 t_1^2 \sigma_\Lambda^2 \Big)e^{2 + \frac{(t_0-t_1)^2 \sigma_X^2}{\sigma_N^2 + (1-\alpha)^2  t_1^2 \sigma_\Lambda^2}} - \sigma_N^2 \right]} \geq  \nonumber \\
&& t_1 e^{1 + \frac{(t_0-t_1)^2 \sigma_X^2}{2[\sigma_N^2 + (1-\alpha)^2  t_1^2 \sigma_\Lambda^2]}};  \label{eq:upper_deterministic2}
\end{eqnarray}
as it was proved in App.~\ref{sec:app2}, 
$t_1 e^{\frac{(t_0-t_1)^2 \sigma_X^2}{2[\sigma_N^2 + (1-\alpha)^2  t_1^2 \sigma_\Lambda^2]}} \geq t_0$
for any $t_1 \geq t_0$, and for any $t_1>0$ smaller than or equal to
a value that goes to $t_0$ whenever HLR $\rightarrow \infty$, and
TNLR $<1$.
Therefore, under those conditions $t_{\textrm{upper}} \geq e t_0$.

Finally, note that a similar strategy, based on a simplified version of 
$L_2(t, \bm z)$, can not be used
for the derivation of a lower-bound, as one can not neglect
either
of both terms in $L_2(t, \bm z)$. Indeed, the second term  is
required in order to
$L_2(t, \bm z) \rightarrow \infty$ when $t\rightarrow 0$, and the first one
should be also considered in order to be able to upper-bound
$L(t, \bm z)$, as it might be negative.
%%%%%%%%%%%%%%%%%%%%%%%%%%%%%%%%%%%%%%%%%%%%%%%%%%%%%%%%%%%%%%%%%%%%%

\subsection{Probabilistic approach}\label{sec:upper_bound_prob}
Trying to reduce the search interval upper-bounds  based on deterministic
approaches proposed in the previous
sections, we can use the
previously introduced  probabilistic modeling of $L(t, \bm Z)$
(valid under the conditions introduced in Sect.~\ref{sec:target_function_t_0})
for $t=t_0$ in order to obtain a new, hopefully smaller, search interval
upper-bound value by relaxing the former deterministic constraints to just be verified with a given high probability. Formally, we can calculate such a point as the smallest positive real scaling factor $t$ such that
for every $t' > t$
the event of $L(t', \bm Z)$ being smaller than $L(t_1, \bm z)$ has
a probability smaller than or equal to $P_{e1}$, i.e.,
\begin{eqnarray}
  t_{\textrm{upper}} = \argmin_{t: t\geq0, \forall t' > t, \textrm{Pr} \left ( L(t', \bm Z) < L(t_1, \bm z) | t_0 = t'\right ) \leq P_{e1}} t. \label{eq:compt_upper}
\end{eqnarray}
Taking into account the probabilistic modeling of $L(t, \bm Z)$
for $t=t_0$, the considered probability can
be calculated as
\begin{eqnarray}
  \textrm{Pr} \left ( L(t, \bm Z) < L(t_1, \bm z) | t = t_0\right ) = 
\int_{-\infty}^{L(t_1, \bm z)} f_{\chi^2_{2n}} \left (x - n \log \Big [ 2 \pi \left ( \sigma_N^2 + (1-\alpha)^2 
  t^2 \sigma_\Lambda^2 \right ) \Big ] \right ) dx, \label{eq:compt_prob}
\end{eqnarray}
with $f_{\chi^2_{2n}}(x)$ the pdf of the $\chi^2$ distribution with $2n$
degrees of freedom; from the last equation it is obvious that
this probability is monotonically decreasing (indeed approaching $0$)
with $t$. Therefore,  the constraint in (\ref{eq:compt_upper}) will
be always well defined
(i.e., it will never define an empty set of feasible $t$'s for any particular
$P_{e1} > 0$ and $L(t_1, \bm z)$), although we will not be able to
use it in order to define a lower-bound counterpart. Moreover, due to the monotonicity
of $\textrm{Pr} \left ( L(t, \bm Z) < L(t_1, \bm z) | t = t_0\right )$,
(\ref{eq:compt_upper}) is equivalent to
\begin{eqnarray}
  t_{\textrm{upper}} = \argmin_{t: t\geq0, \textrm{Pr} \left ( L(t, \bm Z) < L(t_1, \bm z) | t_0 = t\right ) \leq P_{e1}} t. \nonumber
\end{eqnarray}

\begin{figure}[t] 
  \begin{center}
    \includegraphics[width=0.75\linewidth]{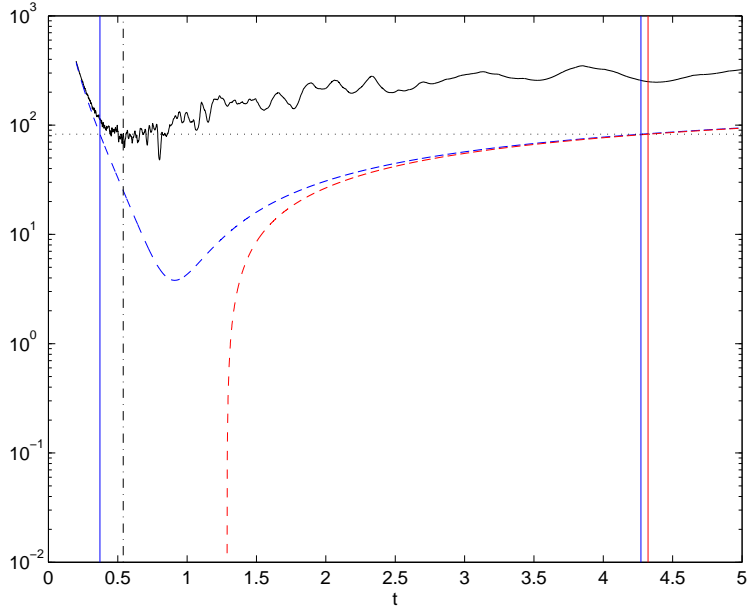}  
    \caption{Comparison between $L(t, \bm z)$ (solid black line),
      $L(t_1, \bm z)$ (dotted black line),
$L_2(t, \bm z)$ (dashed blue line), and $L_4(t, \bm z)$ (dashed red line).
The corresponding search-interval bounds are plotted by using
vertical solid lines with the same color that the function used
for computing them. $t_1$ (vertical dash-dot black line) calculated by using (\ref{eq:approx_min_L1}).
HLR $\approx 34.9765 $ dB, SCR $\approx -1.0618$ dB, TNLR $\approx -3.5736$ dB, $t_0 = 0.8$, 
$\alpha = \alpha_{\textrm{Costa}} \approx 0.5608$, $n = 40$, scalar
quantizer.}\label{fig:bounds3}
  \end{center}
\end{figure}

In Fig.~\ref{fig:bounds3} we can compare the upper-bound obtained
by using the method proposed in this section, and the bounds
obtained by using the deterministic method
proposed in Sects.~\ref{sec:lower_deter_asympt} and 
\ref{sec:upper_deter_asympt}. As expected, the upper-bound derived in this
section is larger than that obtained in Sect.~\ref{sec:upper_deter_asympt},
due to the different definition of $L_2(t,\bm z)$ and $L_4(t, \bm z)$;
nevertheless, one must also take into account that the upper-bound
derived from $L_4(t, \bm z)$ can be calculated by using a closed
formula.

Furthermore, if we denote by $F_{\chi^2_{2n}}(x)$ the distribution function of the
$\chi^2$ distribution with $2n$ degrees of freedom, the mentioned constraint
can be rewritten as
\begin{eqnarray}
  F_{\chi^2_{2n}} \left (L(t_1, \bm z) -  n \log \Big [ 2 \pi \left ( \sigma_N^2 + (1-\alpha)^2 
  t^2 \sigma_\Lambda^2 \right ) \Big ] \right ) \leq P_{e1}, \nonumber
\end{eqnarray}
which due to the strictly monotonically increasing nature of $F_{\chi^2_{2n}}(x)$,
and the strictly monotonically increasing nature of the argument
of the $\log$ function with $t$,
allows one to calculate $t_{\textrm{upper}}$ as the only solution
to
\begin{eqnarray}
 n \log \Big [ 2 \pi \left ( \sigma_N^2 + (1-\alpha)^2 
  t^2 \sigma_\Lambda^2 \right ) \Big ] +   F_{\chi^2_{2n}}^{-1}(P_{e1}) = L(t_1, \bm z) , \label{eq:sec_acomer33}
\end{eqnarray}
which can be shown to be
\begin{eqnarray}
  t_{\textrm{upper}} = \left [\left (\frac{1}{(1-\alpha)^2 \sigma_\Lambda^2} \left [\frac{e^{\frac{L(t_1, \bm z) - F_{\chi^2_{2n}}^{-1}(P_{e1})}{n}}}{2\pi} - \sigma_N^2 \right ]\right )^+\right ] ^{1/2} . \label{eq:t_tocho}
\end{eqnarray}

Similarly, if one decides to use the CLT-based Gaussian approximation 
to the pdf of $L(t, \bm Z)$ given that $t = t_0$ 
(introduced in Sect.~\ref{sec:target_function_t_0}, formulas (\ref{eq:mean_mom_t_0}), (\ref{eq:var_mom_t_0}), and (\ref{eq:var_mom_t_0_mult}), assuming additionally that SCR $\rightarrow 0$ for the scalar quantizer case),
should consider that
the mean and variance of that distribution are $2n$ and $4n$
respectively, so (\ref{eq:t_tocho}) would be replaced by
\begin{eqnarray}
  t_{\textrm{upper}} = \left [\left (\frac{1}{(1-\alpha)^2 \sigma_\Lambda^2} \left [\frac{e^{\frac{L(t_1, \bm z) - 2n + \sqrt{4n}Q^{-1}(P_{e1})}{n}}}{2\pi} - \sigma_N^2 \right ]\right )^+\right ]^{1/2} . \label{eq:t_tocho2}
\end{eqnarray}

\begin{figure}[t] 
  \begin{center}
    \includegraphics[width=0.75\linewidth]{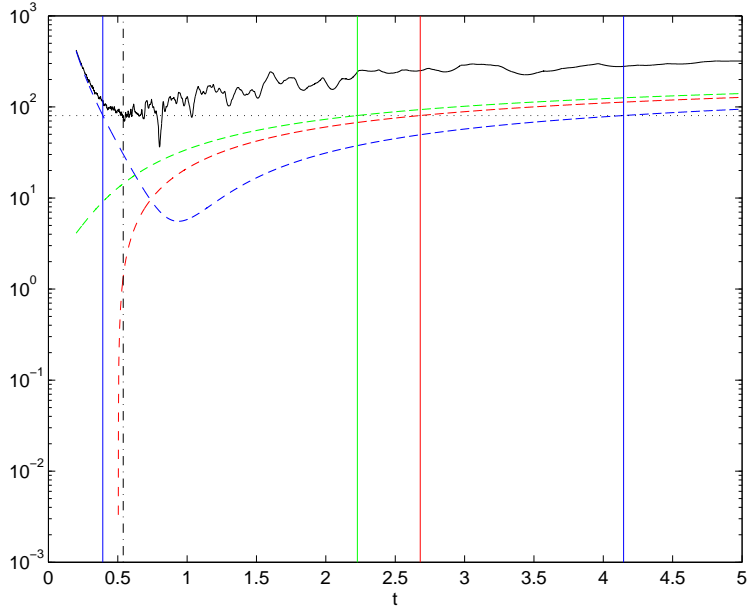}  
    \caption{Comparison between $L(t, \bm z)$ (solid black line),
      $L(t_1, \bm z)$ (dotted black line),
$L_2(t, \bm z)$ (dashed blue line), and the left term of (\ref{eq:sec_acomer33})
for $P_{e1} = 10^{-6}$ (dashed red line), and $P_{e1} = 10^{-3}$ (dashed green line). The corresponding search-interval bounds are plotted by using
vertical solid lines with the same color that the function used
for computing them. $t_1$ (vertical dash-dot black line) calculated by using (\ref{eq:approx_min_L1}).
HLR $\approx 34.9765 $ dB, SCR $\approx -1.0618$ dB, TNLR $\approx -3.5736$ dB, $t_0 = 0.8$, 
$\alpha = \alpha_{\textrm{Costa}} \approx 0.5608$, $n = 40$, scalar
quantizer.}\label{fig:bounds2}
  \end{center}
\end{figure}

In Fig.~\ref{fig:bounds2} we can compare the upper-bounds obtained
by using the method proposed in this section, and the bounds
obtained by using the deterministic method
proposed in Sects.~\ref{sec:lower_deter_asympt} and 
\ref{sec:upper_deter_asympt}. Similarly to Fig.~\ref{fig:bounds1},
the larger $P_{e1}$, the smaller
the search interval upper-bound.

%%%%%%%%%%%%%%%%%%%%%%%%%%%%%%%%%%%%%%%%%%%%%%%%%%%%%%%%%%%%%%%%%%%%%
\subsubsection{Asymptotic value when the number of observations goes to infinity}\label{sec:asym_large1}

For the sake of completeness we will analyze the asymptotic value of
$t_{\textrm{upper}}$ as a function of $t_1$ when $n\rightarrow \infty$; this
analysis will provide us with a theoretical insight of the
search interval upper-bound (and therefore the computational cost) when a large
number of observations are available at the decoder.

First, we will focus on the case where
$t_1\approx t_0$; as it is shown in Sect.~\ref{sec:L_assympt},
\begin{eqnarray}
L(t_1, \bm z) &\approx &2n + \frac{n (t_0-t_1)^2\sigma_X^2}{\sigma_N^2 + (1-\alpha)^2 t_1^2 \sigma_\Lambda^2}
+ n \log\left(2\pi\left ( \sigma_N^2 + (1-\alpha)^2 t_1^2 \sigma_\Lambda^2 \right) \right ), \nonumber
\end{eqnarray}
yielding
\begin{eqnarray}
  t_{\textrm{upper}} \approx \left ( t_1^2 e^{\frac{\sigma_X^2(t_0-t_1)^2}{\sigma_N^2 + (1-\alpha)^2  t_1^2 \sigma_\Lambda^2}} + \frac{\sigma_N^2 \left (e^{\frac{\sigma_X^2(t_0-t_1)^2}{\sigma_N^2 + (1-\alpha)^2  t_1^2 \sigma_\Lambda^2}} - 1\right)}{(1-\alpha)^2 \sigma_\Lambda^2}\right)^{1/2}. \nonumber
\end{eqnarray}
At the sight of the last expression one can check that whenever $t_1\rightarrow
t_0$, then $t_{\textrm{upper}} \rightarrow t_1$, and consequently 
$t_{\textrm{upper}} \rightarrow t_0$. One interesting question to be answered
is if $t_{\textrm{upper}} \geq t_0$ irrespectively of $t_1$.
A sufficient condition for that to happen is that
\begin{eqnarray}
  t_1^2 e^{\frac{\sigma_X^2(t_0-t_1)^2}{\sigma_N^2 + (1-\alpha)^2 t_1^2 \sigma_\Lambda^2}} \geq t_0^2,
\nonumber
\end{eqnarray}
which is discussed in App.~\ref{sec:app2}.
It is interesting to point out the similarities
between the left term of the last expresion and
the square of the last term of (\ref{eq:upper_deterministic2});
in Sect.~\ref{sec:deter_upper_2} a determistic approach was proposed
based on a lossy version of the objective function, in order to be
able to obtain closed formulas for $t_{\textrm{upper}}$. Nevertheless,
the penalty to be paid is also clear: when following that
deterministic lossy approach the obtained upper-bound
is nothing but the upper-bound derived in the current section
multiplied by $e$, consequently increasing  the size of the search
interval.

On the other hand, if we study the case where $t_1$ is not close to $t_0$,
$L(t_1, \bm z)$ is shown in Sect.~\ref{sec:L_assympt} to be
approximately 
\begin{eqnarray}
L(t_1, \bm z) &\approx &n \frac{ t_1^2 \sigma_\Lambda^2}{\sigma_N^2 + (1-\alpha)^2 t_1^2 \sigma_\Lambda^2} + 
n \log\left(2\pi\left ( \sigma_N^2 + (1-\alpha)^2 t_1^2 \sigma_\Lambda^2\right) \right )
+ n \frac{t_0^2}{t_1^2},\nonumber
\end{eqnarray}
yielding
\begin{eqnarray}
  t_{\textrm{upper}} \approx\left ( t_1^2 e^{\frac{t_0^2}{t_1^2} + \frac{ t_1^2 \sigma_\Lambda^2}{\sigma_N^2 + (1-\alpha)^2  t_1^2 \sigma_\Lambda^2} - 2} + \frac{\sigma_N^2 \left (e^{\frac{t_0^2}{t_1^2} + \frac{ t_1^2 \sigma_\Lambda^2}{\sigma_N^2 + (1-\alpha)^2  t_1^2 \sigma_\Lambda^2}-2} - 1\right)}{(1-\alpha)^2 \sigma_\Lambda^2}\right)^{1/2}. \nonumber
\end{eqnarray}

In order to check if also in this case $t_{\textrm{upper}} \geq t_0$ we will
consider the sufficient condition
\begin{eqnarray}
  t_1^2 e^{\frac{t_0^2}{t_1^2}} \geq t_0^2,\nonumber
\end{eqnarray}
or equivalently
\begin{eqnarray}
  \log \left ( \frac{t_1^2}{t_0^2} \right ) \geq - \frac{t_0^2}{t_1^2}, \nonumber
\end{eqnarray}
which is true for any value of $t_1>0$, since, as it is shown in App.~\ref{sec:app3}, $\log(x) \geq - \frac{1}{x}$ for
any $x \geq 0$.

%%%%%%%%%%%%%%%%%%%%%%%%%%%%%%%%%%%%%%%%%%%%%%%%%%%%%%%%%%%%%%%%%%%%%
%%%%%%%%%%%%%%%%%%%%%%%%%%%%%%%%%%%%%%%%%%%%%%%%%%%%%%%%%%%%%%%%%%%%%
\subsection{Partially probabilistic approach}\label{sec:another_possible2}

Please refer to Sect.~\ref{sec:another_possible}.
%%%%%%%%%%%%%%%%%%%%%%%%%%%%%%%%%%%%%%%%%%%%%%%%%%%%%%%%%%%%%%%%%%%%%
%%%%%%%%%%%%%%%%%%%%%%%%%%%%%%%%%%%%%%%%%%%%%%%%%%%%%%%%%%%%%%%%%%%%%
\subsection{Search interval upper-bound calculated
from the variance-based estimator}\label{sec:upper_var_based}

In this section we will follow an approach similar
to that proposed at Sect.~\ref{eq:sec_lowerbound_var_based}
in order to obtain an upper-bound to the search interval
when the variance-based estimator is used as $t_1$.

Similarly to (\ref{eq:lower_var_est}), the upper-bound will be computed
this time as
the smallest $t>0$ such that for every
$(t')^2 \geq t^2$, if 
the square real scaling factor is $(t')^2$, then
the probability of
obtaining a variance-based estimator smaller than or equal to $t_1$,  is smaller than
or equal to $P_{e1}$, i.e., 
\begin{eqnarray}
  t_{\textrm{upper}} &=& \sqrt{\argmin_{t^2: t^2 \geq 0, \forall (t')^2 \geq t^2, \textrm{Pr} \left ( \hat{T^2}_{\textrm{var}} \leq t_1^2 | t_0^2 = (t')^2  \right ) \leq P_{e1}} t^2} . \label{eq:upper_var_est} 
\end{eqnarray}
Based on (\ref{eq:pdf_va_est2}), the probability in (\ref{eq:upper_var_est})
can be calculated as
\begin{eqnarray}
  \textrm{Pr} \left (\hat{T^2}_{\textrm{var}} \leq t_1^2|t_0^2 \right ) \approx Q\left ( \frac{t_0^2}{\sqrt{\frac{2 [(\sigma_X^2 + \alpha^2 \sigma_\Lambda^2) t_0^2  + \sigma_N^2]^2}{n(\sigma_X^2 + \alpha^2 \sigma_\Lambda^2)^2}}} 
\right ) u(t_1^2) + \int_{0}^{t_1^2}
\frac{e^{\frac{- n (\sigma_X^2 + \alpha^2 \sigma_\Lambda^2)^2 (\tau - t_0^2)^2}{4 [(\sigma_X^2 + \alpha^2 \sigma_\Lambda^2)t_0^2 + \sigma_N^2]^2}}}{\sqrt{\frac{4\pi [(\sigma_X^2 + \alpha^2 \sigma_\Lambda^2) t_0^2  + \sigma_N^2]^2}{n(\sigma_X^2 + \alpha^2 \sigma_\Lambda^2)^2}}} d\tau,
\label{eq:cuscus}
\end{eqnarray}
where $u(x)$ stands for the step function, defined as
\begin{eqnarray}
  u(x) \triangleq \left \{ \begin{array}{ll}
    0, &\;\;\; \textrm{ if } x <0, \\
    1, &\;\;\; \textrm{ otherwise}
  \end{array}
  \right . . \nonumber
\end{eqnarray}
Taking into account that $t_1^2 = \hat{t^2}_{\textrm{var}}(\bm z)$ defined
in (\ref{eq:def_va_est2}) is non-negative,  (\ref{eq:cuscus}) can be rewritten as
\begin{eqnarray}
  \textrm{Pr} \left (\hat{T^2}_{\textrm{var}} \leq t_1^2|t_0^2 \right ) \approx  \int_{-\infty}^{t_1^2}
\frac{e^{\frac{- n (\sigma_X^2 + \alpha^2 \sigma_\Lambda^2)^2 (\tau - t_0^2)^2}{4 [(\sigma_X^2 + \alpha^2 \sigma_\Lambda^2)t_0^2 + \sigma_N^2]^2}}}{\sqrt{\frac{4\pi [(\sigma_X^2 + \alpha^2 \sigma_\Lambda^2) t_0^2  + \sigma_N^2]^2}{n(\sigma_X^2 + \alpha^2 \sigma_\Lambda^2)^2}}} d\tau
= Q\left ( \frac{t_0^2 - t_1^2}{\sqrt{\frac{2 [(\sigma_X^2 + \alpha^2 \sigma_\Lambda^2) t_0^2  + \sigma_N^2]^2}{n(\sigma_X^2 + \alpha^2 \sigma_\Lambda^2)^2}}} 
\right ).
\label{eq:cantimpalo_Pe}
\end{eqnarray}
Given that $Q(x)$ is monotonically decreasing with $x$, we will prove that
its argument in (\ref{eq:cantimpalo_Pe}) is monotonically increasing with $t_0^2$ 
in order to show that (\ref{eq:upper_var_est})
has a single solution. Indeed,
\begin{eqnarray}
  \frac{\partial}{\partial t_0^2}\left (
  \frac{t_0^2 - t_1^2}{\sqrt{\frac{2 [(\sigma_X^2 + \alpha^2 \sigma_\Lambda^2) t_0^2  + \sigma_N^2]^2}{n(\sigma_X^2 + \alpha^2 \sigma_\Lambda^2)^2}}}  \right ) = 
  \frac{ \sqrt{n}\left ((\sigma_X^2 + \alpha^2 \sigma_\Lambda^2) t_1^2  + \sigma_N^2 \right)\sigma_X^2}{\sqrt{2} \left ((\sigma_X^2 + \alpha^2 \sigma_\Lambda^2) t_0^2  + \sigma_N^2 \right )^2}. \nonumber
\end{eqnarray}
Based on $\textrm{Pr} \left (\hat{T^2}_{\textrm{var}} \leq t_1^2|t_0^2 \right )$
being monotonically decreasing with $t_0^2$, (\ref{eq:upper_var_est})
is equivalent to
\begin{eqnarray}
  t_{\textrm{upper}} &=&\argmin_{t: t \geq 0, \textrm{Pr} \left ( \hat{T^2}_{\textrm{var}} \leq t_1^2 | t_0^2 =t^2  \right ) \leq P_{e1}} t . \nonumber
\end{eqnarray}

Note that in this case the minimum value of $P_{e1}$ one could expect to
obtain is 
\begin{eqnarray}
  Q\left ( \sqrt{\frac{n}{2}} \right ). \label{eq:cantimpalo_Pe2}
\end{eqnarray}
Similarly to the discussion in Sect.~\ref{eq:sec_lowerbound_var_based},
the bound on (\ref{eq:cantimpalo_Pe}) due to (\ref{eq:cantimpalo_Pe2})
can be dismissed when the number of considered dimensions
is very large, as in that scenario (\ref{eq:cantimpalo_Pe2}) goes
to $0$.

Denoting by $\xi = Q^{-1}(P_{e1})$, and assuming that $P_{e1} \geq Q(\sqrt{n/2})$,
the mentioned only solution is
\begin{eqnarray}
  t_{\textrm{upper}}^2 = \frac{2 \xi^2\sigma_N^2 + n (\sigma_X^2 + \alpha^2 \sigma_\Lambda^2) t_1^2 + \sqrt{2  n} \xi 
  [(\sigma_X^2 + \alpha^2 \sigma_\Lambda^2) t_1^2  + \sigma_N^2]}{(n-2\xi^2)
    (\sigma_X^2+ \alpha^2 \sigma_\Lambda^2)} . \nonumber
\end{eqnarray}

\begin{figure}[t] 
  \begin{center}
    \includegraphics[width=0.75\linewidth]{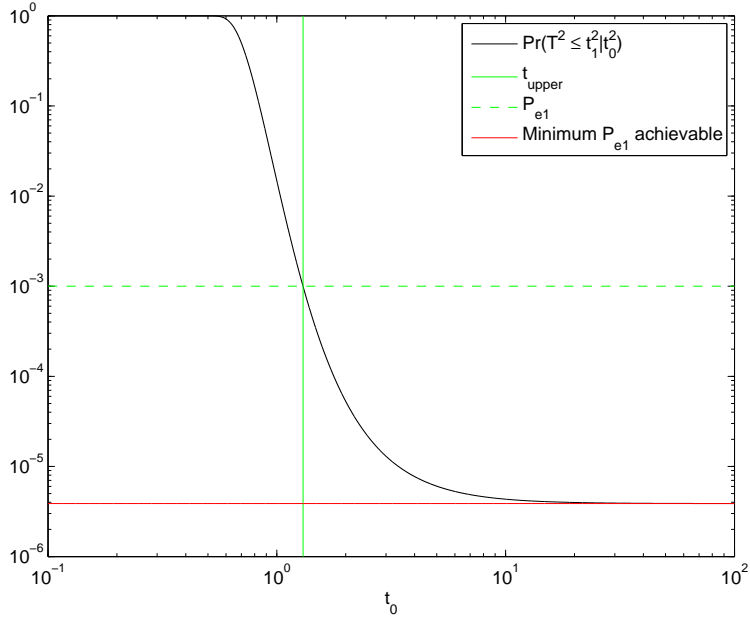}  
    \caption{$\textrm{Pr} \left (\hat{T^2}_{\textrm{var}} \leq t_1^2|t_0^2 \right )$, (\ref{eq:chorizo_Pe2}), and $t_{\textrm{lower}}$. HLR $= 10$ dB, WNR $=-3$ dB, $t_1 = 0.7$, $\alpha = 0.5$, $n = 40$, $P_{e1} = 10^{-3}$. In this case we use the WNR instead 
of the SCR and TNLR the latter quantities depend
on $t_0$, while the WNR does not depend on it.}\label{fig:var_based_upper}
  \end{center}
\end{figure}

The probability studied at (\ref{eq:cantimpalo_Pe}), the lower-bound on it
derived derived in (\ref{eq:cantimpalo_Pe2}), and the value of $t_{\textrm{upper}}$
obtained in the last formula are illustrated in Fig.~\ref{fig:var_based_upper}.

%%%%%%%%%%%%%%%%%%%%%%%%%%%%%%%%%%%%%%%%%%%%%%%%%%%%%%%%%%%%%%%%%%%%%
%%%%%%%%%%%%%%%%%%%%%%%%%%%%%%%%%%%%%%%%%%%%%%%%%%%%%%%%%%%%%%%%%%%%%

\subsection{Discussion on the importance of $t_1$ choice}

Finally, it is worth to point out that although deterministic and
probabilistic methods for the derivation of lower and upper-bounds
can use any initial point $t_1$, the size of the obtained interval (and consequently
the computational cost of the subsequent optimization of the target function
in that interval) will depend on the
choice of $t_1$, as it is evident, for example, at the sight of (\ref{eq:upper_3approx}), (\ref{eq:t_tocho}), or (\ref{eq:t_tocho2}).

%%%%%%%%%%%%%%%%%%%%%%%%%%%%%%%%%%%%%%%%%%%%%%%%%%%%%%%%%%%%%%%%%%%%%
%%%%%%%%%%%%%%%%%%%%%%%%%%%%%%%%%%%%%%%%%%%%%%%%%%%%%%%%%%%%%%%%%%%%%
%%%%%%%%%%%%%%%%%%%%%%%%%%%%%%%%%%%%%%%%%%%%%%%%%%%%%%%%%%%%%%%%%%%%%
%%%%%%%%%%%%%%%%%%%%%%%%%%%%%%%%%%%%%%%%%%%%%%%%%%%%%%%%%%%%%%%%%%%%%

%%%%%%%%%%%%%%%%%%%%%%%%%%%%%%%%%%%%%%%%%%%%%%%%%%%%%%%%%%%%%%%%%%%%%
%%%%%%%%%%%%%%%%%%%%%%%%%%%%%%%%%%%%%%%%%%%%%%%%%%%%%%%%%%%%%%%%%%%%%

\section{Sampling the search-interval}\label{sec:samp_se_int}

As it was already discussed, the main problem
when dealing with the optimization of (\ref{eq:SecEstimator}) is
the presence of non-differentiable points. Nevertheless,
the target function is convex between any two consecutive non-differentiable 
points; therefore,
as a first
approach to the optimization problem one could think of considering all those non-differentiable
points that have a non-neglectable probability and optimizing the target function into the intervals defined
by each two consecutive points. Doing so, the convergence to the value minimizing
$L(t, \bm z)$ can be ensured.

The first question to be answered about this approach is its feasibility.
In other words, how does the number of non-differentiable points to be
considered
increases with the dimensionality of the problem? Obviously, if this number were
not bounded, this would imply a serious implementation problem
to this strategy, as the required resources would be also unlimited.

\begin{figure}[t] 
  \begin{center}
    \includegraphics[width=0.75\linewidth]{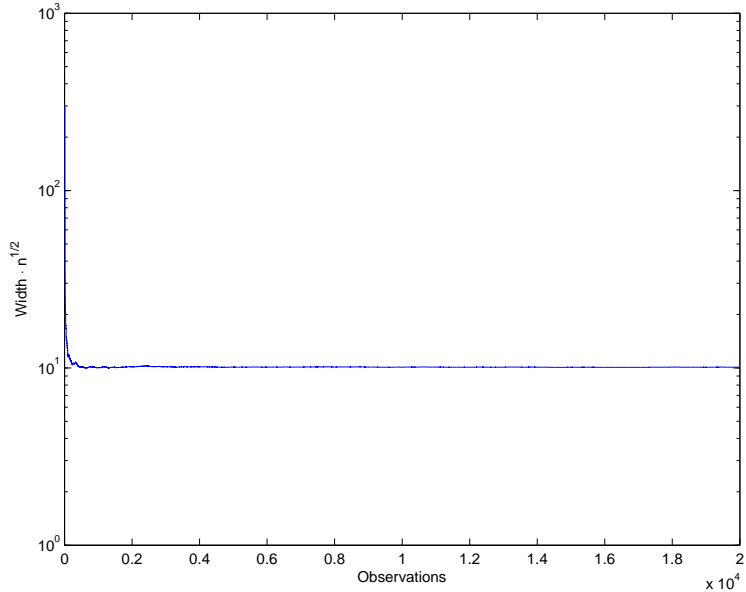}  
    \caption{Width of the interval (obtained by using the variance-based methodology
described above) multiplied by $\sqrt{n}$ versus the dimensionality of the problem $n$. HLR $\approx 25.5630$ dB, SCR $\approx -6.6199$ dB, TNLR $\approx -0.4833$ dB, $\alpha = 0.6$, $t_0 = 0.7$.}\label{fig:cutre1}
  \end{center}
\end{figure}

\begin{figure}[t] 
  \begin{center}
    \includegraphics[width=0.75\linewidth]{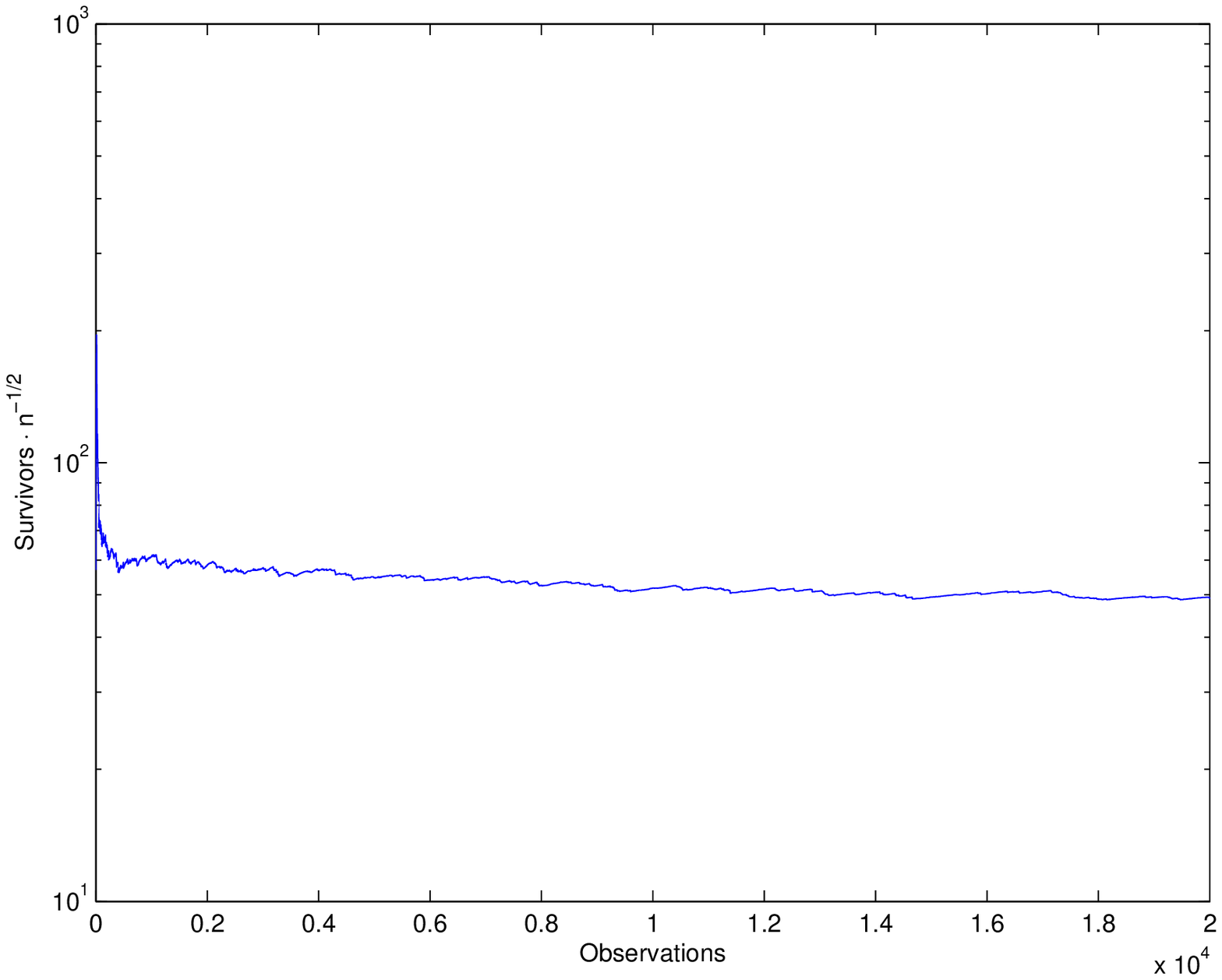}  
    \caption{Number of non-differentiable points in the intervals described above divided by $\sqrt{n}$ versus the dimensionality of the problem $n$. HLR $\approx 25.5630$ dB, SCR $\approx -6.6199$ dB, TNLR $\approx -0.4833$ dB, $\alpha = 0.6$, $t_0 = 0.7$.}\label{fig:cutre2}
  \end{center}
\end{figure}

\begin{figure}[t] 
  \begin{center}
    \includegraphics[width=0.75\linewidth]{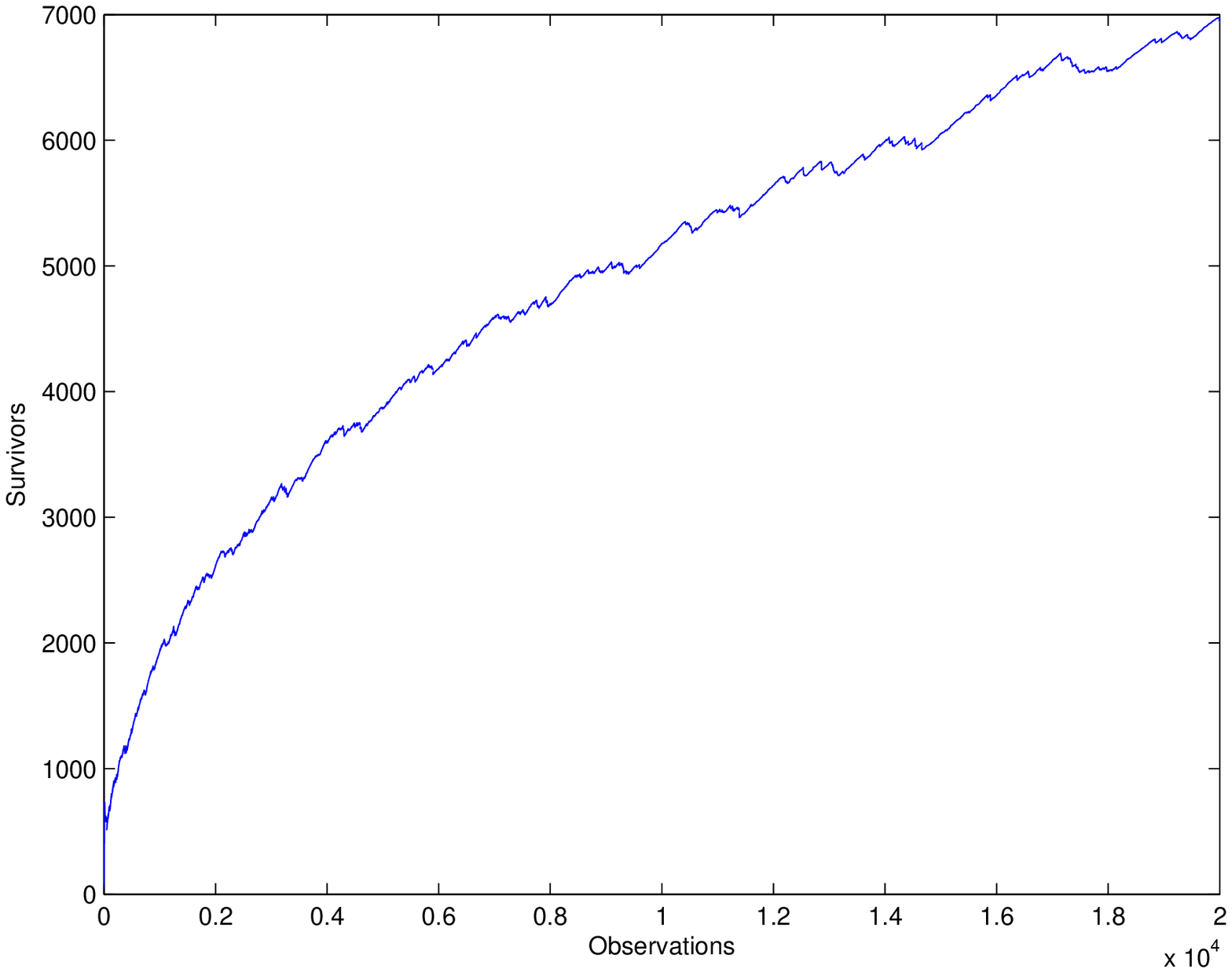}  
    \caption{Number of non-differentiable points in the intervals described above versus the dimensionality of the problem $n$. HLR $\approx 25.5630$ dB, SCR $\approx -6.6199$ dB, TNLR $\approx -0.4833$ dB, $\alpha = 0.6$, $t_0 = 0.7$.}\label{fig:cutre3}
  \end{center}
\end{figure}

In order to answer this question, experiments were performed where
the variance-based
scaling estimator $\hat{t}_{\textrm{var}}(\bm z)$  was used to define  an interval of radius $5$ times
the square root of the CRB of that estimator around the estimated value;
this
interval determines the non-differentiable points that can not
be neglected. As the
estimation variance decreases with $1/n$, the width of that interval will
decrease with $1/\sqrt{n}$. On the other hand, the density of non-differentiable
points increases with $n$. Consequently, the number of points that one
must take into account in the proposed strategy increases in
average with $\sqrt{n}$.
Plots illustrating these results can be found in Figs.~\ref{fig:cutre1}, 
\ref{fig:cutre2}, and \ref{fig:cutre3}.

Given that the approach of considering all the non-differential points has
been shown to be computationally unfeasible, other strategies must be 
proposed in order to explore the search interval seeking
$L(t, \bm z)$ minimum.

Our proposals are based on the study of $L(t, \bm Z)$
main lobe width. Basically, the width of that lobe
depends on the behavior with $t$ of the quantization error Euclidean norm
in the first term of (\ref{eq:SecEstimator}). The value
of the target function is minimum at a point close to $t_0$ (check Sect.~\ref{sec:bias} for further details), as
the variance
of the quantization error is minimized in the neighborhood of
that point. When $t$ deviates
from $t_0$, but it is still close to $t_0$ (following
the conditions introduced at the first part of Sect.~\ref{sec:L_assympt}),
the modulo-$\Lambda$ reduction has not to be considered and
the quantization error variance can be written as
$(t_0-t)^2 \sigma_X^2 + (t - \alpha t_0)^2 \sigma_\Lambda^2 + \sigma_N^2$.
As $t$ deviates from $t_0$, and
due to the presence of the modulo operation, this variance is finally
upper-bounded by $t^2 \sigma_\Lambda^2$; from a statistical point of view, the
target function main
lobe ends when the variance of the modulo-reduced quantization error approaches
this upper-bound. Based on these ideas, we will determine a set of
points in the search interval trying to ensure that at least one of them
is in the target function main lobe. When the number of observations is large,
$L(t, \bm z)$ will tend to be convex in its main lobe, as it
will asymptotically follow the 
characterization introduced in Sect.~\ref{sec:L_assympt}; thus, conventional
or ad-hoc optimization algorithms could be used to minimize that function,
taking as starting points those mentioned above. As long as one of those
points indeed belongs to the target function main lobe, it will lead us
to the minimum  of $L(t, \bm z)$ (based on the asymptotic
convexity of the target function for $n\rightarrow \infty$).

Therefore, the question to be solved is how we can define a minimum
set of points (as the final optimization algorithm considers as starting points all the elements in that set, its computational cost is linearly increased
with the cardinality of the set, so we are interested in defining a minimum
size set) in order to be reasonably sure that one of them will lie on the
main lobe of the target function.
We will define our  {\it candidate point set} (hereafter, {\it candidate set}) 
by exploiting that 
the boundary
of the target function main lobe depends on the quantization
error variance approaching $t^2 \sigma_\Lambda^2$.

\subsection{First approach. Scalar quantizer case} 

The reasoning behind this approach is that for the scalar quantizer case, and similarly for the low-dimensional lattice case, the effect of the modulo-reduction
on the variance of the modularized random vector is rather smooth, in 
the sense that for achieving the mentioned upper-bound $t^2 \sigma_\Lambda^2$,
the variance of the input random vector should be much larger
than that value. This is not the case for the high-dimensional lattice
case, where the input variance needed for saturating the output
variance is just $t^2 \sigma_\Lambda^2$; that is why a specific interval 
search sampling method is proposed for the
high-dimensional good lattice case (please refer to the next section).
Coming back to the scalar quantizer scenario,
as the input vs. output variance function is smooth, one has same 
some margin for considering 
values of $(t_0-t)^2 \sigma_X^2 + (t - \alpha t_0)^2 \sigma_\Lambda^2 + \sigma_N^2$
larger than
$t^2 \sigma_\Lambda^2$, and consequently this threshold will not be
considered by the current search interval sampling method.

Instead, the approach proposed for the low-dimensional lattice
case is based on assuming that
the current point, let us denote it as $t(i)$, is indeed $t_0$; if we want
its nearest points in the candidate set to also belong to its main lobe, upper and lower-bounds
can be defined to those nearest point values in order to the resulting quantization error
variance be close enough to the quantization error variance at the current point
(calculated assuming that it is indeed the real scaling factor), as otherwise
we will probably be outside the main lobe. Mathematically,
the variance of the quantization error at $t(i)$, assuming that it is the real scaling factor,
is given by
\begin{eqnarray}
t^2(i)(1 - \alpha)^2 \sigma_\Lambda^2 + \sigma_N^2, \nonumber
\end{eqnarray}
while that at the nearest point in the candidate set, denoted as $t(i+1)$,
can be calculated as
\begin{eqnarray}
  [t(i+1) - t(i)]^2 + [t(i+1) - \alpha t(i)]^2 \sigma_\Lambda^2 + \sigma_N^2. \nonumber
\end{eqnarray}
Constraining the latter variance to be smaller than or equal to the former
one plus $K_1 t^2(i+1) \sigma_\Lambda^2$ in order to ensure that $t(i+1)$ is still
in the main lobe of the target function, the two following constraints
on $t(i+1)$ are obtained:
\begin{eqnarray}
  t(i+1) &\geq& t_{l,\textrm{ld}} \triangleq
\frac{t(i) \left [\alpha \sigma_\Lambda^2  +  \sigma_X^2 - \sigma_\Lambda
    \sqrt{\sigma_\Lambda^2  [ (1-\alpha)^2 + K_1 (2\alpha - 1)] +  
    K_1 \sigma_X^2 } \right ]}{ \sigma_X^2 +  \sigma_\Lambda^2 (1-K_1)}, 
\label{eq:sampling_lower}\\
  t(i+1) &\leq& t_{u,\textrm{ld}} \triangleq \frac{t(i) \left [\alpha \sigma_\Lambda^2  + \sigma_X^2 + \sigma_\Lambda
    \sqrt{\sigma_\Lambda^2  [ (1-\alpha)^2 + K_1 (2\alpha - 1)] + 
    K_1 \sigma_X^2 } \right ]}{ \sigma_X^2 +  \sigma_\Lambda^2 (1-K_1)}. \label{eq:sampling_upper}
\end{eqnarray}

Obviously $K_1$ plays a trade-off role: the smaller  $K_1$, the more confident
we can be that our candidate set will contain at least one point
at the main lobe, but also the larger the cardinality of that set,
and consequently the larger the computational cost of the subsequent
optimization algorithm.

Whenever explicit reference to other method defining the candidate set
is not provided, we will assume that the procedure used for defining that set
is initialized with the search interval lower-bound $t_{\textrm{lower}}$
and successive
points are computed based on (\ref{eq:sampling_upper}), until 
the computed point being larger than or equal to the search interval
upper-bound $t_{\textrm{upper}}$.

Note that other criteria could be defined based on this variance-based 
idea for setting the candidate set, as for example:
1) starting
the sampling at $t_{\textrm{upper}}$, and use (\ref{eq:sampling_lower})
for computing
$t(i+1)$, or 2) starting at $t_1$, and use both (\ref{eq:sampling_lower})
and (\ref{eq:sampling_upper})
for computing the next candidate points (the latter is indeed used
in Sect.~\ref{sec:progress_dfe}).

\subsection{Second approach. High-dimensional good lattices}\label{sec:sampling_high_dimen}

Contrarily to the low-dimensional lattice case, in the high-dimensional good
lattice case the variance of the modulo-reduced random vector
is equal to the variance of the non-reduced random vector
as long as this value is smaller than or equal to $t^2 \sigma_\Lambda^2$;
once the input variance is larger than this threshold,
the output variance stalls at $t^2 \sigma_\Lambda^2$. Therefore,
in this scenario it will be specially critical to consider points
in the candidate set such that the variance
of the quantization error is smaller than $t^2\sigma_\Lambda^2$;
otherwise, the main lobe will not be longer detectable.

Taking into account the previous discussion, the method
proposed for sampling the search interval will look for
those $t$ such that
\begin{eqnarray}
  (t_0 - t)^2 \sigma_X^2 + (t-\alpha t_0)^2 \sigma_\Lambda^2 + \sigma_N^2
  = t^2 \sigma_\Lambda^2. \label{eq:puntos_corte}
\end{eqnarray}
This equation has two solutions with $t$, namely
\begin{eqnarray}
  t_{l,\textrm{hd}} = \frac{t_0 \left (\sigma_X^2 + \alpha \sigma_\Lambda^2 \right ) - \sqrt{t_0^2 \alpha \sigma_\Lambda^2\left [ \sigma_X^2 \left ( 2-\alpha\right ) + \alpha \sigma_\Lambda^2 \right ]  - \sigma_N^2 \sigma_X^2}}{\sigma_X^2}, \label{eq:sol_low_hd} \\
  t_{u,\textrm{hd}} = \frac{t_0 \left (\sigma_X^2 + \alpha \sigma_\Lambda^2 \right ) + \sqrt{t_0^2 \alpha \sigma_\Lambda^2\left [ \sigma_X^2 \left ( 2-\alpha\right ) + \alpha \sigma_\Lambda^2 \right ]  - \sigma_N^2 \sigma_X^2}}{\sigma_X^2}. \label{eq:sol_up_hd}
\end{eqnarray}
Be aware that both of them go to $t_0$ when HLR $\rightarrow 
\infty$, and TNLR $<1$.
These solutions will be real, consequently defining the real roots of
(\ref{eq:puntos_corte}) we are interested in, when
\begin{eqnarray}
  |t_0| \geq \sqrt{\frac{\sigma_X^2 \sigma_N^2}{\alpha \sigma_\Lambda^2
\left [ \sigma_X^2 \left ( 2-\alpha\right ) + \alpha \sigma_\Lambda^2 \right]}}; \label{eq:condicion_imag}
\end{eqnarray}
otherwise, $t^2 \sigma_\Lambda^2$ will be smaller than the left term
in (\ref{eq:puntos_corte}) for any $t$, indicating that the main lobe
just does not exist. Be aware that when HLR $\rightarrow \infty$, the last condition
is equivalent to
\begin{eqnarray}
  t_0^2 \sigma_\Lambda^2 \alpha (2-\alpha) \geq \sigma_N^2, \nonumber
\end{eqnarray}
which is nothing but TNLR $\leq  1$.

Note that if (\ref{eq:condicion_imag}) is verified, then
$t_{u,\textrm{hd}} \geq t_0$ holds. Furthermore, $t_{l,\textrm{hd}} \leq t_0$ if and only if
\begin{eqnarray}
  \sigma_\Lambda^2 \geq \frac{\sigma_N^2}{\alpha(2-\alpha) t_0^2}, \nonumber
\end{eqnarray}
or equivalently, TNLR $\leq 1$ (independently of HLR).

Consequently, the algorithm we propose in this scenario for 
sampling the search interval is the following:
\begin{itemize}
\item $t(1) = t_{\textrm{lower}}$.
\item $i = 1$.
\item While $(t(i) < t_{\textrm{upper}})$
  \begin{itemize}
    \item If $|t(i)| \geq \sqrt{\frac{\sigma_X^2 \sigma_N^2}{\alpha \sigma_\Lambda^2
\left [ \sigma_X^2 \left ( 2-\alpha\right ) + \alpha \sigma_\Lambda^2 \right]}}$, then
\begin{itemize}
  \item   $t(i+1) = \frac{t(i) \left (\sigma_X^2 + \alpha \sigma_\Lambda^2 \right ) + \sqrt{t(i)^2 \alpha \sigma_\Lambda^2\left [ \sigma_X^2 \left ( 2-\alpha\right ) + \alpha \sigma_\Lambda^2 \right ]  - \sigma_N^2 \sigma_X^2}}{\sigma_X^2}$.
\end{itemize}
else
\begin{itemize}
  \item $t(i+1) = \frac{t(i) \left [\alpha \sigma_\Lambda^2  + \sigma_X^2 + \sigma_\Lambda
    \sqrt{\sigma_\Lambda^2  [ (1-\alpha)^2 + K_1 (2\alpha - 1)] + 
    K_1 \sigma_X^2 } \right ]}{ \sigma_X^2 +  \sigma_\Lambda^2 (1-K_1)}$
\end{itemize}
    \item $i = i + 1$.
  \end{itemize}
\end{itemize}

It is important to note that for the case where the value of $t(i)$ is so small
that if it were the real scaling factor, then no real solution exists
to (\ref{eq:puntos_corte}), i.e., the quantization error will
be uniformly distributed over the considered Voronoi region for
any value of $t$, we have chosen an alternative strategy for
computing $t(i+1)$ based on the procedure proposed in the last section.
Another possible strategy dealing with the case where $t_{\textrm{lower}}  < \sqrt{\frac{\sigma_X^2 \sigma_N^2}{\alpha \sigma_\Lambda^2
\left [ \sigma_X^2 \left ( 2-\alpha\right ) + \alpha \sigma_\Lambda^2 \right]}}$ is setting $t(1) = t_{\textrm{lower}}$,
and $t(2) = \sqrt{\frac{\sigma_X^2 \sigma_N^2}{\alpha \sigma_\Lambda^2
\left [ \sigma_X^2 \left ( 2-\alpha\right ) + \alpha \sigma_\Lambda^2 \right]}}$,
since as long as $t_0$ (or $t(i+1)$)
were smaller than
the right term in (\ref{eq:condicion_imag}), the target function
will not have main lobe, making impossible the search
of $\hat{t}(\bm z)$.

Similarly to the previous method, alternative criteria could be developed
for computing the candidate set based on this idea.
%%%%%%%%%%%%%%%%%%%%%%%%%%%%%%%%%%%%%%%%%%%%%%%%%%%%%%%%%%%%%%%%
%%%%%%%%%%%%%%%%%%%%%%%%%%%%%%%%%%%%%%%%%%%%%%%%%%%%%%%%%%%%%%%%
%%%%%%%%%%%%%%%%%%%%%%%%%%%%%%%%%%%%%%%%%%%%%%%%%%%%%%%%%%%%%%%%
%%%%%%%%%%%%%%%%%%%%%%%%%%%%%%%%%%%%%%%%%%%%%%%%%%%%%%%%%%%%%%%%

\subsection{Future work}\label{sec:sampling_future}
It would be interesting to study the cardinality of the candidate set (and consequently the computational cost of the subsequent optimization algorithm)
as a function of:
\begin{itemize}
\item $\sigma_X^2$
\item $\sigma_N^2$
\item $\sigma_W^2$
\item $K_1$
\item $n$
\item $P_{e1}$
\end{itemize}

%%%%%%%%%%%%%%%%%%%%%%%%%%%%%%%%%%%%%%%%%%%%%%%%%%%%%%%%%%%%%%%%%%%%%
%%%%%%%%%%%%%%%%%%%%%%%%%%%%%%%%%%%%%%%%%%%%%%%%%%%%%%%%%%%%%%%%%%%%%

\section{Optimization algorithms}

Once the candidate set has been determined, we will look for
the minimum of $L(t, \bm z)$, using as starting points of the optimization
algorithm each point in that set; as it was mentioned above,
for large values of $n$ the considered target function
will be convex in the main lobe, so for a large
number of observations one can ensure the global minimum of the
target function to be found. Several choices can be proposed
for this optimization algorithm; next, some of them are
detailed. 

\subsection{Derivative-based optimization}

This method determines for each point in the candidate set a point
where the derivative of the target
function is null, providing as final estimation of $t_0$ that
null-derivative point where the target function is minimized. Namely:

\begin{itemize}
\item For each $t(i)$ in the candidate set:
\begin{itemize}
\item $s_0 = \textrm{sign} \left [L(t(i) + \epsilon_1, \bm z)-  L(t(i), \bm z)\right]$.
In the reported experiments $\epsilon_1 = 10^{-5}$.
\item $s_1 = s_0$.
\item $\textrm{step} = 10^{-3}$.
\item While $(s_1 == s_0)$
\begin{itemize}
  \item $\textrm{step} = 2 \textrm{step}$.
  \item $t_{\textrm{aux}} = |t(i) - s_0 \textrm{step}|$.
  \item $s_1 = \textrm{sign}\left [L(t_{\textrm{aux}} + \epsilon_1, \bm z)-  L(t_{\textrm{aux}}, \bm z)\right]$.
\end{itemize}
\item $t_{l} = \min(t(i), t_{\textrm{aux}})$.
\item $t_{u} = \max(t(i), t_{\textrm{aux}})$.
\item While $((t_{u} - t_{l}) >\epsilon_2)$ (in the reported experiments $\epsilon_2 = 10^{-5}$)
\begin{itemize}
\item $t_{\textrm{aux}} = (t_{u} + t_{l})/2$.
  \item $s_1 = \textrm{sign}\left [L(t_{\textrm{aux}} + \epsilon_1, \bm z)-  L(t_{\textrm{aux}}, \bm z)\right]$.
\item If $(s_1 > 0)$ then
\begin{itemize}
\item  $t_{u} = t_{\textrm{aux}}$.
\end{itemize}
\item else
\begin{itemize}
\item  $t_{l} = t_{\textrm{aux}}$.
\end{itemize}
\end{itemize}
\item L\_value$(i) = L(t_{\textrm{aux}}, \bm z)$.
\item null\_der$(i)= t_{\textrm{aux}}$.
\end{itemize}
\item $i^* = \argmin_i$ L\_value$(i)$.
\item If $L($null\_der$(i^*), \bm z) < L(t_1, \bm z)$ then
\begin{itemize}
\item $\hat{t}(\bm z) = $null\_der$(i^*)$.
 \end{itemize}
\item else
\begin{itemize}
\item $\hat{t}(\bm z) = t_1$.
\end{itemize}
\end{itemize}

{\bf Advantages}:
\begin{itemize}
\item The provided estimate is ensured to be at least a local minimum.
\item Almost exclusively based on evaluating the target function. Additional
logic is very simple.
\end{itemize}

{\bf Drawbacks}:
\begin{itemize}
\item Very high computational cost. The target function must be evaluated at a large number of points for each initial candidate.
\item As it is based on the derivative of the target function, the target
function must be evaluated twice for each point that is really tested.
\item The computational cost will depend on the precision one wants 
to achieve ($\epsilon_2$).
\end{itemize}

\subsection{Decision-Aided optimization}

The next method we proposed to estimate $t_0$ from the candidate set
decodes the received sequence assuming that the initial considered point
is a good approximation to $t_0$, so the decoded centroid
will be indeed the one used at the embedder. From that estimate
of the centroid used at the embedder, the scaling factor
minimizing the distance from the scaled centroid to the received
signal is computed. 

It comprises the following steps:

\begin{itemize}
\item For each $t(i)$ in the candidate set:
\begin{itemize}
  \item $\textrm{centroid} = Q_{\Lambda}\left (\frac{\bm z}{t(i)} -\bm d \right) +\bm  d$.
  \item $t_{\textrm{aux}} = \frac{||\bm z||^2}{\bm z^T \cdot \textrm{centroid}}$.
  \item L\_value$(i) = L(t_{\textrm{aux}}, \bm z)$.
  \item $t_{\textrm{dfe}}(i) = t_{\textrm{aux}}$.
\end{itemize}
\item $i^* = \argmin_i$ L\_value$(i)$.
\item If $L(t_{\textrm{dfe}}(i^*), \bm z) < L(t_1, \bm z)$ then
\begin{itemize}
\item $\hat{t}(\bm z) = t_{\textrm{dfe}}(i^*)$.
 \end{itemize}
\item else
\begin{itemize}
\item $\hat{t}(\bm z) = t_1$.
\end{itemize}
\end{itemize}

{\bf Advantages}:
\begin{itemize}
\item Reduced computational cost. The target function is only evaluated once
per initial candidate point.
\item Additional computational cost, beyond the target function evaluation,
is cheap.
\item The computational cost does not depend on the desired precision.
\end{itemize}

{\bf Drawbacks}:
\begin{itemize}
\item The resulting estimate is not guaranteed to be a local minimum of 
the target function. The target function used for updating
$t(i)$, i.e., the distance from the scaled estimated centroid to the received
signal, is not the target function of our optimization problem (although
under the proposed hypotheses it will be a good approximation).
\end{itemize}

\subsection{Progressively widened DA}\label{sec:progress_dfe}

The two methods proposed so far verify that their computational cost
is proportional to the number of points in the candidate set.
Nevertheless, most of those points will be far from $t_0$, as the
search interval will be usually pessimistic, since the probability
of the real scaling factor being out of that interval will be set to a very
small value. Therefore, a way for reducing the computational
cost of the previous schemes would be to perform the optimization
algorithm first at those points that one assumes to be most likely
close to $t_0$, and check if one can trust on them
for being a good estimate of $t_0$; if one can, the
search can be stopped. In order to do this suitability check,  we will
need some
kind of test or measure informing us about the goodness of that point.
Given that we have a statistical
characterization of the target function both for $t = t_0$ (provided
in Sect.~\ref{sec:target_function_t_0}) and for
$t$ not being close to $t_0$ (following the description
provided in Sect.\ref{sec:pdf_no_eq}), we will exploit these
characterizations for proposing statistical tests that can
help us 
to determine if a given point $t$ is likely to be $t_0$
or not.

Therefore, departing from a initial point that we think that could be
a good estimate of $t_0$, e.g. $t_1$, we will perform an optimization
in order to minimize the value of the target function, obtaining
a value $t_2$. If the new point $t_2$
passes the mentioned statistical test, then the search will be stopped and that point will be provided as our
estimation of $t_0$. Otherwise, we will evaluate the statistical test
for a larger and a smaller value of $t$. Consequently we will progressively
widen the search interval taking as initial point $t_1$ (or any other
point that is assumed to be a good approximation to $t_0$), so
in most cases it will not be
necessary to perform an optimization algorithm for all the points
in the candidate set, but just in those that are considered to be
more likely, allowing to reduce the computational cost of 
the resulting scheme.

Several choices can be proposed for the computation of
$t_2$; given that our final target is to have a good estimate of $t_0$,
we could use the two optimization algorithms proposed in the last
sections in order to compute that point. Indeed, in the subsequent
description of the current method we will focus on the case where
the DA-based scheme is used for computing the mentioned point. In that case, 
the current optimization method can be described for the scalar
quantizer case as:

\begin{itemize}
\item $t_{u} = t_1$
\item $t_{l} = t_1$
\item $i = 1$
\item found$=0$
\item While ((($t_{u} \leq  t_{\textrm{upper}}$) or ($t_{l} \geq t_{\textrm{lower}}$)) and (found$=0$))
  \begin{itemize}
  \item If ($t_{u} \leq t_{\textrm{upper}}$) then
    \begin{itemize}
    \item $\textrm{centroid} = Q_{\Lambda}\left (\frac{\bm z}{t_u} -\bm d \right) + \bm d$.
    \item $t_{\textrm{aux}} = \frac{||\bm z||^2}{\bm z^T \cdot \textrm{centroid}}$.
    \item L\_value$(i) = L(t_{\textrm{aux}}, \bm z)$.
    \item $t_{\textrm{dfe}}(i) = t_{\textrm{aux}}$.
    \item $i = i+1$.
    \item 
      \begin{eqnarray}
        t_u = \frac{t_u \left [\alpha \sigma_\Lambda^2  + \sigma_X^2 + \sigma_\Lambda
            \sqrt{\sigma_\Lambda^2  [ (1-\alpha)^2 + K_1 (2\alpha - 1)] + 
              K_1 \sigma_X^2 } \right ]}{ \sigma_X^2 + \sigma_\Lambda^2 (1-K_1)}. \nonumber
      \end{eqnarray}
    \end{itemize}
  \item If ($t_{l}\geq t_{\textrm{lower}}$) then    
    \begin{itemize}
    \item If $(i\neq2)$ then
      \begin{itemize}
      \item $\textrm{centroid} = Q_{\Lambda}\left (\frac{\bm z}{t_l} - \bm d \right) + \bm d$.
      \item $t_{\textrm{aux}} = \frac{||\bm z||^2}{\bm z^T \cdot \textrm{centroid}}$.
      \item L\_value$(i) = L(t_{\textrm{aux}}, \bm z)$.
      \item $t_{\textrm{dfe}}(i) = t_{\textrm{aux}}$.
      \item $i = i+1$.
      \end{itemize}
    \item 
      \begin{eqnarray}
        t_l = \frac{t_l \left [\alpha \sigma_\Lambda^2  + 12 \sigma_X^2 - \sigma_\Lambda
            \sqrt{\sigma_\Lambda^2  [ (1-\alpha)^2 + K_1 (2\alpha - 1)] + 
              K_1 \sigma_X^2 } \right ]}{ \sigma_X^2 +  \sigma_\Lambda^2 (1-K_1)}, \nonumber
      \end{eqnarray}
      
    \end{itemize}
  \item $i^* = \argmin_i$ L\_value$(i)$.
  \item If $((L(t_{\textrm{dfe}}(i^*), \bm z) < \tau_{t = t_0}(t_{\textrm{dfe}}(i^*), P_{e2}))$ and $
    (L(t_{\textrm{dfe}}(i^*), \bm z) < \tau_{t \neq t_0}(t_{\textrm{dfe}}(i^*), P_{e3})))$ then
    \begin{itemize}
    \item found $=1$
    \end{itemize}
  \end{itemize}
\item $i^* = \argmin_i$ L\_value$(i)$.
\item If $L(t_{\textrm{dfe}}(i^*), \bm z) < L(t_1, \bm z)$ then
  \begin{itemize}
  \item $\hat{t}(\bm z) = t_{\textrm{dfe}}(i^*)$.
  \end{itemize}
\item else
  \begin{itemize}
  \item $\hat{t}(\bm z) = t_1$.
  \end{itemize}
\end{itemize}
where, assuming that the Gaussian approximation can be used,
\begin{eqnarray}
  \tau_{t = t_0}(t, P_{e2})) =   \textrm{E}[L(t, \bm Z)|t = t_0] + 
\sqrt{\textrm{Var}[L(t, \bm Z)|t = t_0]} Q^{-1}(P_{e2}), \nonumber
\end{eqnarray}
and
\begin{eqnarray}
  \tau_{t \neq t_0}(t, P_{e3})) =   \textrm{E}[L(t, \bm Z)|t \neq t_0] + 
\sqrt{\textrm{Var}[L(t, \bm Z)|t \neq t_0]} Q^{-1}(1-P_{e3}), \nonumber
\end{eqnarray}
with $\textrm{E}[L(t, \bm Z)|t = t_0]$ 
taking the value in  (\ref{eq:mean_mom_t_0}), and $\textrm{Var}[L(t, \bm Z)|t = t_0]$
by (\ref{eq:var_mom_t_0}) for the scalar quantizer case and
(\ref{eq:var_mom_t_0_mult}) for the high-dimensional good lattice case;
for $\textrm{E}[L(t, \bm Z)|t \neq t_0]$ and
$\textrm{Var}[L(t, \bm Z)|t \neq t_0]$ we use
\begin{eqnarray}
  \textrm{E}[L(t, \bm Z)|t \neq t_0] = \frac{n t^2 \sigma_\Lambda^2}{
\sigma_N^2 + (1-\alpha)^2 t^2 \sigma_\Lambda^2} + n \log \left [ 2 \pi \left ( \sigma_N^2 + (1-\alpha)^2 
  t^2 \sigma_\Lambda^2 \right ) \right ] + \frac{||\bm z||^2}{\sigma_X^2 t^2} ,\nonumber
\end{eqnarray}
and
\begin{eqnarray}
\textrm{Var}[L(t, \bm Z)|t \neq t_0] &=& \frac{n 144 t^4 \sigma_\Lambda^4 /180 }{\left (\sigma_N^2 + (1-\alpha)^2 t^2 \sigma_\Lambda^2 \right )^2}, \nonumber
\end{eqnarray}
for the scalar quantizer case, while for the high-dimensional good lattice
case
\begin{eqnarray}
\textrm{Var}[L(t, \bm Z)|t \neq t_0] &=& \frac{2 n t^4 \sigma_\Lambda^4 }{\left (\sigma_N^2 + (1-\alpha)^2 t^2 \sigma_\Lambda^2\right )^2}; \nonumber
\end{eqnarray}
note that the mean and variance used for the case $t \neq t_0$ are
not those at (\ref{eq:pdf_t_neq_t0}), as the values there depend on $t_0$,
which is not available when running the optimization algorithm;
therefore, we have preferred to consider the last term in the target function
to be deterministic, as it was also done in Sect.~\ref{sec:another_possible}.

Furthermore, when high-dimensional good lattices are used, one could
replace the  computation of $t_u$ and $t_l$ in the previous description
of the algorithm, which is based on
$t_{u,\textrm{ld}}$
and $t_{l,\textrm{ld}}$, by its counterpart respectively based on 
$t_{u,\textrm{hd}}$
and $t_{l,\textrm{hd}}$, defined in
Sect.~\ref{sec:sampling_high_dimen}, or on some of its
variants depending on the verification of (\ref{eq:condicion_imag}).

{\bf Advantages}:
\begin{itemize}
\item The number of points used for initializing the optimization algorithm
is reduced in comparison with those used by the previously proposed
algorithms.
\end{itemize}

{\bf Drawbacks}:
\begin{itemize}
\item The statistical tests performed for checking the suitability
of the studied points increase the computational cost per
initial point.
\end{itemize}

\subsection{On-line DA}

All the methods presented so far are designed for working with
blocks of $n$ samples of the received signal. Nevertheless, 
practical communications schemes usually require on-line algorithms,
where an algorithm output is generated each time a sample of the received
signal is input. With this target in mind, an on-line version of
the DA-based algorithm presented above was developed. Denoting
by $\bm z^{(j)} = (z_1, \cdots, z_j)$, the proposed scheme can be
summarized as follows:

\begin{itemize}
\item locked $=0$
\item When the $j$-th sample is available:
  \begin{itemize}
  \item If (locked$=0$) then
    \begin{itemize}
    \item Compute $t_1$ using $\bm z^{(j)}$ and a scheme of those introduced in
      Sect.~\ref{sec:t_1_computation}.
    \end{itemize}
  \item else
    \begin{itemize}
    \item $t_1 = t_{\textrm{on-line}}$
    \end{itemize}
  \item $\textrm{centroid} = Q_{\Lambda}\left (\frac{\bm z^{(j)}}{t_1} -\bm d^{(j)} \right) + \bm d^{(j)}$.
  \item $t_3 = \frac{||\bm z^{(j)}||^2}{(\bm z^{(j)})^T \cdot \textrm{centroid}}$.
  \item If $((L(t_3, \bm z^{(j)}) < \tau_{t = t_0}(t_3, P_{e2}))$ and $
    (L(t_3, \bm z^{(j)}) < \tau_{t \neq t_0}(t_3, P_{e3})))$ then
    \begin{itemize}
    \item locked $=1$
    \item $t_{\textrm{on-line}} = t_3$
    \end{itemize}
  \item else 
    \begin{itemize}
      \item Compute the interval search from $t_3$ and $\bm z^{(j)}$, for $P_{e1}$.
        \item Compute the candidate set from the interval search.
      \item For each $t(i)$ in the candidate set:
        \begin{itemize}
        \item $\textrm{centroid} = Q_{\Lambda}\left (\frac{\bm z^{(j)}}{t(i)} -\bm d^{(j)} \right) + \bm d^{(j)}$.
        \item $t_{\textrm{aux}} = \frac{||\bm z^{(j)}||^2}{(\bm z^{(j)})^T \cdot \textrm{centroid}}$.
        \item L\_value$(i) = L(t_{\textrm{aux}}, \bm z^{(j)})$.
        \item $t_{\textrm{dfe}}(i) = t_{\textrm{aux}}$.
        \end{itemize}
      \item $i^* = \argmin_i$ L\_value$(i)$.
      \item If $L(t_{\textrm{dfe}}(i^*), \bm z^{(j)}) < L(t_1, \bm z^{(j)})$ then
        \begin{itemize}
        \item $t_1 = t_{\textrm{dfe}}(i^*)$.
        \end{itemize}
      \item If $((L(t_1, \bm z^{(j)}) < \tau_{t = t_0}(t_1, P_{e4}))$ and $
        (L(t_1, \bm z^{(j)}) < \tau_{t \neq t_0}(t_1, P_{e5})))$ then
        \begin{itemize}
        \item locked $=1$.
        \item $t_{\textrm{on-line}} = t_1$.
        \end{itemize}
      \item else
        \begin{itemize}
        \item locked $=0$.
        \end{itemize}
    \end{itemize}
  \end{itemize}
\item The output estimation is $t_{\textrm{on-line}}$.
\end{itemize}

{\bf Advantages}:
\begin{itemize}
\item Once the system is locked, a very reduced computational cost
is required.
\item Due to the statistical characterization of $L(t, \bm z)$, one
  knows when $t_{\textrm{on-line}}$ is a good estimate.
\item It allows to establish different threshold probabilities depending
on the system being initially locked or not ($P_{e2}$ and $P_{e3}$, versus
$P_{e4}$ and $P_{e5}$, respectively).
\end{itemize}

{\bf Drawbacks}:
\begin{itemize}
\item If the system unlocks, one must define a search interval and
run the optimization algorithm considering each point in the candidate set
as initial point (although a strategy similar to {\it Progressively Widened
DA} might be also adopted); this could make difficult
a real-time implementation where all the received samples were
processed.
\end{itemize}

\subsection{General remark}

Finally, we would like to remark that there is a series of parameters
involved in all the proposed schemes that allow to achieve a trade-off
between computational cost and estimation algorithm performance.
Some of these parameters are $P_{e1}$, $P_{e2}$, $P_{e3}$, $P_{e4}$, $P_{e5}$, 
and $K_1$.  Setting these parameters to different values, one can range from
the basic variance-based estimator to more sophisticated and accurate
estimates. Therefore, the choice of those parameters values will
depend on the considered application scenario.

Furthermore, one could also think of other optimization algorithms
specially designed for working in strongly computationally constrained
scenarios. For example, one could envisage the case where only
a fixed number of candidates can be considered as initial points of the optimization algorithm,
due to real-time computation restrictions; in that case, a possible choice could
be to take as initial point of the search interval sampling 
method $t_1$, and 
then calculate the two sequences of both
larger and smaller candidates, obtained by applying the
formulas derived in Sect.~\ref{sec:samp_se_int}, until the
upper-bounded
number of initial points being achieved.

%%%%%%%%%%%%%%%%%%%%%%%%%%%%%%%%%%%%%%%%%%%%%%%%%%%%%%%%%%%%%%%%%%%%%
%%%%%%%%%%%%%%%%%%%%%%%%%%%%%%%%%%%%%%%%%%%%%%%%%%%%%%%%%%%%%%%%%%%%%

\section{Fisher information analysis}\label{sec:crb_approx1}

For calculating the Fisher information of the proposed scheme, we will use
the pdf approximation in (\ref{eq:multidimen_natural}), whose minus logarithm
is
\begin{eqnarray}
&\frac{1}{2}\left [ \frac{||(\bm z - t\bm d )\moda (t \Lambda)||^2}{ \sigma_N^2 + (1-\alpha)^2  t^2 \sigma_\Lambda^2  } 
  + n \log \left ( 2 \pi \left ( \sigma_N^2 + (1-\alpha)^2 
  t^2 \sigma_\Lambda^2 \right ) \right )\right .\nonumber \\
  & \left .+  \frac{||\bm z||^2}{ \sigma_X^2 t^2}  + n \log \left (2 \pi \sigma_X^2 \right )\right ]; \label{eq:llf1}
\end{eqnarray}
in order to statistically characterize the first term, we will
consider that, as it was shown above, $(\bm z - t\bm d )\moda (t \Lambda) = \Big [ (t_0-t)\bm x + (t-\alpha t_0) \left [(\bm x - \bm d)\moda\Lambda  \right ] + n \Big]\moda(t\Lambda)$, so (\ref{eq:llf1}) can be rewritten as
\begin{eqnarray}
&\frac{1}{2}\left [ \frac{\left \|\big [ (t_0-t)\bm x + (t-\alpha t_0) \left [(\bm x - \bm d)\moda\Lambda  \right ] + \bm n \big]\moda(t\Lambda)\right \|^2}{ \sigma_N^2 + (1-\alpha)^2  t^2 \sigma_\Lambda^2  } 
  + n \log \left ( 2 \pi \left ( \sigma_N^2 + (1-\alpha)^2 
  t^2 \sigma_\Lambda^2 \right ) \right ) \right . \nonumber \\
  &  +  \frac{||(\bm x + \bm w)t_0 + \bm n||^2}{ \sigma_X^2 t^2}  + n \log \left (2 \pi \sigma_X^2 \right )\Bigg ]. \label{eq:llf2}
\end{eqnarray}
In order to calculate the
Fisher information we will evaluate the second derivative of (\ref{eq:llf2})
with respect to $t$ at $t_0$; consequently, we will be only interested
in those values of $t$ in an arbitrarily small neighborhood
around $t_0$.
Thus,  as far as $(1-\alpha) t_0^2 \sigma_\Lambda^2 + \sigma_N^2
<<  t_0^2 \sigma_\Lambda^2$, i.e., as far as TNLR $\rightarrow 0$ for the
low-dimensional lattice case, or
$(1-\alpha) t_0^2 \sigma_\Lambda^2 + \sigma_N^2
<  t_0^2 \sigma_\Lambda^2$, i.e., as far as TNLR $<1$ for the
high-dimensional good lattice case, we can 
neglect the modulo-$\Lambda$ reduction in the first term,
and (\ref{eq:llf2}) can be rewritten as
\begin{eqnarray}
&\frac{1}{2}\left [ \frac{\left \|(t_0-t)\bm x + (t-\alpha t_0) \left [(\bm x - \bm d)\moda\Lambda  \right ] + \bm n \right \|^2}{ \sigma_N^2 + (1-\alpha)^2  t^2 \sigma_\Lambda^2  } 
  + n \log \left ( 2 \pi \left ( \sigma_N^2 + (1-\alpha)^2 
  t^2 \sigma_\Lambda^2 \right ) \right ) \right . \nonumber \\
  &  +  \frac{||(\bm x + \bm w)t_0 + \bm n||^2}{ \sigma_X^2 t^2}  + n \log \left (2 \pi \sigma_X^2 \right )\Bigg ]. \label{eq:llf3}
\end{eqnarray}
Calculating the first derivative of the last expression with respect to $t$,
one obtains
\begin{eqnarray}
&\frac{1}{2}\sum_{j=1}^n \Bigg [-\frac{2 \left( (t_0-t) x_j + (t-\alpha t_0)w_j + n_j\right ) \left (
  \sigma_N^2(x_j - w_j) + (1-\alpha)^2\sigma_\Lambda^2 t(n_j + t_0(x_j - \alpha  w_j)) \right )}{\left ( \sigma_N^2 + (1-\alpha)^2  t^2 \sigma_\Lambda^2 \right)^2} 
\nonumber \\& + 
\frac{2(1-\alpha)^2\sigma_\Lambda^2 t}{ \sigma_N^2 + (1-\alpha)^2  t^2 \sigma_\Lambda^2}
-\frac{2((x_j + w_j)t_0 + n_j)^2}{\sigma_X^2 t^3} \Bigg ], \label{eq:der1_tochisima}
\end{eqnarray}
and the second derivative is
\begin{eqnarray}
&\frac{1}{2}\sum_{j=1}^n  &\Bigg [2\Big(\sigma_N^4(w_j - x_j)^2 + (1-\alpha)^4 \sigma_\Lambda^4 t^2
    \big ((x_j-\alpha  w_j)t_0 +n_j \big) \left (
    (3t_0 - 2t)x_j - (3t_0 \alpha - 2 t) w_j + 3 n_j \right )  \nonumber \\
&& - 
    (1-\alpha)^2 \sigma_\Lambda^2 \sigma_N^2\Big [3 t^2 ( w_j - x_j)^2 + n_j^2
     - 6 t t_0(w_j - x_j) (\alpha w_j - x_j) + t_0^2 (x_j - \alpha w_j)^2
\nonumber \\
&&
     + 2 n_j\big(3t(w_j-x_j) + t_0(x_j - \alpha w_j)\big) \Big ]\Big )\left ( \sigma_N^2 + (1-\alpha)^2\sigma_\Lambda^2 t^2 \right)^{-3}   \nonumber \\
&& + \frac{2(1-\alpha)^2\sigma_\Lambda^2  \sigma_N^2 - 2(1-\alpha)^4 \sigma_\Lambda^4 t^2}{(\sigma_N^2 + (1-\alpha)^2  \sigma_\Lambda^2 t^2)^2 }+ \frac{6\left((x_j + w_j)t_0 + n_j\right)^2}{\sigma_X^2 t^4}
\Bigg ]. \nonumber
\end{eqnarray}
Calculating the mean with respect to $\bm X$, $\bm W$, and $\bm N$, one obtains
\begin{eqnarray}
&& n \Big(\sigma_N^4(\sigma_X^2 + \alpha^2 \sigma_\Lambda^2) + (1-\alpha)^4 \sigma_\Lambda^4 t^2
    \big (t_0(3t_0 -2t)\sigma_X^2 + t_0(3\alpha t_0 - 2t)\alpha^3 \sigma_\Lambda^2 + 3\sigma_N^2 \big)  \nonumber \\
    && - (1-\alpha)^2 \sigma_\Lambda^2 \sigma_N^2 \Big [3t^2 (\sigma_X^2 + \alpha^2 \sigma_\Lambda^2)
+\sigma_N^2 - 6 t t_0(\sigma_X^2 + \alpha^3 \sigma_\Lambda^2) + t_0^2(\sigma_X^2 + 
\alpha^4 \sigma_\Lambda^2)  \Big ] \Big )
\nonumber \\
&&
\left ( \sigma_N^2 + (1-\alpha)^2 \sigma_\Lambda^2 t^2 \right)^{-3}   + n \frac{ (1-\alpha)^2\sigma_\Lambda^2   \sigma_N^2 - (1-\alpha)^4 \sigma_\Lambda^4 t^2}{( \sigma_N^2 + (1-\alpha)^2 \sigma_\Lambda^2t^2)^2 }+ \frac{3 n \left((\sigma_X^2 + \alpha^2 \sigma_\Lambda^2)t_0^2 + \sigma_N^2 \right)}{\sigma_X^2 t^4}; \nonumber  
\end{eqnarray}
evaluating that expression at $t= t_0$,
the result is the Fisher information
\begin{eqnarray}
I(t_0)&=&n \Big(\sigma_N^4(\sigma_X^2 + \alpha^2 \sigma_\Lambda^2) + (1-\alpha)^4 \sigma_\Lambda^4 t_0^2
    \big (t_0^2\sigma_X^2 + t_0^2(3\alpha - 2)\alpha^3 \sigma_\Lambda^2 + 3\sigma_N^2 \big)  \nonumber \\
    && - (1-\alpha)^2\sigma_\Lambda^2 \sigma_N^2 \Big [3t_0^2 (\sigma_X^2 + \alpha^2 \sigma_\Lambda^2)
+\sigma_N^2 - 6 t_0^2(\sigma_X^2 + \alpha^3 \sigma_\Lambda^2) + t_0^2(\sigma_X^2 + 
\alpha^4 \sigma_\Lambda^2)  \Big ] \Big )
\nonumber \\
&&
\left (\sigma_N^2 + (1-\alpha)^2 t_0^2\sigma_\Lambda^2 \right)^{-3}   + n\frac{  (1-\alpha)^2\sigma_\Lambda^2   \sigma_N^2 - (1-\alpha)^4 \sigma_\Lambda^4 t_0^2}{(\sigma_N^2 + (1-\alpha)^2 \sigma_\Lambda^2t_0^2)^2 }+ \frac{3 n \left((\sigma_X^2 + \alpha^2 \sigma_\Lambda^2)t_0^2 + \sigma_N^2 \right)}{\sigma_X^2 t_0^4}. \nonumber  
\end{eqnarray}

In order to get a larger insight into the last formula, and be able
to use it for further theoretical derivations, we will approximate the first
term taking into account that HLR $\rightarrow \infty$, and TNLR $<1$, so
it will asymptotically converge to
\begin{eqnarray}
&&\frac{n \Big( \sigma_X^2 \left (\sigma_N^4 + (1-\alpha)^4 \sigma_\Lambda^4 t_0^4
     + 2(1-\alpha)^2 \sigma_\Lambda^2 \sigma_N^2 t_0^2\right)\Big)}{
\left (\sigma_N^2 + (1-\alpha)^2\sigma_\Lambda^2 t_0^2 \right)^{3}} = 
\frac{n \Big( \sigma_X^2 \left (\sigma_N^2 + (1-\alpha)^2
   \sigma_\Lambda^2 t_0^2\right)^2\Big)}{
\left (\sigma_N^2 + (1-\alpha)^2\sigma_\Lambda^2 t_0^2 \right)^{3}} = \nonumber \\
&&\frac{ n \sigma_X^2 }{
\sigma_N^2 + (1-\alpha)^2 \sigma_\Lambda^2 t_0^2}
, \label{eq:asympt_fisher_info}
\end{eqnarray}
going to infinity under the aforementioned conditions.
Concerning the second term, it can be lower-bounded as 
\begin{eqnarray}
  n\frac{  (1-\alpha)^2\sigma_\Lambda^2   \sigma_N^2 - (1-\alpha)^4 \sigma_\Lambda^4 t_0^2}{(\sigma_N^2 + (1-\alpha)^2 \sigma_\Lambda^2t_0^2)^2 } \geq
  n\frac{- (1-\alpha)^4 \sigma_\Lambda^4 t_0^2}{(\sigma_N^2 + (1-\alpha)^2 \sigma_\Lambda^2t_0^2)^2 } \geq n\frac{- (1-\alpha)^4 \sigma_\Lambda^4 t_0^2}{( (1-\alpha)^2 \sigma_\Lambda^2t_0^2)^2 } = \frac{-n}{t_0^2}; \nonumber
\end{eqnarray}
similarly, it can be upper-bounded as
\begin{eqnarray}
  n\frac{  (1-\alpha)^2\sigma_\Lambda^2   \sigma_N^2 - (1-\alpha)^4 \sigma_\Lambda^4 t_0^2}{(\sigma_N^2 + (1-\alpha)^2 \sigma_\Lambda^2t_0^2)^2 }  \leq 
  n\frac{  (1-\alpha)^2\sigma_\Lambda^2   \sigma_N^2 }{(\sigma_N^2 + (1-\alpha)^2 \sigma_\Lambda^2t_0^2)^2 } \leq   n\frac{(1-\alpha)^2\sigma_\Lambda^2   \sigma_N^2 }{2(1-\alpha)^2\sigma_\Lambda^2   \sigma_N^2 t_0^2} = \frac{n}{2 t_0^2}. \nonumber
\end{eqnarray}
Finally, based on HLR $\rightarrow \infty$, and TNLR $<1$, the last
term will go to $\frac{3n}{t_0^2}$. Therefore, as long as $t_0 > 0$,
HLR $\rightarrow \infty$, and TNLR $ < 1$, 
the Fisher information will asymptotically converge to (\ref{eq:asympt_fisher_info}). In terms of $\sigma_X^2$, $\sigma_W^2$, $\sigma_N^2$, $t_0$, and $\alpha$
(\ref{eq:asympt_fisher_info}) can be written as
\begin{eqnarray}
I(t_0) = \frac{ n \sigma_X^2 }{\sigma_N^2 + \frac{(1-\alpha)^2}{\alpha^2} \sigma_W^2 t_0^2}.\label{eq:crb_quant2}
\end{eqnarray}

It is interesting to note that the last function is monotonically
increasing with $\alpha$.
Nevertheless, in order to TNLR $< 1$ (the relaxed condition on {\bf Hypothesis 3} corresponding to high-dimensional good lattices), the following upper-bound
on the feasible value of $\alpha$ can be established
\begin{eqnarray}
  \alpha \leq \min \left (\frac{2\sigma_W^2 t_0^2}{\sigma_N^2 + \sigma_W^2 t_0^2} -\epsilon, 1 \right )
  \triangleq \alpha_{\textrm{sup-FI}},  \nonumber
\end{eqnarray}
where $\epsilon > 0$ is arbitrarily small, and the left term in the $\min$ function  is nothing but twice the $\alpha$ proposed by Costa, after replacing the 
embedded signal power by its effective value at the receiver;
$\alpha_{\textrm{sup-FI}}$ stands for the supremum of $\alpha$ values
such that TNLR $<1$, which at the same time is arbitrarily close
to the value of $\alpha$ maximizing the Fisher information.

Remember that for the low-dimensional lattice case, the condition on 
TNLR is TNLR $\rightarrow 0$, being much more strict, and constraining
$\alpha$ to be smaller than $\alpha_{\textrm{sup-FI}}$. Consequently, the value
of the Fisher information for the scalar/low-dimensional lattice quantizer
case will be smaller than that achieved for high-dimensional good lattices,
proving the goodness of the latter for this application.

In any case, the difference between the value of
$\alpha$ maximizing the Fisher information
for low-dimensional lattices and $\alpha_{\textrm{sup-FI}}$
will verify:
\begin{itemize}
\item The larger the shaping-gain of the considered lattice, the closer both
values will be, as the fundamental Voronoi region of that lattice will
be more similar to a hypersphere, reducing the probability that the total
noise is modulo-$\Lambda$ reduced.
\item The smaller the effective WNR, $\sigma_W^2 t_0^2/\sigma_N^2$, the
closer both values will be. Indeed, the smaller the effective WNR, the smaller
will be the importance of the AWGN introduced by the channel in the total
noise, and the more important the role of the self-noise; nevertheless, the self-noise
has the same shape that the fundamental Voronoi region of the considered lattice,
so it does not produce modulo-$\Lambda$ reduction.
\end{itemize}

From a geometric point of view, one can see the links between the Fisher
information  and
the curvature (usually defined as the second derivative
of the considered function) of the target function at $t=t_0$.

\subsection{Fisher information based on Erez's approach}

In order to show that the obtained result could be also achieved
by using Erez's lattice-quantization approach, instead of the one followed in
the last sections (more related to Chen and Wornell's interpretation),
in this section we summarize what would happen if Erez's quantizing
strategy were followed.
For the sake of avoiding unnecessary repetition, we will focus on those
aspects that differ from the approach introduced above.

In Erez's approach, the watermarked signal would be calculated as
\begin{eqnarray}
  \bm y = \bm x - [(\alpha \bm x - \bm d) \moda \Lambda], \nonumber
\end{eqnarray}
so
\begin{eqnarray}
  f_{\bm Z|T, K}(\bm z|t, \bm d) &\approx& 
  \frac{|\mathcal{V}(\Lambda)| e^{-\frac{||\bm z||^2}{2 \sigma_X^2 t^2}}}{\left (2 \pi
\sigma_X^2 \right)^{n/2}} \frac{ e^{- \frac{||(\alpha \bm z - t\bm d )\moda (t \Lambda)||^2}{2 \left ( \alpha^2 \sigma_N^2 + (1-\alpha)^2 t^2 \sigma_\Lambda^2 \right ) } }}{\Big [ 2 \pi \left ( \alpha^2 \sigma_N^2 + (1-\alpha)^2 t^2 \sigma_\Lambda^2 \right )  \Big]^{n/2}},  \label{eq:approx_pdf_erez}
\end{eqnarray}
and
\begin{eqnarray}
  \left ( \alpha \bm z - t \bm d \right ) \moda (t \Lambda) &=& \left [\alpha t_0 \left ( \bm x - \Big[(\alpha \bm x - \bm d) \moda \Lambda\Big] \right ) + \alpha \bm n - t \bm d \right] \moda (t \Lambda) \nonumber \\
&=& \left [t_0 \left ( \alpha \bm x - \alpha  
    \Big [\left (\alpha \bm x - \bm d \right ) \moda \Lambda \Big ]   \right) + \alpha \bm n - t \bm d \right] \moda (t \Lambda) \nonumber  \\
&=& \Bigg [(t_0-t) \alpha \bm x  - (t_0 - t) \alpha \Big [ \left ( \alpha \bm x - \bm d \right )
\moda \Lambda \Big] + \alpha \bm n \nonumber \\
&+& t \left (\alpha \bm x - \bm d - \alpha \Big [ \left ( \alpha \bm x - \bm d 
 \right ) \moda \Lambda \Big ] \right)  \Bigg]\moda(t\Lambda) \nonumber \\
&=& \Bigg [(t_0-t) \alpha \bm x  - (t_0 - t) \alpha \Big [ \left ( \alpha \bm x - \bm d \right )
\moda \Lambda \Big] + \alpha \bm n \nonumber \\
&+& t \left ((\alpha \bm x - \bm d)\moda\Lambda - \alpha \Big [ \left ( \alpha \bm x - \bm d 
 \right ) \moda \Lambda \Big ] \right)  \Bigg]\moda(t\Lambda) \nonumber \\
&=& \Bigg [(t_0-t) \alpha \bm x  - (t_0 - t) \alpha \Big [ \left ( \alpha \bm x - \bm d \right )
\moda \Lambda \Big] + \alpha \bm n \nonumber \\
&+& t (1-\alpha) \Big ( \left ( \alpha \bm x - \bm d 
 \right ) \moda \Lambda \Big )   \Bigg]\moda(t\Lambda) \nonumber \\
&=& \left [(t_0-t) \alpha \bm x  + (t - \alpha t_0)  \Big [ \left (\alpha  \bm x - \bm d \right )
\moda \Lambda \Big] + \alpha \bm n \right ]\moda(t\Lambda) \nonumber.
\end{eqnarray}
As it was shown in Sect.~\ref{sec:crb_approx1}, when computing the
derivatives of  the logarithm of the pdf the dominant term will the
that corresponding to $[(t_0-t) \alpha \bm x] \moda(t \Lambda)$, so we will
approximate the first derivative of minus the logarithm of (\ref{eq:approx_pdf_erez}) with respect to $t$
by
\begin{eqnarray}
\sum_{j=1}^n 
\frac{\alpha^2(t-t_0)(\alpha^2\sigma_N^2 + (1-\alpha)^2 t t_0 \sigma_\Lambda^2) x_j^2}{\left (\alpha^2 \sigma_N^2 + (1-\alpha)^2  t^2 \sigma_\Lambda^2 \right)^2},\nonumber
\end{eqnarray}
and the second derivative by
\begin{eqnarray}
&\sum_{j=1}^n &  \Bigg (2\alpha^2\Big [\sigma_N^2\left(\alpha^4\sigma_N^2 - 3(1-\alpha)^2\alpha^2 t^2 \sigma_\Lambda^2  \right ) - 2(1-\alpha)^2 \sigma_\Lambda^2t(-3\alpha^2\sigma_N^2 + (1-\alpha)^2 t^2 \sigma_\Lambda^2)t_0 \nonumber \\
&&  + (1-\alpha)^2 \sigma_\Lambda^2 (-\alpha^2\sigma_N^2 + 3(1-\alpha)^2 t^2 \sigma_\Lambda^2 )t_0^2 \Big ]x_j^2 \Bigg)\left (\alpha^2 \sigma_N^2 + (1-\alpha)^2 t^2 \sigma_\Lambda^2 \right)^{-3}. \nonumber
\end{eqnarray}
Taking the mean of the last expression and replacing $t$ by $t_0$, one obtains
\begin{eqnarray}
&  n\Bigg (2\alpha^2\Big [\sigma_N^2\left(\alpha^4\sigma_N^2 - 3(1-\alpha)^2\alpha^2 \sigma_\Lambda^2 t_0^2 \right ) - 2(1-\alpha)^2 \sigma_\Lambda^2 t_0(-3\alpha^2\sigma_N^2 + (1-\alpha)^2 t_0^2 \sigma_\Lambda^2 )t_0 \nonumber \\
&  + (1-\alpha)^2\sigma_\Lambda^2(-\alpha^2\sigma_N^2 + 3(1-\alpha)^2 t_0^2 \sigma_\Lambda^2)t_0^2 \Big ]\sigma_X^2 \Bigg)\left (\alpha^2 \sigma_N^2 + (1-\alpha)^2 t_0^2 \sigma_\Lambda^2 \right)^{-3}= \nonumber\\
& \frac{n\alpha^2 \sigma_X^2}{\alpha^2 \sigma_N^2 + (1-\alpha)^2  t_0^2 \sigma_\Lambda^2}; \nonumber
\end{eqnarray}
therefore, the Fisher information in terms of $\sigma_X^2$, $\sigma_W^2$, $\sigma_N^2$, $t_0$, and $\alpha$ is given by
\begin{eqnarray}
I(t_0) =   \frac{n\alpha^2 \sigma_X^2}{\alpha^2 \sigma_N^2 + (1-\alpha)^2 \sigma_W^2 t_0^2}, \label{eq:crb_erez}
\end{eqnarray}
providing the same result that was previously described in (\ref{eq:crb_quant2}).

%%%%%%%%%%%%%%%%%%%%%%%%%%%%%%%%%%%%%%%%%%%%%%%%%%%%%%%%%%%%%%%%%%%%%
%%%%%%%%%%%%%%%%%%%%%%%%%%%%%%%%%%%%%%%%%%%%%%%%%%%%%%%%%%%%%%%%%%%%%
%%%%%%%%%%%%%%%%%%%%%%%%%%%%%%%%%%%%%%%%%%%%%%%%%%%%%%%%%%%%%%%%%%%%%
%%%%%%%%%%%%%%%%%%%%%%%%%%%%%%%%%%%%%%%%%%%%%%%%%%%%%%%%%%%%%%%%%%%%%
\section{Bias analysis}\label{sec:bias}

At the sight of (\ref{eq:der1_tochisima}), it is evident that the
derivation of $\hat{t}(\bm z)$, even when one just considers a
small neighborhood around $t_0$, is a tough question. Indeed, one
must find the roots of a six-order polynomial. Furthermore, in order
to compute the bias, one should average over the pdfs of
$\bm X$, $\bm W$, and $\bm N$ those cumbersome expressions. Therefore,
some simplifications will be required in order
to be able to derive insightful results. In this report, the bias
analysis will be based on the following assumptions:
\begin{itemize}
  \item $n\rightarrow \infty$: this will allow us to replace the square Euclidean
    norm of a random vector by its variance.
  \item We will constrain our analysis to the high-dimensional good lattice
    case, as in that case, as long as TNLR $<1$ the modulo-lattice reduction
    can be neglected.
  \item As it was mentioned before in this report,  we will assume
that the variance of the modulo-reduced version of the
total noise vector is equal to the variance the total noise vector
as long as the latter is smaller than or equal to $\sigma_\Lambda^2 t^2$, i.e.,
TNLR $<1$, and equal to $\sigma_\Lambda^2 t^2$ otherwise.
\end{itemize}

Therefore, based on $\sigma_E^2(t) \triangleq (t_0 - t)^2 \sigma_X^2 + (t-\alpha t_0)^2 \sigma_\Lambda^2 + \sigma_N^2$, our study of the bias will be splitted into two different
scenarios:
\begin{itemize}
  \item  $\sigma_E^2(t) < \sigma_\Lambda^2 t^2 $, or equivalentl, 
    $t\in (t_{l,\textrm{hd}}, t_{u,\textrm{hd}})$, based on the results
derived in Sect.~\ref{sec:sampling_high_dimen} (formulas (\ref{eq:sol_low_hd}) and (\ref{eq:sol_up_hd})).
  \item   $\sigma_E^2(t) \geq \sigma_\Lambda^2 t^2 $, or equivalently
    $t \leq t_{l,\textrm{hd}}$, or $t\geq t_{u,\textrm{hd}}$.
\end{itemize}

Obviously,
\begin{eqnarray}
  \min_{t\in (t_{l,\textrm{hd}}, t_{u,\textrm{hd}})} \frac{L(t, \bm z)}{n} \leq \frac{L(t_0, 
\bm z)}{n},\nonumber
\end{eqnarray}
since, as it was proved in Sect.~\ref{sec:sampling_high_dimen},
whenever TNLR $\leq 1$, $t_0 \in (t_{l,\textrm{hd}}, t_{u,\textrm{hd}})$.
Therefore, if 
\begin{eqnarray}
  \frac{L(t_0, \bm z)}{n} \leq \min_{t \in (0, t_{l,\textrm{hd}}] \bigcup
    [t_{u,\textrm{hd}}, \infty)} \frac{L(t, \bm z)}{n}, \nonumber
\end{eqnarray}
it would be clear that the minimum of the target function would
be achieved in $(t_{l,\textrm{hd}}, t_{u,\textrm{hd}})$, and we could
focus our analysis in that interval. 

We base our comparison on HLR $\rightarrow \infty$, and TNLR $ <1$, so
\begin{eqnarray}
  \frac{L(t_0, \bm z)}{n} = 2 +\log \left [2\pi \left (\sigma_N^2 + (1-\alpha)^2 \sigma_\Lambda^2 t_0^2 \right )
\right], \nonumber
\nonumber
\end{eqnarray}
and, for $t \in (0, t_{l,\textrm{hd}}] \bigcup
    [t_{u,\textrm{hd}}, \infty)$,
\begin{eqnarray}
    \frac{L(t, \bm z)}{n} \approx \frac{\sigma_\Lambda^2 t^2}{\sigma_N^2 + (1-\alpha)^2 \sigma_\Lambda^2 t^2}
  +\log \left [2\pi \left (\sigma_N^2 + (1-\alpha)^2 \sigma_\Lambda^2 t^2 \right)
\right] + \frac{t_0^2 }{ t^2}; \nonumber
\end{eqnarray}
it is straigforward to see that the result of substracting the first
formula to the second one is
\begin{eqnarray}
  \frac{t_0^2}{t^2}  -2 + \frac{ \sigma_\Lambda^2 t^2}{\sigma_N^2 + (1-\alpha)^2 t^2 \sigma_\Lambda^2} + \log \left (\frac{\sigma_N^2 + (1-\alpha)^2 t^2 \sigma_\Lambda^2}{\sigma_N^2 + (1-\alpha)^2  t_0^2 \sigma_\Lambda^2} \right ), \nonumber
\end{eqnarray}
which is nothing but $f(t, \sigma_\Lambda^2)$, defined in 
(\ref{eq:eq_another3}). In Sect.~\ref{sec:asym_large2} we have already
proved that this function is non-negative for any $t$,
confirming that the minimum of the target function will
be achieved in $(t_{l,\textrm{hd}}, t_{u,\textrm{hd}})$

Therefore, we will focus our analysis on the
interval $t\in (t_{l,\textrm{hd}}, t_{u,\textrm{hd}})$.
Based on the assumptions mentioned at the beginning of this section,
$L(t, \bm z)/n$ can be
approximated by
\begin{eqnarray}
  \frac{L(t, \bm z)}{n} \approx \frac{(t_0-t)^2\sigma_X^2 + (t-\alpha t_0)^2 \sigma_\Lambda^2 + \sigma_N^2}{\sigma_N^2 + (1-\alpha)^2 \sigma_\Lambda^2 t^2} 
  +\log \left [2\pi \left (\sigma_N^2 + (1-\alpha)^2 \sigma_\Lambda^2 t^2 \right)
\right] + \frac{\left (\sigma_X^2 + \alpha^2 \sigma_\Lambda^2 \right )t_0^2 
+ \sigma_N^2}{\sigma_X^2 t^2}, \label{eq:bias_approx_L}
\end{eqnarray}
whose derivative with respect to $t$ is
\begin{eqnarray}
  &&\frac{2\sigma_N^2\left [\left(2-\alpha \right )\alpha \sigma_\Lambda^2 + \sigma_X^2 \right ] t + 2\left (\alpha \sigma_\Lambda^2 + \sigma_X^2 \right )\left (
    (1-\alpha)^2\sigma_\Lambda^2t^2 -\sigma_N^2 \right )t_0 - 2(1-\alpha)^2
    \sigma_\Lambda^2\left (\sigma_X^2 + \alpha^2 \sigma_\Lambda^2\right )t t_0^2}
  {\left ( \sigma_N^2 + (1-\alpha)^2\sigma_\Lambda^2 t^2\right )^2} \nonumber\\
&&+\frac{2(1-\alpha)^2\sigma_\Lambda^2 t}{\sigma_N^2 + (1-\alpha)^2 \sigma_\Lambda^2 t^2 } + \frac{2\left [\left (\sigma_X^2 + \alpha^2 \sigma_\Lambda^2 \right ) t_0^2 + \sigma_N^2\right ]}{\sigma_X^2 t^3}. \label{eq:bias_der_complete}
\nonumber
\end{eqnarray}
In order to find the points where the derivative is null, we can multiply
the last formula by $\sigma_X^2 t^3\left ( \sigma_N^2 + (1-\alpha)^2\sigma_\Lambda^2 t^2\right )^2$, obtaining a polynomial that can be written as
$a_0 + a_1 t + a_2 t^2 + a_3 t^3 + a_4 t^4 + a_5 t^5 + a_6 t^6$,
where
\begin{eqnarray}
  a_0 &=&-2\sigma_N^4 \left [ \left (\sigma_X^2 + \alpha^2 \sigma_\Lambda^2 \right ) t_0^2  + \sigma_N^2\right ], \nonumber \\
  a_1 &=& 0, \nonumber \\
  a_2 &=& -4 (1-\alpha)^2 \sigma_\Lambda^2 \sigma_N^2 \left [ \left (\sigma_X^2 + \alpha^2 \sigma_\Lambda^2 \right ) t_0^2 + \sigma_N^2 \right ], \nonumber \\
  a_3 &=& -2\sigma_N^2  \sigma_X^2 \left ( \sigma_X^2 + \alpha \sigma_\Lambda^2 \right )t_0, \nonumber \\
  a_4 &=& -2 \Bigg ( - \sigma_N^2\sigma_X^4 + (1-\alpha)^4 \alpha^2 \sigma_\Lambda^6
  t_0^2 + \sigma_\Lambda^2 \sigma_X^2 \left [ \left (1-\alpha \right )^2 \sigma_X^2t_0^2 -
    \sigma_N^2 \right] \nonumber \\ &+& (1-\alpha)^2 \sigma_\Lambda^4  \left [(1-\alpha)^2\sigma_N^2 + (1-2\alpha + 2\alpha^2) \sigma_X^2t_0^2 \right ] \Bigg ), \nonumber \\
  a_5 &=& 2(1-\alpha)^2 \sigma_\Lambda^2 \sigma_X^2\left (\alpha \sigma_\Lambda^2 + \sigma_X^2  \right )t_0, \nonumber \\
  a_6 &=& 2(1-\alpha)^4 \sigma_\Lambda^4 \sigma_X^2; \nonumber
\end{eqnarray}
therefore, using Descartes' sign rule, the mentioned derivative will
be null at most at one positive $t$.

\begin{figure}[t] 
  \begin{center}
    \includegraphics[width=0.75\linewidth]{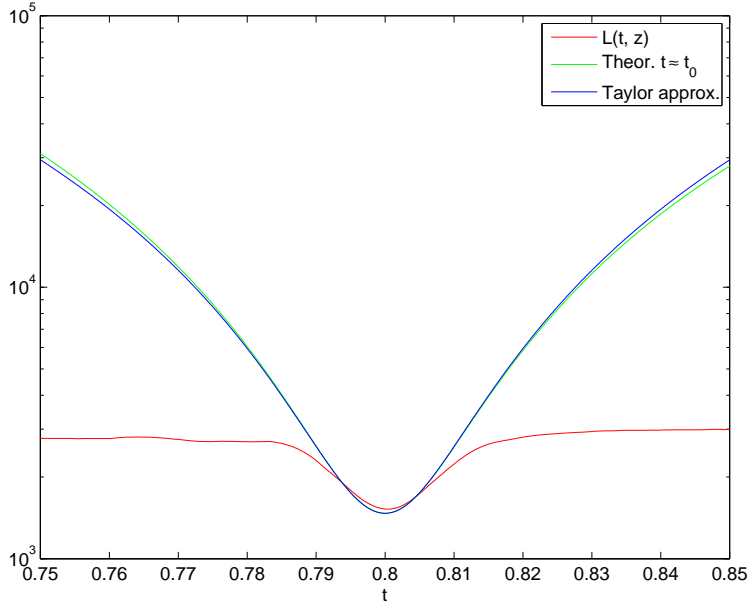}  
    \caption{Comparison between $L(t, \bm z)$, (\ref{eq:bias_approx_L}) and the
order $2$ Taylor series expansion. HLR $\approx 34.9765 $ dB, SCR $\approx -1.0618$ dB, TNLR $\approx -3.5736$ dB, $t_0 = 0.8$, 
$\alpha = \alpha_{\textrm{Costa}} \approx 0.5608$, $n = 1000$, scalar
quantizer.}\label{fig:Taylor}
  \end{center}
\end{figure}

As the solution of the previous polynomial is indeed pretty cumbersome,
for any $t_0 > \epsilon$, with $\epsilon > 0$,
we will calculate the order $2$ Taylor series expansion around $t_0$,
i.e., $b_0 + b_1(t-t_0) + b_2(t-t_0)^2/2$,
based on the reasonable assumption that the minimum of $L(t, \bm z)$
will be close to $t_0$. Take into account that this assumption
is based on  $t\in (t_{l,\textrm{hd}}, t_{u,\textrm{hd}})$,
HLR $\rightarrow \infty$, and TNLR $< 1$; as it was mentioned in
Sect.~\ref{sec:sampling_high_dimen}, under those conditions
both $t_{l,\textrm{hd}}$, and $t_{u,\textrm{hd}}$ go to $t_0$.
The resulting Taylor series expansion will be optimized.
Consequently, we will evaluate
$L(t, \bm z)/n$ and
(\ref{eq:bias_der_complete}) at $t=t_0$, yielding
\begin{eqnarray}
  b_0 &=& 2 + \frac{\sigma_N^2 + \alpha^2 \sigma_\Lambda^2t_0^2}{\sigma_X^2t_0^2}
  +\log \left [2\pi \left (\sigma_N^2 + (1-\alpha)^2 \sigma_\Lambda^2 t_0^2
\right ) \right ], \nonumber\\
  b_1 &=& \frac{2(1-\alpha) \sigma_\Lambda^2 t_0}{\sigma_N^2 + (1-\alpha)^2
    \sigma_\Lambda^2t_0^2} - \frac{2\left [\left (\sigma_X^2 + \alpha^2 \sigma_\Lambda^2
    \right)t_0^2 + \sigma_N^2 \right ]}{\sigma_X^2 t_0^3}. \nonumber
\end{eqnarray}
Concerning the order two coefficient, we will calculate the second derivative
of (\ref{eq:bias_approx_L}), yielding
\begin{eqnarray}
 && 2 \Bigg (\sigma_N^4 \sigma_X^2 - (1-\alpha)^4 \sigma_\Lambda^6t^2 \left [ \left(1-\alpha \right)^2t^2 + 2\alpha t t_0 -3 \alpha^2 t_0^2 \right ] \nonumber \\
 && + \left (1-\alpha \right)^2 \sigma_\Lambda^4 \left (3 \left (-2 \alpha \right)\alpha \sigma_N^2
    t^2 - 2t \left [\left(1-\alpha \right)^2 \sigma_X^2 t^2 - 3 \alpha \sigma_N^2 \right ]t_0 + \left [3 \left(1-\alpha\right )^2 \sigma_X^2 t^2 - \alpha^2 \sigma_N^2
      \right ]t_0^2 \right )  \nonumber \\&+&
    \sigma_\Lambda^2 \sigma_N^2 \left [\sigma_N^2 
  - (1-\alpha)^2 \sigma_X^2 \left (3 t^2 - 6t t_0 +t_0^2 \right ) \right ]
\Bigg ) \Bigg (\sigma_N^2 + (1-\alpha)^2 \sigma_\Lambda^2 t^2 \Bigg )^{-3}
+ \frac{6 \left [ \left (\sigma_X^2 + \alpha^2 \sigma_\Lambda^2 \right)t_0^2 + 
    \sigma_N^2 \right ]}{\sigma_X^2 t^4}, \nonumber
\end{eqnarray}
that evaluated at $t=t_0$ gives
\begin{eqnarray}
  b_2 = \frac{4(1-\alpha)^2\sigma_\Lambda^2 \sigma_N^2 + 2 \left [ \left(2\alpha^2 - 1 \right)\sigma_\Lambda^2 + \sigma_X^2 \right ] \left(\sigma_N^2 + (1-\alpha)^2
\sigma_\Lambda^2t_0^2 \right )}{\left (\sigma_N^2 + (1-\alpha)^2 \sigma_\Lambda^2
    t_0^2 \right )^2} + \frac{6 \left [\left (\sigma_X^2 + \alpha^2
      \sigma_\Lambda^2 \right )t_0^2 + \sigma_N^2 \right ]}{\sigma_X^2 t_0^4}.
\nonumber
\end{eqnarray}
An example of the resulting approximation can be found in Fig.~\ref{fig:Taylor}.

Obviously, the minimum will be achieved at $t_{\textrm{min}} = t_0 - \frac{b_1}{b_2}$.
In order to gain theoretical insight, we will use that
HLR $\rightarrow \infty$, and TNLR $<1$, or equivalently for fixed $\sigma_\Lambda^2$ and $\sigma_N^2$,
$\sigma_X^2 \rightarrow \infty$, for calculating
\begin{eqnarray}
  \lim_{\sigma_X^2 \rightarrow \infty} -\frac{\sigma_X^2 b_1}{b_2} = 
  \frac{\sigma_N^2
    + (\alpha^2 - \alpha) \sigma_\Lambda^2 t_0^2}{t_0} = 
  \frac{\sigma_N^2}{t_0} - \frac{(1-\alpha)\sigma_W^2 t_0}{\alpha}.
  \nonumber
\end{eqnarray}
Therefore, the bias in that asymptotic case has been shown to converge to
\begin{eqnarray}
  b(t_0) = \frac{1}{\sigma_X^2}\left [ \frac{\sigma_N^2}{t_0} - \frac{(1-\alpha)\sigma_W^2 t_0}{\alpha} \right ].\label{eq:bias_compact}
\end{eqnarray}
It is easy to see that under the assumptions of HLR $\rightarrow \infty$,
and TNLR $< 1$,
the bias nullifies for
\begin{eqnarray}
  \alpha_{\textrm{no-bias}} \triangleq \frac{\sigma_W^2 t_0^2}{\sigma_N^2 + \sigma_W^2 t_0^2},
\nonumber
\end{eqnarray}
which is nothing but the $\alpha$ derived by Costa, after replacing the 
embedded signal power by its effective value at the receiver.

In the case where $t_0 \rightarrow 0$ the Taylor series expansion of order 2 
can not be performed, as there does not exist a neighborhood of $t_0$
where the target function is differentiable.
In any case, it is easy to see that the derivative of 
(\ref{eq:bias_approx_L}) evaluated at $t_0 = 0$ is
\begin{eqnarray}
  \frac{2 \left (t^2  - \frac{\sigma_N^2}{\sigma_X^2} \right )}{t^3};\nonumber
\end{eqnarray}
therefore, in that case the bias will be
\begin{eqnarray}
  b(t_0) = \sqrt{\frac{\sigma_N^2}{\sigma_X^2}}. \nonumber
\end{eqnarray}

%%%%%%%%%%%%%%%%%%%%%%%%%%%%%%%%%%%%%%%%%%%%%%%%%%%%%%%%%%%%%%%%%%%%%
%%%%%%%%%%%%%%%%%%%%%%%%%%%%%%%%%%%%%%%%%%%%%%%%%%%%%%%%%%%%%%%%%%%%%
%%%%%%%%%%%%%%%%%%%%%%%%%%%%%%%%%%%%%%%%%%%%%%%%%%%%%%%%%%%%%%%%%%%%%
%%%%%%%%%%%%%%%%%%%%%%%%%%%%%%%%%%%%%%%%%%%%%%%%%%%%%%%%%%%%%%%%%%%%%
\section{Total MSE lower-bound and $\alpha$ optimization}\label{sec:total_mse}

In the previous sections we have studied both the Fisher information
and the bias of our estimator, obtaining close formulas in both cases
for the high-dimensional good lattice case, when HLR $\rightarrow \infty$,
and TNLR $< 1$. Combining both expressions, 
the MSE of the proposed estimator can be lower-bounded by
\begin{eqnarray}
\textrm{E}\left \{ \left (t_0 - \hat{t}(\bm z) \right )^2\right \} \geq
\frac{ \left [1 + \frac{\partial b(t_0)}{\partial t_0} \right ]^2}{I(t_0)} + b(t_0)^2. \label{eq:mse_total_mola}
\end{eqnarray}
We will focus our analysis on the non-degenerated case where $t_0$ does not
go to $0$. In this scenario, $I(t_0)$ is given by (\ref{eq:crb_quant2}),
and $b(t_0)$ by (\ref{eq:bias_compact}); thus, the derivative
of $b(t_0)$ with respect to $t_0$ is 
\begin{eqnarray}
\frac{\partial b(t_0)}{\partial t_0} = \frac{-1}{\sigma_X^2 t_0^2}
\left [ \left (1-\alpha \right) \alpha \sigma_\Lambda^2 t_0^2 + \sigma_N^2 
\right]; \nonumber
\end{eqnarray}
if TNLR $< 1$ and HLR $\rightarrow \infty$, then the last expression
goes to $0$, so it can be neglected in (\ref{eq:mse_total_mola}).

Therefore, we will write
\begin{eqnarray}
\textrm{E}\left \{ \left (t_0 - \hat{t}(\bm z) \right )^2\right \} \geq
\frac{1}{I(t_0)} + b(t_0)^2 =
\frac{\sigma_N^2 + \frac{(1-\alpha)^2}{\alpha^2} \sigma_W^2 t_0^2}{ n \sigma_X^2 }
 + \left ( \frac{1}{\sigma_X^2}\left [ \frac{\sigma_N^2}{t_0} - \frac{(1-\alpha)\sigma_W^2 t_0}{\alpha} \right ] \right )^2. \label{eq:mse_total_mola2}
\end{eqnarray}
It is worth pointing out that the Fisher information contribution decreases with 
$n \sigma_X^2$, while the bias one decreases with $\sigma_X^4$.
Therefore, the dominant effect will depend on the ratio
$\frac{\sigma_X^2}{n}$: the larger it is, the more dominant the Fisher
information will be; the smaller it is, the more important the
bias contribution. 

\subsection{$\alpha$ optimization}

In the previous sections we also derived the value of $\alpha$
maximizing the Fisher information $\alpha_{\textrm{sup-FI}}$, and
minimizing the bias effect $\alpha_{\textrm{no-bias}}$; given that
both values are different, when trying to optimize both functions
simultaneously, i.e. when minimizing (\ref{eq:mse_total_mola2}),
a trade-off between the optimal $\alpha$ corresponding
to both contributions must be achieved. In that sense, and taking
into account the discussion in the previous paragraph, one 
would expect the obtained $\alpha$ to be closer to 
$\alpha_{\textrm{sup-FI}}$ when $\frac{\sigma_X^2}{n}$ is large,
and closer to $\alpha_{\textrm{no-bias}}$ when that ratio is small.

Indeed, the derivative of the total MSE lower-bound with respect
to $\alpha$ is
\begin{eqnarray}
\frac{\partial \left (\frac{1}{I(t_0)} + b(t_0)^2 \right )}{\partial \alpha} = 
\frac{2 \sigma_W^2\left [n \alpha \sigma_N^2 - (1-\alpha) \left (n \sigma_W^2 + 
    \sigma_X^2 \right)t_0^2 \right]}{n \alpha^3 \sigma_X^4}, \nonumber
\end{eqnarray}
which is null at
\begin{eqnarray}
  \alpha = \frac{n\sigma_W^2 t_0^2 + \sigma_X^2 t_0^2}{n \sigma_N^2 + n\sigma_W^2 t_0^2 + \sigma_X^2 t_0^2}, \label{eq:cansau}
\end{eqnarray}
where the target function has its only local (and consequently global)
minimum.
Be aware that when $\frac{\sigma_X^2}{n} \rightarrow \infty$ (and
based on HLR $\rightarrow \infty$, and TNLR $<1$), 
(\ref{eq:cansau}) goes to $1$, while if $\frac{\sigma_X^2}{n}
\rightarrow 0$, (\ref{eq:cansau}) goes to $\alpha_{\textrm{no-bias}}$
(i.e., $\alpha_{\textrm{Costa}}$).
Taking into account that $\alpha$ must be smaller than
$\alpha_{\textrm{sup-FI}}$ in order to TNLR $<1$, the feasible value
of $\alpha$ minimizing the total MSE lower-bound is finally given by
\begin{eqnarray}
  \alpha_{\textrm{opt}} \triangleq \min \left ( \frac{n\sigma_W^2 t_0^2 + \sigma_X^2 t_0^2}{n \sigma_N^2 + n\sigma_W^2 t_0^2 + \sigma_X^2 t_0^2}, \frac{2\sigma_W^2 t_0^2}{\sigma_N^2 + \sigma_W^2 t_0^2} -\epsilon\right), \label{eq:alpha_opt_mola}
\end{eqnarray}
for $\epsilon > 0$ arbitrarily small.

As expected, when $\frac{\sigma_X^2}{n} \rightarrow 0$, 
$\alpha_{\textrm{opt}} \rightarrow \alpha_{\textrm{no-bias}}$, while
when $\frac{\sigma_X^2}{n} \rightarrow \infty$, 
$\alpha_{\textrm{opt}} \rightarrow \alpha_{\textrm{sup-FI}}$.
In the intermediate cases, $\alpha_{\textrm{opt}}$ will balance the effect of
both contributions. Although obvious at the sight of (\ref{eq:alpha_opt_mola}),
it is interesting to mention that the optimal $\alpha$ value
depends on this scenario both on $\sigma_X^2$ and $n$, contrarily
to Costa's original proposal.

Finally, when $\alpha_{\textrm{opt}}$ is used, if the left argument of the $\min$ function is active in 
(\ref{eq:alpha_opt_mola}), one has that 
\begin{eqnarray}
  \textrm{E}\left \{ \left (t_0 - \hat{t}(\bm z) \right )^2\right \} \geq
\frac{\sigma_N^2 \left (\frac{1}{n} + \frac{\sigma_N^2}{\left (n \sigma_W^2+ \sigma_X^2 \right)t_0^2} \right)}{\sigma_X^2};\nonumber
\end{eqnarray}
if HLR $\gg 1$ and TNLR $<1$, then we can approximate the right term by
\begin{eqnarray}
\frac{\sigma_N^2}{n\sigma_X^2}. \label{eq:var_alpha_cota}
\end{eqnarray}
On the other hand, if the active argument is the second one, then
\begin{eqnarray}
  \textrm{E}\left \{ \left (t_0 - \hat{t}(\bm z) \right )^2\right \} \geq
  \frac{\left (n\sigma_W^2 + \sigma_X^2 \right) \left (\sigma_N^2 + \sigma_W^2t_0^2 \right)^2}{4 n \sigma_W^2 \sigma_X^4 t_0^2}. \nonumber
\end{eqnarray}

Finally, for further comparison we also find useful to derive the total
MSE lower-bound when $\alpha = \alpha_{\textrm{no-bias}}$, which is
\begin{eqnarray}
  \textrm{E}\left \{ \left (t_0 - \hat{t}(\bm z) \right )^2\right \} \geq
\frac{\sigma_N^2 \left ( \sigma_N^2 + \sigma_W^2 t_0^2 \right)}{n \sigma_W^2
  \sigma_X^2 t_0^2}, \nonumber
\end{eqnarray}
based on TNLR $<1$, the right term is smaller than
\begin{eqnarray}
  \frac{\sigma_N^2 /\alpha^2}{n\sigma_X^2}. \label{eq:var_alpha_costa}
\end{eqnarray}
%%%%%%%%%%%%%%%%%%%%%%%%%%%%%%%%%%%%%%%%%%%%%%%%%%%%%%%%%%%%%%%%%%%%%
%%%%%%%%%%%%%%%%%%%%%%%%%%%%%%%%%%%%%%%%%%%%%%%%%%%%%%%%%%%%%%%%%%%%%
%%%%%%%%%%%%%%%%%%%%%%%%%%%%%%%%%%%%%%%%%%%%%%%%%%%%%%%%%%%%%%%%%%%%%
%%%%%%%%%%%%%%%%%%%%%%%%%%%%%%%%%%%%%%%%%%%%%%%%%%%%%%%%%%%%%%%%%%%%%
\section{Fundamental bounds}

The bounds derived in this section will not
depend on the particular estimation scheme that one is considering.
Therefore, we will obtain an estimation error variance lower-bound to any
estimation scheme where
\begin{itemize}
\item the channel state $\bm X$, follows a $\mathcal{N}(\bm 0, \sigma_X^2 I_{n\times n})$.
\item the transmitter, who anticausally knows $\bm X$ but can not choose it, can produce a signal $\bm W$ of power $\sigma_W^2 << \sigma_X^2$, which is added to $\bm X$.
\item  the resulting
signal is scaled
by an unknown parameter $t_0$, and added AWGN $\bm N$ independent of
both $\bm X$ and $\bm W$, with power $\sigma_N^2$.
\item the decoder has not access to $\bm X$, but only to $t_0(\bm X + \bm W) + \bm N$.
\end{itemize}

In that scenario, and given that $\frac{\sigma_W^2}{\sigma_X^2}\rightarrow 0$, one can assume
that $\bm Y = \bm X + \bm W$ will be also asymptotically Gaussian, as $\bm W$ will not
significantly modify the distribution of $\bm X$. Additionally,
$(\sigma_X - \sigma_W)^2 \leq \sigma_Y^2 \leq (\sigma_X + \sigma_W)^2$,
% la correlación normalizada está acotada superiorment por 1.
and
whenever $\bm W$ is constrained to be independent of $\bm X$,
$\sigma_Y^2 = \sigma_X^2 + \sigma_W^2$.

Therefore, in the Gaussian framework we can bound the performance of such
a scheme by that of a scheme where the transmitter can choose to his/her will
$\bm X$ (as long as the Gaussian constraint is respected), and then
design $\bm W$ accordingly, and the chosen
value $\bm Y$ is also known at the receiver side; this new scenario
corresponds to the case where {\it pilot signals} are used. In that framework,
the pdf of the received signal is
\begin{eqnarray}
  f_{Z|Y, t_0}(\bm z|\bm y, t)= \frac{e^{\frac{-||\bm z - t \bm y||^2}{2\sigma_N^2}}}{\left (2\pi \sigma_N^2 \right )^{n/2}},\nonumber
\end{eqnarray}
since, as it was mentioned before, the receiver will know what was
the trasmitted signal $\bm y$.
It is straightforward to see that in this scenario the CRB equals to 
\begin{eqnarray}
  \frac{\sigma_N^2}{n \sigma_Y^2}. \nonumber
\end{eqnarray}

Having the last expression in mind, we will envisage two different scenarios
providing different values of $\sigma_Y^2$; first, if complete freedom on choosing
$\bm W$ is available, one has that
\begin{eqnarray}
  \textrm{E}\{(t_0 - \hat{t}(\bm Z))^2\} \geq \sigma^2_{\textrm{bound1}} \triangleq
  \frac{\sigma_N^2}{n (\sigma_X + \sigma_W)^2}; \nonumber
\end{eqnarray}
on the other hand, if $\bm W$ is required to be independent of $\bm X$, then
\begin{eqnarray}
  \textrm{E}\{(t_0 - \hat{t}(\bm Z))^2\} \geq \sigma^2_{\textrm{bound2}} \triangleq
  \frac{\sigma_N^2}{n (\sigma_X^2 + \sigma_W^2)}. \nonumber
\end{eqnarray}
Obviously $\sigma^2_{\textrm{bound1}} \leq \sigma^2_{\textrm{bound2}}$; nevertheless,
when $\frac{\sigma_W^2}{\sigma_X^2} \rightarrow 0$, both expressions are asymptotically
equivalent.

%%%%%%%%%%%%%%%%%%%%%%%%%%%%%%%%%%%%%%%%%%%%%%%%%%%%%%%%%%%%%%%%%%%%%
%%%%%%%%%%%%%%%%%%%%%%%%%%%%%%%%%%%%%%%%%%%%%%%%%%%%%%%%%%%%%%%%%%%%%
%%%%%%%%%%%%%%%%%%%%%%%%%%%%%%%%%%%%%%%%%%%%%%%%%%%%%%%%%%%%%%%%%%%%%
%%%%%%%%%%%%%%%%%%%%%%%%%%%%%%%%%%%%%%%%%%%%%%%%%%%%%%%%%%%%%%%%%%%%%
\section{Intuitive insights}\label{sec:intuit_insight}

In this section we try to provide an intuitive insight on 
the total MSE lower-bound obtained in Sect.~\ref{sec:total_mse}, and
its comparison with the fundamental lower-bound derived
in the previous section.

First, we would like to emphasize the similarities between 
(\ref{eq:var_alpha_cota}) and (\ref{eq:var_alpha_costa}),  and the CRB obtained for the case where
the transmitter can choose the state of the channel, i.e., when pilot signals are used, $\frac{\sigma_N^2}{ n(\sigma_X + \sigma_W)^2}$. Based on HLR $\rightarrow \infty$, $(\sigma_X + \sigma_W)^2 \approx \sigma_X^2$. Therefore, 
the lower-bound in (\ref{eq:var_alpha_costa}), which is non-minimal, as
the optimal value of $\alpha$ is not used, differs from the fundamental bound
just by the division by $\alpha^2$; furthermore, (\ref{eq:var_alpha_cota})
asymptotically converge to that fundamental bound.
Thus, the advantage of the proposed quantization-based
scheme being able not only to reduce the host signal interference, but
indeed to use it for helping in the estimation
of the scalar multiplicative factor.
Similarly,
the signal variance term that appears at the denominator of both expressions
is slightly reduced when quantization-based techniques are used
in comparison with the pilot-based strategy. In any case, the variance used by the
pilot-based strategy will be $(\sigma_X + \sigma_W)^2$, while the variance
used by our quantization-based scheme is just $\sigma_W^2$; the advantage
of our scheme could be also studied in terms of the increased payload
(as in pilot-based schemes the considered signals
does not convey information).

But, how it is possible
that similar performance is achieved by our quantization-based
scaling estimation scheme and by the pilot-based strategies?
If we assume that the estimator of the proposed quantization-based
scheme were aided by a genie, who would
say what was the centroid used at the embedder,
both schemes would be equivalent, except
for two facts already mentioned (the increase in the total noise variance,
and the decrease of transmitted power).
Although the scheme we are considering is obviously non-genie-based,
it is indeed implicitly based on correctly estimating the centroid used
at the embedder. This fact is reflected on the derivation of the
total MSE lower-bound when assuming that the modulo-$\Lambda$ reduction can be neglected, which
is only possible if the estimated centroid is that used at the
embedder.

From the last discussion one can glimpse the importance of being
able to find a $t$ close enough to $t_0$ so the total noise at $t$, i.e.,
$(t_0-t)\bm x + (t-\alpha t_0)[(\bm x - \bm d)\moda\Lambda] + \bm n$
is not modulo-reduced, so the used centroid can be
correctly estimated. It is easy to figure out that the closer
$\alpha$ to $\frac{2\sigma_W^2 t_0^2}{\sigma_N^2 + \sigma_W^2 t_0^2}$,
the smaller the size of the interval of $t$ around $t_0$ for which
the modulo reduction can be neglected. Indeed, this discussion
is also related to our
study of the search interval sampling. 
Therefore, if high-dimensional good lattices were
used, and the value of $\alpha$ approached $\alpha_{\textrm{sup-FI}}$,
then the distance between consecutive candidate points should
be arbitrarily small, in order to the variance of the total noise at
the new location be still smaller than the second moment per dimension
of the lattice. Consequently, the correct estimation of the
centroid at the embedder is achieved at the cost of
increasing the computational cost of the
estimation algorithm.
This comment also makes evident the trade-off that should
be achieved between estimation algorithm performance, in terms
of estimation error variance, and the required computational cost.

Finally, we would like to emphasize that
the larger the HLR, the smaller the total MSE lower-bound,
but the larger the computational cost, as the
sampling of the search interval should be thinner.

%%%%%%%%%%%%%%%%%%%%%%%%%%%%%%%%%%%%%%%%%%%%%%%%%%%%%%%%%%%%%%%%%%%%%
%%%%%%%%%%%%%%%%%%%%%%%%%%%%%%%%%%%%%%%%%%%%%%%%%%%%%%%%%%%%%%%%%%%%%
%%%%%%%%%%%%%%%%%%%%%%%%%%%%%%%%%%%%%%%%%%%%%%%%%%%%%%%%%%%%%%%%%%%%%
%%%%%%%%%%%%%%%%%%%%%%%%%%%%%%%%%%%%%%%%%%%%%%%%%%%%%%%%%%%%%%%%%%%%%
\section{Experimental results}

In this section we summarize the main results obtained by considering multidimensional lattices. Specifically, we quantize the host signal by using the coarse lattice proposed in \cite{Sun}; the reported results were obtained by considering constraint length $7$, and the generator polynomial proposed in \cite{Larsen} (in octal notation, $(133, 171)$), which yields a shaping gain of $1.26$ dB. Note that other choices, for example based on the coarse lattice used in \cite{Erez05}, are possible. In all the plots in this section {\it Simplified bound} stands for the value of (\ref{eq:mse_total_mola2}).

\begin{figure}[t] 
  \begin{center}
    \includegraphics[width=0.75\linewidth]{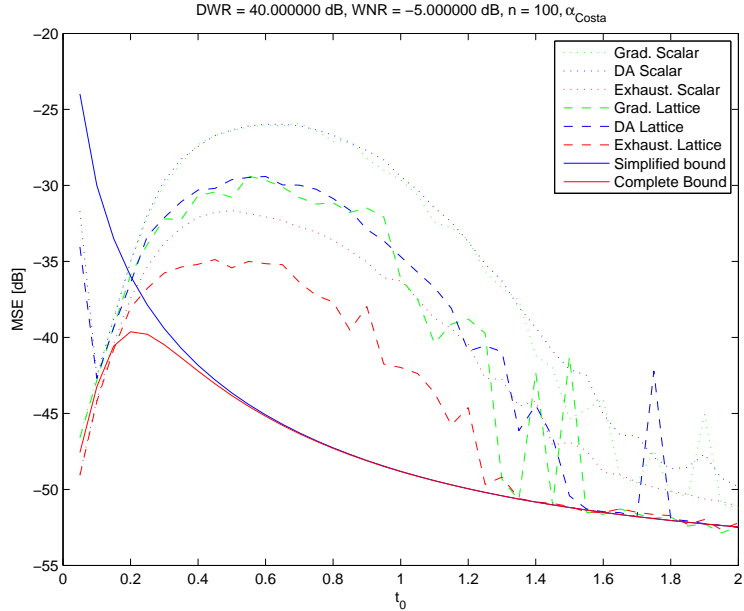} 
    \caption{MSE obtained for different estimation algorithms, both for scalar quantizer and multidimensional lattices. DWR = $40$ dB, WNR = $-5$ dB, $n=100$, $\alpha_{\textrm{Costa}}$.}\label{fig:mse1}
  \end{center}
\end{figure}

\begin{figure}[t] 
  \begin{center}
    \includegraphics[width=0.75\linewidth]{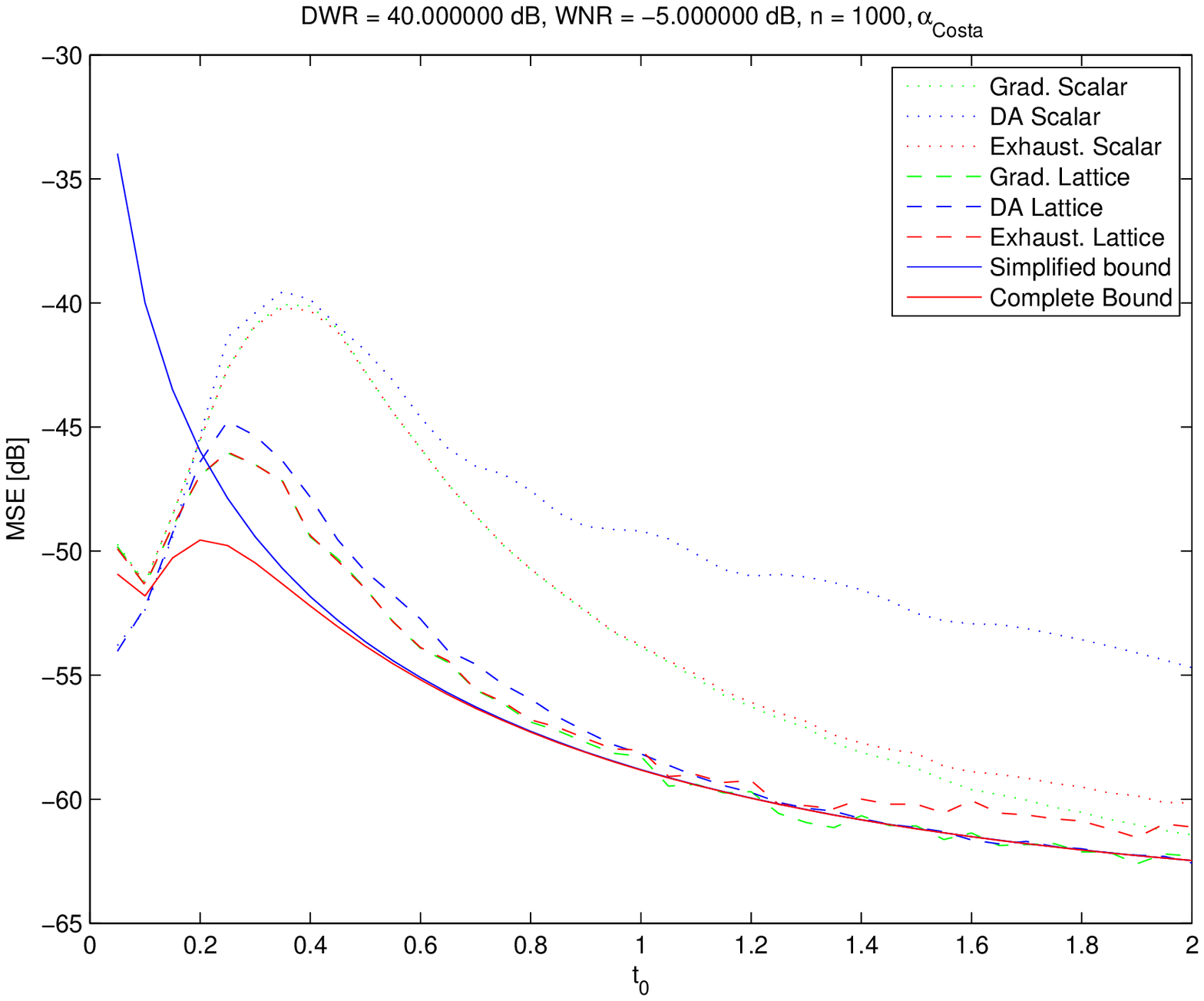}
    \caption{MSE obtained for different estimation algorithms, both for scalar quantizer and multidimensional lattices. DWR = $40$ dB, WNR = $-5$ dB, $n=1000$, $\alpha_{\textrm{Costa}}$.}\label{fig:mse2}
  \end{center}
\end{figure}

\begin{figure}[t] 
  \begin{center}
    \includegraphics[width=0.75\linewidth]{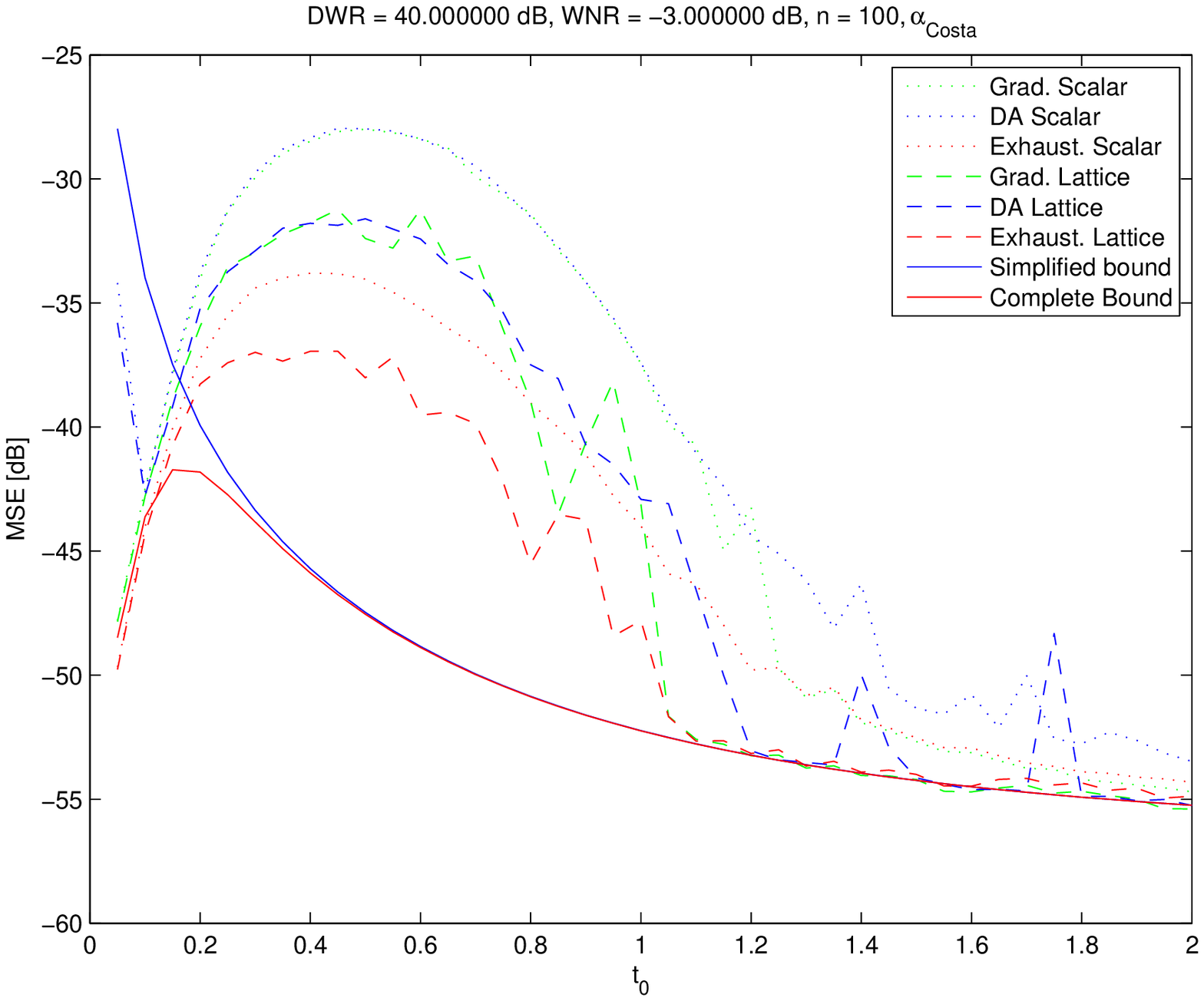} 
    \caption{MSE obtained for different estimation algorithms, both for scalar quantizer and multidimensional lattices. DWR = $40$ dB, WNR = $-3$ dB, $n=100$, $\alpha_{\textrm{Costa}}$.}\label{fig:mse3}
  \end{center}
\end{figure}

\begin{figure}[t] 
  \begin{center}
    \includegraphics[width=0.75\linewidth]{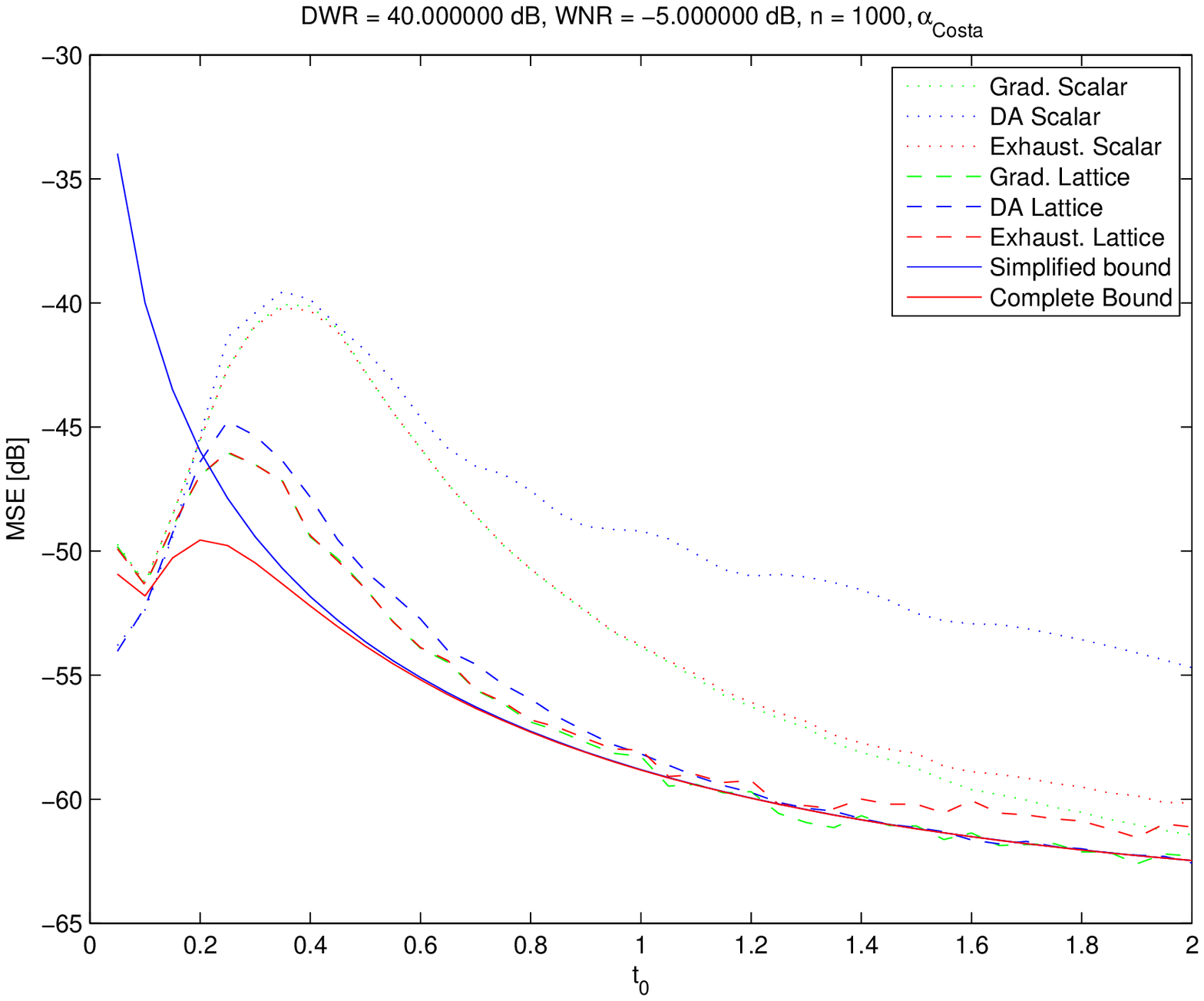}
    \caption{MSE obtained for different estimation algorithms, both for scalar quantizer and multidimensional lattices. DWR = $40$ dB, WNR = $-3$ dB, $n=1000$, $\alpha_{\textrm{Costa}}$.}\label{fig:mse4}
  \end{center}
\end{figure}

\begin{figure}[t] 
  \begin{center}
    \includegraphics[width=0.75\linewidth]{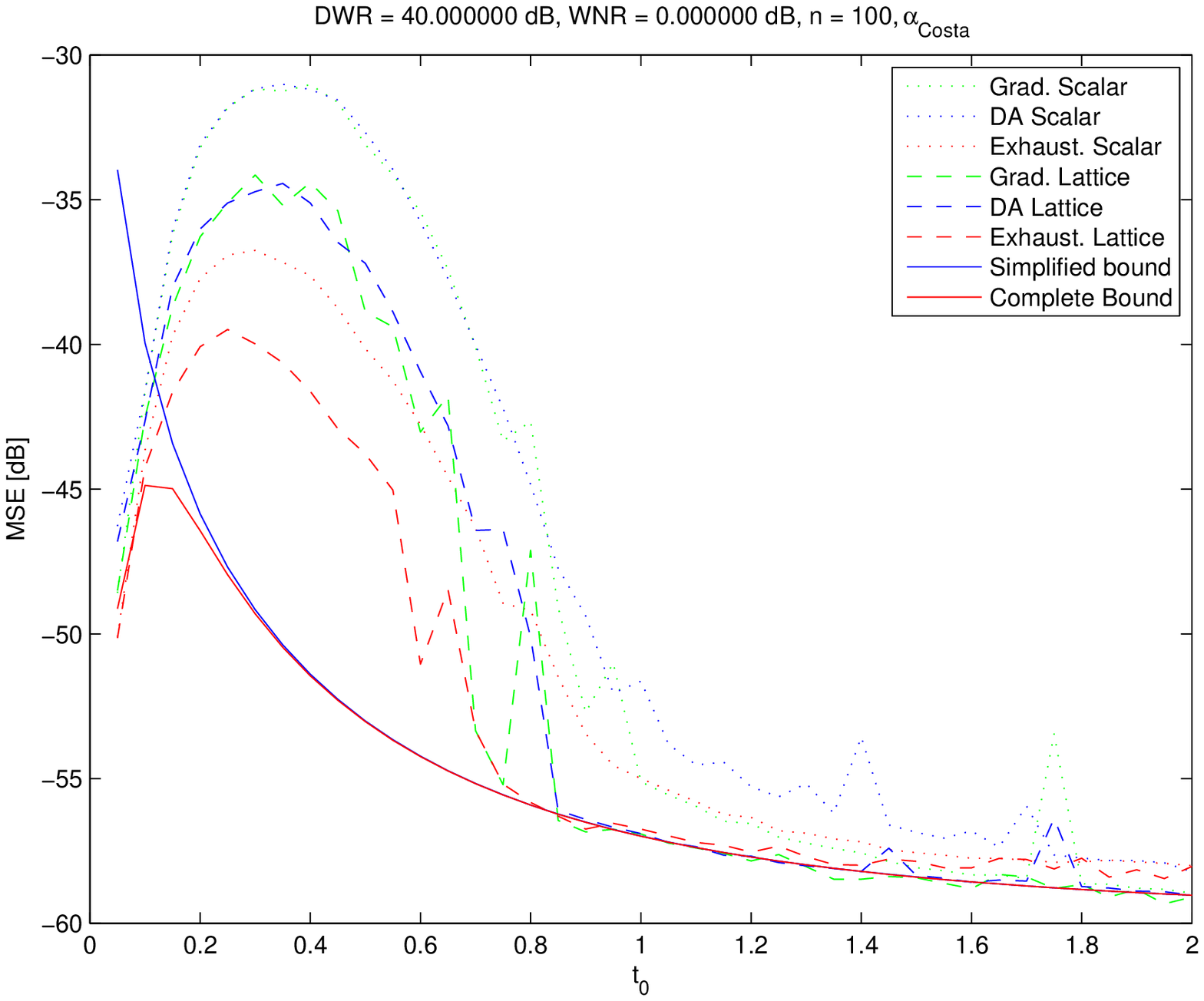}  
    \caption{MSE obtained for different estimation algorithms, both for scalar quantizer and multidimensional lattices. DWR = $40$ dB, WNR = $0$ dB, $n=100$, $\alpha_{\textrm{Costa}}$.}\label{fig:mse5}
  \end{center}
\end{figure}

\begin{figure}[t] 
  \begin{center}
    \includegraphics[width=0.75\linewidth]{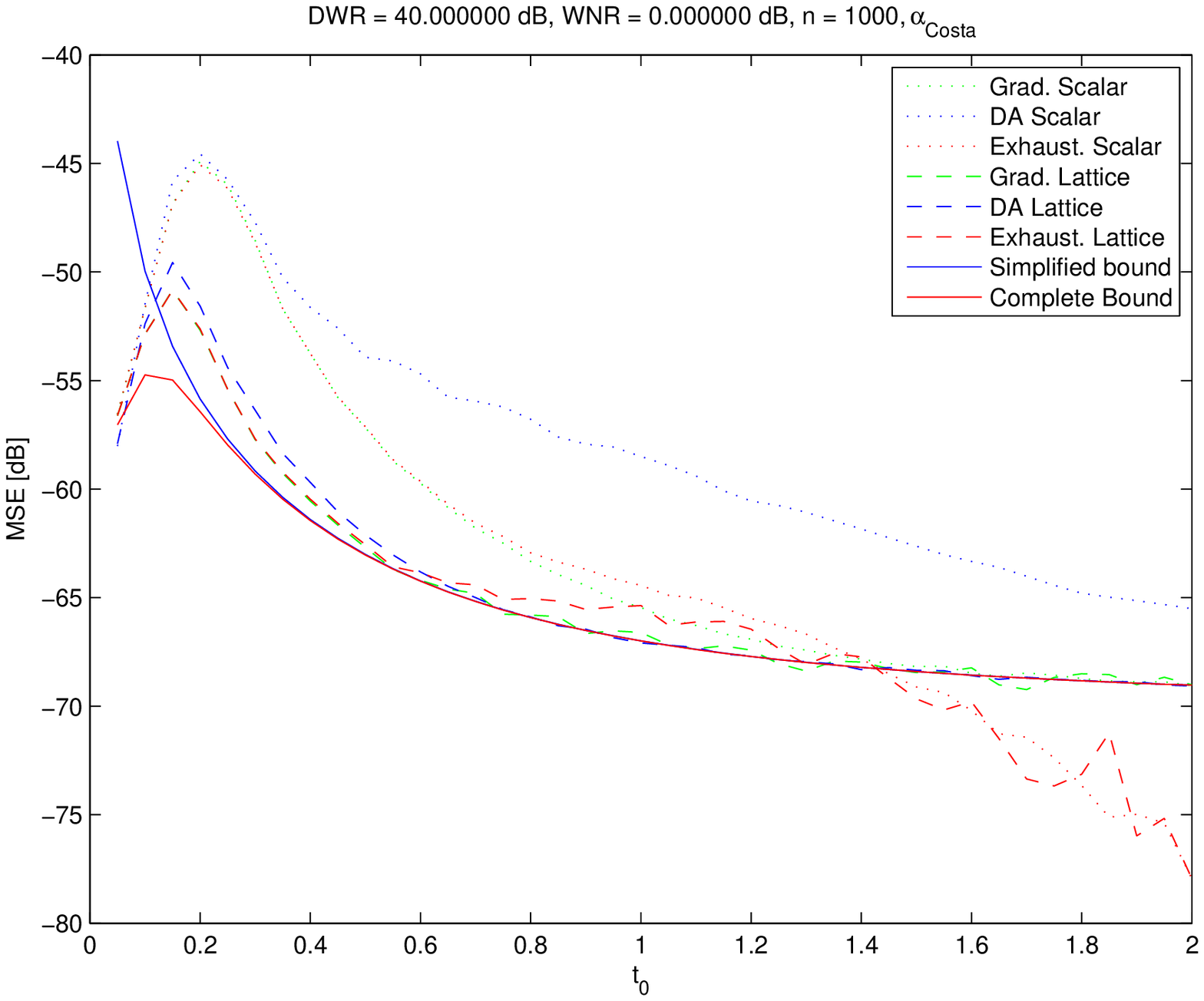}
    \caption{MSE obtained for different estimation algorithms, both for scalar quantizer and multidimensional lattices. DWR = $40$ dB, WNR = $0$ dB, $n=1000$, $\alpha_{\textrm{Costa}}$.}\label{fig:mse6}
  \end{center}
\end{figure}

\begin{figure}[t] 
  \begin{center}
    \includegraphics[width=0.75\linewidth]{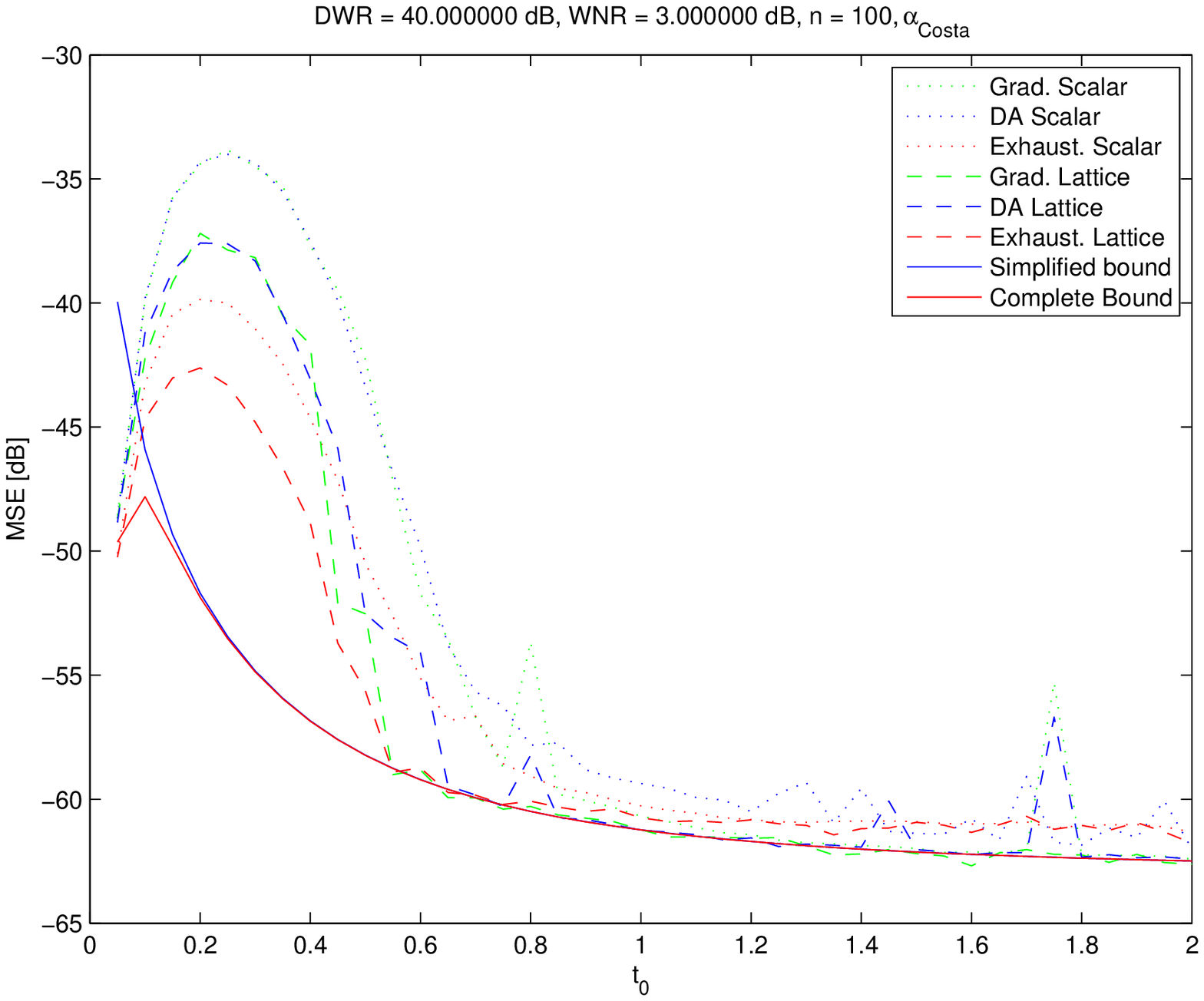}  
    \caption{MSE obtained for different estimation algorithms, both for scalar quantizer and multidimensional lattices. DWR = $40$ dB, WNR = $3$ dB, $n=100$, $\alpha_{\textrm{Costa}}$.}\label{fig:mse7}
  \end{center}
\end{figure}

\begin{figure}[t] 
  \begin{center}
    \includegraphics[width=0.75\linewidth]{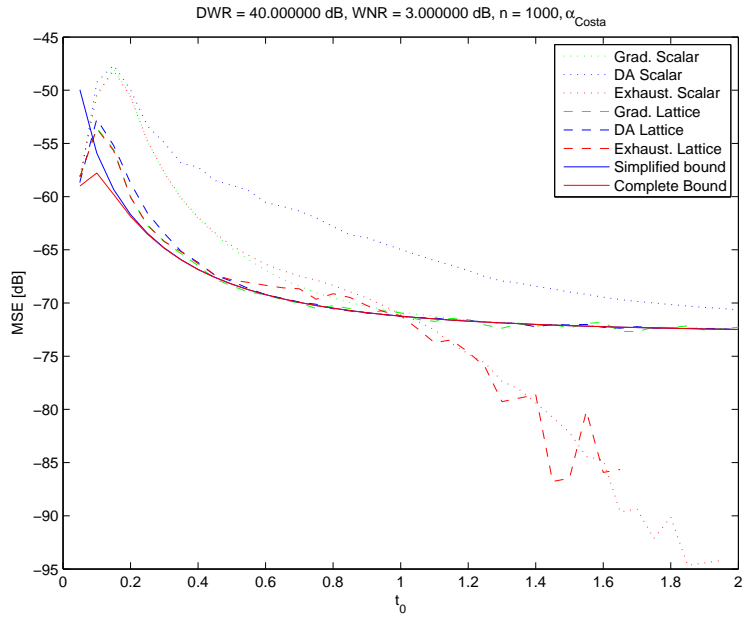}
    \caption{MSE obtained for different estimation algorithms, both for scalar quantizer and multidimensional lattices. DWR = $40$ dB, WNR = $3$ dB, $n=1000$, $\alpha_{\textrm{Costa}}$.}\label{fig:mse8}
  \end{center}
\end{figure}

First, in Figs.~\ref{fig:mse1}-\ref{fig:mse8} we show the MSE results as a function of $t_0$ for different estimation algorithms, for both scalar quantizers and the multidimensional lattice proposed in \cite{Sun}. Note that in some cases the plots show an MSE obtained by exhaustive search which is smaller than the corresponding MSE bound; this is due to the fact that the dominant effect in those cases is the exhaustive search quantization.

In general, all those plots show the advantage of considering multidimensional lattices instead of scalar quantizers, specially for large values of $n$, where one can really take advantage of the shaping of the multidimensional lattices, as well as a very good agreement between the theoretical bounds and the experimental results.

\clearpage

\begin{figure}[t] 
  \begin{center}
    \includegraphics[width=0.75\linewidth]{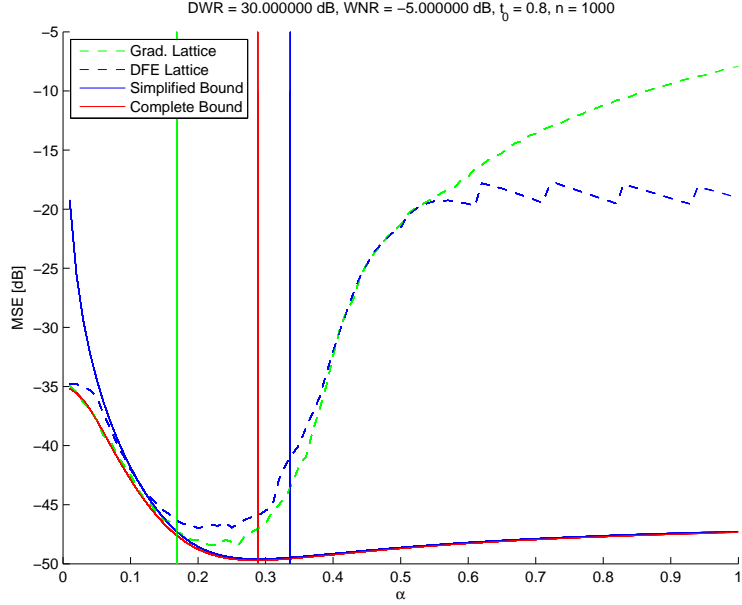}  
    \caption{MSE as a function of $\alpha$ obtained for different estimation algorithms and multidimensional lattices. DWR = $30$ dB, WNR = $-5$ dB, $n=1000$.}\label{fig:mse9}
  \end{center}
\end{figure}

\begin{figure}[t] 
  \begin{center}
    \includegraphics[width=0.75\linewidth]{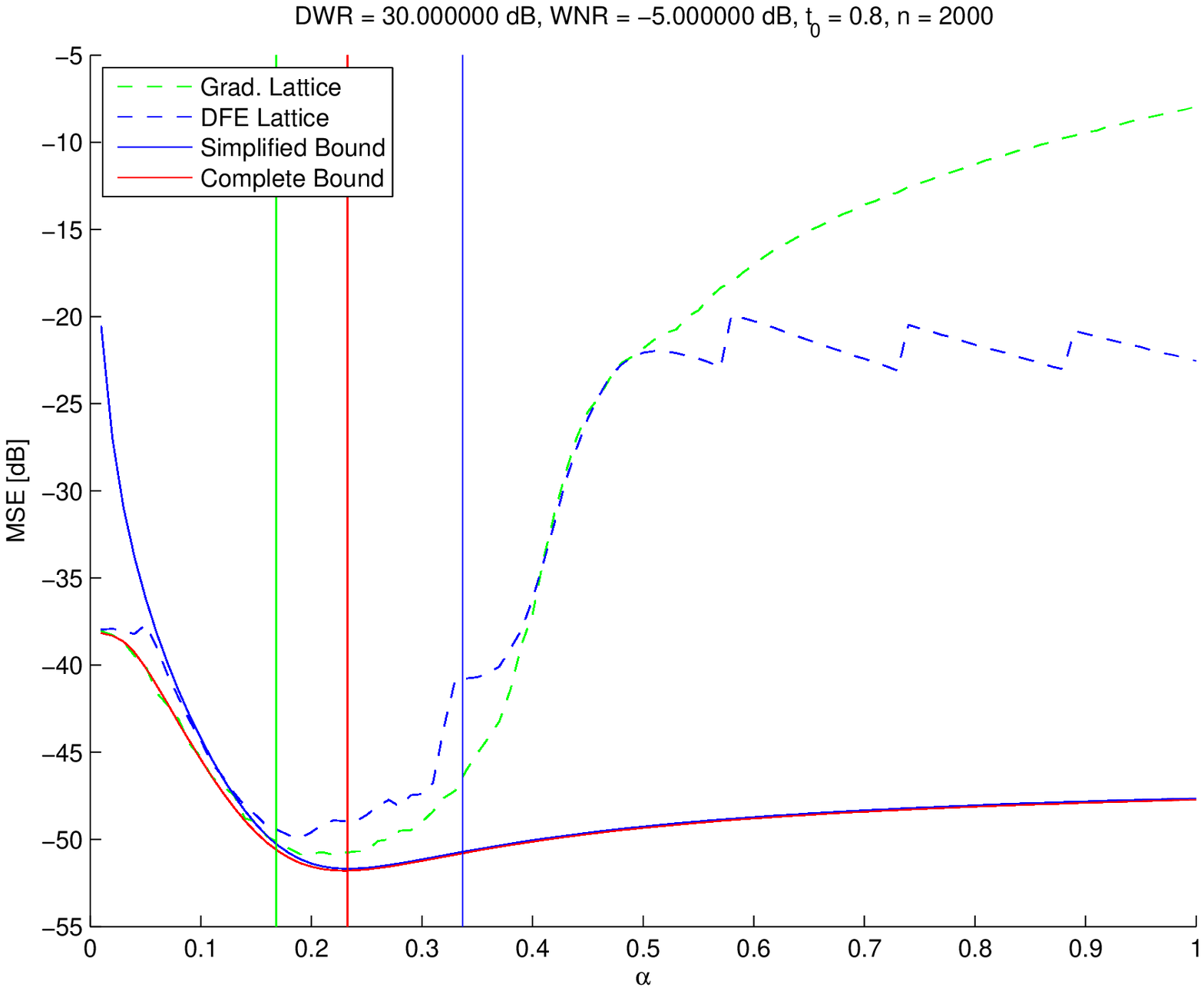}  
    \caption{MSE as a function of $\alpha$ obtained for different estimation algorithms and multidimensional lattices. DWR = $30$ dB, WNR = $-5$ dB, $n=2000$.}\label{fig:mse10}
  \end{center}
\end{figure}

\begin{figure}[t] 
  \begin{center}
    \includegraphics[width=0.75\linewidth]{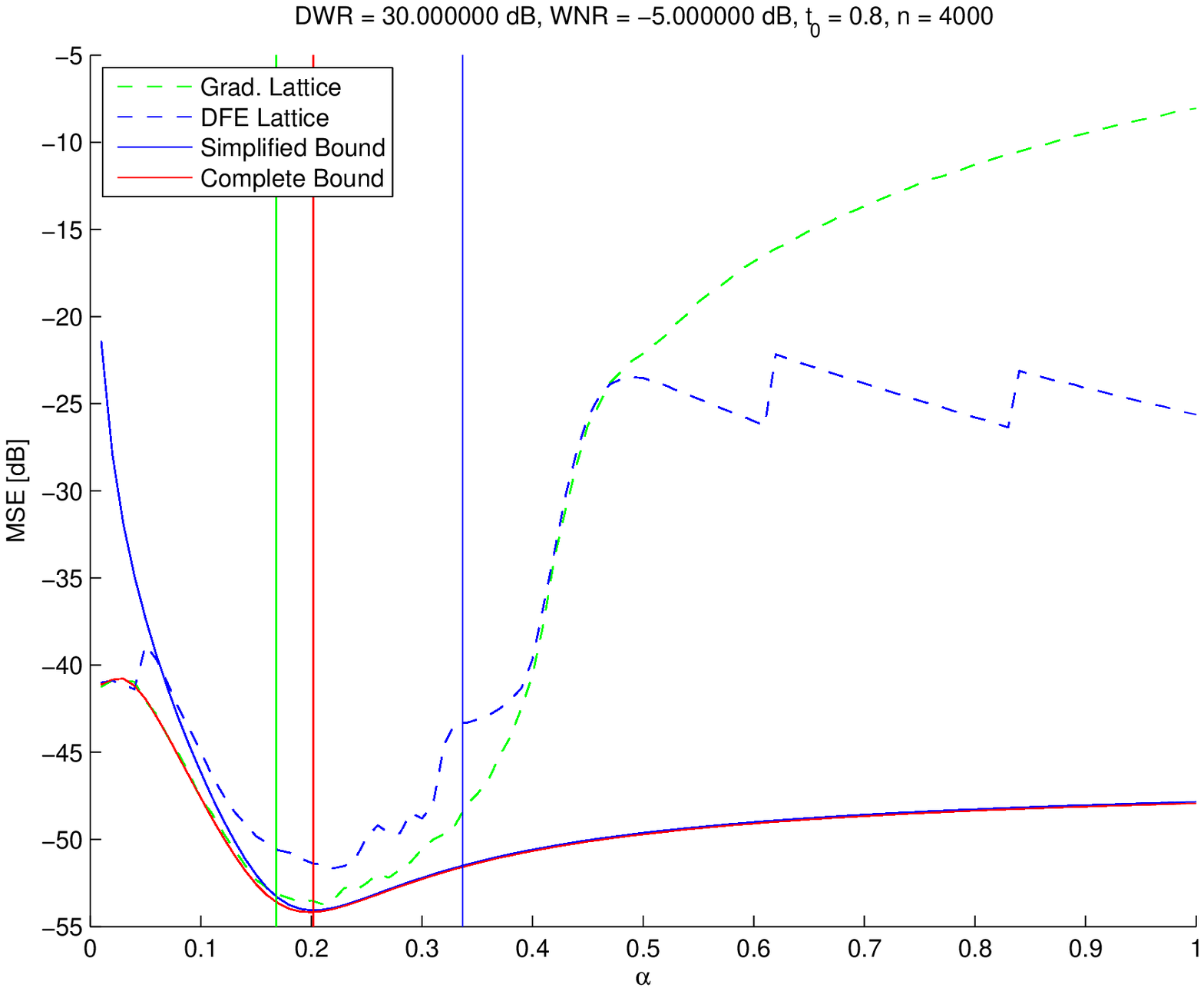}  
    \caption{MSE as a function of $\alpha$ obtained for different estimation algorithms and multidimensional lattices. DWR = $30$ dB, WNR = $-5$ dB, $n=4000$.}\label{fig:mse11}
  \end{center}
\end{figure}

\begin{figure}[t] 
  \begin{center}
    \includegraphics[width=0.75\linewidth]{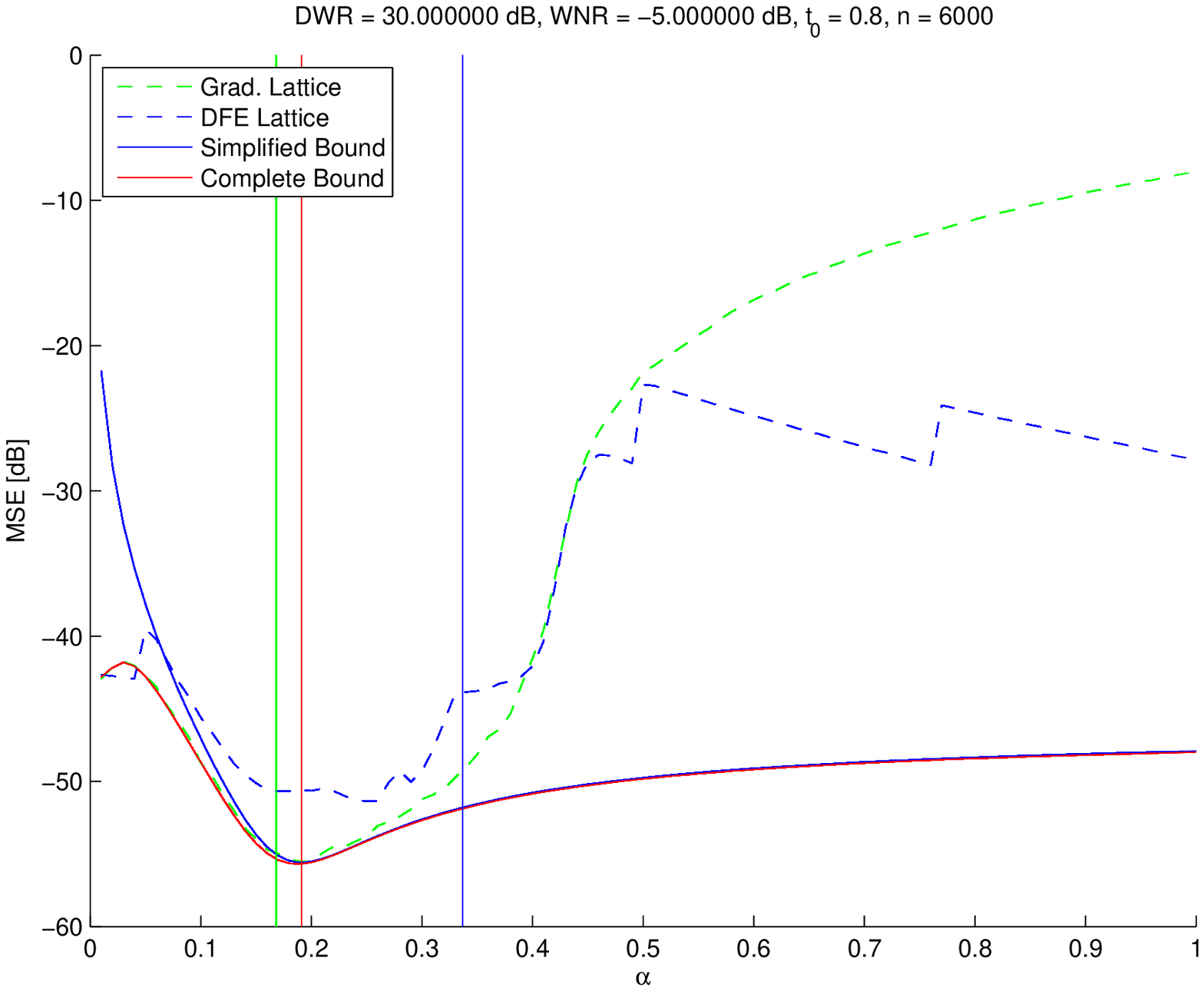}  
    \caption{MSE as a function of $\alpha$ obtained for different estimation algorithms and multidimensional lattices. DWR = $30$ dB, WNR = $-5$ dB, $n=6000$.}\label{fig:mse12}
  \end{center}
\end{figure}

\begin{figure}[t] 
  \begin{center}
    \includegraphics[width=0.75\linewidth]{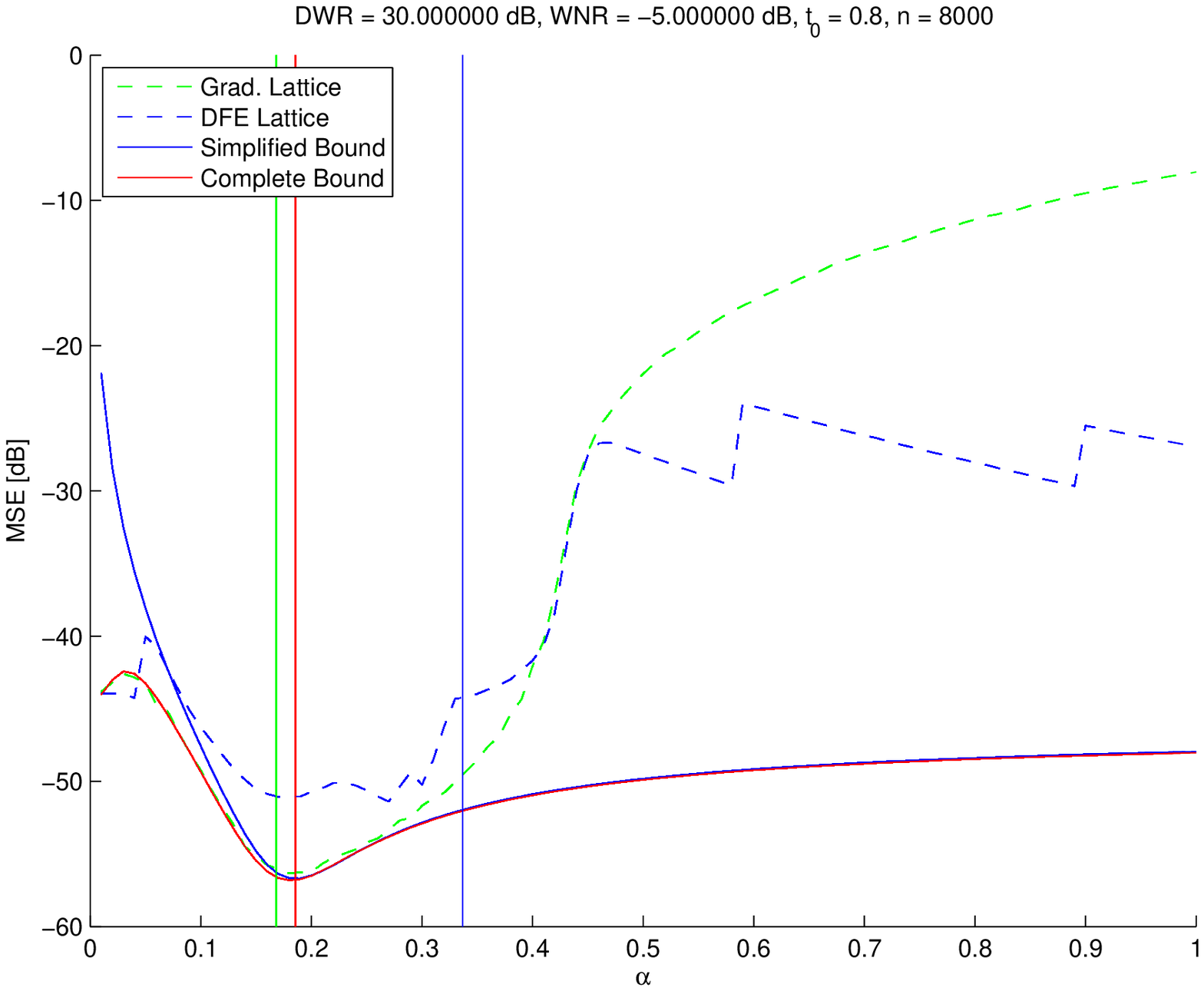}  
    \caption{MSE as a function of $\alpha$ obtained for different estimation algorithms and multidimensional lattices. DWR = $30$ dB, WNR = $-5$ dB, $n=8000$.}\label{fig:mse13}
  \end{center}
\end{figure}

\begin{figure}[t] 
  \begin{center}
    \includegraphics[width=0.75\linewidth]{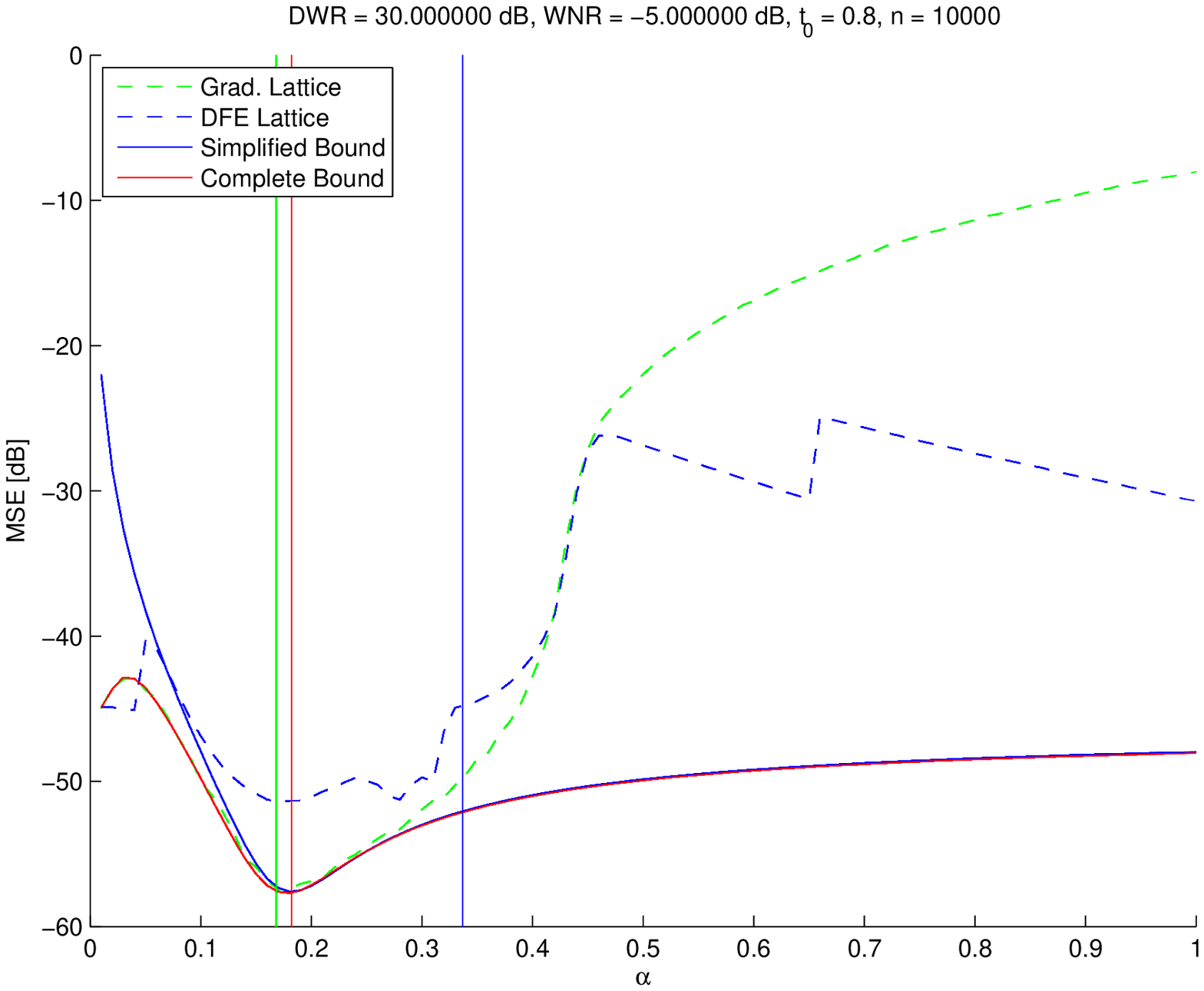}  
    \caption{MSE as a function of $\alpha$ obtained for different estimation algorithms and multidimensional lattices. DWR = $30$ dB, WNR = $-5$ dB, $n=10000$.}\label{fig:mse14}
  \end{center}
\end{figure}

\begin{figure}[t] 
  \begin{center}
    \includegraphics[width=0.75\linewidth]{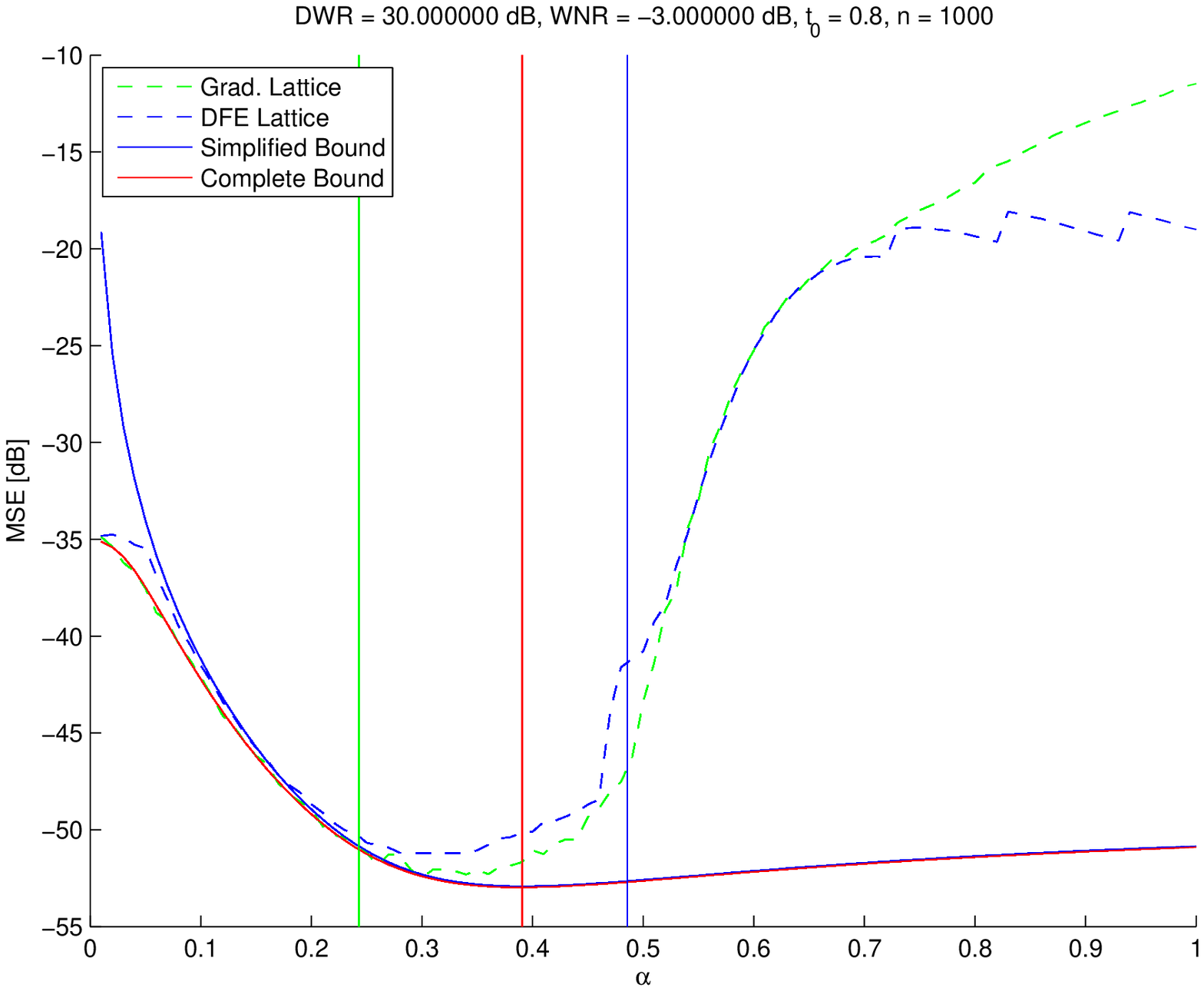}  
    \caption{MSE as a function of $\alpha$ obtained for different estimation algorithms and multidimensional lattices. DWR = $30$ dB, WNR = $-3$ dB, $n=1000$.}\label{fig:mse15}
  \end{center}
\end{figure}

\begin{figure}[t] 
  \begin{center}
    \includegraphics[width=0.75\linewidth]{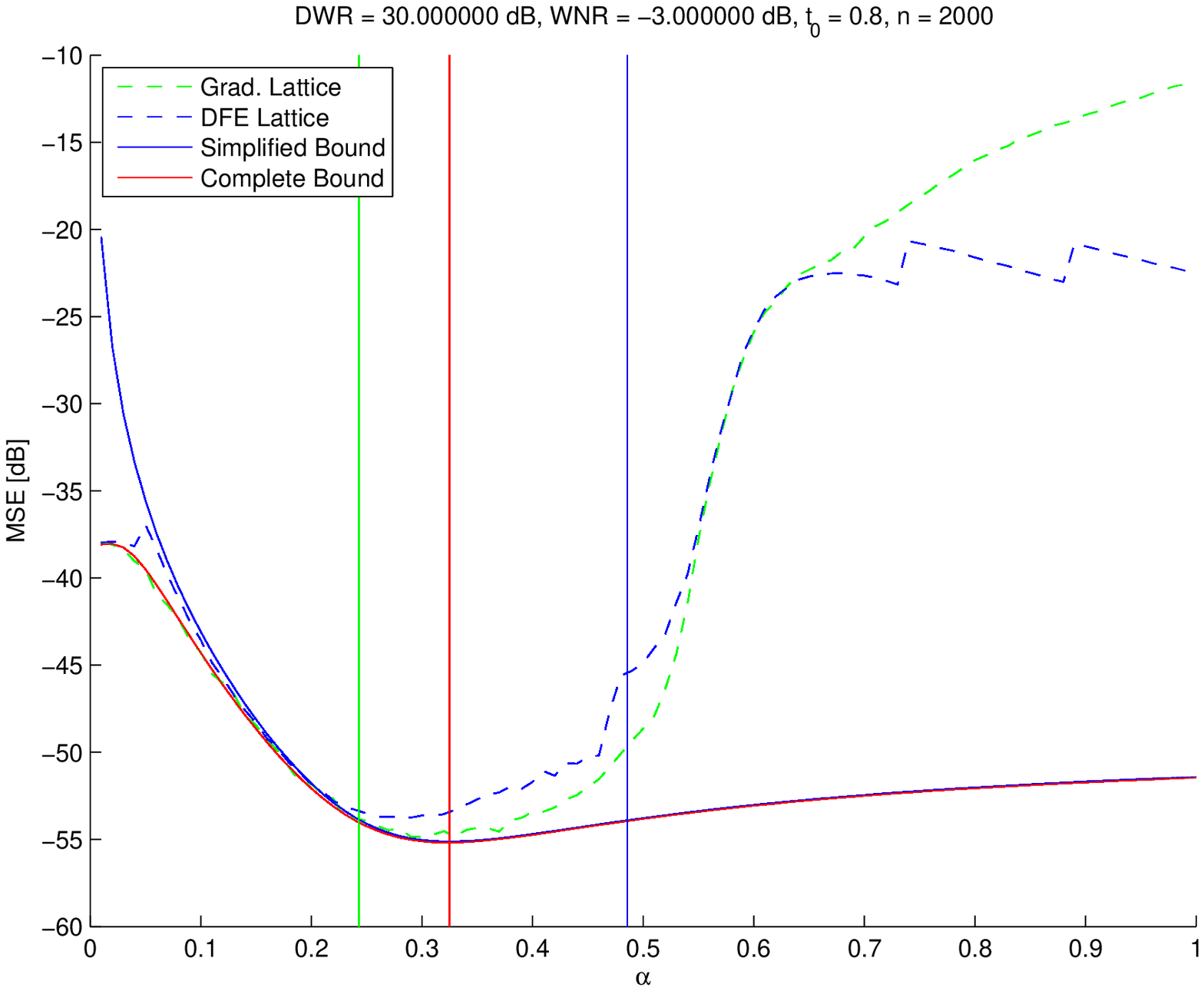}  
    \caption{MSE as a function of $\alpha$ obtained for different estimation algorithms and multidimensional lattices. DWR = $30$ dB, WNR = $-3$ dB, $n=2000$.}\label{fig:mse16}
  \end{center}
\end{figure}

\begin{figure}[t] 
  \begin{center}
    \includegraphics[width=0.75\linewidth]{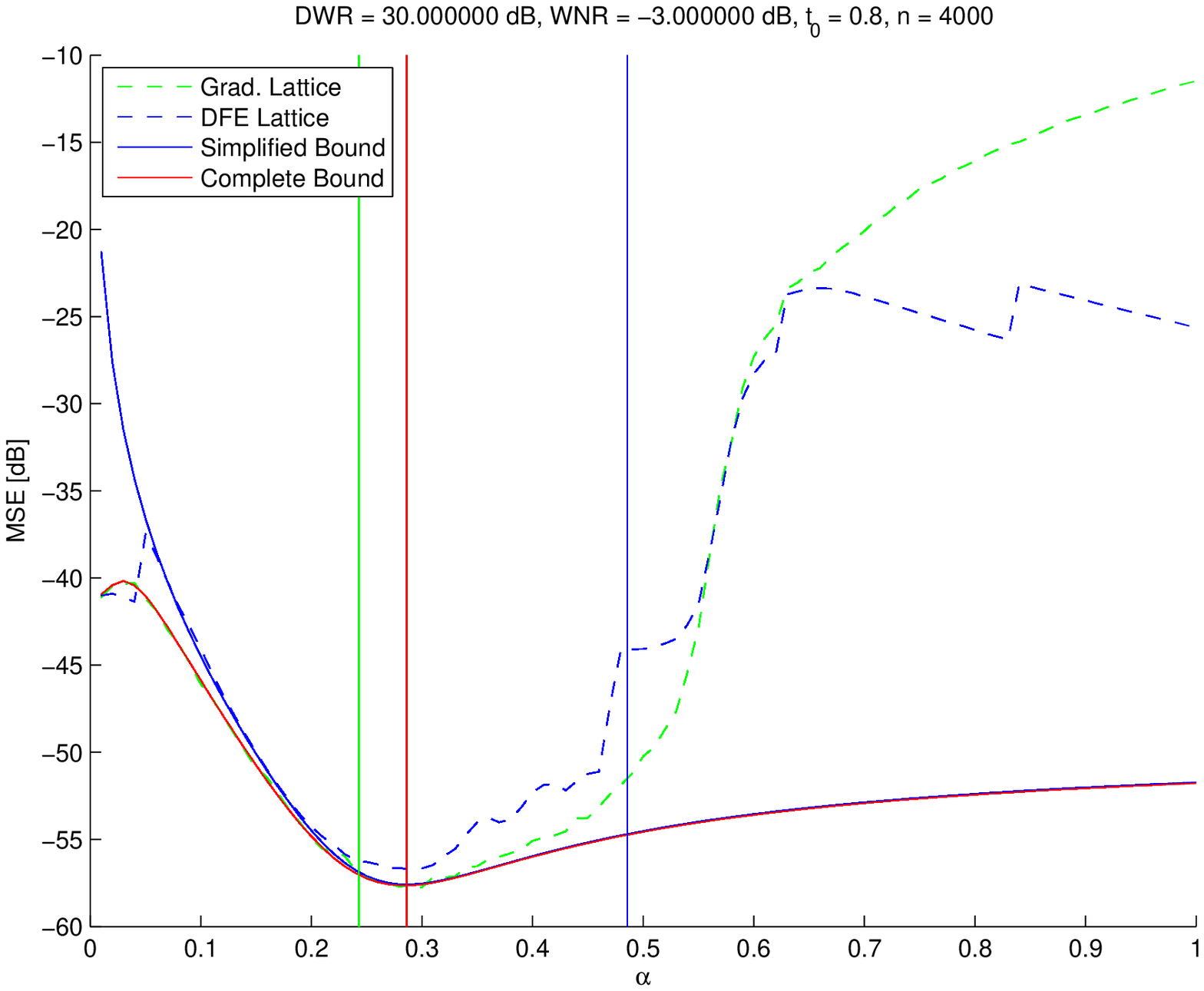}  
    \caption{MSE as a function of $\alpha$ obtained for different estimation algorithms and multidimensional lattices. DWR = $30$ dB, WNR = $-3$ dB, $n=4000$.}\label{fig:mse17}
  \end{center}
\end{figure}

\begin{figure}[t] 
  \begin{center}
    \includegraphics[width=0.75\linewidth]{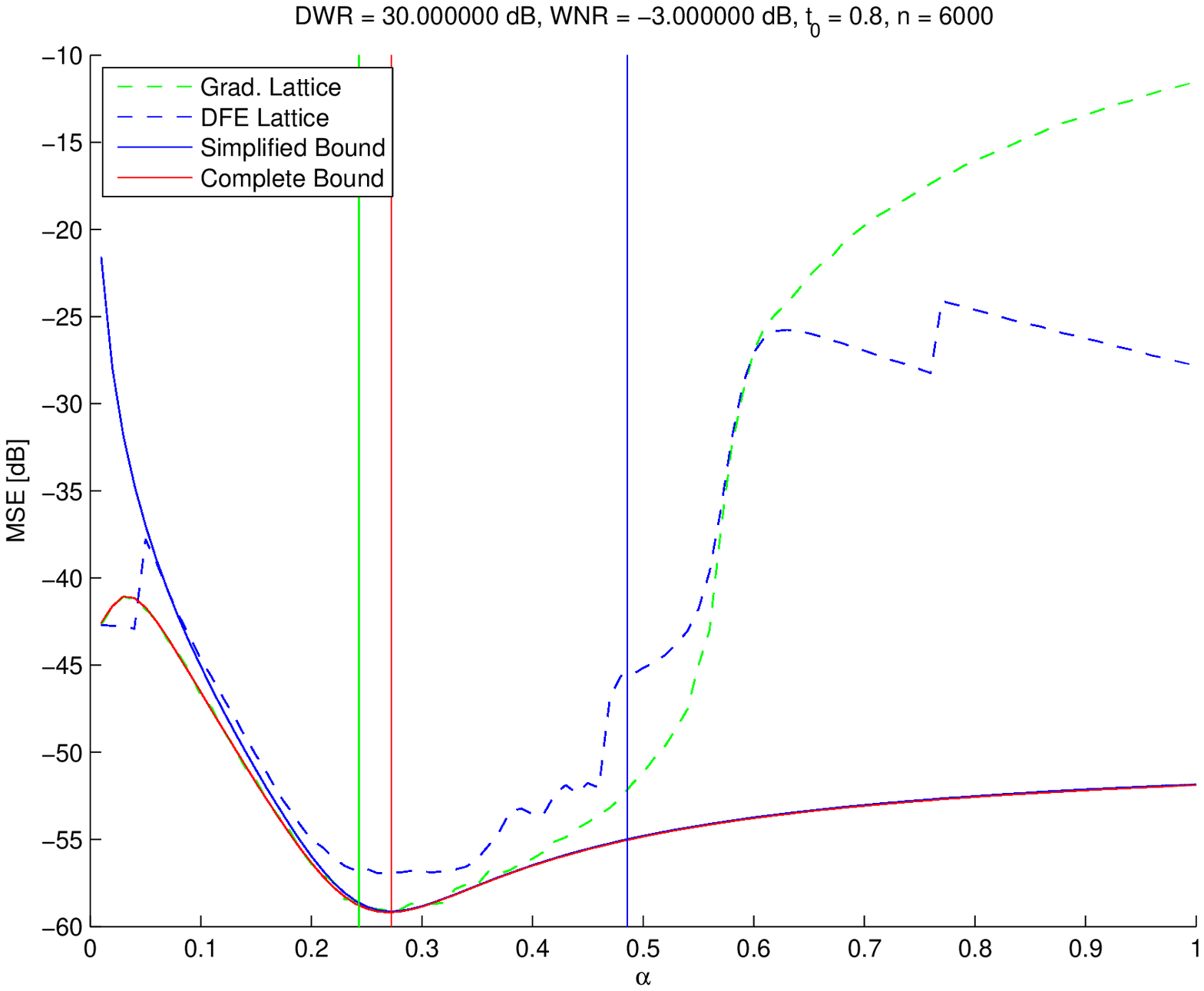}  
    \caption{MSE as a function of $\alpha$ obtained for different estimation algorithms and multidimensional lattices. DWR = $30$ dB, WNR = $-3$ dB, $n=6000$.}\label{fig:mse18}
  \end{center}
\end{figure}

\begin{figure}[t] 
  \begin{center}
    \includegraphics[width=0.75\linewidth]{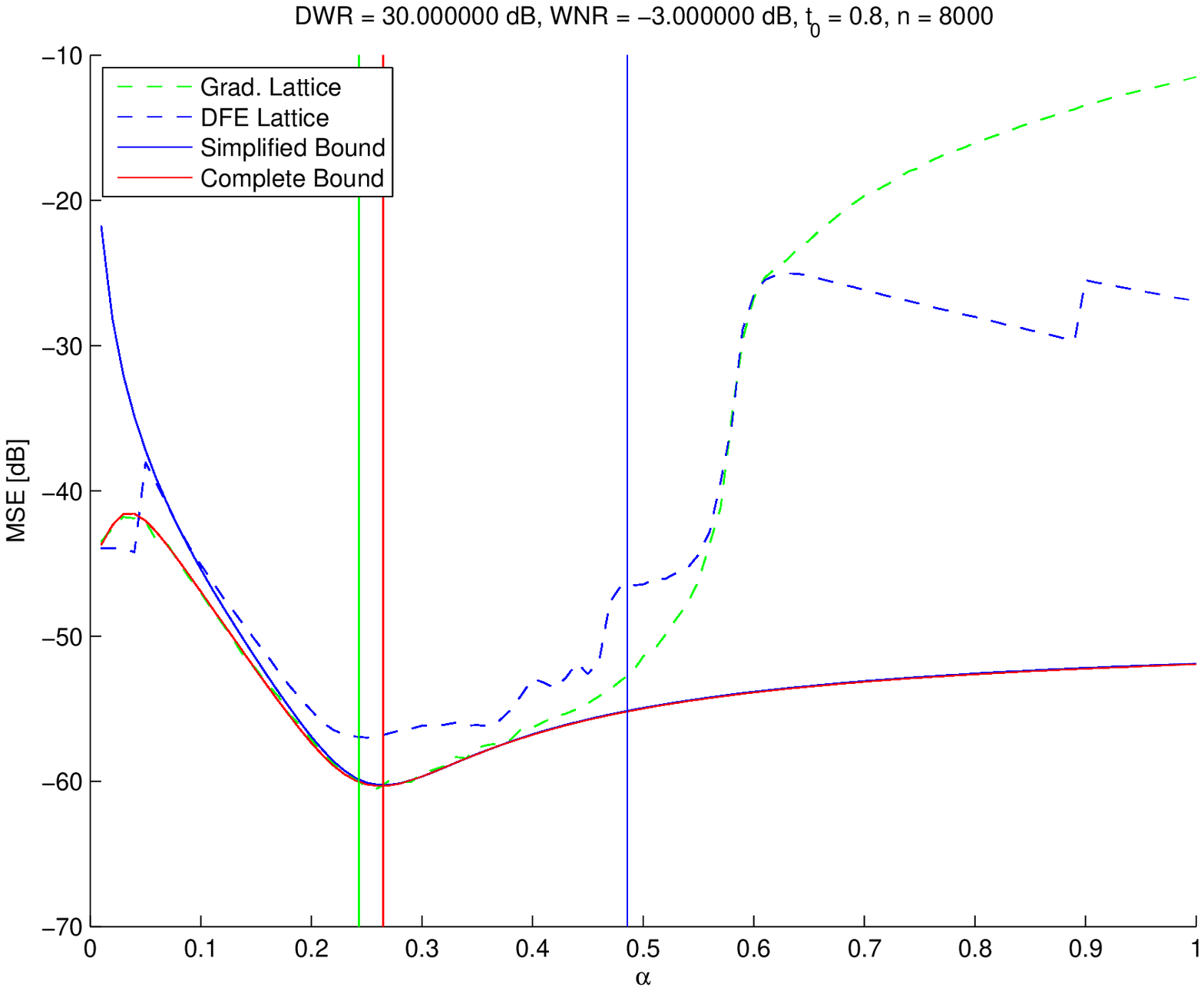}  
    \caption{MSE as a function of $\alpha$ obtained for different estimation algorithms and multidimensional lattices. DWR = $30$ dB, WNR = $-3$ dB, $n=8000$.}\label{fig:mse19}
  \end{center}
\end{figure}

\begin{figure}[t] 
  \begin{center}
    \includegraphics[width=0.75\linewidth]{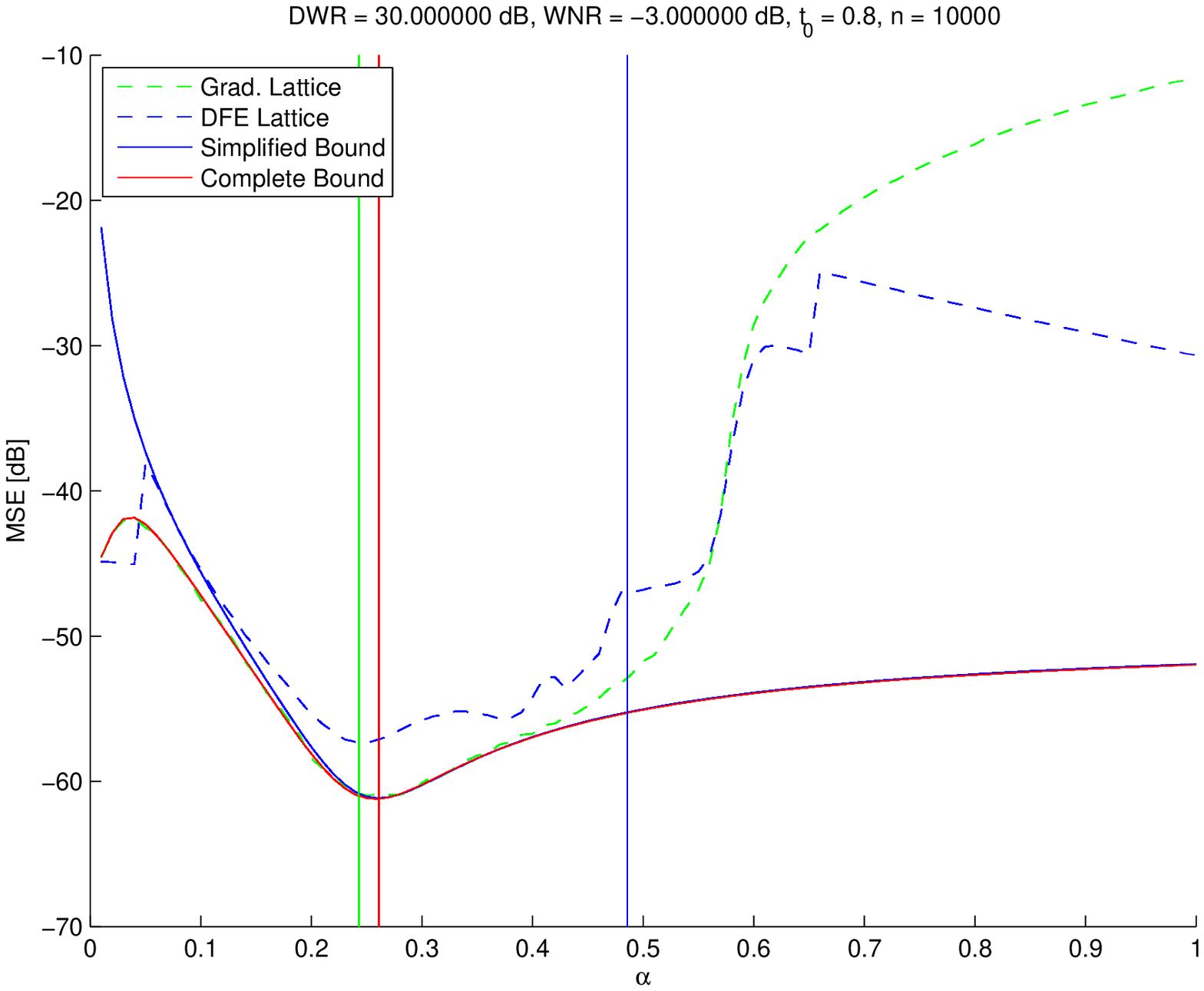}  
    \caption{MSE as a function of $\alpha$ obtained for different estimation algorithms and multidimensional lattices. DWR = $30$ dB, WNR = $-3$ dB, $n=10000$.}\label{fig:mse20}
  \end{center}
\end{figure}

\begin{figure}[t] 
  \begin{center}
    \includegraphics[width=0.75\linewidth]{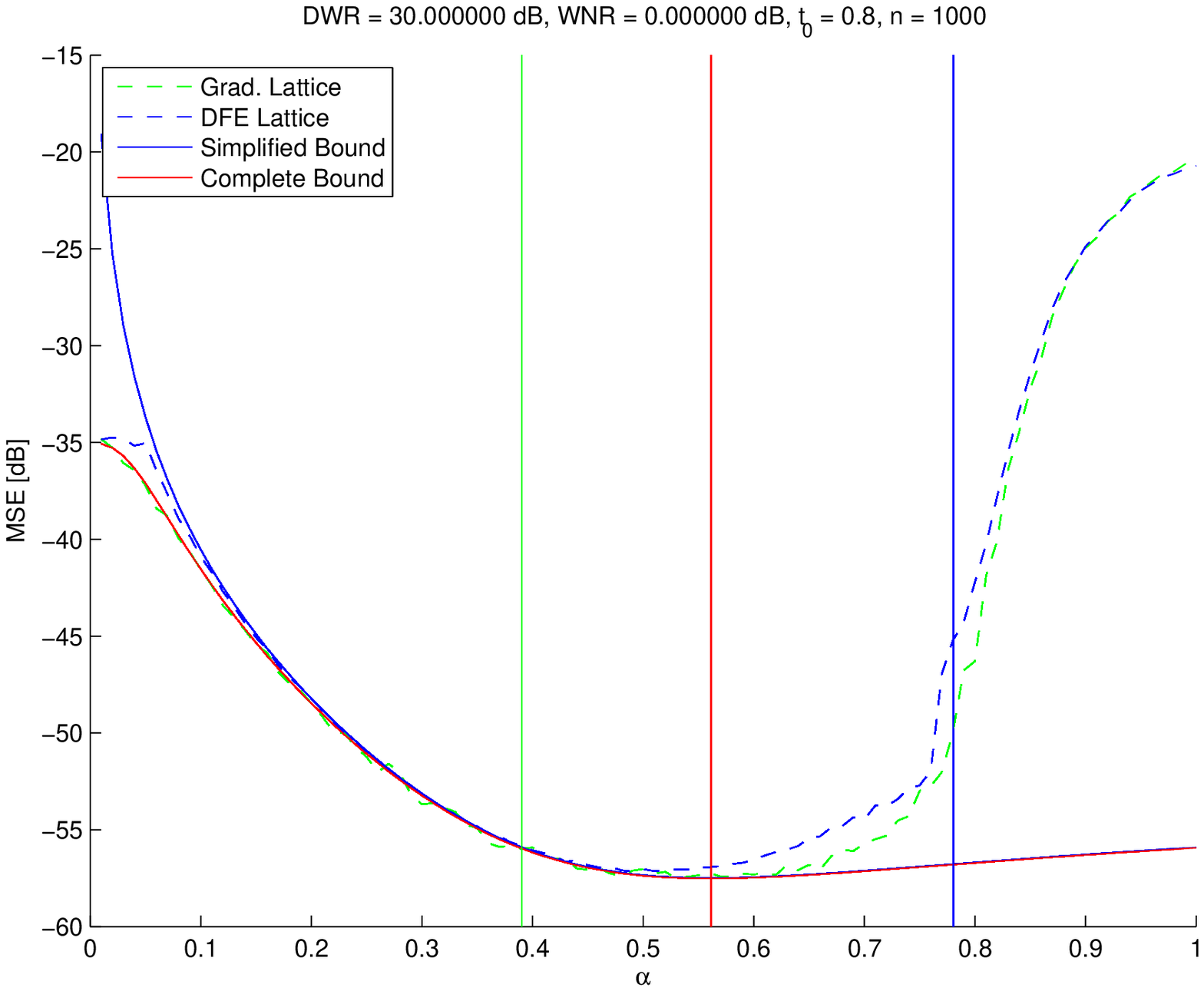}  
    \caption{MSE as a function of $\alpha$ obtained for different estimation algorithms and multidimensional lattices. DWR = $30$ dB, WNR = $0$ dB, $n=1000$.}\label{fig:mse21}
  \end{center}
\end{figure}

\begin{figure}[t] 
  \begin{center}
    \includegraphics[width=0.75\linewidth]{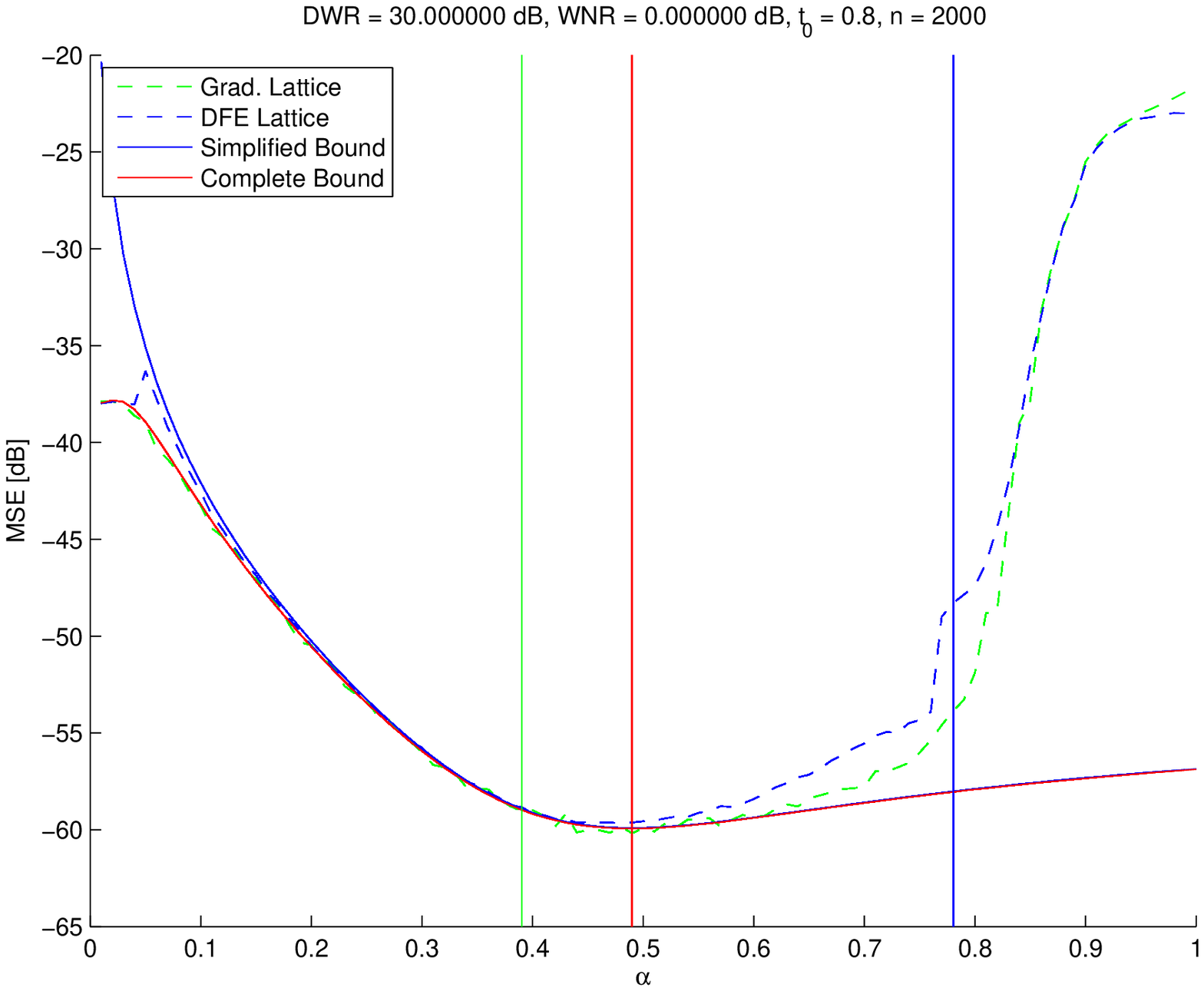}  
    \caption{MSE as a function of $\alpha$ obtained for different estimation algorithms and multidimensional lattices. DWR = $30$ dB, WNR = $0$ dB, $n=2000$.}\label{fig:mse22}
  \end{center}
\end{figure}

\begin{figure}[t] 
  \begin{center}
    \includegraphics[width=0.75\linewidth]{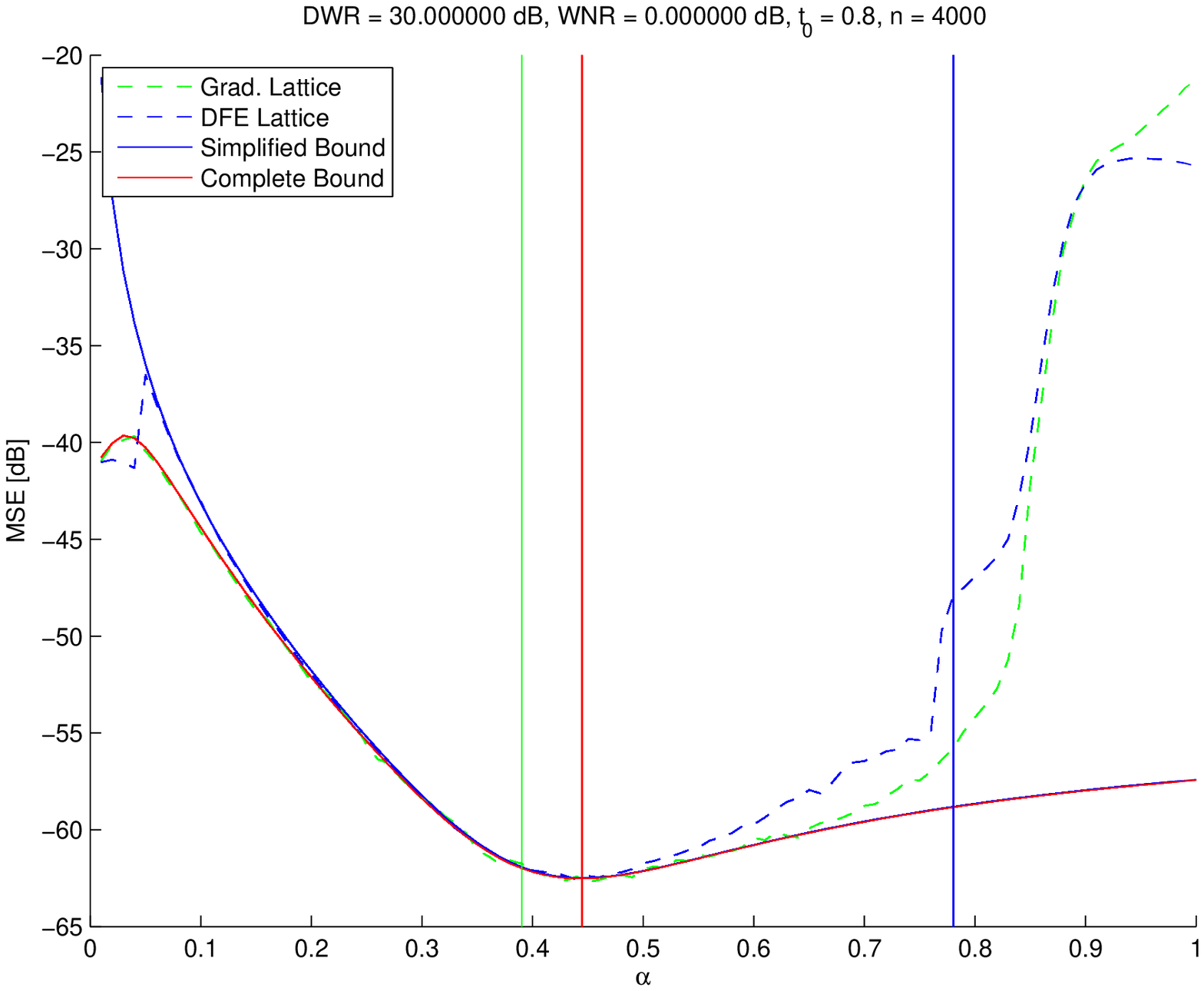}  
    \caption{MSE as a function of $\alpha$ obtained for different estimation algorithms and multidimensional lattices. DWR = $30$ dB, WNR = $0$ dB, $n=4000$.}\label{fig:mse23}
  \end{center}
\end{figure}

\begin{figure}[t] 
  \begin{center}
    \includegraphics[width=0.75\linewidth]{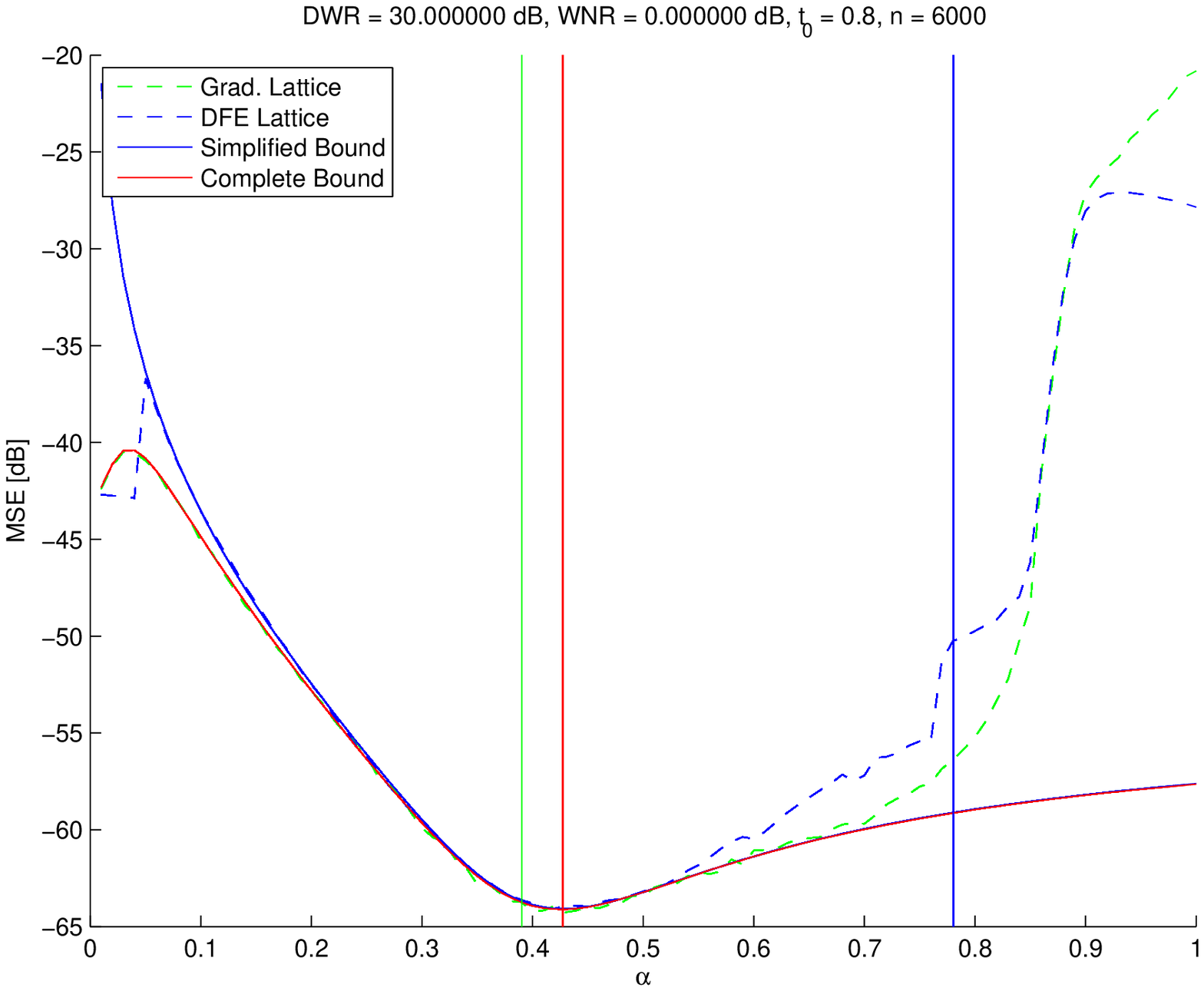}  
    \caption{MSE as a function of $\alpha$ obtained for different estimation algorithms and multidimensional lattices. DWR = $30$ dB, WNR = $0$ dB, $n=6000$.}\label{fig:mse24}
  \end{center}
\end{figure}

\begin{figure}[t] 
  \begin{center}
    \includegraphics[width=0.75\linewidth]{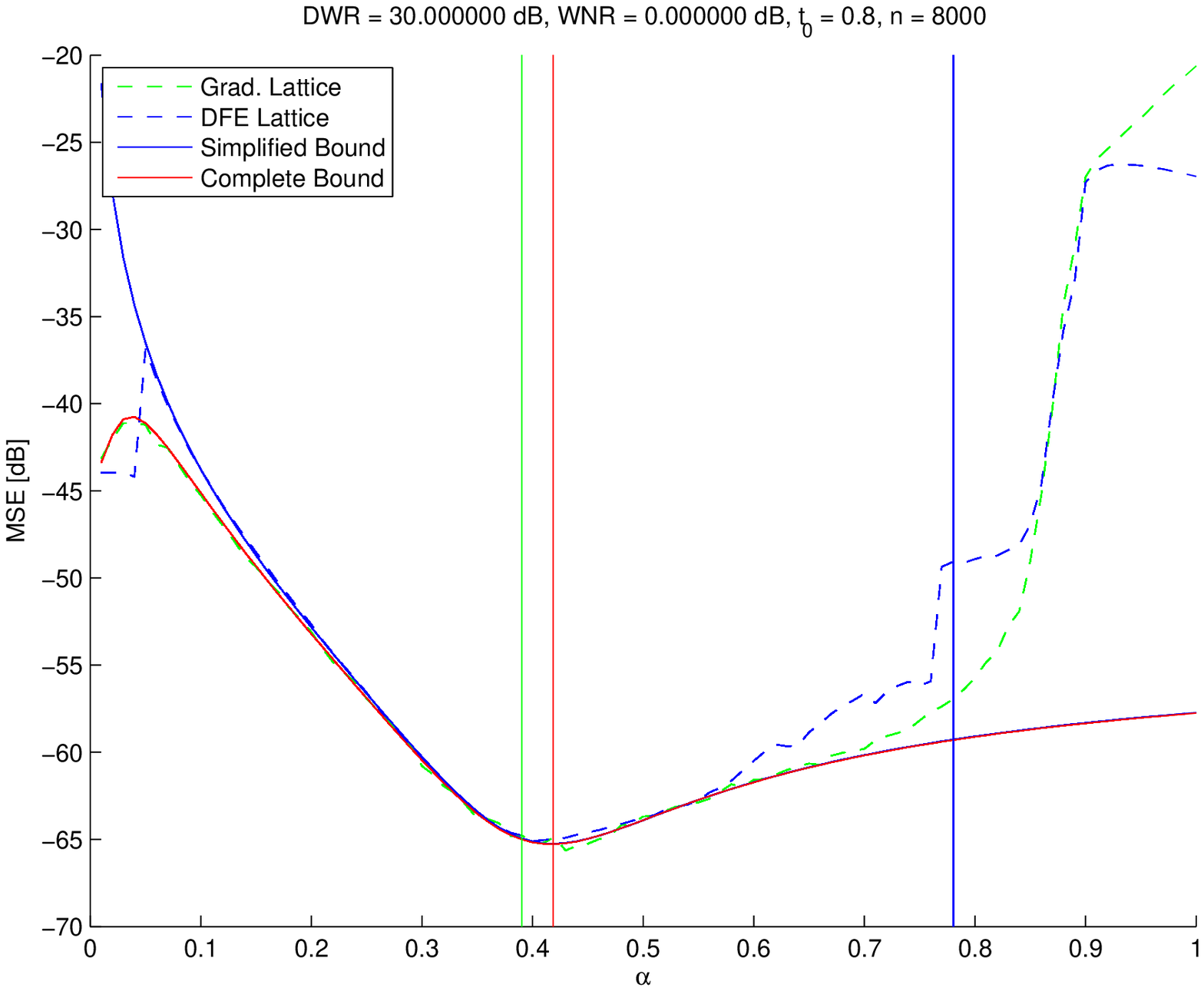}  
    \caption{MSE as a function of $\alpha$ obtained for different estimation algorithms and multidimensional lattices. DWR = $30$ dB, WNR = $0$ dB, $n=8000$.}\label{fig:mse25}
  \end{center}
\end{figure}

\begin{figure}[t] 
  \begin{center}
    \includegraphics[width=0.75\linewidth]{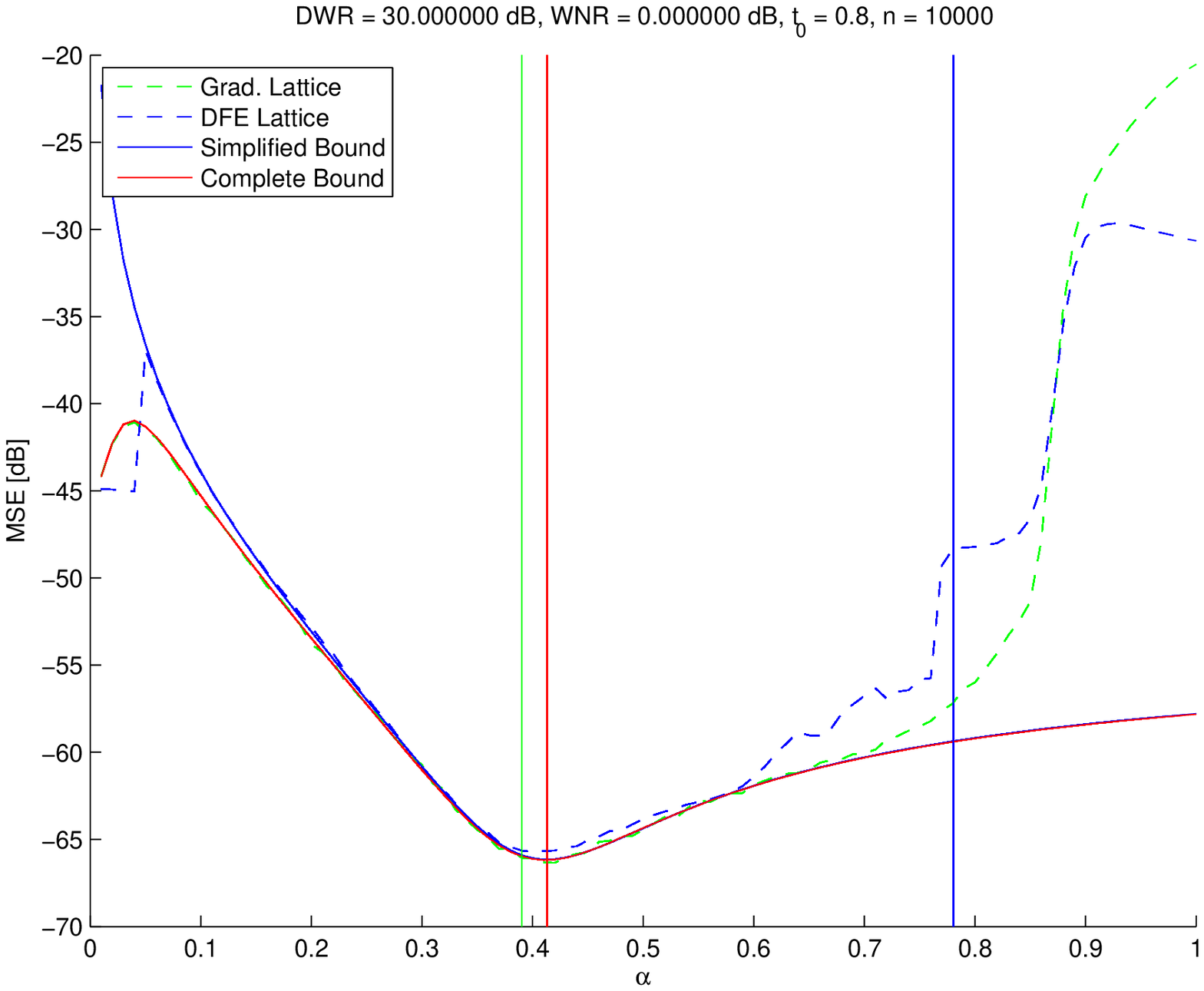}  
    \caption{MSE as a function of $\alpha$ obtained for different estimation algorithms and multidimensional lattices. DWR = $30$ dB, WNR = $0$ dB, $n=10000$.}\label{fig:mse26}
  \end{center}
\end{figure}

\begin{figure}[t] 
  \begin{center}
    \includegraphics[width=0.75\linewidth]{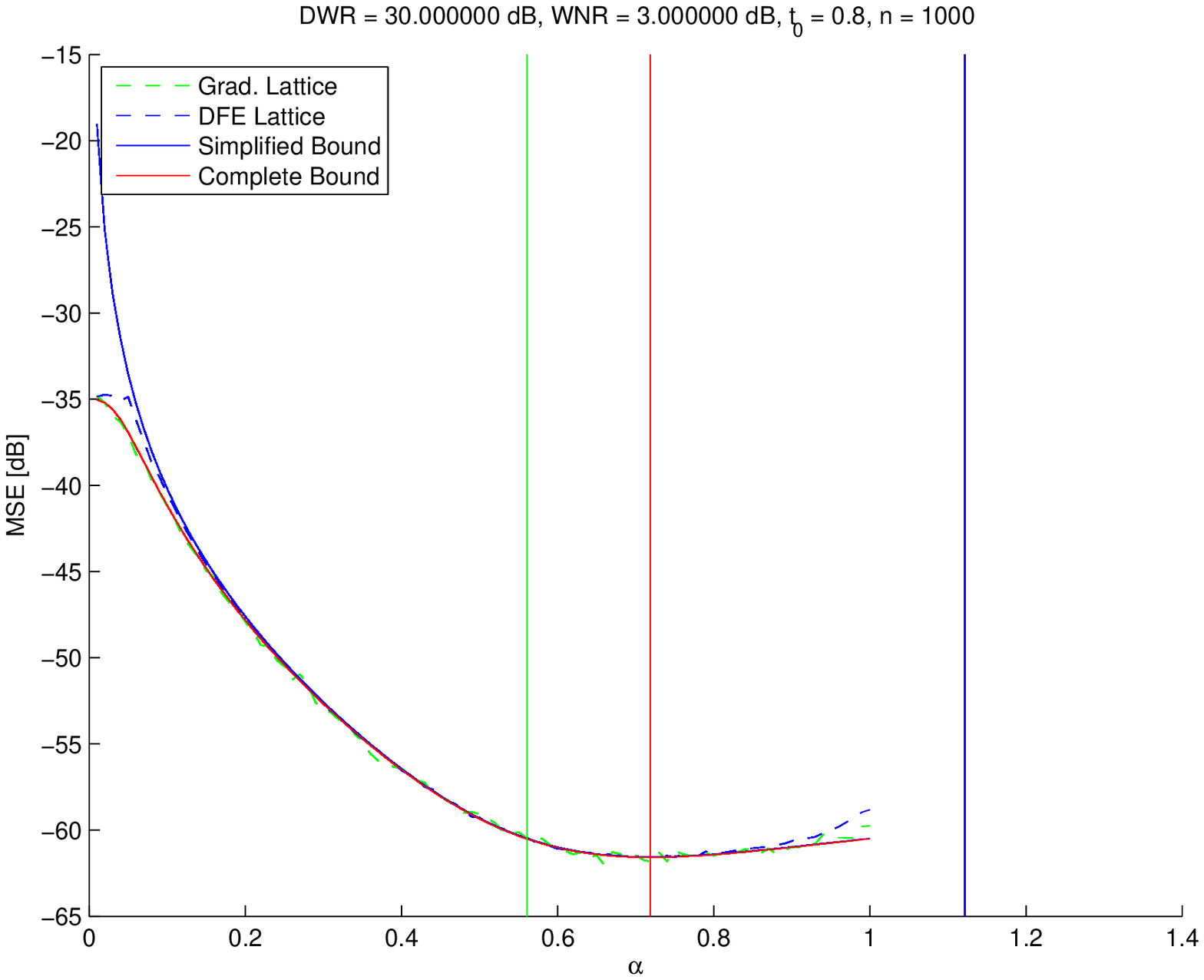}  
    \caption{MSE as a function of $\alpha$ obtained for different estimation algorithms and multidimensional lattices. DWR = $30$ dB, WNR = $3$ dB, $n=1000$.}\label{fig:mse27}
  \end{center}
\end{figure}

\begin{figure}[t] 
  \begin{center}
    \includegraphics[width=0.75\linewidth]{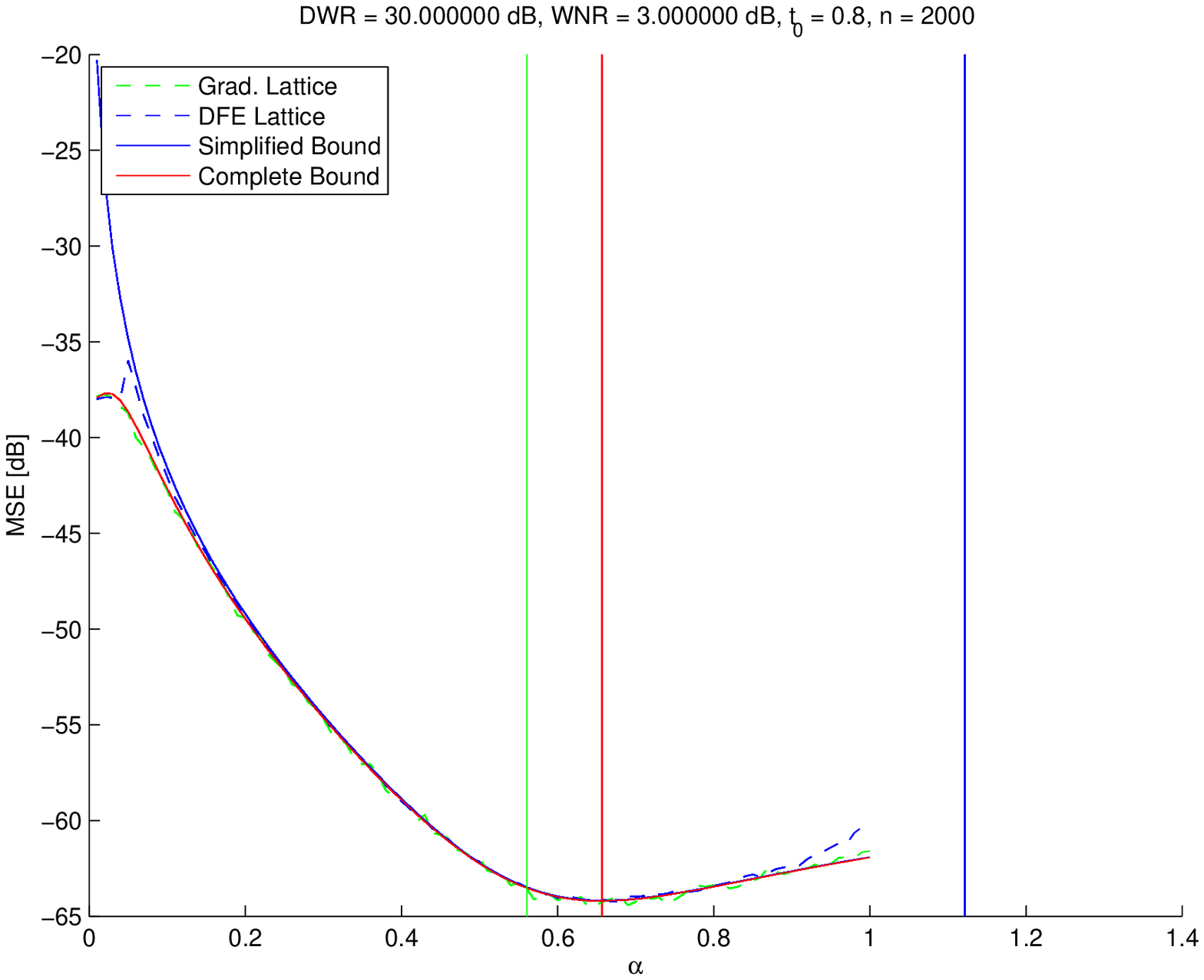}  
    \caption{MSE as a function of $\alpha$ obtained for different estimation algorithms and multidimensional lattices. DWR = $30$ dB, WNR = $3$ dB, $n=2000$.}\label{fig:mse28}
  \end{center}
\end{figure}

\begin{figure}[t] 
  \begin{center}
    \includegraphics[width=0.75\linewidth]{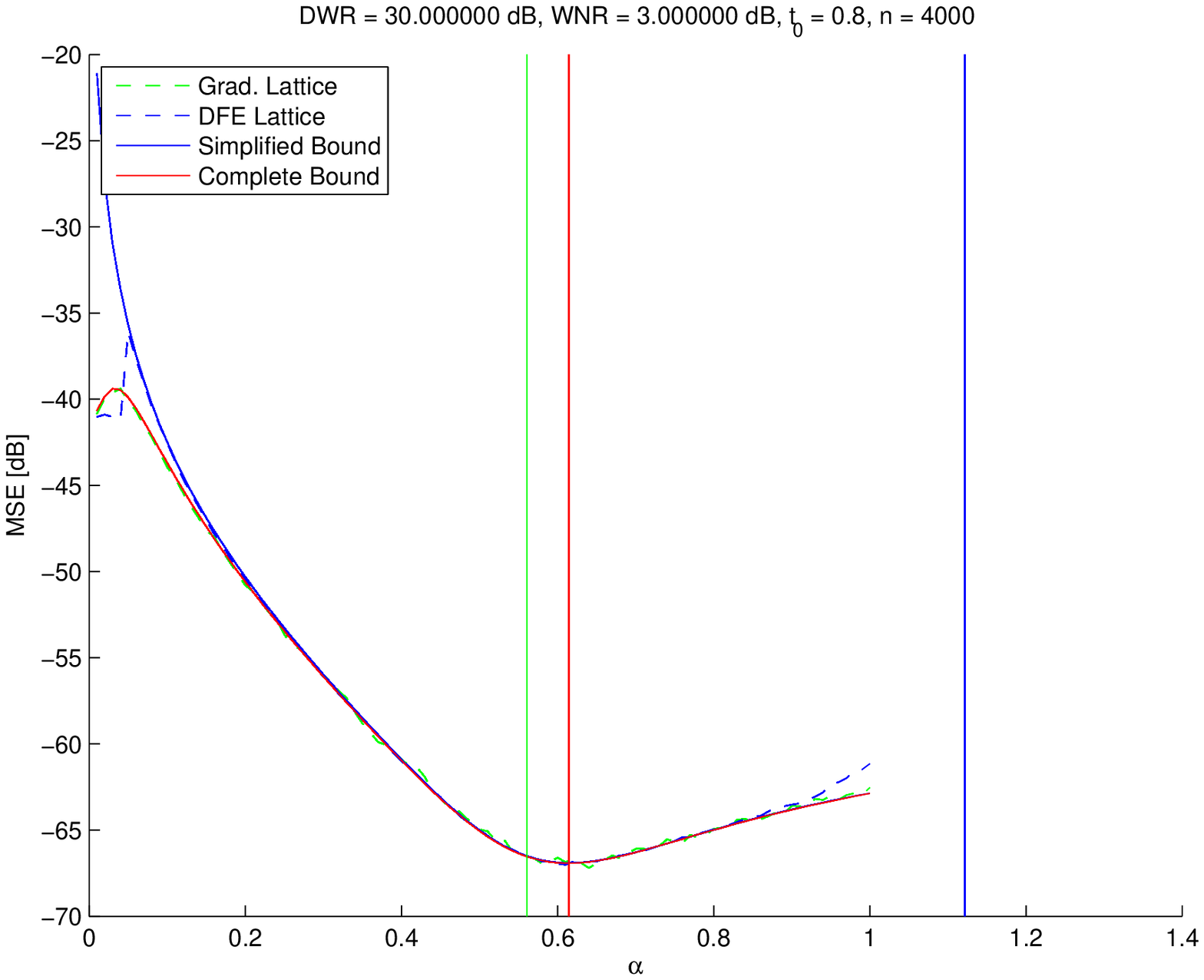}  
    \caption{MSE as a function of $\alpha$ obtained for different estimation algorithms and multidimensional lattices. DWR = $30$ dB, WNR = $3$ dB, $n=4000$.}\label{fig:mse29}
  \end{center}
\end{figure}

\begin{figure}[t] 
  \begin{center}
    \includegraphics[width=0.75\linewidth]{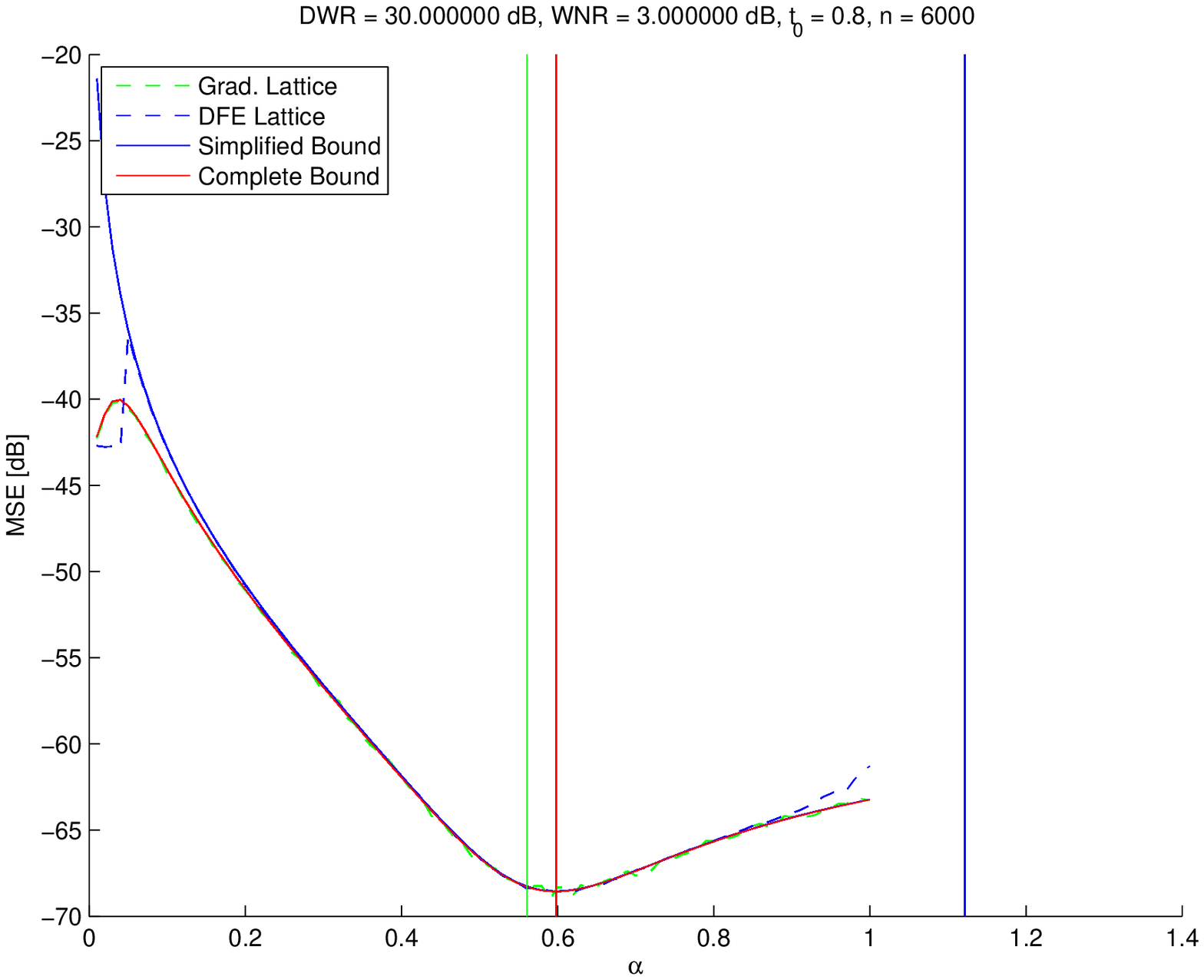}  
    \caption{MSE as a function of $\alpha$ obtained for different estimation algorithms and multidimensional lattices. DWR = $30$ dB, WNR = $3$ dB, $n=6000$.}\label{fig:mse30}
  \end{center}
\end{figure}

\begin{figure}[t] 
  \begin{center}
    \includegraphics[width=0.75\linewidth]{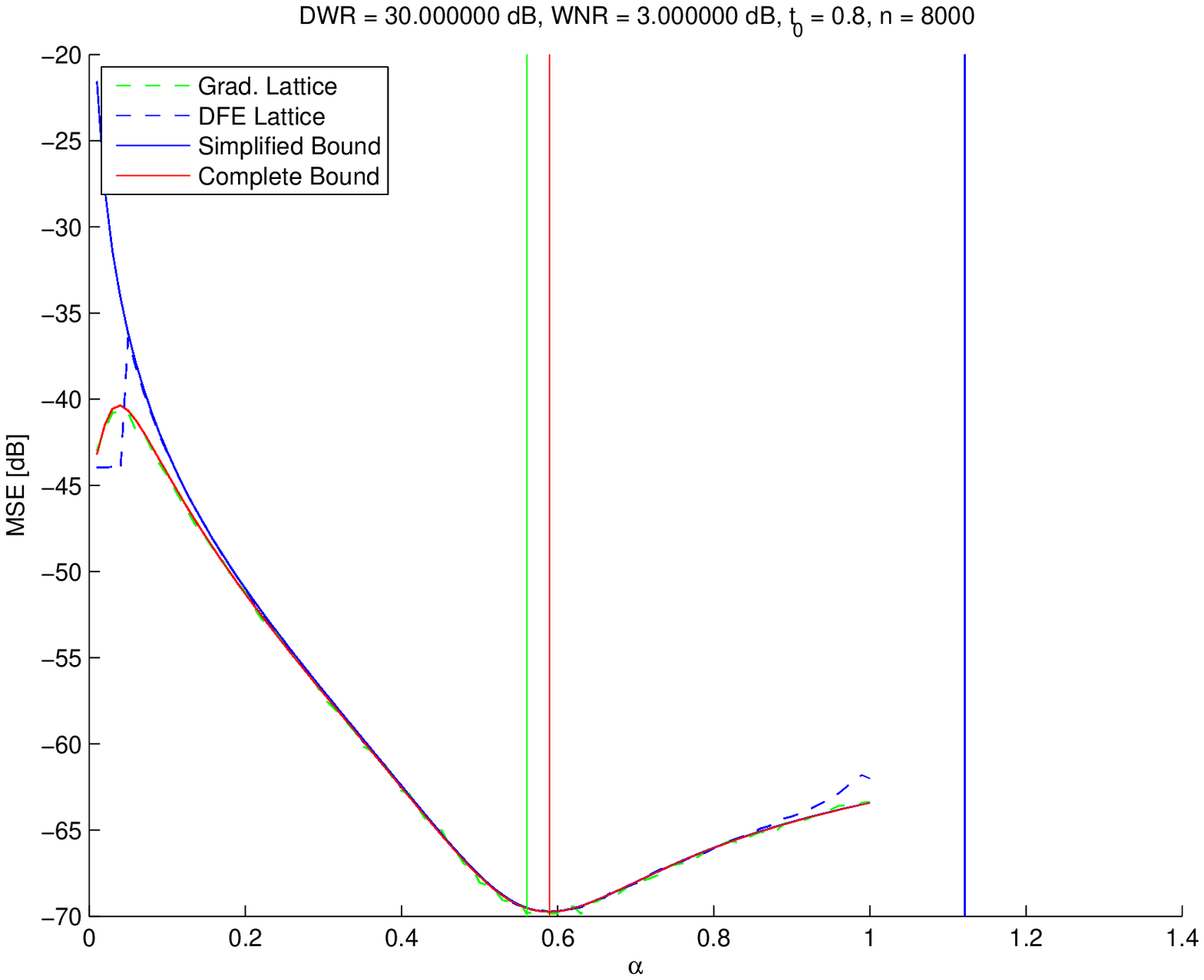}  
    \caption{MSE as a function of $\alpha$ obtained for different estimation algorithms and multidimensional lattices. DWR = $30$ dB, WNR = $3$ dB, $n=8000$.}\label{fig:mse31}
  \end{center}
\end{figure}

\begin{figure}[t] 
  \begin{center}
    \includegraphics[width=0.75\linewidth]{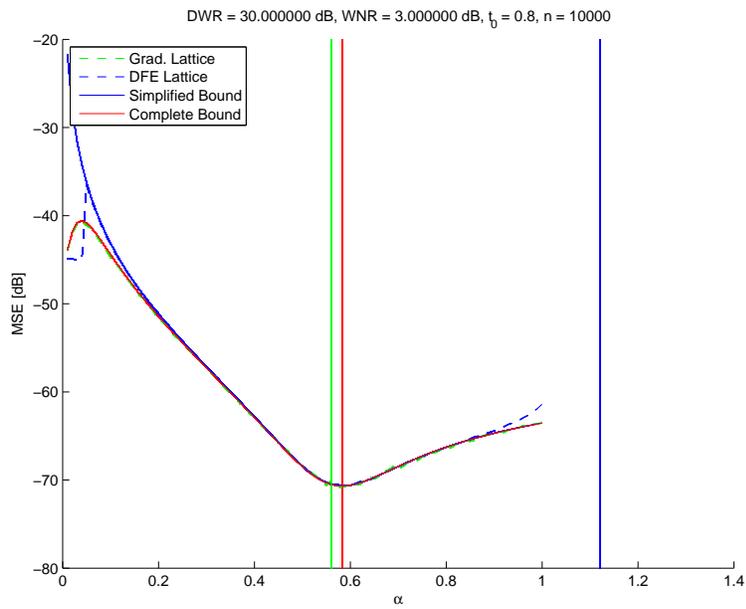}  
    \caption{MSE as a function of $\alpha$ obtained for different estimation algorithms and multidimensional lattices. DWR = $30$ dB, WNR = $3$ dB, $n=10000$.}\label{fig:mse32}
  \end{center}
\end{figure}

A second set of plots are those in Figs.~\ref{fig:mse9}-\ref{fig:mse32}, where we show the behavior of the MSE as a function of $\alpha$, in order to illustrate the correctness of the discussion in Sect.~\ref{sec:total_mse}. In particular, we can see how the optimal value of $\alpha$ depends on the ratio $\sigma_X^2/n$, as a trade-off between the CRB component of the MSE and the bias of the estimation must be achieved. These plots also show the $\alpha_{\textrm{no-bias}}$ value (remember that this is the $\alpha$ derived by Costa for the communications problem) as a solid green line, $\alpha_{\textrm{sup-FI}}$ as a solid blue line, and $\alpha_{\textrm{opt}}$ as a solid red line, illustrating that indeed the optimal value of $\alpha$ is virtually $\alpha_{\textrm{opt}}$. Furthermore, and contrarily to what happens in the communications framework, the location of the optimal value of $\alpha$ (i.e., approximately $\alpha_{\textrm{opt}}$) depends on $n$.
We would like to emphasize the outstanding agreement between theoretical and experimental results show in these plots, as well as the expected significant increase in the MSE for those values of $\alpha$ larger than $\alpha_{\textrm{sup-FI}}$.

%%%%%%%%%%%%%%%%%%%%%%%%%%%%%%%%%%%%%%%%%%%%%%%%%%%%%%%%%%%%%%%%%%
%%%%%%%%%%%%%%%%%%%%%%%%%%%%%%%%%%%%%%%%%%%%%%%%%%%%%%%%%%%%%%%%%%

\clearpage

\clearpage

\clearpage
%%%%%%%%%%%%%%%%%%%%%%%%%%%%%%%%%%%%%%%%%%%%%%%%%%%%%%%%%%%%%%%%%%
%%%%%%%%%%%%%%%%%%%%%%%%%%%%%%%%%%%%%%%%%%%%%%%%%%%%%%%%%%%%%%%%%%

\clearpage

\clearpage

\clearpage
%%%%%%%%%%%%%%%%%%%%%%%%%%%%%%%%%%%%%%%%%%%%%%%%%%%%%%%%%%%%%%%%%%
%%%%%%%%%%%%%%%%%%%%%%%%%%%%%%%%%%%%%%%%%%%%%%%%%%%%%%%%%%%%%%%%%%

\clearpage

\clearpage

\clearpage

%%%%%%%%%%%%%%%%%%%%%%%%%%%%%%%%%%%%%%%%%%%%%%%%%%%%%%%%%%%%%%%%%%
%%%%%%%%%%%%%%%%%%%%%%%%%%%%%%%%%%%%%%%%%%%%%%%%%%%%%%%%%%%%%%%%%%

\clearpage

\clearpage

\clearpage

%%%%%%%%%%%%%%%%%%%%%%%%%%%%%%%%%%%%%%%%%%%%%%%%%%%%%%%%%%%%%%%%%%
%%%%%%%%%%%%%%%%%%%%%%%%%%%%%%%%%%%%%%%%%%%%%%%%%%%%%%%%%%%%%%%%%%
%%%%%%%%%%%%%%%%%%%%%%%%%%%%%%%%%%%%%%%%%%%%%%%%%%%%%%%%%%%%%%%%%%
%%%%%%%%%%%%%%%%%%%%%%%%%%%%%%%%%%%%%%%%%%%%%%%%%%%%%%%%%%%%%%%%%%
\appendix
\section{Proof of $\log(x) \geq - \frac{1}{x}$ for
any $x \geq 0$}\label{sec:app3}

\begin{figure}[t] 
  \begin{center}
    \includegraphics[width=0.75\linewidth]{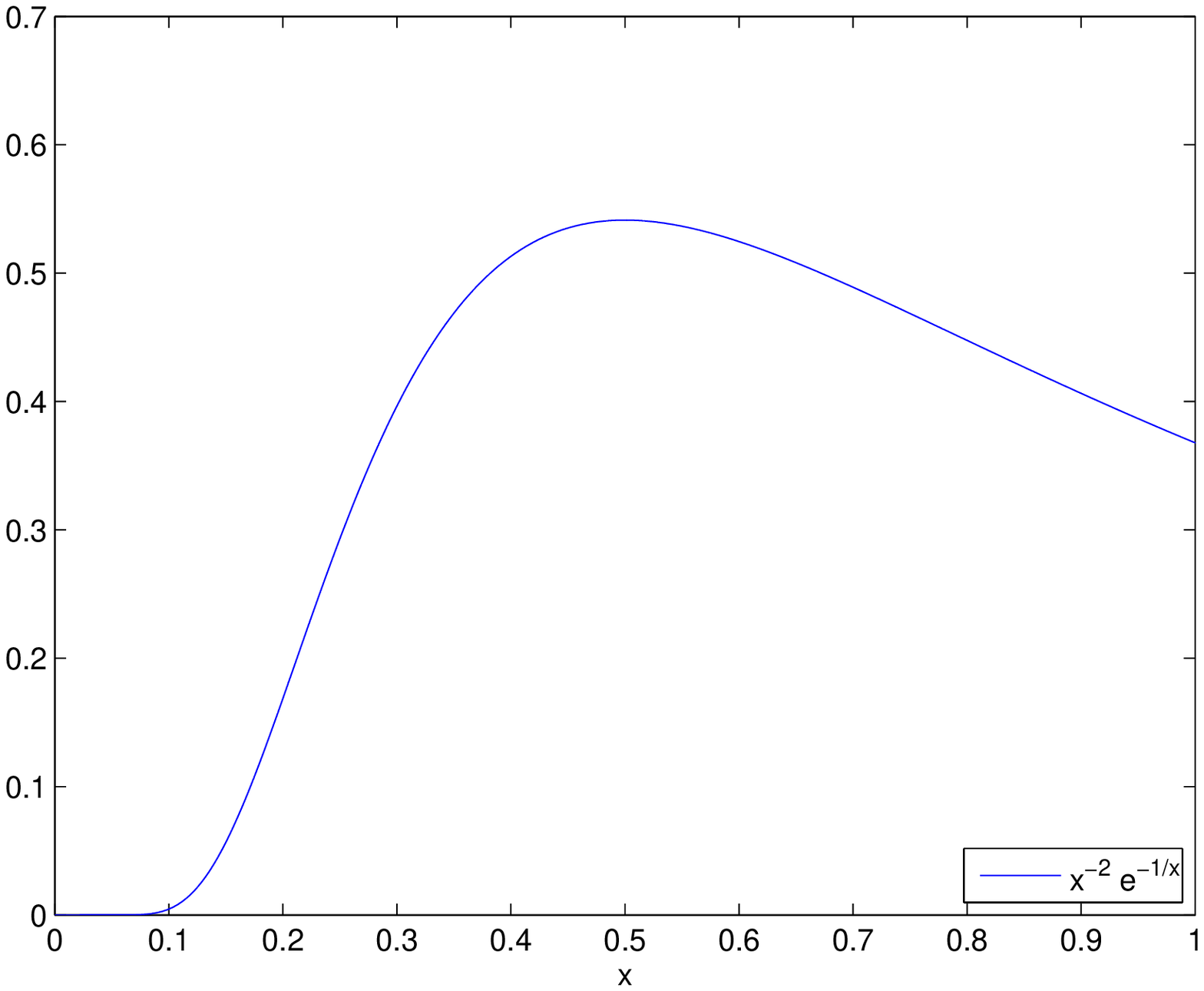}  
    \caption{$\frac{e^{\frac{-1}{x}}}{x^2}$.}\label{fig:app2}
  \end{center}
\end{figure}

Due to the monotonically increasing nature with $x$ of $e^x$, it is
obvious that the target inequality can be rewritten as
\begin{eqnarray}
  x \geq e^{\frac{-1}{x}}, \textrm{   for any  } x \geq 0. \label{eq:app31}
\end{eqnarray}
Given that at $x=0$ both sides of the inequality take the same value, and
that they are analytical functions, a sufficient condition for
(\ref{eq:app31}) to be verified is that the derivative with respect to
$x$ of the left side of (\ref{eq:app31}) is larger than or equal to the
right side for every $x \geq 0$, i.e.,
\begin{eqnarray}
  1 \geq \frac{e^{\frac{-1}{x}}}{x^2}, \textrm{   for every  } x \geq 0. \label{eq:app32}
\end{eqnarray}
Indeed, the derivative of the right side of (\ref{eq:app32}) with respect to
$x$ is
\begin{eqnarray}
  \frac{e^{\frac{-1}{x}}(1-2x)}{x^4}, \nonumber
\end{eqnarray}
so the right side of (\ref{eq:app32}) will have its global maximum located at
$x = 1/2$ (as it is shown in Fig.~\ref{fig:app2}), where it will take the value $4 e^{-2} \approx 0.5413 < 1$, proving
the initial inequality.

%%%%%%%%%%%%%%%%%%%%%%%%%%%%%%%%%%%%%%%%%%%%%%%%%%%%%%%%%%%%%%%%%%
%%%%%%%%%%%%%%%%%%%%%%%%%%%%%%%%%%%%%%%%%%%%%%%%%%%%%%%%%%%%%%%%%%

\section{A useful inequality}\label{sec:app2}

In this appendix we will prove that for any $t_1 \geq t_0$, or
$t_1 \in (\epsilon, t_0 - \epsilon)$, with $\epsilon > 0$ 
going to $0$ for HLR $\rightarrow \infty$ and TNLR $<1$,
the following inequality holds
\begin{eqnarray}
  t_1^2 e^{\frac{\sigma_X^2(t_0-t_1)^2}{\sigma_N^2 + (1-\alpha)^2 t_1^2 \sigma_\Lambda^2}} \geq t_0^2,
\nonumber
\end{eqnarray}
or equivalently
\begin{eqnarray}
  \log \left( \frac{t_1^2}{t_0^2} \right ) \geq  \frac{-\sigma_X^2(t_0-t_1)^2}{\sigma_N^2 + (1-\alpha)^2  t_1^2 \sigma_\Lambda^2}.\label{eq:condi13}
\end{eqnarray}
This inequality  is obviously true for $t_1 \geq t_0$. For the case $t_1 < t_0$ we will analyze the derivative with respect to $t_1$ of 
\begin{eqnarray}
  \log \left( \frac{t_1^2}{t_0^2} \right ) + \frac{\sigma_X^2(t_0-t_1)^2}{\sigma_N^2 + (1-\alpha)^2  t_1^2 \sigma_\Lambda^2}, \label{eq:condi14}
\end{eqnarray}
which is nothing but
\begin{eqnarray}
  \frac{2}{t_1} - \frac{ 2\sigma_X^2(t_0-t_1) \left (\sigma_N^2 + (1-\alpha)^2 t_0 t_1\sigma_\Lambda^2 \right )}{\left ( \sigma_N^2 + (1-\alpha)^2 t_1^2 \sigma_\Lambda^2 \right )^2} . \label{eq:condi15}
\end{eqnarray}
If we multiply the last expression by $\left ( \sigma_N^2 + (1-\alpha)^2 t_1^2 \sigma_\Lambda^2 \right )^2 t_1$, then the result can be written like
$a_0 + a_1 t_1 + a_2 t_1^2 + a_3 t_1^3 + a_4 t_1^4$, where
\begin{eqnarray}
  a_0 &=& 2\sigma_N^4, \nonumber \\
  a_1 &=& -2\sigma_N^2 \sigma_X^2 t_0, \nonumber \\
  a_2 &=& 2\sigma_N^2 \sigma_X^2  + 2(1-\alpha)^2\sigma_\Lambda^2 \left (2\sigma_N^2 - \sigma_X^2 t_0^2 \right ), \nonumber\\
  a_3 &=& 2(1-\alpha)^2 \sigma_\Lambda^2 \sigma_X^2 t_0, \nonumber \\
  a_4 &=& 2(1-\alpha)^4 \sigma_\Lambda^4; \nonumber 
\end{eqnarray}
using Descartes' sign rule one gets that the number of points where
the derivative of (\ref{eq:condi14}) is null for $t_1 > 0$ is at most $2$.
Additionally, based on (\ref{eq:condi14}) evaluated at $t_1 = t_0$ being
null, and (\ref{eq:condi15}) evaluated at the same point being
positive, it is obvious that there is an interval of points
just to the left of $t_1 = t_0$ where (\ref{eq:condi14}) is negative.
Furthermore, (\ref{eq:condi14}) at $t_1 \rightarrow 0$ is also
negative. Taking into account that (\ref{eq:condi15}) is null at most at $2$ points, it is clear that these
two intervals (one on the left of $t_0$, and the other on the right of $0$)
will contain the only points $t_1 > 0$ where (\ref{eq:condi13})
is not verified. Be aware that so far we have not proved that these
intervals were disjoint, so it could be the case that (\ref{eq:condi13}) 
were  indeed negative for $t_1\in[0, t_0)$.

In order to bound the width of both intervals, we will take into account
that
\begin{eqnarray}
  \log \left( \frac{t_1^2}{t_0^2} \right )+  \frac{\sigma_X^2(t_0-t_1)^2}{\sigma_N^2 + (1-\alpha)^2  t_1^2 \sigma_\Lambda^2} \geq   \log \left( \frac{t_1^2}{t_0^2} \right ) +  \frac{\sigma_X^2(t_0-t_1)^2}{\sigma_N^2 + (1-\alpha)^2  t_0^2 \sigma_\Lambda^2}, \label{eq:condi16}
\end{eqnarray}
for any $t_1 < t_0$, and will focus on the left term.
If one considers the following two points
\begin{eqnarray}
  t_{1,u} &=& \frac{1}{2} \left ( t_0  + \sqrt{\frac{t_0}{\sigma_X}}
  \sqrt{\sigma_X t_0 - 4 \sqrt{\sigma_N^2 + (1-\alpha)^2\sigma_\Lambda^2
      t_0^2}} \right ), \nonumber\\
  t_{1,l} &=& \frac{1}{2} \left ( t_0  - \sqrt{\frac{t_0}{\sigma_X}}
  \sqrt{\sigma_X t_0 - 4 \sqrt{\sigma_N^2 + (1-\alpha)^2\sigma_\Lambda^2
      t_0^2}} \right ), \nonumber 
\end{eqnarray}
where $t_{1,u} < t_0$, and $t_{1,l} > 0$,
then the evaluation of (\ref{eq:condi16}) at them can be written as
$\log(x_u) + \frac{1}{x_u}$ and $\log(x_l) + \frac{1}{x_l}$, respectively,
where
\begin{eqnarray}
  x_{u} = \frac{\left ( t_0  + \sqrt{\frac{t_0}{\sigma_X}}
  \sqrt{\sigma_X t_0 - 4 \sqrt{\sigma_N^2 + (1-\alpha)^2\sigma_\Lambda^2
      t_0^2}} \right )^2}{4t_0^2}, \nonumber\\
  x_{l} = \frac{\left ( t_0  - \sqrt{\frac{t_0}{\sigma_X}}
  \sqrt{\sigma_X t_0 - 4 \sqrt{\sigma_N^2 + (1-\alpha)^2\sigma_\Lambda^2
      t_0^2}} \right )^2}{4t_0^2}. \nonumber
\end{eqnarray}
Given that both $x_u$ and $x_l$ are non-negative, by using
App.~\ref{sec:app3} we have proved that (\ref{eq:condi16})
evaluated at $t_{1,u}$ and at $t_{1,l}$ is positive. Furthermore, 
for TNLR $<1$,
\begin{eqnarray}
  \lim_{\textrm{HLR}\rightarrow \infty} t_{1,u} &=& t_0, \nonumber \\
  \lim_{\textrm{HLR}\rightarrow \infty} t_{1,l} &=& 0, \nonumber 
\end{eqnarray}
proving the desired result.

\begin{figure}[t] 
  \begin{center}
    \includegraphics[width=0.75\linewidth]{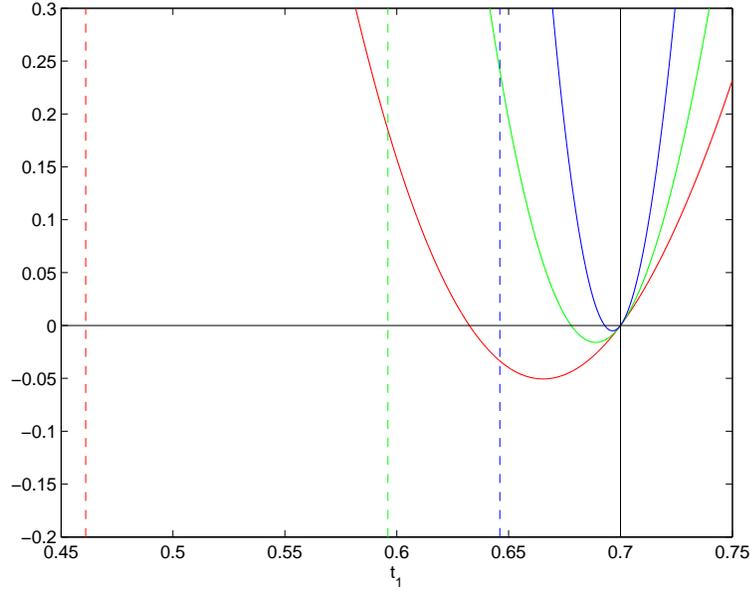}  
    \caption{Comparison  of (\ref{eq:condi14}) for different values of 
      HLR (red $10$ dB, green $15$ dB, blue $20$ dB). SCR $\approx 0.0980$ dB, TNLR $\approx -2.9616$ dB, $t_0 = 0.7$, 
      $\alpha = \alpha_{\textrm{Costa}} \approx 0.4944$. The vertical colored dashed
    lines stand for the values of $t_{1,u}$ derived by using
    (\ref{eq:t1_asymp}), showing that they converge to $t_0$ (vertical black
dashed line) as HLR $\rightarrow \infty$.}\label{fig:app11}
  \end{center}
\end{figure}

Fig.~\ref{fig:app11} illustrates the behavior of the considered
function for $t_1$ smaller than but close to $t_0$. As expected, the
larger the HLR, the closer to $t_0$ the considered function
takes positive values. Additionally, the corresponding values
of $t_{1,u}$ are plotted for the sake of completeness.

\begin{figure}[t] 
  \begin{center}
    \includegraphics[width=0.75\linewidth]{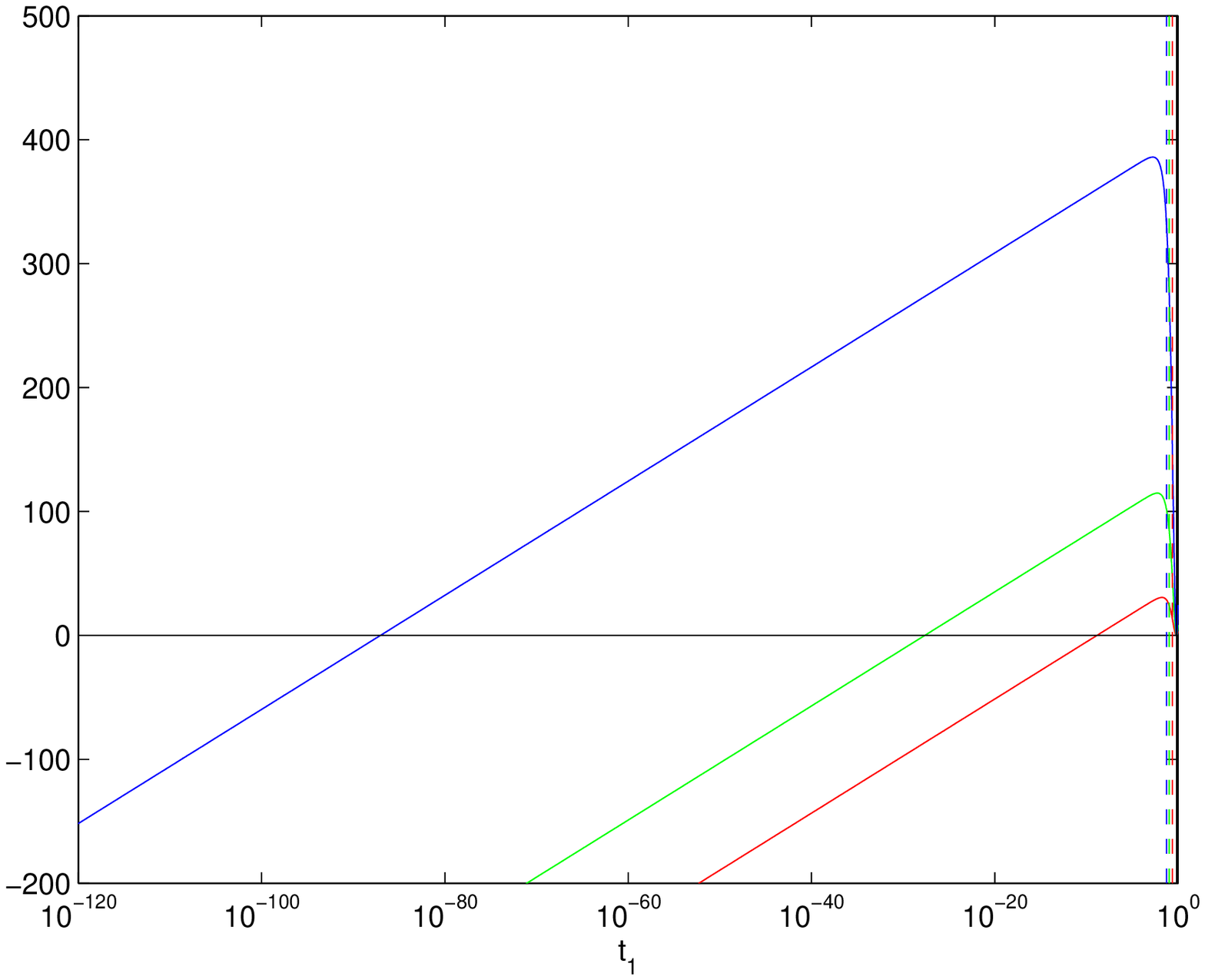}  
    \caption{Comparison  of (\ref{eq:condi14}) for different values of 
      HLR (red $10$ dB, green $15$ dB, blue $20$ dB). SCR $\approx 0.0980$ dB, TNLR $\approx -2.9616$ dB, $t_0 = 0.7$, 
      $\alpha = \alpha_{\textrm{Costa}} \approx 0.4944$. The vertical colored dashed
    lines stand for the values of $t_{1,l}$ derived by using
    (\ref{eq:t1_asymp}), showing that they converge to $0$ 
as HLR $\rightarrow \infty$.}\label{fig:app12}
  \end{center}
\end{figure}

\begin{figure}[t] 
  \begin{center}
    \includegraphics[width=0.75\linewidth]{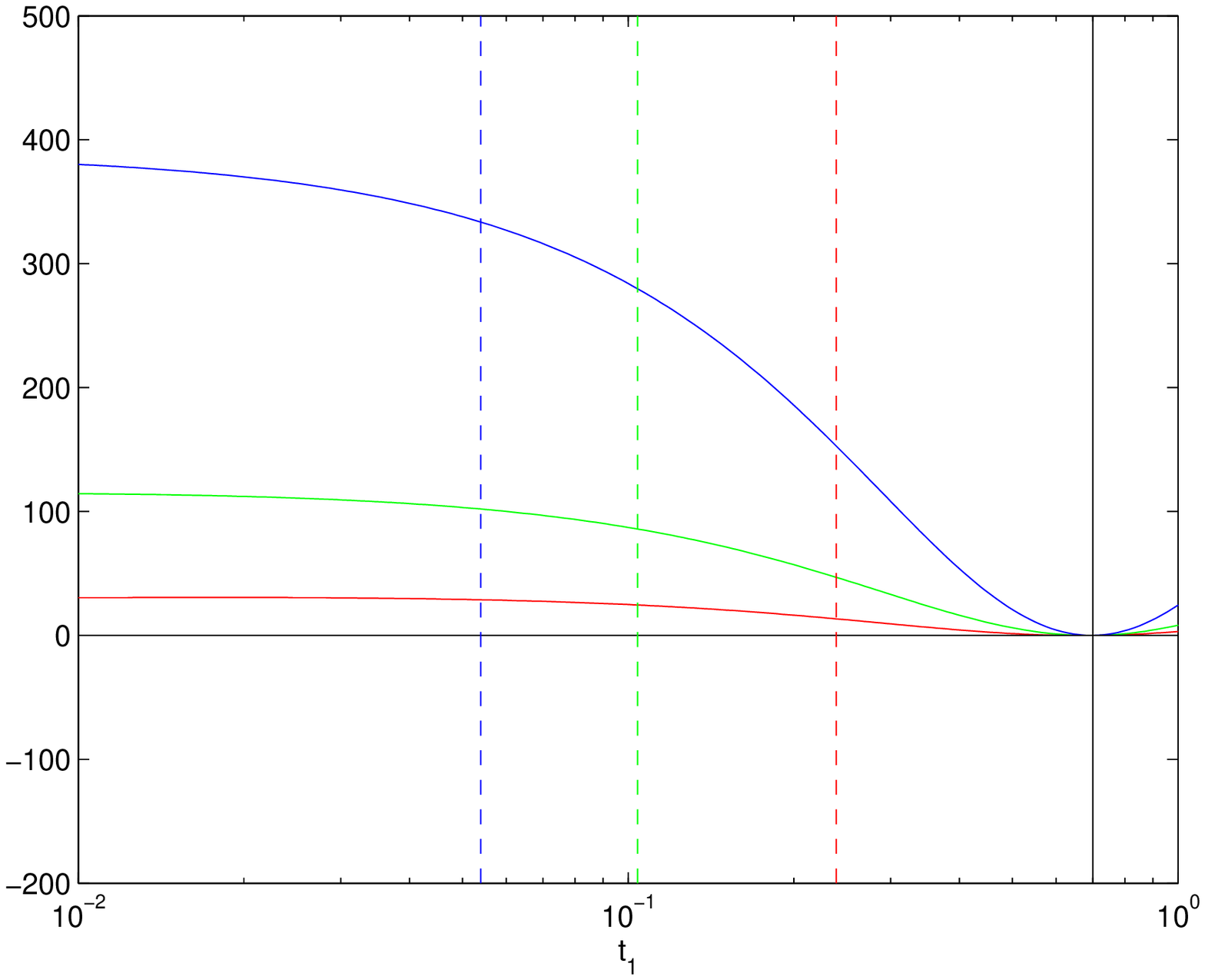}  
    \caption{Comparison  of (\ref{eq:condi14}) for different values of 
      HLR (red $10$ dB, green $15$ dB, blue $20$ dB). SCR $\approx 0.0980$ dB, TNLR $\approx -2.9616$ dB, $t_0 = 0.7$, 
      $\alpha = \alpha_{\textrm{Costa}} \approx 0.4944$. The vertical colored dashed
    lines stand for the values of $t_{1,l}$ derived by using
    (\ref{eq:t1_asymp}), showing that they converge to $0$ 
as HLR $\rightarrow \infty$.}\label{fig:app13}
  \end{center}
\end{figure}

Similarly, Figs.~\ref{fig:app12} and \ref{fig:app13} illustrate the behavior
of the considered
function for values of $t_1$ close to $0$. Again, the shown
behavior is that predicted by the performed theoretical analysis.
In that sense, the
larger the HLR, the closer to $0$ the considered function
takes positive values. Additionally, the corresponding values
of $t_{1,l}$ are plotted for the sake of completeness.

\bibliographystyle{plain}
\bibliography{biblio}

\end{document}